\documentclass[11pt]{book}
\usepackage{titlesec}
\usepackage{hyperref}
\usepackage{simplewick}
\usepackage[utf8]{inputenc}
\usepackage[T1]{fontenc}
\usepackage[italian]{babel}
\addto\captionsitalian{}
\addto\captionsitalian{}
\addto\captionsitalian{}

\titleclass{\subsubsubsection}{straight}[\subsection]
\usepackage[utf8]{inputenc}
\usepackage{amsmath,amsfonts,amssymb,stackengine,graphicx}
\title{appendix H2}
\author{Mario Meo}
\date{August 2025}

\usepackage[utf8]{inputenc}
\newif\iffigs\figstrue

\usepackage[title]{appendix}
\usepackage{epsfig,latexsym}
\usepackage{hyperref}
\usepackage{amsmath}
\usepackage{verbatim}
\usepackage{color}
\usepackage{mathrsfs}
\usepackage{slashed}
\usepackage{amssymb}

\DeclareMathAlphabet{\mathpzc}{OT1}{pzc}{m}{it}

 \csname
@addtoreset\endcsname{equation}{section}

\def\gz0{\gamma^{0}}

 \def\det{{\rm det\,}}

\def\beq{\begin{equation}}
\def\eeq{\end{equation}}
\def\bea{\begin{eqnarray}}
\def\eea{\end{eqnarray}}
\def\ba{\begin{array}}
\def\ea{\end{array}}
\def\bec{\begin{center}}
\def\ec{\end{center}}
\def\ba{\begin{align}}
\def\ena{\end{align}}

\def\12{\frac{1}{2}}

\newcounter{subsubsubsection}[subsubsection]
\renewcommand\thesubsubsubsection{\thesubsubsection.\arabic{subsubsubsection}}
\titleformat{\subsubsubsection}
  {\normalfont\normalsize\bfseries}{\thesubsubsubsection}{1em}{}
\titlespacing*{\subsubsubsection}
{0pt}{3.25ex plus 1ex minus .2ex}{1.5ex plus .2ex}

\makeatletter
\renewcommand\paragraph{\@startsection{paragraph}{5}{\z@}%
  {3.25ex \@plus1ex \@minus.2ex}%
  {-1em}%
  {\normalfont\normalsize\bfseries}}
\renewcommand\subparagraph{\@startsection{subparagraph}{6}{\parindent}%
  {3.25ex \@plus1ex \@minus .2ex}%
  {-1em}%
  {\normalfont\normalsize\bfseries}}
\def\toclevel@subsubsubsection{4}
\def\toclevel@paragraph{5}
\def\toclevel@paragraph{6}
\def\l@subsubsubsection{\@dottedtocline{4}{7em}{4em}}
\def\l@paragraph{\@dottedtocline{5}{10em}{5em}}
\def\l@subparagraph{\@dottedtocline{6}{14em}{6em}}
\makeatother

\title{Primordial Non-Gaussianity from a String-Inspired Cosmology}
\author{Mario Meo}
\date{}
\usepackage[numbers,sort&compress]{natbib}
\begin{document}
\begin{titlepage}
\begin{figure}[!htb]
    \centering
    \includegraphics[keepaspectratio=true,scale=0.5]{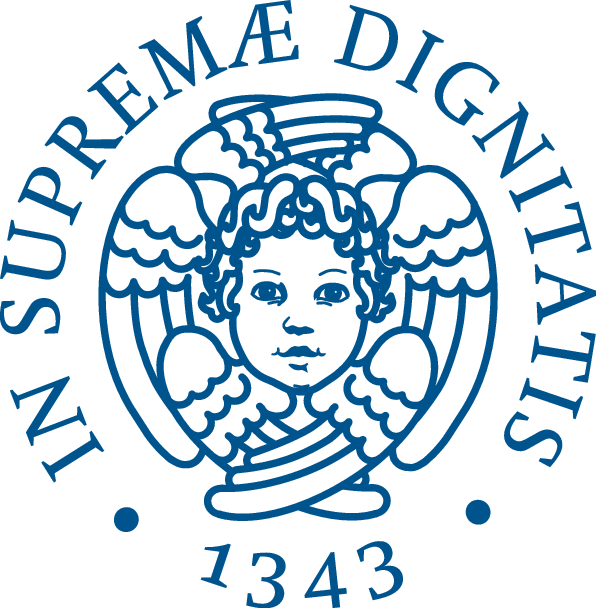}
\end{figure}

\begin{center}
    \LARGE{UNIVERSITÀ DI PISA}
    \vspace{5mm}
    \\ \large{DIPARTIMENTO DI FISICA}
    \vspace{5mm}
    \\ \LARGE{Corso di Laurea Magistrale in Fisica}
\end{center}

\vspace{15mm}
\begin{center}
    {\LARGE{\bf Primordial Non-Gaussianity from a String-Inspired Cosmology}}
\end{center}
\vspace{30mm}

\begin{minipage}[t]{0.47\textwidth}
	{\large{Relatori:}{\normalsize\vspace{3mm}
	\bf\\ \large{Augusto Sagnotti\vspace{2mm}\\Guilherme Leite Pimentel}}}
\end{minipage}
\hfill
\begin{minipage}[t]{0.47\textwidth}\raggedleft
	{\large{Candidato:}{\normalsize\vspace{3mm} \bf\\ \large{Mario Meo\\ }}}
\end{minipage}

\vspace{30mm}
\hrulefill
\\\centering{\large{ANNO ACCADEMICO 2024/2025}}

\end{titlepage}
\cleardoublepage
\thispagestyle{empty}

\vspace*{\fill}  % spinge verso il centro verticale

\begin{flushright} % allinea a destra
    {\Large \emph{To my parents, my sister Ilaria, and my girlfriend Grazia.}}
\end{flushright}

\vspace*{\fill}  % riempie lo spazio fino al fondo

\cleardoublepage
\chapter*{Acknowledgments}
I wish to thank my primary advisor, Professor Augusto Sagnotti, for his support, availability, and the dedication with which he guided me throughout these two years at the Scuola Normale, serving as both a mentor and a guide during a crucial phase of my academic growth and my development as a scientist.

I would also like to sincerely thank my co-advisor, Professor Guilherme Pimentel, for introducing me to the fascinating world of cosmology, which has inspired the direction of my future academic career.

\chapter*{Introduction and Summary}
The quest for a consistent theory unifying quantum mechanics and gravity has guided much of the development of theoretical physics during the final portion of the last century. \cite{Goroff:1985th,GoroffSagnotti1986}. Classical General Relativity, though successful on macroscopic scales, leads to uncontrollable ultraviolet divergences when treated perturbatively in a quantum field theoretic setting~\cite{tHooftVeltman1974,GoroffSagnotti1986}. This failure suggests that the gravitational interaction cannot be consistently described within the point-particle paradigm.

A conceptual shift occurred in the late 1960s with the discovery of the Veneziano amplitude~\cite{Veneziano1968}, originally proposed as a phenomenological model for hadronic resonances. It soon became clear that this amplitude, and its closed-string counterpart, the Shapiro–Virasoro formula~\cite{ShapiroVirasoro1969}, admit an elegant interpretation in terms of the dynamics of one-dimensional extended objects: strings. The so-called ``dual models'' thus gradually evolved into what is now known as string theory.

The following decade was marked by the construction of the fermionic string~\cite{Ramond1971,NeveuSchwarz1971}, and the realization that string theory inevitably includes gravity in its low-energy spectrum~\cite{ScherkSchwarz1974,Yoneya1974}. These developments transformed string theory from a phenomenological tool into a serious candidate for a fundamental theory of all interactions.

A significant milestone came with the introduction of supersymmetry in string theory~\cite{Gliozzi:1977}, which eventually led to five consistent superstring models in ten dimensions: type I, type IIA, type IIB, and the two heterotic theories with gauge groups \(SO(32)\) and \(E_8 \times E_8\)~\cite{GreenSchwarzWitten1987,Gross1985,Gross1986}. The discovery of anomaly cancellation mechanisms~\cite{GreenSchwarz1984} and the formulation of modular-invariant GSO projections laid the foundation for a perturbatively consistent framework.

In the mid-1980s, string theory made contact with four-dimensional physics through compactification schemes involving Calabi–Yau manifolds~\cite{Candelas1985} and orbifolds~\cite{DixonHarveyVafaWitten1985}. These approaches allowed for the emergence of chiral matter and \( \mathcal{N}=1 \) supersymmetry in four dimensions, essential features for realistic model building.

A deeper understanding of the theory's non-perturbative aspects emerged in the 1990s, particularly with the identification of D-branes as fundamental dynamical objects carrying Ramond–Ramond charges~\cite{Polchinski1995}. This insight sparked the so-called ``second string revolution,'' which revealed a rich web of dualities relating the various superstring theories and M-theory~\cite{Witten1995,HullTownsend1995}.

Despite its elegance, a persistent challenge in string theory is the vacuum selection problem. The landscape of consistent string backgrounds includes an enormous number of vacua, many of which are metastable or supersymmetry-breaking~\cite{BoussoPolchinski2000,DouglasKachru2007}. Supersymmetry breaking, while essential to connect string theory to our non-supersymmetric low-energy world, often destabilizes the moduli and introduces runaway directions in scalar potentials.

This is particularly relevant in cosmology, where scalar fields—such as the dilaton and various moduli—can play a dynamical role in the evolution of the Universe. The early Universe is, in many respects, the best laboratory we have to probe fundamental physics. During its earliest moments, it operated at energy scales that can approach those of string theory, and therefore may carry imprints of its dynamics~\cite{Linde1990,Guth2007}.

One of the most striking features of the cosmic microwave background (CMB) is its near scale-invariance and Gaussianity. Yet, anomalies remain. In particular, the "lack of power at low multipoles" in the CMB angular power spectrum—especially the quadrupole and octupole—has prompted investigations into non-standard initial conditions and pre-inflationary physics~\cite{Planck2018,Contaldi2003}.

This motivates the study of models beyond the minimal slow-roll paradigm. In recent years, non-supersymmetric string models have offered new perspectives. Supersymmetry breaking at the string scale can induce non-trivial potentials for scalar fields, typically with exponential behavior~\cite{Sugimoto1999,Antoniadis1999,Angelantonj2000,Dudas:2000ff}. These potentials give rise to the \textit{climbing scalar} mechanism~\cite{DudasKitazawaPatilSagnotti:2012,KitazawaSagnotti:2014}, a novel scenario in which the scalar field is forced to climb up a steep string--induced potential as it emerges from the initial singularity.

The climbing phase precedes a turning point, after which the scalar descends and can potentially drive inflation \cite{Dudas:2010climbing,DudasKitazawaPatilSagnotti:2012,Martin2014} in the presence of milder contributions. This pre-inflationary epoch can leave distinct imprints in cosmological observables. In particular, the climbing dynamics suppresses long-wavelength modes in the primordial spectrum, naturally leading to a low-\( \ell \) cutoff that resonates with the CMB power suppression~\cite{DudasKitazawaPatilSagnotti:2012,GruppusoKitazawaMandolesiNatoliSagnotti2016}. The suppression induced by the climbing phase thus appears to leave tangible signs in the two-point function. However, this distinctive dynamics can leave additional imprints in the bispectrum, encoded in the three-point correlation functions. This possibility motivates the present work, which is devoted to a detailed analysis of the non-Gaussian features arising in this framework in the three--point amplitude for scalar perturbations.

This thesis presents a detailed investigation of some cosmological implications of the climbing scalar mechanism. After deriving the framework and analyzing its effects on the two-point correlation function, we focus on how the climbing dynamics could affect the standard computation of primordial non-Gaussianities, originally developed by Maldacena~\cite{maldacena2003non,collins2011primordial}. 

In Chapter 1, we review the structure of ten-dimensional string spectra, covering both supersymmetric and tachyon-free non-supersymmetric models, emphasizing how supersymmetry breaking leads to exponential potentials. Chapter 2 develops the effective field theory for the climbing scalar in a cosmological setting, analyzing the dynamics and computing the associated slow-roll parameters. In Chapter 3, we study cosmological perturbations. We compute the two-point function in both the standard and climbing scenarios, demonstrating a suppression at low momenta regulated by a characteristic scale \(\Delta\) \cite{DudasKitazawaPatilSagnotti:2012,KitazawaSagnotti:2014,GruppusoKitazawaMandolesiNatoliSagnotti2016}. Maldacena’s cubic action is then derived to prepare for the analysis of non-Gaussianities. Chapter 4 is devoted to constructing a framework for their computation, considering two distinct resolutions of the conformal-time singularity: one via a persistent de Sitter phase, and another through a bouncing cosmology, which is ultimately adopted. These ought to emerge from a complete string treatment, which is needed to grant finite results, since the initial singularity gives rise to a divergent result. The resulting non-Gaussianities \cite{meo2025preinflationarynongaussianities} are shown to comprise two components: one originating from the post-bounce dynamics and another solely from the turning point of the climbing phase. While the first contribution is small and oscillates around Maldacena's result, the second can be sizable and observationally relevant. In particular, for \(63 < N < 65\) inflationary e-folds, the predicted \(f_{\mathrm{NL}}\) lies within current bounds and may be detectable in future CMB surveys.

This work focuses on scalar amplitudes. However, we are aware of the possibility of computing amplitudes involving tensorial gravitational modes. This appears to be a natural future development of the present study, since the overall structure of the computation remains essentially the same. Nevertheless, we believe that the most interesting features arise from the scalar non-Gaussianities, which are the main focus of the present analysis.
\tableofcontents
\chapter{ Strings and Supersymmetry Breaking}
This chapter introduces the key string theory concepts necessary for developing the \textit{Climbing Scalar} model, the main focus of this work. Rather than covering the full scope of string theory, it lays a theoretical foundation for the analysis that follows.

We first review the bosonic string and its quantization, then we present the main supersymmetric string theories in 10d: Type IIA, Type IIB, Type I, and heterotic strings with gauge groups $SO(32)$ and $E_8 \times E_8$. These serve as standard, well-understood vacua where supersymmetry is preserved.

We then focus on three tachyon-free, non-supersymmetric models: the heterotic $SO(16) \times SO(16)$, and two orientifold models with gauge groups $U(32)$ and $USp(32)$. These offer concrete examples of stable vacua without supersymmetry.

All spectra are constructed using the $SO(8)$ character formalism, effective in the light-cone gauge, where physical states fall into representations of $SO(8)$. This framework makes it easier to distinguish between supersymmetric and non-supersymmetric configurations and to identify \textit{brane supersymmetry breaking}\cite{Freedman:1976xh,Deser:1976eh,Freedman:2012zz,Sagnotti:1987Cargese,Pradisi:1989zz,Horava:1989fv,Horava:1989bg,BianchiSagnotti:1990,BianchiSagnotti:1991a,Bianchi1992, Sagnotti:1992note, Dudas:2000review,Angelantonj:2002ct, AngelantonjFlorakis:2024,Aldazabal1999}—a mechanism where supersymmetry is preserved in the bulk but broken on D-branes or orientifold planes.

Supersymmetry breaking modifies the low-energy effective action, introducing scalar potentials and changing the vacuum structure—especially relevant in weak coupling and low curvature regimes. These effects are crucial for understanding how stable, non-supersymmetric string vacua can lead to cosmologically viable models.

By the end of the chapter, we will have assembled all the ingredients needed to understand how \textit{brane supersymmetry breaking}\cite{Freedman:1976xh,Deser:1976eh,Freedman:2012zz,Sagnotti:1987Cargese,Pradisi:1989zz,Horava:1989fv,Horava:1989bg,BianchiSagnotti:1990,BianchiSagnotti:1991a,Bianchi1992, Sagnotti:1992note, Dudas:2000review,Angelantonj:2002ct, AngelantonjFlorakis:2024,Aldazabal1999} can give rise to effective scalar dynamics with exponential potentials —core feature of the \textit{Climbing Scalar} model.

\section{The Polyakov Expansion}
This section introduces the Feynman path integral formalism as a natural framework for quantizing string theory. Transition amplitudes are computed as sums over all possible histories weighted by the phase factor:
\beq
    \exp\left(\frac{i S}{\hbar}\right),
\eeq
where $S$ is the classical action evaluated along a world-sheet— a two-dimensional surface interpolating between initial and final string configurations. In string theory, this replaces the sum over point-particle trajectories with a sum over world-sheet geometries.

\begin{figure}[h]
    \centering
    \includegraphics[width=0.35\linewidth]{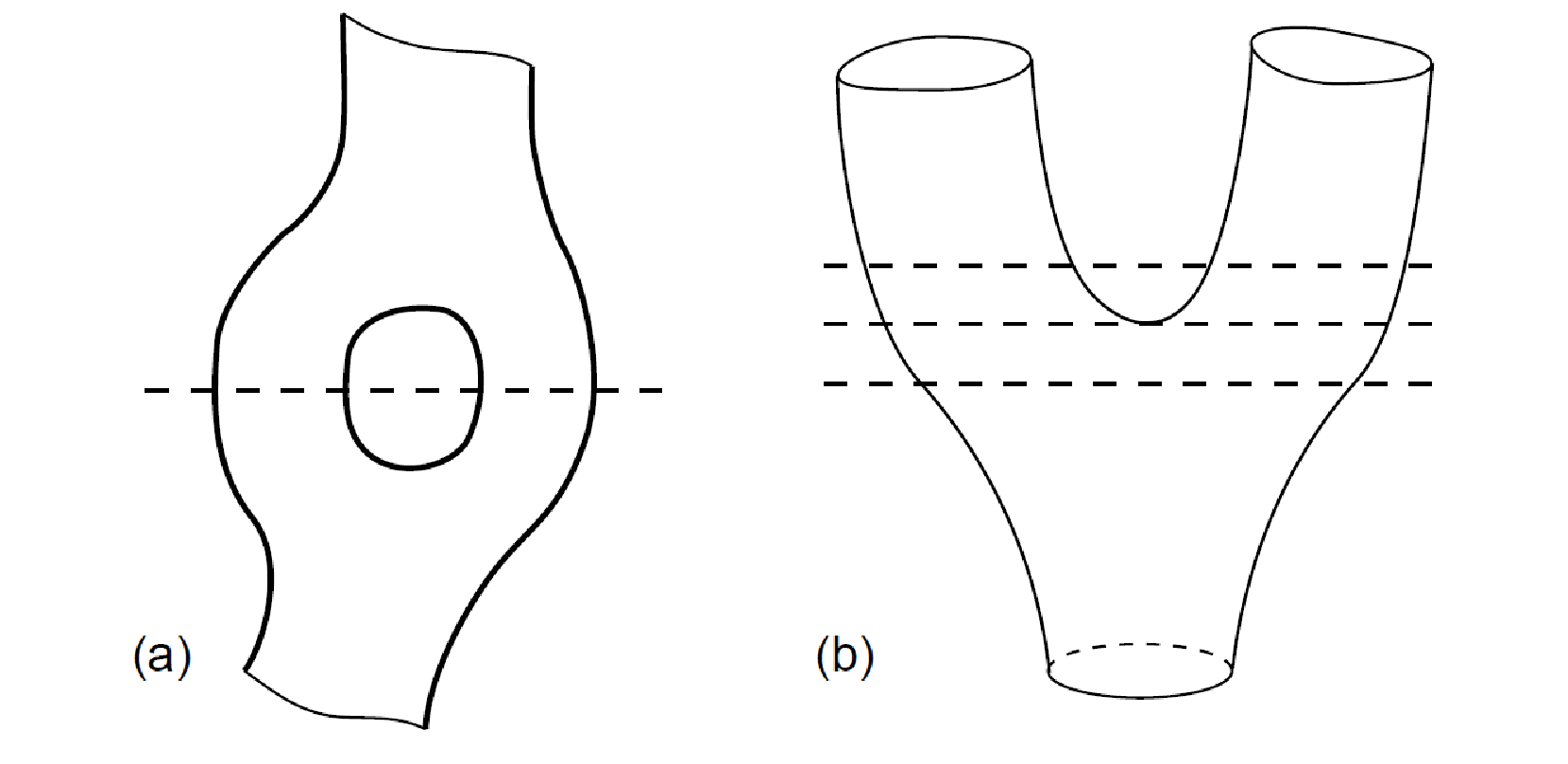}
    \caption{(a) Quantum correction to open string propagation. (b) Decay of one
    closed string into two. Time evolves upward. Source:~\cite{Polchinski:1998rq}.}
    \label{fig:str1}
\end{figure}

Unlike point-particle theories, where interactions must be added explicitly, string interactions are inherently encoded in the topology of the world-sheet. For instance, a world-sheet that splits or merges describes string emission, absorption, or scattering. In the case of closed strings, fundamental processes include one string splitting into two or two strings merging into one, as illustrated in Fig.~\ref{fig:str1}.

All known fundamental interactions—gravity, gauge forces, and Yukawa couplings—emerge from different vibrational modes of the same fundamental object: the closed string. For example, the graviton arises from a specific massless mode, while gauge bosons and scalars correspond to other string excitations. Thus, string theory provides a unified description in which all forces stem from geometric processes on world-sheets.

For open strings, similar topological interactions occur: splitting, joining, and transitions to/from closed strings. Crucially, these do not occur at localized spacetime points; their localization is frame-dependent and determined solely by the global world-sheet topology. This non-locality is a key feature that softens ultraviolet divergences, making string theory finite at high energies.

The classification of consistent string theories relies on the types of world-sheet topologies allowed in the path integral. Two types of boundaries are relevant: source boundaries which defines initial/final string states; endpoint boundaries, corresponding to open string endpoints in spacetime.
Focusing on endpoint boundaries, four classes of free string theories emerge:
\begin{enumerate}
    \item Closed oriented string theory: oriented, boundaryless world-sheets.
    \item Closed unoriented string theory: all boundaryless world-sheets, regardless of orientation.
    \item Oriented open and closed string theory: oriented world-sheets with boundaries.
    \item Unoriented open and closed string theory: all (oriented and unoriented) world-sheets with boundaries.
\end{enumerate}

We now proceed to construct the path integral over string world-sheets by integrating over the fields appearing in the Polyakov action \cite{GreenSchwarzWitten1987,Polchinski:1998rq,Angelantonj:2002ct}, starting from the bosonic string model. The Polyakov action reads:
\beq
    S = S_X + \lambda \chi \,,
\eeq
\beq
    S_X = \frac{1}{4\pi \alpha'} \int_{M} d^2 \sigma \, g^{1/2} g^{ab} \partial_a X^\mu \partial_b X_\mu \quad , \quad
    \chi = \frac{1}{4\pi} \int_{M} d^2 \sigma \, g^{1/2} R + \frac{1}{2\pi} \int_{\partial M} ds \, k \,.
\eeq

Here, $X^\mu$ is a $d$-vector embedding the string in flat Minkowski spacetime, depending on world-sheet coordinates $(\sigma_1, \sigma_2)$. The metric $g_{ab}$ on the world-sheet is taken with Euclidean signature for convergence in the path integral. $R$ is the Ricci scalar of $g_{ab}$, and $k$ the extrinsic curvature (see \ref{appendix:ADM}) on the boundary $\partial M$.

Following the conventions of~\cite{Polchinski:1998rq,Polchinski:1998rr}, the string tension $T$ relates to the Regge slope $\alpha'$ by
\beq
    T = \frac{1}{2\pi \alpha'} \,.
\eeq
The action is invariant under:
\begin{itemize}
    \item The $d$-dimensional Poincaré group (isometries of embedding spacetime),
    \item Two-dimensional world-sheet diffeomorphisms,
    \item Weyl transformations:
    \beq
        X'^\mu(\tau, \sigma) = X^\mu(\tau, \sigma) \,, \quad
        g'_{ab}(\tau, \sigma) = e^{2\omega(\tau, \sigma)} g_{ab}(\tau, \sigma) \,.
    \eeq
\end{itemize}

From the world-sheet perspective, $X^\mu$ behave as massless scalar fields minimally coupled to $g_{ab}$, and Poincaré invariance acts as an internal symmetry at fixed world-sheet coordinates. The term $\chi$ is a gravitational contribution including the Gibbons-Hawking-York boundary term (see ~\ref{appendix:ADM}), where $k$ denotes the extrinsic curvature and $ds$ the proper length element on the boundary.

The Euler characteristic term $\chi$ depends only on topology; in two dimensions the Einstein–Hilbert action does not propagate metric dynamics. This follows because the variation under a local Weyl rescaling satisfies $\delta (g^{1/2} R) \propto g^{1/2} \nabla^2 \omega \,$,
a total derivative ensuring Weyl invariance for closed world-sheets. For surfaces with boundaries, the boundary term involving $k$ is essential to preserve this invariance.

In the path integral, the factor $e^{-\lambda \chi}$ weights contributions by topology: adding a boundary strip (corresponding to an open string emission/reabsorption) reduces $\chi$ by one and multiplies the amplitude by $e^\lambda$, while adding a handle (a closed string loop) reduces $\chi$ by two and contributes $e^{2\lambda}$. Thus, the string coupling constants relate as
\beq
    g_o^2 \sim g_c \sim e^{\lambda} \,,
\eeq
linking open and closed string couplings to the topological term coefficient $\lambda$.

This construction lays the groundwork for perturbative string theory, with $X^\mu$ fields describing string fluctuations and the Euler term controlling the strength of string interactions through world-sheet topology.
The parameter $\lambda$ in the Polyakov action might appear as a free parameter, seemingly contradicting the well-known statement that string theory contains no free parameters. This puzzle is resolved by introducing the \textbf{dilaton} field $\Phi(X)$, a scalar coupled to the world-sheet through the term
\beq
    \frac{1}{4\pi} \int_M d^2 \sigma \, g^{1/2} R \, \Phi(X) \quad .
\eeq
The $\beta$-functions of the two-dimensional world-sheet theory yield spacetime equations where $\Phi$ appears only through its derivatives, implying invariance under constant shifts $\Phi \to \Phi + \text{const}$. Such shifts modify the world-sheet action by a term proportional to the Euler characteristic, leaving local properties like Weyl invariance unaffected.

In particular, a flat metric background with constant dilaton $\Phi = \Phi_0$
is exactly Weyl-invariant for any constant $\Phi_0$. Identifying $\Phi_0 = \lambda$
shows that the previously introduced parameter $\lambda$ corresponds to the constant background value of the dilaton, $\lambda = \langle \Phi \rangle $. 
Thus, $e^\lambda$ sets the string coupling strength, but different values of $\lambda$ represent different backgrounds within a single theory, not distinct string theories.

String scattering amplitudes simplify when external sources are sent to infinity, reducing the problem to computing S-matrix elements for asymptotic string states. In this limit, world-sheet external legs become infinite cylinders. Using complex coordinates \( w \sim w + 2\pi \) with \( \text{Im}(w) \in [-2\pi t, 0] \), the limit \( t \to \infty \) corresponds to free propagation.

\begin{figure}[h]
    \centering
    \includegraphics[width=0.35\linewidth]{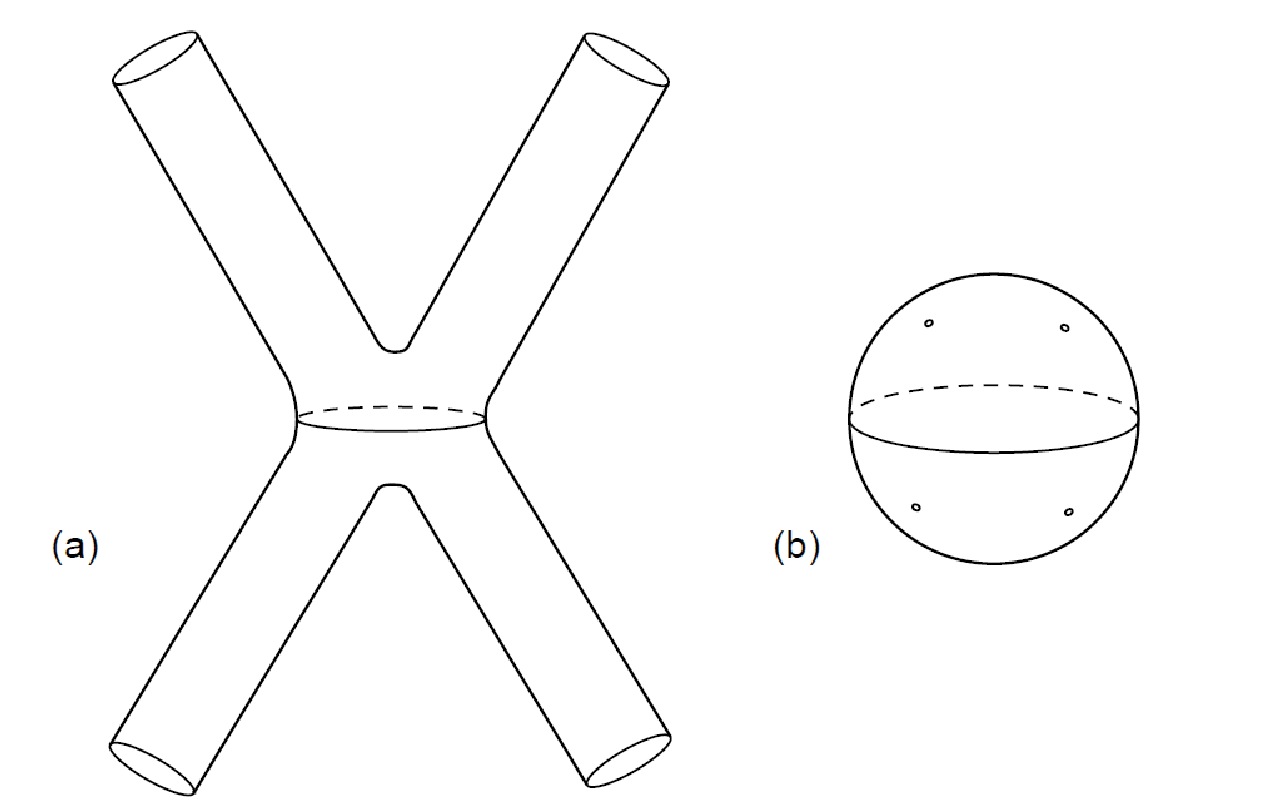}
    \caption{(a) Scattering of four closed strings with sources at \( x_0 = \pm \infty \). (b) Conformally equivalent picture as a sphere with punctures. Taken from~\cite{Polchinski:1998rq}.}
    \label{fig:str2}
\end{figure}

A conformal map \( z = e^{-i w} \) transforms each cylinder into a small circle near the origin on the complex plane. As \( t \to \infty \), these shrink to point insertions, yielding a sphere with punctures (see Fig.~\ref{fig:str2}, from~\cite{Polchinski:1998rq}).

Via the state--operator correspondence, each asymptotic state becomes a vertex operator \( V_j(k) \). The S-matrix is then:
\begin{align}
    S_{j_1 \dots j_n}(k_1, \dots, k_n) 
    = \sum_{\text{c.t.}} \int \frac{[dX\,dg]}{\text{Diff} \times \text{Weyl}} \, e^{-S_X - \lambda \chi} \times \prod_{i=1}^n \int d^2\sigma_i\, \sqrt{g(\sigma_i)}\, V_{j_i}(k_i, \sigma_i) \quad.
\end{align}
The sum runs over connected compact world-sheets (c.t. stands for compact topologies), with insertions in the interior (closed strings) or boundary (open strings). Oriented surfaces are classified by genus \( g \) and number of boundaries \( b \), e.g., disk: \( (0,1) \), annulus: \( (0,2) \), pair of pants: \( (0,3) \).

Unoriented surfaces include cross-caps $c$ \cite{Angelantonj:2002ct}(e.g., projective plane: \( (0,0,1) \), Möbius strip: \( (0,1,1) \), Klein bottle: \( (0,0,2) \), torus with cross-cap: \( (1,0,1) \) or \( (0,0,3) \)), built via identifications like \( z' = -1/\bar{z} \). Minimal classification allows only one of \( g \), \( c \) non-zero. All surfaces satisfy the Euler characteristic:
\beq
    \chi = 2 - 2g - b - c \quad.
\eeq
These topological data determine diagram weights in string perturbation theory. Unlike field theory, vacuum amplitudes in string theory---constrained by geometry and infinite spectra---encode essential physical information about the full theory, due to their geometrical constraints.

\section{The Bosonic String}
To prepare for the construction of vacuum amplitudes in string theory, we begin with the simplest case: the closed, oriented bosonic string. As a warm-up, it is useful to recall the analogous situation in quantum field theory. Consider a scalar field \( \phi \) with mass \( M \) in \( D \) dimensions, described by the action:
\beq
    S = \int d^D x \left( \frac{1}{2} \partial_\mu \phi \, \partial^\mu \phi - \frac{1}{2} M^2 \phi^2 \right) \quad .
\eeq

Performing a Wick rotation to Euclidean space leads to the Euclidean path integral:
\beq
    e^{-\Lambda} = \int [D\phi] \, e^{-S_E} \propto \det{}^{-1/2}(-\Delta + M^2) \quad ,
\eeq
where \( S_E \) is the Euclidean action and \( \Delta \) is the Laplacian operator. 

The mass dependence of the determinant can be extracted using the identity:
\beq
    \log \det A = -\int_\epsilon^\infty \frac{dt}{t} \, \text{tr}(e^{-tA}) \quad ,
\eeq
where \( \epsilon \) is a UV cutoff, and \( t \) is the Schwinger proper time  \cite{Angelantonj:2000ct}.

We now recall that momentum eigenstates diagonalize the kinetic operator in field theory. Thus, the vacuum energy can be written performing the gaussian integral over $p$ as:
\beq
    \Lambda = -\frac{V}{2 (4\pi)^{D/2}} \int_\epsilon^\infty \frac{dt}{t^{D/2 + 1}} e^{-t M^2} \quad ,
\eeq
where \( V \) is the spacetime volume. 
For a Dirac fermion, one obtains the same expression with a positive sign and a factor \( 2^{[D/2]} \), due to Grassmann integration. In general, for arbitrary spin, the result takes the form:
\beq
    \Lambda = -\frac{V}{2 (4\pi)^{D/2}} \int_\epsilon^\infty \frac{dt}{t^{D/2 + 1}} \, \text{Str}(e^{-t M^2}) \quad ,
\eeq
where \( \text{Str} \) is the supertrace over physical degrees of  \cite{Angelantonj:2002ct}.

This formalism extends naturally to string theory. In the critical bosonic string with \( d = 26 \), the mass spectrum is:
\beq
    M^2 = \frac{2}{\alpha'} (L_0 + \bar{L}_0 - 2) \quad ,
    \label{eq:mass}
\eeq
with \( L_0 = \bar{L}_0 \) due to level matching. Substituting into the previous expression and implementing level matching condition including an apposite $\delta$-function gives:

\beq
    \Lambda = -\frac{V}{2(4\pi)^{13}} \int_{-1/2}^{1/2} ds \int_\epsilon^\infty \frac{dt}{t^{14}} \, \text{Tr}\left( e^{-(2/\alpha')(L_0 + \bar{L}_0 - 2)t} e^{2\pi i (L_0 - \bar{L}_0)s} \right) \quad .
\eeq
Defining the complex Schwinger parameter \( \tau = \tau_1 + i\tau_2 = s + i \frac{t}{\pi \alpha'} \) , and letting $q = e^{2\pi i \tau} \, \, \, \, \bar{q} = e^{-2\pi i \bar{\tau}}$,
this expression can finally be restricted to the fundamental domain \( \mathcal{F} \), exploiting modular invariance:
\beq
    \Lambda = -\frac{V}{2(4\pi^2 \alpha')^{13}} \int_{\mathcal{F}} \frac{d^2 \tau}{\tau_2^2} \, \frac{1}{\tau_2^{12}} \, \text{Tr} \left( q^{L_0 - 1} \bar{q}^{\bar{L}_0 - 1} \right) \quad .
\eeq
This one-loop vacuum amplitude for the closed bosonic string encodes the full spectrum and is finite after appropriate regularization, thanks to modular invariance.

Once we restrict to closed oriented strings, the vacuum amplitude at one loop corresponds to a torus diagram, with its Teichmüller parameter naturally identified with the complex Schwinger parameter \( \tau \). Since the torus has vanishing Euler characteristic (\( \chi = 0 \)), it gives the leading contribution to the Polyakov expansion. However, not every value of \( \tau \) in the strip \( \{-\tfrac{1}{2} < \tau_1 < \tfrac{1}{2}, \, \tau_2 > \epsilon\} \) represents a physically distinct surface.

\begin{figure}
    \centering
    \includegraphics[width=0.35\linewidth]{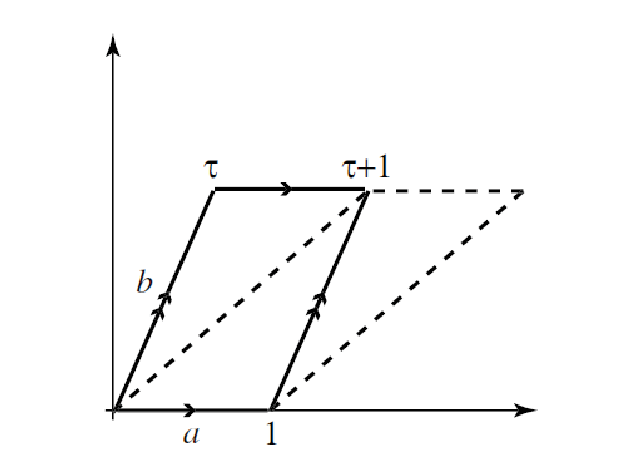}
    \caption{The torus as a periodic lattice. This picture is taken from~\cite{Angelantonj:2002ct}.}
    \label{fig:torus}
\end{figure}

As shown in Fig.~\ref{fig:torus}, a torus can be represented as a parallelogram in the complex plane with opposite sides identified. By rescaling, one edge can be made horizontal and of unit length, so that the shape of the torus is fully encoded in a single complex parameter \( \tau = \tau_1 + i \tau_2 \), with \( \tau_2 > 0 \). This parameter characterizes the complex structure of the surface.

However, many such values of \( \tau \) correspond to equivalent tori. In fact, the modular group \( PSL(2,\mathbb{Z}) \) acts on \( \tau \) as:
\beq
    \tau \rightarrow \frac{a \tau + b}{c \tau + d} \quad \text{with} \quad ad - bc = 1 \,, \quad a,b,c,d \in \mathbb{Z} \quad ,
\eeq
and all values related by this transformation are physically indistinguishable. The group is generated by the transformations:
\beq
    T: \tau \rightarrow \tau + 1 \,, \qquad S: \tau \rightarrow -\frac{1}{\tau} \quad ,
\eeq
which satisfy the relation \( S^2 = (ST)^3 \). The \( T \) transformation modifies the slanted side of the torus, while \( S \) interchanges the vertical and horizontal directions.

\begin{figure}
    \centering
    \includegraphics[width=0.35\linewidth]{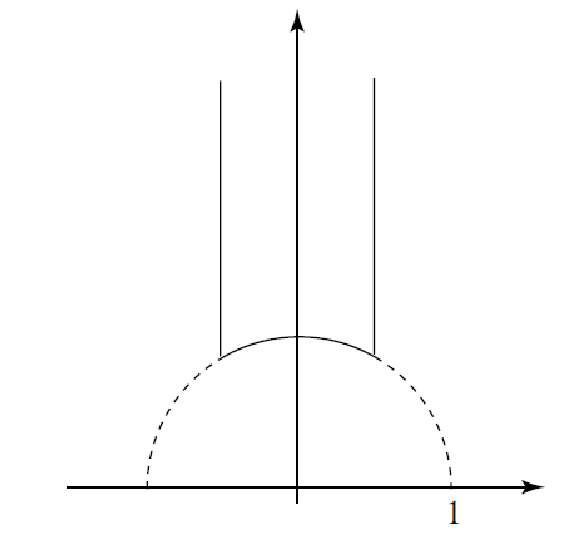}
    \caption{Fundamental domain for the torus. This picture is taken from~\cite{Angelantonj:2002ct}.}
    \label{fig:torus1}
\end{figure}

The physically inequivalent tori are thus parametrized by the values of \( \tau \) in a fundamental domain of the modular group, typically taken to be:
\beq
    \mathcal{F} = \left\{ -\frac{1}{2} < \tau_1 \le \frac{1}{2}, \quad |\tau| \ge 1 \right\} \quad ,
\eeq
as shown in Fig.~\ref{fig:torus1}. Restricting the integration domain to \( \mathcal{F} \) avoids overcounting and automatically imposes an ultraviolet cutoff at the string scale. This leads us to define the one-loop vacuum amplitude, or torus partition function, for the closed bosonic string:
\beq
    \mathcal{T} = \int_{\mathcal{F}} \frac{d^2\tau}{\tau_2^2} \, \frac{1}{\tau_2^{12}} \, \text{Tr}\left( q^{L_0 - 1} \bar{q}^{\tilde{L}_0 - 1} \right) \quad .
    \label{eq:torus1}
\eeq
This expression applies to any oriented closed string model, once the appropriate form of the Virasoro generators \( L_0 \) and \( \tilde{L}_0 \) is specified by the string spectrum.

We can now compute the torus amplitude of the bosonic string explicitly. For each transverse space-time direction, the Virasoro operator is given in terms of oscillator modes and each individual mode contributes to the trace as a geometric series:
\beq
    L_0 = \sum_n n a_n^\dagger a_n \quad \to
    \quad \text{Tr}(q^{n a_n^\dagger a_n}) = \sum_{k=0}^\infty q^{nk} = \frac{1}{1 - q^n} \quad .
\eeq
Collecting all contributions, the full one-loop amplitude becomes:
\beq
    \mathcal{T} = \int_{\mathcal{F}} \frac{d^2 \tau}{\tau_2^2} \frac{1}{\tau_2^{12}} \frac{1}{|\eta(\tau)|^{48}} \quad ,
\eeq
where \( \eta(\tau) \) is the Dedekind function:
\beq
    \eta(\tau) = q^{1/24} \prod_{n=1}^\infty (1 - q^n) \quad .
\eeq
The integrand is modular invariant. This follows from the known modular properties of \( \eta(\tau) \)~\cite{Angelantonj:2002ct,Fay1973,Mumford1983,Erdelyi1953,WhittakerWatson1927}:
\beq
    T: \eta(\tau+1) = e^{i\pi/12} \eta(\tau) \,, \qquad S: \eta(-1/\tau) = \sqrt{-i\tau} \, \eta(\tau) \quad ,
\eeq
which imply that \( \tau_2^{1/2} |\eta(\tau)|^2 \) is invariant. Each transverse coordinate contributes an invariant factor, independently of the total central charge.

The factor \( \tau_2^{-12} \) arises from integrating over 24 transverse momenta
so the amplitude can be rewritten as:
\beq
    \mathcal{T} = (\alpha')^{12} \int_{\mathcal{F}} \frac{d^2\tau}{\tau_2^2} \int d^{24}p \left| \frac{q^{\alpha' p^2 / 4}}{\eta(\tau)^{24}} \right|^2 \quad ,
\eeq
highlighting a continuum of ground states \( |p\rangle \), each accompanied by a tower of excitations. In conformal field theory terms, each tower corresponds to a Verma module~\cite{Angelantonj:2002ct,Belavin:1984vu,Friedan:1985ge,Ginsparg:1988ui,Cardy:1988,ItzyksonDrouffe1989v1,ItzyksonDrouffe1989v2}, with associated character:
\beq
    \chi_i(q) = \text{Tr}(q^{L_0 - c/24})_i = q^{h_i - c/24} \sum_k d_k q^k \quad ,
\eeq
where \( d_k \) counts excitations at level \( h_i + k \). The general torus amplitude thus takes the form:
\beq
    \mathcal{T} = \int_{\mathcal{F}} \frac{d^2\tau}{\tau_2^2} \sum_{i,j} \bar{\chi}_i(\bar{q}) X_{ij} \chi_j(q) \quad ,
\eeq
with \( X_{ij} \in \mathbb{Z} \), determined by spin-statistics. In the bosonic string, this structure is realized as an integral over momentum modes, with each character given by:
\beq
    \chi_p(q) = q^{\alpha' p^2 / 4} \eta(\tau)^{-24} \quad .
\eeq

The low-energy spectrum can be extracted from the expansion of the integrand, focusing on terms with equal powers of \( q \) and \( \bar{q} \), in accordance with level matching:
\bea
    &&\frac{1}{q \bar{q}} \prod_{n=1}^\infty \left( \frac{1}{(1 - q^n)^{24}} \right) \left( \frac{1}{(1 - \bar{q}^n)^{24}} \right)
    \sim \frac{1}{q \bar{q}} + 24^2 + \mathcal{O}(q \bar{q}) \quad .
\eea
The first term corresponds to the state \( |0 \tilde{0} \rangle \), with no left or right excitations. The mass is computed via the mass formula~\eqref{eq:mass}, yielding:
\beq
    M^2 = -\frac{4}{\alpha'} < 0 \quad ,
\eeq
indicating a tachyon, a sign of vacuum instability~\cite{GreenSchwarzWitten1987,Polchinski:1998rq,Zwiebach2004}. This pathology reflects the absence of fermions and supersymmetry in the bosonic string.

The next term, \( 24^2 \), comes from states of the form \( \alpha^i_{-1} \tilde{\alpha}^j_{-1} |0 \tilde{0} \rangle \), describing massless
as required by Lorentz invariance in the critical dimension \( d = 26 \). These massless modes decompose into: a symmetric tensor \( h_{\mu \nu} \), corresponding to the \textbf{graviton}, an antisymmetric tensor \( B_{\mu \nu} \), a scalar \( \phi \), that is once again the \textbf{dilaton}.
These fields account for the \( (24)^2 \) physical degrees of freedom found in the expansion. This procedure can be systematically applied to uncover the full tower of excited string states.

\subsection{The Orientifold Construction}
We now discuss the simplest example of an orientifold, or open descendant~\cite{Sagnotti:1987Cargese,Pradisi:1989zz,Horava:1989fv,Horava:1989bg,BianchiSagnotti:1990,BianchiSagnotti:1991a,Bianchi1992,Sagnotti:1992note,Dudas:2000review,Angelantonj:2002ct,AngelantonjFlorakis:2024}, where world-sheet parity $\Omega$ is used to project the closed string spectrum. Recall that the low-lying bosonic closed string spectrum starts with a tachyonic scalar, followed by massless states from $\alpha_{-1}^i \tilde{\alpha}_{-1}^j |0, \tilde{0}\rangle$: a traceless symmetric tensor, a scalar (the trace), and an antisymmetric tensor. 

These states and all excitations have definite symmetry under interchange of left $\alpha$ and right $\tilde{\alpha}$ oscillators, since $\Omega$ is preserved by the action and quantization, and satisfies $\Omega^2=1$. This splits the spectrum into two subsets with eigenvalues $\pm 1$. However, because string states can scatter, the product of two odd states produces even states, so the only consistent choice is to keep the $\Omega$-invariant states, eliminating, in particular, the massless antisymmetric tensor. Thus, the original massless level of $24 \times 24 = 576$ states reduces to $24 (24+1)/2 = 300$ states after projection.

To reflect this, the torus amplitude contribution must be halved and supplemented by an additional term where left and right modes are identified, described by the \textit{Klein-bottle amplitude}. This amplitude represents a vacuum diagram of a closed string undergoing orientation reversal and can be written as a trace over states with insertion of $\Omega$:
\beq
    \mathcal{K} = \frac{1}{2} \int_{\mathcal{F}_K} \frac{d^2 \tau}{\tau_2^2} \frac{1}{\tau_2^{12}} \, \text{tr}\left(q^{L_0 - 1} \bar{q}^{\bar{L}_0 - 1} \Omega \right)
    \quad .
\eeq
Explicitly, using $\Omega |L,R\rangle = |R,L\rangle$ and orthonormality,
\beq
    \sum_{L,R} \langle L,R| q^{L_0 - 1} \bar{q}^{\bar{L}_0 - 1} \Omega |L,R\rangle 
    = \sum_L \langle L,L| (q \bar{q})^{L_0 - 1} |L,L \rangle
    \quad ,
\eeq
where the restriction to diagonal states $|L,L\rangle$ effectively identifies $L_0$ and $\bar{L}_0$. This construction thus describes bosonic, closed but unoriented strings.

The Klein bottle is the second surface with Euler characteristic $\chi=0$: the torus has $h=1,b=0,c=0$, while the Klein bottle has $h=0,b=0,c=2$. It cannot be embedded in three-dimensional Euclidean space without self-intersections. There are two fundamental polygons for the Klein bottle and one for the doubly covering torus. 

The first polygon, with sides $1$ and $i \tau_2$, differs from the torus in that its horizontal sides have opposite orientations and $\tau$ is purely imaginary. The Klein bottle arises from its covering torus with Teichmüller parameter $2i \tau_2$ by supplementing lattice translations with the anticonformal involution $z \to 1 - \bar{z} + i \tau_2$, where $\tau_2$ plays the role of the proper world-sheet time for closed string propagation.

The second polygon halves the horizontal side and doubles the vertical one, keeping the area fixed. This alternative representation shows the Klein bottle as a tube ending on two crosscaps, with horizontal sides now oriented the same way, and the vertical sides identified pairwise by translations of half their lengths.

The Klein bottle is thus the natural projection counterpart of the torus (sharing $\chi=0$) but without orientation. The amplitude depends on $2i \tau_2$, the modulus of the covering torus, and the integration domain covers the entire positive imaginary $\tau$-axis, as the involution breaks modular invariance down to a finite subgroup.

Performing the trace for the bosonic string yields:
\beq
    \mathcal{K} = \frac{1}{2} \int_0^\infty \frac{d \tau_2}{\tau_2^{14}} \frac{1}{\eta^{24}(2 i \tau_2)}
    \quad .
    \label{eq:KB}
\eeq
It is useful to compare the $q$-expansions of the torus and Klein-bottle integrands, retaining only terms with equal powers of \(q\) and \(\bar{q}\) corresponding to physical states. Aside from \(\tau_2\) powers, the integrands read:
\bea
    \mathcal{T} &\sim& (q\bar{q})^{-1} + 24^2 + \cdots \quad , \nonumber \\
    \mathcal{K} &\sim& \frac{1}{2} \left( (q\bar{q})^{-1} + 24 + \cdots \right) \quad ,
\eea
showing that the projected spectrum is obtained by halving \(\mathcal{T}\) and adding \(\mathcal{K}\). Using \(t=2\tau_2\), the Klein-bottle amplitude becomes
\beq
    \mathcal{K} = \frac{2^{13}}{2} \int_0^\infty \frac{dt}{t^{14}} \frac{1}{\eta^{24}(i t)}
    \quad .
\eeq
The Klein bottle admits two natural "times": vertical \(\tau_2\) for the direct-channel loop amplitude, and horizontal \(l = 1/t\) for the transverse-channel tree amplitude
\beq
    \tilde{\mathcal{K}} = \frac{2^{13}}{2} \int_0^\infty dl \frac{1}{\eta^{24}(i l)}
    \quad ,
    \label{eq:Ktad}
\eeq
obtained from Eq.~\eqref{eq:KB} via an \(S\) modular transformation.

Including open strings requires the annulus and M\"obius strip amplitudes. The annulus polygon has horizontal sides identified and vertical sides as boundaries, fixed by involutions $z \to -\bar{z}, \quad z \to 2 - \bar{z}$.

The direct-channel annulus amplitude written in terms of $\eta$ and $t=\tau_2/2$ is
\beq
    \mathcal{A} = \frac{N^2 2^{-13}}{2} \int_0^\infty \frac{dt}{t^{14}} \frac{1}{\eta^{24}(i t)}
    \quad .
\eeq
with \(N^2\) from Chan-Paton multiplicities~\cite{Angelantonj:2002ct,Paton:1969je,cp2,cp3,cp4}. 
The integrand expansion in powers of \(\sqrt{q}\) starts as
\beq
    \mathcal{A} \sim \frac{N^2}{2} \left( (\sqrt{q})^{-1} + 24 + \cdots \right)
    \quad .
\eeq
The transverse-channel amplitude, with \(\ell = 1/t\), is
\beq
    \tilde{\mathcal{A}} = \frac{N^2 2^{-13}}{2} \int_0^\infty dl \frac{1}{\eta^{24}(i l)}
    \quad .
    \label{eq:Annolus}
\eeq
This result can be obtained from the annulus amplitude written in terms of the previous time via an $S$ modular transformation.

The M\"obius strip, is characterized by a polygon whose horizontal sides have opposite orientation, and the vertical sides correspond to different portions of a single boundary. The proper time $\tau_2$ parametrizes the propagation of an open string along the strip.

An equivalent representation of the surface can be obtained by halving the horizontal side and doubling the vertical one. In this view, the strip appears as a tube connecting a boundary to a crosscap, where the involution $z \to 1 - \bar{z} + i\tau_2$
defines the crosscap identification.

The complex modulus of the doubly covering torus becomes
\beq
    \tau = \frac{1}{2} + \frac{i}{2} \tau_2
    \quad ,
\eeq
and the modular transformation connecting the direct and transverse channels is
\beq
    P: \quad \frac{1}{2} + i\, \frac{\tau_2}{2} \longmapsto \frac{1}{2} + i\, \frac{1}{2\tau_2}
    \quad ,
\eeq
which is generated by $P = T S T^2 S$.
Using \( S^2 = (ST)^3 \), this implies $P^2 = S^2 = (ST)^3$.

Since $\tau_1 = \frac{1}{2}$, the M\"obius amplitude is not manifestly real, unlike $\mathcal{K}$ and $\mathcal{A}$. To handle this, a real basis of characters is introduced:
\beq
    \hat{\chi}_i\left(i\tau_2 + \frac{1}{2}\right) = q^{h_i - c/24} \sum_k (-1)^k d_k^{(i)} q^k
    \quad ,
\eeq
with $q = e^{-2\pi \tau_2}$. These differ from standard characters by phases \( e^{-i\pi(h_i - c/24)} \), and redefine the transformation as:
\beq
    P = T^{1/2} S T^2 S T^{1/2}
    \quad .
\eeq

From standard modular relations,
\beq
    S^2 = (ST)^3 = C \quad \Rightarrow \quad P^2 = C
    \quad ,
\eeq
where \( C \) is the charge conjugation matrix.

The M\"obius amplitude for the open bosonic string then reads:
\beq
    \mathcal{M} = \frac{\epsilon N}{2} \int_0^\infty \frac{d\tau_2}{\tau_2^{14}} \cdot \frac{1}{\hat{\eta}^{24}\left( \tfrac{1}{2} i \tau_2 + \tfrac{1}{2} \right)}
    \quad ,
\eeq
and its expansion yields:
\beq
    \mathcal{M} \sim \frac{\epsilon N}{2} \left( (\sqrt{q})^{-1} - 24 + \cdots \right)
    \quad .
\eeq

Comparing to the annulus, \( \epsilon = +1 \) gives \( \frac{N(N - 1)}{2} \) massless vectors, corresponding to an orthogonal gauge group, while \( \epsilon = -1 \) (with even \( N \)) gives a symplectic group.

In the transverse channel, after the change \( \tau_2 \to 1/t \), the modular transformation \( P \) yields:
\beq
    \hat{\eta}\left( \frac{i}{2t} + \frac{1}{2} \right) = \sqrt{t} \hat{\eta}\left( \frac{it}{2} + \frac{1}{2} \right)
    \quad ,
\eeq
leading to ($l = t/2)$:
\beq
    \tilde{\mathcal{M}} = 2 \frac{\epsilon N}{2} \int_0^\infty dl \frac{1}{\hat{\eta}^{24}\left( il + \frac{1}{2} \right)}
    \quad .
    \label{eq:Mobius}
\eeq
The additional factor of two in eq.~\eqref{eq:Mobius} reflects the vacuum-channel combinatorics: since \( \tilde{\mathcal{M}} \) describes a tube with one hole and one crosscap, its amplitude includes a combinatorial factor of two compared to \( \tilde{\mathcal{K}} \) and \( \tilde{\mathcal{A}} \).

Among the four amplitudes with vanishing Euler characteristic, only the torus amplitude \( \mathcal{T} \) is UV-safe due to modular invariance. The Klein bottle, annulus, and Möbius amplitudes have UV divergences from regions near the real axis.

In the transverse channel, these appear as IR divergences at large \( \ell \), due to exchange of tachyonic or massless states. A state of mass \( M \) contributes
\beq
    \int_0^\infty d\ell\, e^{-M^2 \ell} = \frac{1}{M^2} \quad,
\eeq
which diverges for \( M = 0 \). Massive states contribute only up to \( \ell \sim 1/M^2 \), so removing massless divergences introduces a natural string-scale cutoff.

The divergence is due to the massless dilaton. In the \( \ell \to \infty \) limit, the propagator
diverges for \( M = 0 \), \( p = 0 \), while the residue yields finite one-point functions on boundaries and crosscaps. Since boundary terms scale with the Chan–Paton factor \( N \), cancellation occurs only for specific values of \( N \).
The divergent parts of eqs.~\eqref{eq:Ktad}, \eqref{eq:Annolus}, and \eqref{eq:Mobius} combine as:
\begin{equation}
    \tilde{\mathcal{K}} + \tilde{\mathcal{A}} + \tilde{\mathcal{M}} \;\propto\; \frac{1}{2} \left( 2^{13} + 2^{-13} N^2 - 2 \epsilon N \right)
    = \frac{2^{-13}}{2} \left(N - \epsilon \,2^{13} \right)^2
    \quad ,
\end{equation}
which vanishes for \( N = 2^{13} = 8192 \) and \( \epsilon = +1 \), giving the Chan–Paton gauge group: $\text{SO}(8192)$.
This gives our first example of a \emph{tadpole condition}. Even if not required for consistency here, not satisfying it would result in a non-vanishing potential for the dilaton:
\begin{equation}
    V(\phi) \propto (N - \epsilon 2^{13}) \int d^{26}x\, \sqrt{-g}\, e^{-\phi}
    \quad ,
\end{equation}
where \( g \) is the space–time metric. In more complex settings, tadpole conditions eliminate gauge or gravitational anomalies~\cite{Polchinski:1988,Aldazabal:1999,Scrucca:1999,Scrucca:2000,Bianchi:2000}.

Tadpole contributions arise from surfaces with one boundary or crosscap (disk or projective disk), which have \( \chi = 1 \), explaining the accompanying \( e^{-\phi} \) factor.

Finally, it should be emphasized that vacuum contributions emerge solely from the open-string sector. This completes the construction of the \( SO(8192) \) model of unoriented bosonic open and closed strings.
\section{Ten--Dimensional Superstrings}
This section constructs all known supersymmetric 10d string models, explains how to extract their spectra, and introduces supersymmetry breaking (to be expanded later).

The starting point is the supersymmetric generalization of the string action (still in Euclidean signature)
\beq
    S=\frac{1}{4\pi \alpha'}\int_{M}d^2 \sigma g^{1/2}\left[g^{ab}\partial_aX^\mu \partial_bX_\mu+i\bar{\psi}^\mu \gamma^a \nabla_a \psi_\mu +i\bar{\chi}_a\gamma^b\gamma^a\psi^\mu \left(\partial_b X_\mu - \frac{i}{4}\bar{\chi}_b\psi_\mu\right)\right]
    \quad ,
\eeq
where $\psi^\mu$ are 2d Majorana spinors (superpartners of $X^\mu$), and $\chi_a$ is the gravitino. Exploiting the symmetries of the action:
\beq
    g_{ab}=\Lambda(\sigma)\delta_{ab} \qquad \chi_a = \gamma_a \chi (\sigma)
    \quad ,
\eeq
leads to the simplified gauge-fixed action:
\beq
    S=\frac{1}{4 \pi \alpha'}\int d^2 \sigma \left(\partial^a X^\mu \partial _a X_\mu+i \bar{\psi} ^\mu \gamma^a \partial_a \psi_\mu \right)
    \quad .
\eeq

All features remain as in the bosonic case, except the critical dimension becomes \( d = 10 \). Open and closed strings differ as before. Since Noether currents contain even powers of \( \psi^\mu \), they are periodic for both antiperiodic (NS) and periodic (R) fermions. For closed strings, four sectors appear: NS--NS and R--R (bosons), NS--R and R--NS (fermions). Open strings, having only one chirality, yield two: NS (bosons) and R (fermions) \cite{GreenSchwarzWitten1987,GreenSchwarzWitten1987Vol2,Lust:1989,Kiritsis:1998,Polchinski:1998rr}.

The Virasoro generators receive bosonic and fermionic contributions \cite{Angelantonj:2002ct}:
\beq
    L_m = \frac{1}{2} \sum_n : \alpha^i_{m-n} \alpha^i_n : + \frac{1}{2} :\sum_r \left( r - \frac{m}{2} \right)  \psi^i_{m-r} \psi^i_r : + \delta_{m,0} \, a
    \quad ,
\eeq
with \( r \) half-integer in NS, integer in R. The normal-ordering constant \( a \) is:
NS: \( a = -\frac{1}{16}(D-2) \), R: \( a = 0 \). This gives the mass formulae:
\bea
    M^2 &=& \frac{2}{\alpha'}\left[N_B+\bar{N}_B+N_F+\bar{N}_F-\frac{D-2}{8}\right] \qquad  \text{(NS)} \nonumber \\
    M^2 &=& \frac{2}{\alpha'}\left[N_B+\bar{N}_B+N_F+\bar{N}_F\right] \qquad  \text{(R)} \nonumber \\
    && \text{with} \qquad N_B + \bar{N}_B = N_F + \bar{N}_F \quad .
\eea

In the NS sector, fermions have no zero modes. The vacuum is a tachyonic scalar; the first excitation, \( \psi^i_{-1/2} \), gives a massless transverse vector if $\frac{1}{2} - \frac{1}{16}(D - 2)$
vanishes, which determines the critical dimension \( D = 10 \).

For the partition function, the fermionic oscillator trace is:
\beq
    \mathrm{tr} \left( q^{\sum_r r \, \psi^{\dagger}_r \psi_r} \right) = \prod_r \mathrm{tr} (q^{r \, \psi^{\dagger}_r \psi_r}) = \prod_r (1 + q^r)^8
    \quad ,
\eeq
with \( r \) half-integer (NS) or integer (R). Including the bosonic contribution:
\beq
    \text{Tr}(q^{L_0}) = \frac{\prod\limits_{m=1}^\infty (1 + q^{m - 1/2})^8}{q^{1/2} \prod\limits_{m=1}^\infty (1 - q^m)^8} \quad \text{(NS)}
    \qquad
    \text{Tr}(q^{L_0}) = 16 \frac{\prod\limits_{m=1}^{\infty} (1 + q^m)^8}{\prod\limits_{m=1}^{\infty} (1 - q^m)^8} \quad \text{(R)}
\eeq

Here the factor 8 counts transverse directions. The \( q^{1/2} \) in the NS sector reflects the tachyon, absent in the R sector. The factor 16 in the R sector comes from the degeneracy of the vacuum: $\{ \psi^i_0, \psi^j_0 \} = 2 \delta^{ij}$, 
so the vacuum forms a 16-dimensional representation of the SO(8) Clifford algebra.

The torus amplitude \eqref{eq:torus1} becomes:
\beq
    \mathcal{T}=\int_{\mathcal{F}}\frac{d^2 \tau}{\tau_2^2}\frac{1}{\tau_2^4}\text{Tr}\left(q^{N_B}\bar{q}^{\bar{N}_B}q^{N_F}\bar{q}^{\bar{N}_F}q^{-a_R}\bar{q}^{-a_L}\right)
    \quad ,
\eeq
with: $a_L=a_R=1/2 \quad \text{(NS)}$ , $
a_L=a_R=0 \quad \text{(R)} \quad .$

To construct a consistent spectrum, one must project out states with incorrect spin-statistics and remove tachyons. This leads to the GSO projection \cite{Gliozzi:1976,Gliozzi:1977}, which enforces modular invariance and physical consistency by retaining only states with the correct worldsheet fermion number parity.

The GSO projection in the NS sector is implemented by inserting \(\frac{1 - (-1)^F}{2}\) inside the trace, where \(F\) is the world-sheet fermion number, reversing the sign of contributions from states with an odd number of fermionic oscillators:

\beq
    \mathrm{tr} \left( \frac{1 - (-1)^F}{2} \, q^{L_0} \right)
    = \frac{ \prod\limits_{m=1}^{\infty} (1 + q^{m - 1/2})^8 - \prod\limits_{m=1}^{\infty} (1 - q^{m - 1/2})^8 }{2q^{1/2} \prod\limits_{m=1}^{\infty} (1 - q^m)^8}
    \quad ,
\eeq
where the insertion of \( (-1)^F \), with \( F \) the world-sheet fermion number, reverses the sign of all contributions associated with odd numbers of fermionic oscillators. 

In the R sector, the projection uses \(\frac{1 - \Gamma_{11}(-1)^F}{2}\), where \(\Gamma_{11}\) (the 10d analogue of \(\gamma_5\)) imposes definite chirality, halving the states:

\beq
    \mathrm{tr} \left( \frac{1 - \Gamma_{11}(-1)^F}{2} \, q^{L_0} \right)
    = \frac{16}{2}\frac{\prod^{\infty}_{m=1}(1+q^m)^8}{\prod^{\infty}_{m=1}(1-q^m)^8} 
    \quad .
\eeq

The full torus amplitude sums over the four sectors:

\beq
    \mathcal{T}= \mathcal{T}_{NS-NS}+\mathcal{T}_{R-R}+\mathcal{T}_{NS-R}+\mathcal{T}_{R-NS}
    \quad ,
\eeq
where
\bea
    \mathcal{T}_{\text{NS--NS}} &=& 
    \int \frac{d^2 \tau}{\tau_2^6}  \frac{1}{4}
    \left| 
    \frac{ \prod\limits_{k=1}^\infty \left(1 + q^{k - \frac{1}{2}}\right)^8 
        - \prod\limits_{k=1}^\infty \left(1 - q^{k - \frac{1}{2}}\right)^8 }
        { q^{1/2} \prod\limits_{k=1}^\infty \left(1 - q^k\right)^8 } 
    \right|^2 \nonumber  \quad ,\\
    \mathcal{T}_{\text{R--R}} &=& 
    \int \frac{d^2 \tau}{\tau_2^6} \frac{1}{4}
    \left| 
    \frac{ 16 \prod\limits_{k=1}^\infty \left(1 + q^k\right)^8 }
         { \prod\limits_{k=1}^\infty \left(1 - q^k\right)^8 } 
    \right|^2 \nonumber \quad ,\\
    \mathcal{T}_{\text{NS--R}} &=& 
    - \int \frac{d^2 \tau}{\tau_2^6} \frac{1}{4} \frac{1}{q^{1/2}} 
    \left( 
    \prod\limits_{k=1}^\infty \left(1 + q^{k - \frac{1}{2}}\right)^8 
    - \prod\limits_{k=1}^\infty \left(1 - q^{k - \frac{1}{2}}\right)^8 
    \right) 
    \frac{ -16 \cdot \prod\limits_{k=1}^\infty \left(1 + \bar{q}^k\right)^8 }
        {\left| \prod\limits_{k=1}^\infty \left(1 - q^k\right)^8 \right|^2}  \nonumber  \quad ,\\
    \mathcal{T}_{\text{R--NS}} &=& 
    - \int \frac{d^2 \tau}{\tau_2^6} \frac{1}{4} \frac{16}{\bar{q}^{1/2}} 
    \left( 
    \prod\limits_{k=1}^\infty \left(1 + q^k\right)^8 
    \right) 
    \frac{ \prod\limits_{k=1}^\infty \left(1 + \bar{q}^{k - \frac{1}{2}}\right)^8 
         - \prod\limits_{k=1}^\infty \left(1 - \bar{q}^{k - \frac{1}{2}}\right)^8 }
        {\left| \prod\limits_{k=1}^\infty \left(1 - q^k\right)^8 \right|^2} 
        \nonumber \\ \quad &&.
\eea

To handle the GSO projection effectively, it is convenient to use SO(8) level-one characters. These characters classify the independent sectors with definite spin-statistics on both the world-sheet and in spacetime, exploiting the transverse SO(8) symmetry in ten dimensions. This method will be introduced briefly in the next section.
\subsection{SO(8) Characters}
Let us introduce the \textit{Jacobi theta functions}~\cite{WhittakerWatson1927,Erdelyi1953,Fay1973,Mumford1983},that for our convenience we define by by the infinite products:
\bea
    &&\vartheta \begin{bmatrix} \alpha \\ \beta \end{bmatrix}(z | \tau) = 
    e^{2\pi i \alpha(z + \beta)} q^{\frac{1}{2} \alpha^2}
    \prod_{n=1}^{\infty} (1 - q^n)
    \prod_{n=1}^{\infty} \left(1 + q^{n + \alpha - \frac{1}{2}} e^{2\pi i(z + \beta)} \right)
    \times \nonumber \\&& \times \left(1 + q^{n - \alpha - \frac{1}{2}} e^{-2\pi i(z + \beta)} \right)\quad .
\eea
These \(\vartheta\) functions have well-known modular transformation properties under \(T\) and \(S\), which will not be detailed here but can be found in \cite{Angelantonj:2002ct}.

For our purposes, the fermions \(\psi^i\) are periodic or antiperiodic, so it suffices to consider the \emph{theta-constants} at vanishing argument \(z=0\) with \(\alpha, \beta = 0, \frac{1}{2}\). Among these, \(\vartheta_1 = \vartheta \begin{bmatrix} \frac{1}{2} \\ \frac{1}{2} \end{bmatrix}\) vanishes identically at zero, while the other three are non-zero and commonly denoted \(\vartheta_2, \vartheta_3, \vartheta_4\).

These satisfy:
\beq
    \frac{ \vartheta_2^4(0|\tau) }{ \eta^{12}(\tau) } =
    16\frac{ \prod\limits_{m=1}^\infty (1 + q^m)^8 }{ \prod\limits_{m=1}^\infty (1 - q^m)^8 }
    \quad , \quad
    \frac{ \vartheta_3^4(0|\tau) }{ \eta^{12}(\tau) } =
    \frac{ \prod\limits_{m=1}^\infty (1 + q^{m-\frac{1}{2}})^8 }{ q^{1/2} \prod\limits_{m=1}^\infty (1 - q^m)^8 }
    \quad , \quad
    \frac{ \vartheta_4^4(0|\tau) }{ \eta^{12}(\tau) } =
    \frac{ \prod\limits_{m=1}^\infty (1 - q^{m-\frac{1}{2}})^8 }{ q^{1/2} \prod\limits_{m=1}^\infty (1 - q^m)^8 }
    \quad .
\eeq

Turning to SO(8) representations, the first two characters are defined as:
\bea
    O_8 &=& \frac{\vartheta_3^4 + \vartheta_4^4}{2\, \eta^4} = q^{-1/6}(1 + 28q + \cdots) \quad , \nonumber \\
    V_8 &=& \frac{\vartheta_3^4 - \vartheta_4^4}{2\, \eta^4} = q^{-1/6}(8q^{1/2} + 64q^{3/2} + \cdots) \quad .
\eea  
These correspond to an orthogonal decomposition of the NS sector: \(O_8\) contains even fermion excitations starting with the tachyon, while \(V_8\) contains odd excitations starting with the massless vector. These are associated with the singlet and vector conjugacy classes of the SO(8) weight lattice.

To complete the set, two more characters \(S_8\) and \(C_8\) are introduced for the R sector, built from \(\vartheta_2\) and the vanishing \(\vartheta_1\):
\bea
    S_8 &=& \frac{ \vartheta_2^4 + \vartheta_1^4 }{2 \eta^4} = 8q^{1/3}(1 + 8q + \cdots) \quad , \\
    C_8 &=& \frac{ \vartheta_2^4 - \vartheta_1^4 }{2 \eta^4} = 8q^{1/3}(1 + 8q + \cdots) \quad .
\eea
These describe orthogonal portions of the R spectrum corresponding to the two spinors of opposite chirality. Massive modes mix chiralities, enabling a full description of spinors in the string spectrum.

The famous Jacobi identity,
\beq
    \vartheta_3^4 - \vartheta_4^4 - \vartheta_2^4 = 0
    \quad ,
\eeq
implies that the full torus amplitude,
\beq
    \mathcal{T} = \int \frac{d^2 \tau}{\tau_2^6} \left| \frac{\vartheta_3^4(0|\tau) - \vartheta_4^4(0|\tau) - \vartheta_2^4(0|\tau)}{\eta^{12}(\tau)} \right|^2
    \quad ,
\eeq
contains equal numbers of bosonic and fermionic excitations at all mass levels, ensuring space-time supersymmetry~\cite{Gliozzi:1976}.
SO characters can be defined in the same way also for generic $n$. They form representations of the modular group with unitary \(S\) and diagonal \(T\) matrices; their explicit forms and modular properties are described in detail in \cite{Angelantonj:2002ct}.

In particular, for the physically relevant SO(8), the \(T\) matrix acts on \(\{O_8, V_8, S_8, C_8\}\) as:
\beq
    T = \mathrm{diag}(-1,\, 1,\, 1,\, 1)
    \quad .
\eeq
The matrix \(P\), related to the transverse channel, shares the same diagonal form. This structure reflects the resolution of the ambiguity in distinguishing the spinor sectors \(S\) and \(C\), despite the vanishing of \(\vartheta_1\) at zero argument, a subtlety well understood within two-dimensional Conformal Field Theory~\cite{Angelantonj:2002ct}.
\subsection{Type IIA and Type IIB Superstrings}
Consistent 10-dimensional spectra of oriented closed strings are built by imposing modular invariance on torus amplitudes of the form:
\beq
    \mathcal{T} = \chi X \chi
    \quad ,
\eeq
where \( \chi \) are characters and \( X \) implements the GSO projection. Modularity requires:
\beq
    S^\dagger X S = X, \qquad T^\dagger X T = X
    \quad ,
\eeq

As already discussed, supersymmetry imposes an equal number of bosonic and fermionic degrees of freedom at each mass level. In terms of SO(8) characters, this is equivalent to:
\beq
    V_8 = S_8 = C_8
    \quad ,
    \label{eq:jaeq}
\eeq
a form of Jacobi's identity. As a consequence, supersymmetric torus amplitudes must vanish. Any non-vanishing contribution (aside from special exceptions~\cite{Kachru1999, Harvey1999, Angelantonj1999}) signals a backreaction on spacetime.

There are four consistent modular invariant torus amplitudes:
\beq
    \mathcal{T}_{\text{IIA}} = \int_{\mathcal{F}}\frac{d^2 \tau}{\tau_2^2} \frac{(\bar{V}_8 - \bar{S}_8)(V_8 - C_8)}{\tau_2^4 \eta^8 \bar{\eta}^8}
    \quad , \quad
    \mathcal{T}_{\text{IIB}} = \int_{\mathcal{F}}\frac{d^2 \tau}{\tau_2^2} \frac{|V_8 - S_8|^2}{\tau_2^4 \eta^8 \bar{\eta}^8}
    \quad ,
    \label{eq:Type2}
\eeq
\beq
    \mathcal{T}_{\text{0A}} = \int_{\mathcal{F}}\frac{d^2 \tau}{\tau_2^2} \frac{|O_8|^2 + |V_8|^2 + \bar{S}_8 C_8 + \bar{C}_8 S_8}{\tau_2^4 \eta^8 \bar{\eta}^8}
    \quad , \quad
    \mathcal{T}_{\text{0B}} = \int_{\mathcal{F}}\frac{d^2 \tau}{\tau_2^2} \frac{|O_8|^2 + |V_8|^2 + |S_8|^2 + |C_8|^2}{\tau_2^4 \eta^8 \bar{\eta}^8}
    \quad .
\eeq

Focusing on the supersymmetric models, we note that IIB is worldsheet-parity symmetric but has chiral spacetime content, while IIA is worldsheet-parity asymmetric but spacetime non-chiral. All spectra are tachyon-free, and the first states are massless as can be recognized expanding the amplitude.

From these, the massless particle content can be identified:
\beq
\begin{array}{ll}
\textbf{Type IIA} \quad &
\textbf{Type IIB} \quad \\
\\[0.3em]
\begin{cases}
\text{NS-NS:} \; g_{\mu \nu}, B_{\mu \nu}, \phi \\
\text{R-R:} \; A_{\mu}, C_{\mu \nu \rho} \\
\text{NS-R, R-NS:} \; \psi_\mu^L, \psi_\mu^R, \psi^L, \psi^R
\end{cases}
&
\begin{cases}
\text{NS-NS:} \; g_{\mu \nu}, B^{(1)}_{\mu \nu}, \phi^{(1)} \\
\text{R-R:} \; \phi^{(2)}, B_{\mu \nu}^{(2)}, D^+_{\mu \nu \rho \sigma} \\
\text{NS-R, R-NS:} \; \psi_\mu^{L(1)}, \psi_\mu^{L(2)}, \psi^{R(1)}, \psi^{R(2)}
\end{cases}
\end{array}
\quad .
\label{eq:IIspec}
\eeq

The four-form \( D^+_{\mu \nu \rho \sigma} \) is subject to a self-duality constraint. These massless sectors are accompanied by towers of massive excitations with masses proportional to \( 1/\alpha' \). The Type IIA spectrum completes the non-chiral \((1,1)\) supergravity multiplet in 10d, while Type IIB corresponds to the chiral \((2,0)\) multiplet.
\paragraph{0A, 0B strings}
\leavevmode\\
As previously noted, modular invariance also allows for two additional consistent oriented closed string models: Type 0A and 0B. Their torus amplitudes were already presented. These models are not supersymmetric, as they do not vanish under Jacobi’s identity, and they both contain a tachyon, signaled by the isolated \( O_8 \bar{O}_8 \) term.

Despite being purely bosonic, their low-energy spectra are well-defined:
\beq
\begin{array}{ll}
\textbf{Type 0A} &
\textbf{Type 0B} \quad \\
\\[0.3em]
\begin{cases}
\text{NS-NS:} \; T, g_{\mu \nu}, B_{\mu \nu}, \phi \\
\text{R-R:} \; A_{\mu}^{(1)}, A_\mu^{(2)}, C^{(1)}_{\mu \nu \rho}, C^{(2)}_{\mu \nu \rho}
\end{cases}
&
\begin{cases}
\text{NS-NS:} \; T, g_{\mu \nu}, B^{(1)}_{\mu \nu}, \phi^{(1)} \\
\text{R-R:} \; \phi^{(2)}, \phi^{(3)}, B_{\mu \nu}^{(2)}, B_{\mu \nu}^{(3)}, D_{\mu \nu \rho \sigma}
\end{cases}
\end{array}
\quad .
\label{eq:0spectrum}
\eeq

The four-form \( D_{\mu \nu \rho \sigma} \) is unconstrained. These models will be further discussed in the context of non-supersymmetric strings.
\subsection{Heterotic Superstrings}
Although not directly related to the climbing scalar phenomenon, for completness we have to introduce the ten-dimensional supersymmetric \textit{heterotic string}~\cite{Gross1985,Gross1986} combines right-moving superstring modes with left-moving bosonic string modes. Its name reflects the independence of left- and right-movers: the left sector is bosonic (26d), the right sector supersymmetric (10d).

To ensure spacetime supersymmetry and eliminate tachyons, 10 of the 26 left-moving bosonic coordinates are identified with spacetime coordinates \(X^\mu\) and their fermionic partners \(\psi^\mu\), \(\mu=0,\dots,9\), while the remaining 16 left-moving bosons are compactified.

Compactification \cite{Polchinski:1998rq,Polchinski:1998rr,Zwiebach2004,GreenSchwarzWitten1987,GreenSchwarzWitten1987Vol2,BeckerBeckerSchwarz2006} curls extra dimensions on a compact internal manifold, adding quantized momenta and windings that enrich the spectrum and induce internal gauge symmetries.

Modular invariance requires the internal compactification lattice to be 16-dimensional, even, and self-dual:
\[
\Lambda = \Lambda^* = \{ w \in \mathbb{R}^{16} \mid w \cdot v \in \mathbb{Z}, \forall v \in \Lambda \}.
\]
This ensures tachyon removal, modular invariance, and quantum consistency.

Even unimodular lattices exist only in dimensions multiple of 8; at 16d there are two~\cite{BeckerBeckerSchwarz2006}: \(\Gamma_{E_8 \times E_8}\) (two \(E_8\) root lattices) and \(\Gamma_{D_{16}^+}\) (an extension of \(D_{16}\)), corresponding to gauge groups \(E_8 \times E_8\) and \(\text{Spin}(32)/\mathbb{Z}_2\). These give the only two consistent ten-dimensional supersymmetric heterotic strings: \(E_8 \times E_8\) heterotic (HE), \(\text{Spin}(32)/\mathbb{Z}_2\) heterotic (HO).
Their one-loop vacuum amplitudes, after GSO projections on the right-moving sector, take the modular invariant forms:
\beq
    \mathcal{T}_{\text{HE}} = \int_{\mathcal{F}}\frac{d^2 \tau}{\tau_2^2} \frac{(V_8-S_8)(\bar{O}_{16}+\bar{S}_{16})^2}{\tau_2^4 \eta^8 \bar{\eta}^8} \qquad , \qquad
    \mathcal{T}_{\text{HO}} = \int_{\mathcal{F}}\frac{d^2 \tau}{\tau_2^2} \frac{(V_8-S_8)(\bar{O}_{32}+\bar{S}_{32})}{\tau_2^4 \eta^8 \bar{\eta}^8}
    \quad .
\eeq
These amplitudes encode contributions from spacetime bosons, fermions, and internal gauge degrees of freedom. Level-matching conditions impose strict constraints on physical states, with the left-moving sector starting at zero mass level and the right-moving sector at level \(-1\). The characters \(O_{2n}\), \(V_{2n}\), \(S_{2n}\), and \(C_{2n}\) contribute specific shifts to the mass levels consistent with this counting.

In the HE model, four internal fermionic sectors generate massless level-matched states forming the adjoint representation of \(E_8 \times E_8\). The HO model has a simpler massless spectrum arising solely from the \(\bar{O}_{32}\) sector, whose antisymmetric NS oscillator combinations build the adjoint representation of \(SO(32)\).

In both theories, the combination of a single right-moving bosonic oscillator with the left-moving sector produces the \((1,0)\) supergravity multiplet. Thus, the low-energy spectra are summarized as:
\beq
\begin{array}{ll}
\textbf{HE model} & \textbf{HO model} \quad \\
\\[0.3em]
\left(g_{\mu \nu}, B_{\mu \nu}, \phi, \psi_\mu^L, \psi^R\right) \oplus [E_8 \times E_8 (A_\mu, \lambda_L)] &
\left(g_{\mu \nu}, B_{\mu \nu}, \phi, \psi_\mu^L, \psi^R\right) \oplus [SO(32) (A_\mu, \lambda_L)]
\end{array}
\quad .
\label{eq:hspectrum}
\eeq
These correspond to \((1,0)\) supergravity coupled to supersymmetric Yang--Mills theory with gauge group \(E_8 \times E_8\) for HE and \(SO(32)\) for HO.
\subsection{Type I Superstring}
Before introducing orientifolds, we note a direct relation between Type IIA and Type IIB strings in ten dimensions via an \textit{orbifold construction} without compactification \cite{GreenSchwarzWitten1987Vol2,Polchinski:1998rr,Angelantonj:2002ct}. Orbifolding by \( (-1)^{F_R} \) on Type IIB projects out states odd under this operator and adds twisted sectors to ensure modular invariance. The partition function changes as
\beq
    \mathcal{T}_{\text{IIB}} \rightarrow \frac{1}{2} \mathcal{T}_{\text{IIB}} + \frac{1}{2} \int_{\mathcal{F}} \frac{d^2 \tau}{\tau_2^2} \frac{(V_8 - S_8)(\bar{V}_8 + \bar{S}_8)}{\tau_2^4 |\eta(\tau)|^8} \quad ,
\eeq
which alone is not modular invariant. Including the modular images
\beq
    \frac{1}{2} \int_{\mathcal{F}} \frac{d^2 \tau}{\tau_2^2} \frac{(V_8 - S_8)(\bar{O}_8 - \bar{C}_8)}{\tau_2^4 |\eta(\tau)|^8} \quad , \quad
    \frac{1}{2} \int_{\mathcal{F}} \frac{d^2 \tau}{\tau_2^2} \frac{(V_8 - S_8)(-\bar{O}_8 - \bar{C}_8)}{\tau_2^4 |\eta(\tau)|^8} \quad ,
\eeq
restores modular invariance and reproduces the Type IIA partition function (eq.~\eqref{eq:Type2}). Thus, Type IIA arises as an orbifold of Type IIB by \( (-1)^{F_R} \), and vice versa with \( (-1)^{F_L} \).

\begin{center}
    ***
\end{center}

The \( SO(32) \) Type I superstring is an open descendant of Type IIB with supercharacter \( V_8 - S_8 \). The direct-channel one-loop amplitudes are:
\bea
    \mathcal{K} &=& \frac{1}{2} \int_0^\infty \frac{d \tau_2}{\tau_2^6} \frac{(V_8 - S_8)(2 i \tau_2)}{\eta^8(2 i \tau_2)} \quad , \quad
    \mathcal{A} = \frac{N^2}{2} \int_0^\infty \frac{d \tau_2}{\tau_2^6} \frac{(V_8 - S_8)(\frac{1}{2} i \tau_2)}{\eta^8(\frac{1}{2} i \tau_2)} \quad , \nonumber \\ 
    \mathcal{M} &=& \frac{\epsilon N}{2} \int_0^\infty \frac{d \tau_2}{\tau_2^6} \frac{(\hat{V}_8 - \hat{S}_8)(\frac{1}{2} i \tau_2 + \frac{1}{2})}{\hat{\eta}^8(\frac{1}{2} i \tau_2 + \frac{1}{2})} \quad ,
\eea
where \( \epsilon = \pm 1 \) fixes the gauge group. The Klein-bottle projection symmetrizes the NS–NS sector removing the massless two-form, antisymmetrizes the R–R sector removing the second scalar and self-dual four-form, yielding a closed spectrum consistent with minimal \( \mathcal{N} = (1,0) \) supergravity: graviton, R–R two-form, dilaton, left-handed gravitino, right-handed spinor.

The massless open sector forms a \( (1,0) \) super-Yang–Mills multiplet with gauge group \( SO(N) \) for \( \epsilon = -1 \) or \( USp(N) \) for \( \epsilon = +1 \), with spectrum
\beq
    \begin{cases}
        \text{NS-NS}: (g_{\mu\nu}, \phi), \quad \text{R-R}: (B_{\mu\nu}), \\
        \text{NS-R, R-NS}: (\psi_\mu^L, \psi^R)
    \end{cases}
    \quad \oplus \quad (A_\mu, \lambda) \quad .
\eeq

In the transverse channel, amplitudes read:
\bea
    \tilde{\mathcal{K}} &=& \frac{2^5}{2} \int_0^\infty dl \frac{(V_8 - S_8)(i l)}{\eta^8(i l)} \quad , \quad
    \tilde{\mathcal{A}} = \frac{2^{-5} N^2}{2} \int_0^\infty dl \frac{(V_8 - S_8)(i l)}{\eta^8(i l)} \quad , \quad \nonumber \\ 
    \tilde{\mathcal{M}} &=& \frac{2 \epsilon N}{2} \int_0^\infty dl \frac{(\hat{V}_8 - \hat{S}_8)(i l + \frac{1}{2})}{\hat{\eta}^8(i l + \frac{1}{2})} \quad .
\eea
The tadpole cancellation condition,
\beq
    \frac{2^5}{2} + \frac{2^{-5} N^2}{2} + \frac{2 \epsilon N}{2} = \frac{2^{-5}}{2} (N + 32 \epsilon)^2 = 0 \quad ,
\eeq
requires \( N=32 \) and \( \epsilon = -1 \), fixing the gauge group to \( SO(32) \). The cancellation of the \( S_8 \) tadpole is key for the Green–Schwarz anomaly cancellation~\cite{GreenSchwarz1984} and supersymmetry also cancels the \( V_8 \) tadpole.

This setup corresponds to 32 \( D9 \)-branes filling spacetime and an \( O9^+ \)-orientifold plane. \( Dp \)-branes and \( Op \)-planes couple to R–R \( (p+1) \)-form fields \( C_{p+1} \), carry tension and charge: branes have positive charge and tension; orientifold planes can be \( O^- \) (positive) or \( O^+ \) (negative). Anti-branes and anti-planes exist with opposite charge.

R–R tadpoles enforce charge conservation and anomaly cancellation, NS–NS tadpoles couple to the dilaton affecting vacuum energy and can yield \textbf{dilaton tadpoles}. Non-derivative R–R couplings consistent with zero-momentum tadpoles appear in~\cite{Bianchi1992,DiVecchia1997}.

In the supersymmetric Type I \( SO(32) \) string both R–R and NS–NS tadpoles cancel, but extensions to non-supersymmetric cases with residual dilaton tadpoles exist and will be discussed later.
\begin{center}
    ***
\end{center}
The five consistent ten-dimensional superstring theories --- type IIA, type IIB, type I, and the heterotic models based on $SO(32)$ and $E_8 \times E_8$ --- together with eleven-dimensional supergravity form the basis of \textbf{M-theory}, a unifying framework proposed in the mid-1990s~\cite{Witten1995}. These theories are linked by a network of dualities: T-duality connects type IIA and IIB via circle compactification; S-duality relates type I to the heterotic $SO(32)$ and maps type IIB to itself; U-dualities arise in lower dimensions from combinations of both. Eleven-dimensional supergravity emerges as the low-energy limit of M-theory and connects to type IIA via $S^1$ compactification, and to heterotic $E_8 \times E_8$ via $S^1/\mathbb{Z}_2$.

\begin{figure}[ht]
\centering
\begin{tabular}{ccc}
%\mbox{graphic} & \mbox{table} \\
\includegraphics[width=50mm]{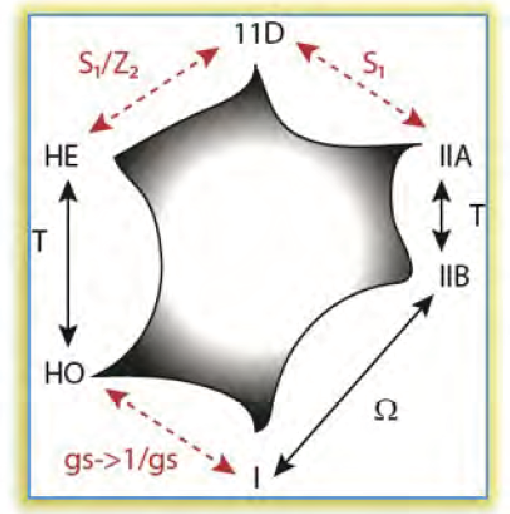} \quad  &
\includegraphics[width=70mm]{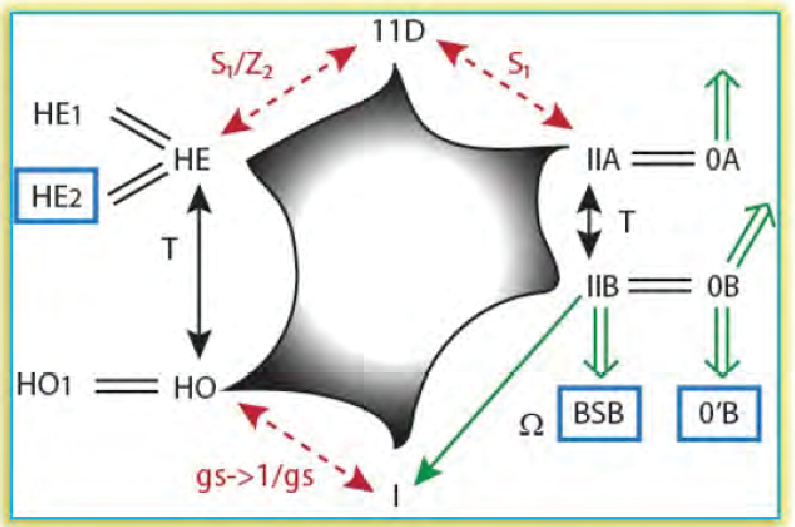}  \\
\end{tabular}
 \caption{\small Left panel: The duality hexagon for ten-dimensional supersymmetric superstrings; Right panel: The larger duality diagram including the ten-dimensional nonsupersymmetric superstrings. The green lines identify orientifold projections, the blue boxes identify the three nontachyonic models: $HE_2$ stands for the $SO(16) \times SO(16)$ model of~\cite{AlvarezGaume1986}, $0'B$ for the $U(32)$ model of~\cite{Sagnotti1995,Sagnotti1997}, and BSB for Sugimoto’s model in~\cite{Sugimoto1999,Antoniadis1999,Angelantonj2000b,Aldazabal1999,Angelantonj2000b} . This picture is taken from \cite{MouradSagnotti2021}.}
\label{fig:hexagons}
\end{figure}

While this duality web elegantly unifies the supersymmetric theories, it does not exhaust the space of consistent string constructions. If one relaxes supersymmetry while retaining tree-level tachyon absence and world-sheet consistency, a distinct class of ten-dimensional \textbf{non-supersymmetric} string models emerges. Though disconnected from M-theory, these models offer a valuable laboratory for exploring novel phenomena --- most notably, the \textit{climbing scalar} mechanism, which is central to this work.

The next section presents a systematic overview of all known tachyon-free, non-supersymmetric string theories in ten dimensions, examining their constructions, spectra, and consistency constraints, and culminating in the discussion of the climbing scalar phenomenon that will occupy the remainder of this thesis.
\section{10d Non-supersymmetric Models}

The five consistent supersymmetric string theories in ten dimensions form the core of the M-theory web, yet this duality framework (see Fig.~\ref{fig:hexagons}) captures only a limited portion of the full string landscape. Enforcing modular invariance, anomaly cancellation, and world-sheet consistency reveals additional ten-dimensional models that, despite lacking spacetime supersymmetry, remain perturbatively consistent.

These non-supersymmetric constructions obey the same string-theoretic criteria and provide a top-down setting for studying supersymmetry breaking, beyond phenomenological assumptions. Most contain tachyons, signaling vacuum instabilities, but three models are tachyon-free: the $SO(16) \times SO(16)$ heterotic string~\cite{AlvarezGaume1986}, the $0'B$ orientifold with $U(32)$ gauge group~\cite{Sagnotti1995,Sagnotti1997}, and the Sugimoto model~\cite{Sugimoto1999,Antoniadis1999,Angelantonj2000b,Aldazabal1999}, where supersymmetry is broken via \textit{brane supersymmetry breaking}\cite{Freedman:1976xh,Deser:1976eh,Freedman:2012zz,Sagnotti:1987Cargese,Pradisi:1989zz,Horava:1989fv,Horava:1989bg,BianchiSagnotti:1990,BianchiSagnotti:1991a,Bianchi1992, Sagnotti:1992note, Dudas:2000review,Angelantonj:2002ct, AngelantonjFlorakis:2024,Aldazabal1999}.

These models offer consistent mechanisms of supersymmetry breaking and a natural framework for dynamical phenomena such as the \textit{climbing scalar}, central to the following analysis.

We now turn to a systematic description of these three non-supersymmetric, tachyon-free string models, focusing on their spectra, consistency conditions, and construction principles.

\subsection{\texorpdfstring{$SO(16) \times SO(16)$} \ \ Model}
Another consistent ten-dimensional, tachyon-free model is the non-supersymmetric heterotic string with gauge group $SO(16) \times SO(16)$, constructed analogously to the supersymmetric heterotic strings \cite{DixonHarvey1986}. Its torus amplitude reads:
\bea
    \mathcal{T}_{SO(16) \times SO(16)}&=&\int_{\mathcal{F}}\frac{d^2 \tau}{\tau_2^2} \, \Bigg[\frac{O_8\left(\bar{V}_{16}\bar{C}_{16}+\bar{C}_{16}\bar{V}_{16}\right)+V_8\left(\bar{O}_{16}\bar{O}_{16}+\bar{S}_{16}\bar{S}_{16}\right)}{\tau_2^4 \, \eta^8 \,\bar{\eta}^8}- \nonumber \\ &+&\frac{-S_8\left(\bar{O}_{16}\bar{S}_{16}+\bar{S}_{16}\bar{O}_{16}\right)-C_8\left(\bar{V}_{16}\bar{V}_{16}+\bar{C}_{16}\bar{C}_{16}\right)}{\tau_2^4 \, \eta^8 \,\bar{\eta}^8} \, \Bigg]
    \quad .
\eea

This is the only one of the three tachyon-free, non-supersymmetric models not built via orientifold projection. Its massless spectrum is:
\begin{center}
    \textbf{$SO(16) \times SO(16)$}
\end{center}
\beq
    \left(g_{\mu \nu}, \, B_{\mu \nu}, \, \phi \right) \,  \oplus \, A_\mu^{(120,1) \, \oplus \, (1, 120) } \, \oplus \, \psi^{L \, \, (128,1) \, \oplus \, (1, 128)} \, \oplus \, \psi^{R \, \, (16, 16)}
    \quad .
    \label{eq:so16crssso16}
\eeq

No gravitino appears, as expected in non-supersymmetric spectra, yet no tachyons are present: the $O_8$ ground state is level-mismatched with the right-moving sector, whose excitations start $3/2$ units above. This model can be seen as a $\mathbb{Z}_2$ orbifold of the two supersymmetric heterotic theories.

As a first instance of a general issue with broken supersymmetry, the partition function~\eqref{eq:so16crssso16} does not vanish even after applying~\eqref{eq:jaeq}. The resulting positive vacuum energy, first computed in~\cite{AlvarezGaume1986}, induces an $O(1/\alpha')$ curvature, invalidating the flat-space background and requiring an effective low-energy (super)gravity description.

This effective approach, shared by the remaining two models, will provide the framework for the climbing scalar dynamics introduced after their analysis.
\subsection{\texorpdfstring{$Usp(32)$} \ \ Model}
Having reviewed the type I string construction, we now consider Sugimoto’s non-supersymmetric model~\cite{Sugimoto1999}, built via a modified orientifold projection of type IIB with a crucial sign reversal on the orientifold plane charge. Reflection coefficients are assigned differently to NS–NS and R–R states in $\tilde{\mathcal{A}}$ and $\tilde{\mathcal{M}}$, yielding
\bea
    \tilde{\mathcal{A}} &=& \frac{2^{-5}}{2} \int \frac{dl}{\eta^8} \left[ (n_+ + n_-)^2 V_8 - (n_+ - n_-)^2 S_8 \right] \quad , \nonumber \\
    \tilde{\mathcal{M}} &=& 2 \int \frac{dl}{\hat{\eta}^8} \left[ \epsilon_{\text{NS}} (n_+ + n_-) \hat{V}_8 - \epsilon_{\text{R}} (n_+ - n_-) \hat{S}_8 \right] \quad .
\eea
The direct-channel amplitudes become
\bea
    \mathcal{A} &=& \frac{1}{2} \int_0^\infty \frac{d \tau_2}{\tau_2^6 \, \eta^8} \left[ (n_+^2 + n_-^2)(V_8 - S_8) + 2n_+ n_- (O_8 - C_8) \right] \quad , \nonumber \\
    \mathcal{M} &=& \frac{1}{2} \int_0^\infty \frac{d\tau_2}{\tau_2^6 \, \hat{\eta}^8} \left[ \epsilon_{\text{NS}} (n_+ + n_-) \hat{V}_8 - \epsilon_{\text{R}} (n_+ - n_-) \hat{S}_8 \right] \quad .
\eea
Two types of Chan–Paton charges $n_+$ and $n_-$ define sectors with like charges exhibiting the usual GSO projection involving $V_8$ and $S_8$, while sectors with unlike charges involve the opposite projection, featuring the tachyonic $O_8$ and $C_8$ modes~\cite{Angelantonj:2002ct}.

The two tadpole conditions are:
\bea
    \text{NS--NS:} \quad & 32 + \epsilon_{\text{NS}} (n_+ + n_-) = 0\quad , \nonumber\\
    \text{R--R:} \quad & 32 + \epsilon_{\text{R}} (n_+ - n_-) = 0\quad .
\eea
These admit the supersymmetric $SO(32)$ solution with $\epsilon_{\text{NS}} = \epsilon_{\text{R}} = -1$, $n_+ = 32$, $n_- = 0$. Relaxing the NS–NS tadpole condition allows an infinite family of solutions, e.g.\ $\epsilon_{\text{R}} = \epsilon_{\text{NS}} = -1$ with $n_+ - n_- = 32$. In spacetime, $n_+$ and $n_-$ count D9-branes and anti-D9-branes, while $\epsilon_{\text{NS}} = \pm 1$ corresponds to different O-planes. The Möbius amplitude $\tilde{\mathcal{M}}$ encodes these signs, as its NS sector is sensitive to relative brane and O-plane tensions.

Configurations with both $n_+, n_- \neq 0$ have tachyonic instabilities from brane–antibrane attraction~\cite{Sen1998a, Sen1998b, Sen1998c, Bergman1998, Sen1998d, Sen1998e, Sen1998f, Lerda2000}. For $n_+ = 0$, Sugimoto’s non-supersymmetric model~\cite{Sugimoto1999} arises with $\epsilon_{\text{NS}} = \epsilon_{\text{R}} = +1$ and $n_- = 32$, involving anti-D9-branes and $O9^-$ planes. The Möbius amplitude is
\beq
    \mathcal{M} = -\frac{1}{2} N \int_0^{\infty} \frac{d\tau_2}{\tau_2^2} \, \frac{-V_8 - S_8}{\tau_2^4 \, \hat{\eta}^8}\left[ i\tau_2/2 + 1/2 \right].
\eeq
This changes the gauge group to $USp(32)$, with $\frac{N(N+1)}{2}$ gauge bosons and $\frac{N(N-1)}{2}$ fermions in the antisymmetric representation ($N=32$). The antisymmetric contains a singlet, the \textit{goldstino}, signaling non-linear supersymmetry breaking.

The shift from $O_-$ to $O_+$, with positive tension and charge, forces the presence of anti-D-branes, leaving a net vacuum tension. This induces a tree-level dilaton potential, enabling a non-supersymmetric open string spectrum coupled to a supersymmetric bulk. Supersymmetry is non-linearly realized on the branes~\cite{Volkov1973,Samuel1983a,Samuel1983b,Samuel1984}, and the dilaton tadpole appears as the leading term in the Volkov–Akulov goldstino action~\cite{Dudas:2000ff,Blumenhagen:2000wh}.

The amplitude $\mathcal{M}$ is UV divergent, but its transverse channel reveals an IR origin due to a massless zero-momentum propagator and a disk NS–NS tadpole. Resolving this requires shifting to a new vacuum solving the equations of motion, accessible only via low-energy effective field theory. Understanding how charged or uncharged D-branes adjust to this background remains open~\cite{Dudas:2001pf,Dudas:2001wd}.

The dilaton tadpole forbids a maximally symmetric Minkowski vacuum, but an $SO(1,8)$ symmetric warped background exists~\cite{Dudas:2000ff,Blumenhagen:2000wh}. Although tachyon-free, residual attraction between $O^-$ planes and anti-D9-branes persists.

Supersymmetry breaking in $USp(32)$ stems from coexisting individually BPS but mutually non-BPS objects. BPS states saturate mass bounds preserving some supersymmetry and stability; here, despite explicit breaking, BPS structures still govern the spectrum and dynamics.

The breaking scale is at the string scale, challenging the low-energy supergravity approach. Compared to $SO(32)$, the bosonic excitation spectrum shifts by one level. The vacuum energy is infinite due to IR effects, corresponding to the dilaton runaway potential ("dilaton tadpole"), whose detailed form will be discussed in the EFT section.
\subsection{0'B Model}
There is another tachyon-free model in ten dimensions~\cite{Sagnotti1995,Sagnotti1997}, an orientifold of the tachyonic 0B model studied in~\cite{Armoni:1999xy,Angelantonj:2000ct,Armoni:2003er,Armoni:2003ug,Armoni:2005ps,Armoni:2018qcd3}. This model leads to several open descendants with chiral spectra and shows that the Klein-bottle projection is not unique.

Starting from the $0B$ unoriented projection defined by the Klein bottle amplitude
\beq
    \mathcal{K}_1 = \frac{1}{2}(O_8 + V_8 - S_8 - C_8)\quad,
\eeq
which symmetrizes the NS--NS sector and antisymmetrizes the R--R sector, the closed spectrum includes a tachyon, graviton, dilaton, and two R--R two-forms. The transverse Klein bottle amplitude is
\beq
    \tilde{\mathcal{K}}_1 = \frac{2^{6}}{2} V_8\quad,
\eeq
indicating only the NS--NS sector propagates.

The annulus and Möbius amplitudes are
\bea
    \mathcal{A}_1 &=& \frac{1}{2}(n_o^2 + n_v^2 + n_s^2 + n_c^2)V_8 + (n_o n_v + n_s n_c)O_8 \nonumber \\
    && - (n_v n_s + n_o n_c)S_8 - (n_v n_c + n_o n_s)C_8,
\eea
\beq
    \mathcal{M}_1 = -\frac{1}{2}(n_o + n_v + n_s + n_c)\hat{V}_8\quad.
\eeq
Tadpole cancellation conditions impose
\bea
    n_o + n_v + n_s + n_c &=& 64, \nonumber \\
    n_o - n_v - n_s + n_c &=& 0, \nonumber \\
    n_o - n_v + n_s - n_c &=& 0,
\eea
yielding $n_o = n_v$ and $n_s = n_c$ and gauge group $SO(n_o) \times SO(n_v) \times SO(n_s) \times SO(n_c)$.
The open-string spectrum is chiral but free of irreducible gauge anomalies.

Two additional inequivalent Klein-bottle projections exist~\cite{Sagnotti1995}:
\beq
    \mathcal{K}_2 = \frac{1}{2}(O_8 + V_8 + S_8 + C_8), \quad \mathcal{K}_3 = \frac{1}{2}(-O_8 + V_8 + S_8 - C_8).
\eeq
The second leads to a tachyonic closed sector but a chiral open sector with unitary gauge groups, while the third, known as the $0'B$ model, features a tachyon-free, chiral closed-string spectrum due to a self-dual R--R four-form from the $|S_8|^2$ term.

The annulus and Möbius amplitudes for the $0'B$ model read:
\beq
    \mathcal{A}_3 = -\frac{1}{2}(n_o^2 + n_v^2 + n_s^2 + n_c^2)C_8 - (n_o n_v + n_s n_c)S_8 + (n_v n_s + n_o n_c)V_8 + (n_v n_c + n_o n_s)O_8,
\eeq
\beq
    \mathcal{M}_3 = \frac{1}{2}(n_o - n_v - n_s + n_c) \hat{C}_8,
\eeq
with transverse amplitudes
\bea
    \tilde{\mathcal{A}}_3 &=& \frac{2^{-6}}{2} \left[(n_o + n_v + n_s + n_c)^2 V_8 - (n_o + n_v - n_s - n_c)^2 O_8 \right. \nonumber \\
    && \left. - (n_o - n_v - n_s + n_c)^2 C_8 + (n_o - n_v + n_s - n_c)^2 S_8 \right],
\eea
\beq
    \tilde{\mathcal{M}}_3 = \frac{2}{2} (n_o - n_v - n_s + n_c) \hat{C}_8.
\eeq

Reinterpreting Chan--Paton factors as $n_v = n$, $n_s = \bar{n}$, $n_o = m$, $n_c = \bar{m}$ yields:
\bea
    \mathcal{A}_3 &=& -\frac{1}{2}(n^2 + \bar{n}^2 + m^2 + \bar{m}^2)C_8 + (n \bar{n} + m \bar{m})V_8 \nonumber \\
    && - (m n + \bar{m} \bar{n})S_8 + (m \bar{n} + \bar{m} n)O_8,
\eea
\beq
    \mathcal{M}_3 = \frac{1}{2}(m + \bar{m} - n - \bar{n}) \hat{C}_8,
\eeq
with tadpole cancellation imposing $m = 32 + n$.

Setting $n=0$ leads to $m=32$ and a fully tachyon-free, chiral model with gauge group effectively $SU(32)$. The anomalous $U(1)$ factor acquires a mass via a Green–Schwarz mechanism and decouples~\cite{Sagnotti1995}.

The final amplitudes of the $0'B$ model are:
\beq
    \mathcal{K}_{0'B} = \frac{1}{2} \int_0^{\infty} \frac{d \tau_2}{\tau_2^2} \frac{-O_8 + V_8 + S_8 - C_8}{\tau_2^4 \eta^8} [2i \tau_2],
\eeq
\beq
    \mathcal{A}_{0'B} = \int_0^{\infty} \frac{d \tau_2}{\tau_2^2} \frac{N \bar{N} V_8 - \tfrac{1}{2} (N^2 + \bar{N}^2) C_8}{\tau_2^4 \eta^8} [i \tau_2 / 2],
\eeq
\beq
    \mathcal{M}_{0'B} = - \frac{N + \bar{N}}{2} \int_0^{\infty} \frac{d \tau_2}{\tau_2^2} \frac{\hat{C}_8}{\tau_2^4 \hat{\eta}^8} [i \tau_2 / 2 + 1/2],
\eeq
and their transverse-channel counterparts
\beq
    \widetilde{\mathcal{K}}_{0'B} = -\frac{2^{6}}{2} \int_0^\infty dl \frac{C_8}{\hat{\eta}^8} [i l],
\eeq
\beq
    \widetilde{\mathcal{A}}_{0'B} = \frac{2^{-6}}{2} \int_0^\infty dl \left[ (N + \bar{N})^2 (V_8 - C_8) - (N - \bar{N})^2 (O_8 - S_8) \right] \frac{1}{\eta^8} [i l],
\eeq
\beq
    \widetilde{\mathcal{M}}_{0'B} = -2 \frac{N + \bar{N}}{2} \int_0^\infty dl \frac{\hat{C}_8}{\hat{\eta}^8} [i l + 1/2],
\eeq
with consistency requiring $N = \bar{N}$ to eliminate the $S_8$ contribution and ensure anomaly cancellation.

The spectrum of the model is:
\begin{center}
    \textbf{0'B}
\end{center}
\beq
    \left(g_{\mu \nu}, \, B_{\mu \nu}, \, \phi , \, a, \, D^+_{\mu \nu \rho \sigma}\right) \, \oplus \, A_\mu^{496 \, \oplus \, \overline{496}}\, \oplus \,\psi_L^{(128,1) \, \oplus \, (1, 128)}
    \quad .
\eeq

All three models discussed admit a geometric interpretation involving D9-branes and O9-planes in type 0B string theory, where the R--R sector is doubled. D9-branes and O9-planes can carry different combinations of R--R charges, leading to various configurations, including branes and anti-branes. The structure of O-planes in each model is inferred from the transverse Klein-bottle amplitude.

The first model contains equal numbers of two types of O9-planes and uses standard projections. The second model modifies the Klein-bottle projection with the operato $\Omega_2 = \Omega (-1)^{F_L}$
which alters the Möbius strip amplitude and leads to complex combinations of Chan--Paton charges. The resulting branes are interpreted as superpositions of the original ones, and the open string sector becomes chiral, with a gauge group \( U(n_b) \times U(n_f) \), free of anomalies.

The third model, known as the \( 0'B \) theory, employs a different projection $\Omega_3 = \Omega (-1)^{F_L}$
also yielding complex brane combinations. In the configuration with \( n = 0 \) and \( m = 32 \), the model is fully tachyon-free and chiral. The gauge group starts as \( U(32) \), but the anomalous \( U(1) \) decouples via a Green--Schwarz mechanism, leaving an effective \( SU(32) \) gauge symmetry.

As in other non-supersymmetric constructions, the system induces a backreaction on the vacuum due to both quantum (torus) and classical (brane tension) effects. Supersymmetry breaking in the torus is a quantum effect and may be softened by cancellations among mass levels~\cite{Kachru1999,Harvey1999,Angelantonj1999}. On the other hand, the open and unoriented sectors involve static extended objects whose mutual attraction can lead to instabilities, particularly between branes and antibranes. Though equilibrium configurations are hard to realize, orientifolds remain non-dynamical and help prevent tachyon formation. Despite their attractive interactions with branes, these setups appear to be consistent in time-dependent backgrounds such as those in cosmology~\cite{Kitazawa:2017dbrane}.
\subsection{Low--Energy Actions}
In this final section, we present the low-energy effective field theory (EFT) setup relevant to the analysis of the climbing scalar phenomenon. This involves the supergravity actions corresponding to the non-supersymmetric string models previously discussed.

We begin by reviewing the construction of the low-energy action for the universal bosonic sector, consisting of the dilaton $\Phi$, the graviton $G_{\mu \nu}$, and the Kalb--Ramond two-form $B_{\mu \nu}$. In backgrounds dominated by gravity, spacetime is curved with a characteristic curvature radius $R_c$, and derivatives of the metric scale as $R_c^{-1}$. The effective coupling becomes dimensionless when scaled with the string length $\alpha'^{1/2}$, giving $\alpha'^{1/2}R_c^{-1}$.

If $R_c \gg \alpha'^{1/2}$, the curvature is small on the string scale, allowing the use of low-energy effective field theory. In this regime, massive string modes are not excited, and the dynamics is captured by the massless fields only. String theory determines the structure of the EFT via consistency conditions like Weyl invariance on the worldsheet.

To ensure this consistency, recall that the energy-momentum tensor \( T_{ab} \) is defined as the variation of the path integral with respect to the worldsheet metric \( g_{ab} \):
\beq
    \delta \langle \ldots \rangle_g = -\frac{1}{4\pi} \int d^2\sigma \, \sqrt{g(\sigma)} \, \delta g^{ab}(\sigma) \, \langle T_{ab}(\sigma) \ldots \rangle_g \quad . 
    \label{eq:variation_path_integral}
\eeq

Neglecting boundary terms, a Weyl variation yields:
\beq
    \delta_W \langle \ldots \rangle_g = -\frac{1}{2\pi} \int d^2\sigma \, \sqrt{g(\sigma)} \, \delta \omega(\sigma) \, \langle T^a_{\ a}(\sigma) \ldots \rangle_g
    \quad ,
\label{eq:weyl_variation}
\eeq
and Weyl invariance requires the stress-energy tensor to be traceless.
While the classical theory satisfies this condition, quantum corrections may induce a trace anomaly due to regularization. However, such anomalies must preserve the remaining symmetries, such as diffeomorphism and target-space Poincaré invariance, and must vanish in flat backgrounds.

To proceed, we briefly introduce the concept of vertex operators, which are crucial for the construction of the effective action. The discussion here is schematic and follows~\cite{Polchinski:1998rr}, intended only to define the objects appearing in the next equations.

\paragraph{Basic Properties of Vertex Operators}
\leavevmode \par \noindent

In string theory, vertex operators are constructed via the state--operator correspondence, typically on a flat worldsheet. Extending this to curved worldsheets in the Polyakov path integral formalism requires operators compatible with local diffeomorphism and Weyl invariance.

Renormalized operators \( [F]_r \) cancel short-distance singularities by generalizing normal ordering:
\beq
    [F]_r = \exp\left( \frac{1}{2} \int d^2\sigma \, d^2\sigma' \, \Delta(\sigma, \sigma') \frac{\delta}{\delta X^\mu(\sigma)} \frac{\delta}{\delta X_\mu(\sigma')} \right) F
    \quad ,
\eeq
where
\beq
    \Delta(\sigma, \sigma') = \frac{\alpha'}{2} \ln d^2(\sigma, \sigma')
    \quad ,
\eeq
and \( d(\sigma, \sigma') \) is the geodesic distance on the worldsheet. This ensures manifest diffeomorphism invariance; the Weyl variation reads:
\beq
    \delta_W [F]_r = [\delta_W F]_r + \frac{1}{2} \int d^2\sigma \, d^2\sigma' \, \delta_W \Delta(\sigma, \sigma') \frac{\delta}{\delta X^\mu(\sigma)} \frac{\delta}{\delta X_\mu(\sigma')} [F]_r
    \quad .
\eeq

The simplest vertex operator is the tachyon:
\beq
    V_0 = 2 g_c \int d^2\sigma \, \sqrt{g} \, e^{i k \cdot X(\sigma)}
    \quad .
\eeq
Under a Weyl rescaling \( g_{ab} \to e^{2\omega(\sigma)} g_{ab} \), near coincident points:
\beq
    d^2(\sigma, \sigma') \approx (\sigma - \sigma')^2 e^{2\omega(\sigma)} \quad \Rightarrow \quad \delta_W \Delta(\sigma, \sigma) = \alpha' \delta\omega(\sigma)
    \quad .
\eeq
The Weyl variation of \( V_0 \) becomes:
\beq
    \delta_W V_0 = 2 g_c \int d^2\sigma \, \sqrt{g} \left( 2 \delta\omega(\sigma) - \frac{\alpha'}{2} k^2 \delta\omega(\sigma) \right) [e^{i k \cdot X}]_r
    \quad ,
\eeq
where \( g_c \) is the closed string coupling constant. Vanishing of this variation for arbitrary \( \delta \omega \) requires: $k^2 = \frac{4}{\alpha'}$,
the on-shell tachyon condition.

Next, the two-derivative vertex operator reads
\beq
    V_1 = \frac{g_c}{\alpha'} \int d^2\sigma \, \sqrt{g} \left[ \left( g^{ab} s_{\mu\nu} + i \varepsilon^{ab} a_{\mu\nu} \right) [\partial_a X^\mu \partial_b X^\nu e^{i k \cdot X}]_r + \alpha' \phi R [e^{i k \cdot X}]_r \right]
    \quad ,
    \label{eq:vertex operator}
\eeq
with \( s_{\mu\nu} \) symmetric, \( a_{\mu\nu} \) antisymmetric, and \( \varepsilon^{ab} \) normalized by \( \sqrt{g} \varepsilon^{12} = 1 \).

The Weyl variation involves derivatives of \( \delta \omega \). Using expansions at coincident points:
\beq
    \partial_a \delta_W \Delta(\sigma, \sigma') \Big|_{\sigma' = \sigma} = \frac{1}{2} \alpha' \partial_a \delta\omega(\sigma)
    \quad ,
\eeq
\beq
    \partial_a \partial'_b \delta_W \Delta(\sigma, \sigma') \Big|_{\sigma' = \sigma} = \frac{1 + \gamma}{2} \alpha' \nabla_a \partial_b \delta\omega(\sigma)
    \quad ,
\eeq
\beq
    \nabla_a \partial_b \delta_W \Delta(\sigma, \sigma') \Big|_{\sigma' = \sigma} = -\frac{\gamma}{2} \alpha' \nabla_a \partial_b \delta\omega(\sigma)
    \quad ,
\eeq
with \( \gamma = -\frac{2}{3} \).

Using the identity
\beq
    [\nabla^2 X^\mu e^{i k \cdot X}]_r = \frac{i \alpha' \gamma}{4} k^\mu R [e^{i k \cdot X}]_r
    \quad ,
\eeq
and integrating by parts, the Weyl variation of \( V_1 \) is
\beq
    \delta_W V_1 = \frac{g_c}{2} \int d^2\sigma \, \sqrt{g} \, \delta\omega(\sigma) \left[ (g^{ab} S_{\mu\nu} + i \varepsilon^{ab} A_{\mu\nu}) [\partial_a X^\mu \partial_b X^\nu e^{i k \cdot X}]_r + \alpha' F R [e^{i k \cdot X}]_r \right]
    \quad ,
    \label{eq:variationofvertex}
\eeq
where
\beq
    S_{\mu\nu} = -k^2 s_{\mu\nu} + k_\nu k^\omega s_{\mu\omega} + k_\mu k^\omega s_{\nu\omega} - (1 + \gamma) k_\mu k_\nu s^\omega_{\ \omega} + 4 k_\mu k_\nu \phi
    \quad ,
\eeq
\beq
    A_{\mu\nu} = -k^2 a_{\mu\nu} + k_\nu k^\omega a_{\mu\omega} - k_\mu k^\omega a_{\nu\omega}
    \quad ,
\eeq
\beq
    F = (\gamma - 1)k^2 \phi + \frac{1}{2} \gamma k^\mu k^\nu s_{\mu\nu} - \frac{1}{4} \gamma(1 + \gamma)k^2 s^\nu_{\ \nu}
    \quad .
\eeq
\begin{center}
    ***
\end{center}
In the limit where \( B_{\mu\nu} \) and \( \Phi \) are small and \( G_{\mu\nu} \approx \eta_{\mu\nu} \), the string action in a background of massless bosonic fields can be written as
\beq
    S_\sigma = S_p - V_1 + \cdots,
\eeq
with \( S_p \) the flat Polyakov action and \( V_1 \) the vertex operator in \eqref{eq:vertex operator}. The background fields are expanded as
\bea
    G_{\mu\nu}(X) = \eta_{\mu\nu} - 4\pi g_c s_{\mu\nu} e^{i k \cdot X}, \quad
    B_{\mu\nu}(X) = -4\pi g_c a_{\mu\nu} e^{i k \cdot X}, \quad
    \Phi(X) = -4\pi g_c \phi e^{i k \cdot X}.
\eea
Using the Weyl variation \eqref{eq:variationofvertex} with a renormalization scheme where \( \gamma = 0 \), the trace of the energy-momentum tensor becomes
\beq
    T^a_{\ a} = -\frac{1}{2\alpha'} \beta^G_{\mu\nu} g^{ab} \partial_a X^\mu \partial_b X^\nu - \frac{i}{2\alpha'} \beta^B_{\mu\nu} \varepsilon^{ab} \partial_a X^\mu \partial_b X^\nu - \frac{1}{2} \beta^\Phi R,
\eeq
where the beta functions to linear order are
\bea
    \beta^G_{\mu\nu} &\approx& -\frac{\alpha'}{2} \left( \partial^2 \chi_{\mu\nu} - \partial_\nu \partial^\omega \chi_{\mu\omega} - \partial_\mu \partial^\omega \chi_{\nu\omega} + \partial_\mu \partial_\nu \chi^\omega_{\ \omega} \right) + 2\alpha' \partial_\mu \partial_\nu \Phi, \label{eq:anomalies1}\\
    \beta^B_{\mu\nu} &\approx& -\frac{\alpha'}{2} \partial^\omega H_{\omega\mu\nu}, \label{eq:anomalies2}\\
    \beta^\Phi &\approx& \frac{D - 26}{6} - \frac{\alpha'}{2} \partial^2 \Phi. \label{eq:anomalies3}
\eea
Here,
\beq
    H_{\omega\mu\nu} = \partial_\omega B_{\mu\nu} + \partial_\mu B_{\nu\omega} + \partial_\nu B_{\omega\mu}
\eeq
is the field strength of the antisymmetric tensor. The constant term in \(\beta^\Phi\) is the critical dimension anomaly that vanishes for \(D=26\) (and similarly in superstring theories at their critical dimension).

Including higher order terms up to two spacetime derivatives, the full beta functions become
\bea
    \beta^G_{\mu\nu} &=& \alpha' R_{\mu\nu} + 2\alpha' \nabla_\mu \nabla_\nu \Phi - \frac{\alpha'}{4} H_{\mu\lambda\omega} H_\nu^{\ \lambda\omega} + \mathcal{O}(\alpha'^2), \\
    \beta^B_{\mu\nu} &=& -\frac{\alpha'}{2} \nabla^\omega H_{\omega\mu\nu} + \alpha' \nabla^\omega \Phi H_{\omega\mu\nu} + \mathcal{O}(\alpha'^2), \\
    \beta^\Phi &=& \frac{D - 26}{6} - \frac{\alpha'}{2} \nabla^2 \Phi + \alpha' \nabla^\omega \Phi \nabla_\omega \Phi - \frac{\alpha'}{24} H_{\mu\nu\lambda} H^{\mu\nu\lambda} + \mathcal{O}(\alpha'^2),
\eea
where all terms are now covariant under spacetime diffeomorphisms. Note the distinction between spacetime Ricci tensor \( R_{\mu\nu} \) and the worldsheet one \( R_{ab} \).

Weyl invariance requires vanishing beta functions,
\beq
    \beta^G_{\mu\nu} = \beta^B_{\mu\nu} = \beta^\Phi = 0,
    \label{eq:fieldeq}
\eeq
which correspond to physically meaningful spacetime field equations: \(\beta^G_{\mu\nu}=0\) generalizes Einstein’s equations including the dilaton and antisymmetric tensor sources; \(\beta^B_{\mu\nu}=0\) governs the antisymmetric tensor dynamics; and \(\beta^\Phi=0\) constrains the dilaton and dimensionality.

These field equations can be derived from the low-energy effective spacetime action
\beq
    S = \frac{1}{2\kappa_0^2} \int d^D x\, (-G)^{1/2} e^{-2\Phi} \left[ R - \frac{1}{12} H_{\mu\nu\lambda} H^{\mu\nu\lambda} + 4\, \partial_\mu \Phi \partial^\mu \Phi + \mathcal{O}(\alpha') \right],
\eeq
specialized to critical dimension \(D\). The normalization \(\kappa_0\) is unphysical and can be absorbed by rescaling \(\Phi\).

Performing a spacetime Weyl rescaling $G_{\mu\nu} \to e^{2\omega(x)} G_{\mu\nu}$,
with $\omega(x) = \frac{2(\Phi_0 - \Phi)}{D-2}$,
and shifting \(\Phi \to \Phi - \Phi_0\), the action transforms into the Einstein frame:
\beq
    S = \frac{1}{2\kappa^2} \int d^D x\, (-G)^{1/2} \left[ R - \frac{1}{12} e^{-\frac{8\Phi}{D-2}} H_{\mu\nu\lambda} H^{\mu\nu\lambda} - \frac{4}{D-2} \partial_\mu \Phi \partial^\mu \Phi + \mathcal{O}(\alpha') \right],
\eeq
where $\kappa = \kappa_0 e^{\Phi_0}$
is the physical gravitational coupling. For \(D=4\), this relates to Newton’s constant as
\beq
    \kappa = \sqrt{8\pi G_N} = \frac{\sqrt{8\pi}}{M_P} \approx (2.43 \times 10^{18}~\text{GeV})^{-1}.
\eeq

The dilaton appears only through derivatives or as an overall factor \(e^{-2\Phi}\), consistent with its shift symmetry and insensitivity to adding a constant to \(\Phi\) on the Weyl anomaly.

Up to now, we focused on the bosonic string , but for cosmological models arising from SUSY breaking, it is consistent and justified to truncate to the NS-NS bosonic sector and impose $H_{\mu\nu\rho} = 0$
to preserve spatial homogeneity and isotropy, assuming an FRW metric

So the climbing scalar dynamics is driven by the dilaton \(\Phi\), retaining only the graviton since we can ignore the two-form piece

The simplified action reads
\beq
    S = \frac{1}{2\kappa^2} \int d^{10} x\, (-G)^{1/2} \left[R  - \frac{1}{2} \partial_\mu\Phi \partial^\mu\Phi + \mathcal{O}(\alpha') \right] \quad .
\eeq

The key ingredient is the vacuum energy from SUSY breaking, absent in fully supersymmetric models but present here.

In the \(SO(16) \times SO(16)\) model, SUSY breaking occurs at the torus level, adding a term coupling as \(e^{-\chi \Phi}\) with \(\chi=0\), yielding the string-frame action
\beq
    S_{SO(16) \times SO(16)} = \frac{1}{2\kappa^2} \int d^{10} x\, (-G)^{1/2} \left[e^{-2\Phi} \left[ R + 4\, \partial_\mu \Phi\, \partial^\mu \Phi \right]-T\right]
    \quad ,
    \label{eq:SO16stringframe}
\eeq
and Einstein-frame action
\beq
    S_{SO(16) \times SO(16)} = \frac{1}{2\kappa^2} \int d^{10} x\, (-G)^{1/2} \left[R  - \frac{1}{2} \partial_\mu\Phi \partial^\mu\Phi - T \, e^{\frac{5}{2}\Phi}\right]
    \quad .
    \label{eq:SO16einsteinframe}
\eeq

In the \(Usp(32)\) and \(0'B\) models~\cite{Sugimoto1999,Sagnotti1995,Sagnotti1997}, the vacuum energy arises from the \(\overline{\mathrm{D9}} - \mathrm{O9}_+\) system at disk level (\(\chi=1\)), resulting in
\beq
    S_{Usp(32) / 0'B} = \frac{1}{2\kappa^2} \int d^{10} x\, (-G)^{1/2} \left[e^{-2\Phi} \left[ R + 4\, \partial_\mu \Phi\, \partial^\mu \Phi \right]-Te^{- \Phi}\right]
    \quad ,
    \label{eq:noncompactlag}
\eeq
and Einstein-frame form
\beq
    S_{Usp(32) / 0'B} = \frac{1}{2\kappa^2} \int d^{10} x\, (-G)^{1/2} \left[R  - \frac{1}{2} \partial_\mu\Phi \partial^\mu\Phi - T \, e^{\frac{3}{2}\Phi}\right]
    \quad .
\eeq

The difference is not crucial since the climbing phenomenon universally occurs for exponential weights above \(3/2\), satisfied by all three SUSY breaking models. The crucial potential term, called the \textit{hard exponential}, that we will consider in the following is
\beq
    V_1(\Phi)=T \, e^{\frac{3}{2} \Phi} \quad .
\eeq
To connect string theory with four-dimensional cosmology, six extra dimensions must be compactified. A Kaluza–Klein reduction of the ten-dimensional action on an FRW background leads to an effective $d$-dimensional Einstein-frame action involving the dilaton $\Phi$ and the internal volume scalar $\sigma$.

After a field redefinition to canonically normalized scalars $\Phi_t$ and $\Phi_s$, only $\Phi_t$ couples to the potential:
\beq
S_d = \frac{1}{2 \kappa_d^2} \int d^d x \sqrt{-G} \left[ R - \frac{1}{2} (\partial \Phi_t)^2 - T \, e^{\sqrt{\frac{2(d-1)}{d - 2}} \, \Phi_t} + \dots \right],
\eeq
where $\Phi_s$ is assumed stabilized. The resulting exponential potential drives the \textit{climbing scalar} behavior, robust even in the presence of higher-curvature corrections and axionic partners~\cite{DudasKitazawaPatilSagnotti:2012,KitazawaSagnotti:2014,FreSagnottiSorin:2013_integrable}.

Specializing to $d=4$ and rescaling the dilaton to canonical cosmological normalization, $\Phi \rightarrow \sqrt{2} \kappa_4 \phi$, one obtains:
\beq
S = \int d^4 x \sqrt{-G} \left[ \frac{M_{pl}^2}{2} R - \frac{1}{2} (\partial \phi)^2 - M_{pl}^2 \frac{T}{2} \, e^{\sqrt{6} \, M_{pl}^{-1} \, \phi} \right],
\eeq
which is the standard form for the climbing scalar potential in 4D cosmology.

Additional branes in lower-dimensional vacua (e.g., $D5$, $\overline{D5}$, non–BPS $D3$) generate extra exponential terms in the potential~\cite{Sugimoto1999,Antoniadis1999,Aldazabal1999,Angelantonj2000b,Dudas:2001wd,Sen1998b,Dudas:2010climbing}, assuming $\Phi_s$ is stabilized.

Generically, a brane of type $(p,\alpha)$ contributes a term with exponent~\cite{Bergshoeff:2011ee,Bergshoeff:2011zk,Bergshoeff:2012pm}:
\beq
    \gamma = \frac{1}{12} (p + 9 - 6\alpha) \quad .
\eeq
Typical values of $\gamma$ include $\gamma = \frac{1}{2}$ from non-BPS $D3$ branes in $d=4$ and $\gamma = \frac{1}{12}$ from NS5 branes. These terms can naturally produce two-exponential potentials. The second will be important to support the slow-roll phase of inflation.

Finally our effective action for the climbing scalar becomes:
\beq
    S=\frac{1}{2 \kappa^2}\int d^4 x \sqrt{-g}\left[R-\frac{1}{2}(\partial \phi)^2 - V(\phi)\right]
    \quad ,
    \label{eq:firstEFTfinal}
\eeq

In cosmological normalization, with $\kappa^{-1} \sim M_{pl}=1$:

\beq
    S=\int d^4 x \sqrt{-g}\left[\frac{R}{2}-\frac{1}{2}g^{\mu \nu}\partial_\mu \phi \partial_\nu \phi-V(\phi)\right]
    \quad ,
    \label{eq:secondEFTfinal}
\eeq
with
\beq
    V(\phi)=\frac{T}{2}\left(e^{\sqrt{6} \, \phi}+e^{\sqrt{6} \, \gamma \, \phi}\right)
    \quad ,
    \label{eq:secondpotential}
\eeq
where $\gamma < 1$ and typically of order $\sim O(0.1)$. The second exponential will be referred to as the \textit{mild exponential}. Different factors at the exponent derives from the compactification.

We conclude this section by emphasizing some conceptual and technical caveats when working with effective field theories. While E.F.T.s are valuable tools derived under assumptions of weak coupling and low curvature, their reliability becomes questionable in regimes where these assumptions fail—particularly in early-universe scenarios involving strong coupling or high curvature.

This issue is especially relevant for non-supersymmetric but tachyon-free string models, where the absence of supersymmetry leads to weaker constraints and potentially significant corrections beyond the low-energy approximation. As a result, while effective actions can reproduce interesting phenomena such as the \emph{climbing scalar}, it is important to remain cautious: these features may arise in regimes where the effective description no longer holds, casting doubt on their robustness.

The climbing scalar mechanism, which occurs when the scalar potential becomes sufficiently steep, has attracted interest for its role in pre-inflationary cosmology and initial condition generation. However, since this behavior typically emerges near regions of high curvature or large string coupling, one must question whether it truly reflects full string dynamics or is simply an artifact of extrapolating E.F.T.s beyond their validity.

Despite these limitations, effective actions may still capture hints of deeper string-theoretic structures. Therefore, while they remain our best available tools, they should be used with care. In the next section, we will study the \emph{climbing scalar} in detail, analyzing its dynamics and possible cosmological implications.
\chapter{The Climbing Scalar}

As discussed earlier, string-inspired cosmological models with brane-induced supersymmetry breaking give rise to exponential scalar potentials, often leading to the \emph{climbing scalar} phenomenon, where the scalar field initially climbs a steep potential before descending~\cite{Dudas:2010climbing}. This arises when the slope exceeds a critical value, preventing immediate slow-roll and inducing a pre-inflationary climbing phase~\cite{Dudas:2010climbing,FreSagnottiSorin:2013_integrable}.

If this phase occurs shortly before observable modes exit the horizon, it may imprint CMB features such as power suppression at low multipoles and mild oscillations~\cite{DudasKitazawaPatilSagnotti:2012,GruppusoKitazawaMandolesiNatoliSagnotti2016}. Analytic solutions from integrable models clarify climbing and inflationary dynamics~\cite{FreSagnottiSorin:2013_integrable}.

This chapter analyzes single and double exponential potentials, the evolution of slow-roll parameters, and variations of the scenario. It concludes by setting the stage for the analysis of cosmological perturbations.

The discussion is based on seminal works on the climbing scalar~\cite{Dudas:2010climbing,DudasKitazawaPatilSagnotti:2012,KitazawaSagnotti:2014,Sagnotti2013,GruppusoKitazawaMandolesiNatoliSagnotti2016,Sagnotti:2013bsb_inflation,KitazawaSagnotti2015,FerraraSagnotti2015,KitazawaSagnotti2015MPLA,CondeescuDudas2013,MouradSagnotti20212} and related studies on exponential dynamics~\cite{FreSagnottiSorin:2013_integrable,PelliconiSagnotti2021,Russo2004,TownsendWohlfarth2004,EmparanGarriga2003,TownsendWohlfarth2003,Cicoli2014}.
\section{One--Exponential Dynamics: the Climbing Phenomenon} 
We begin by considering the low-energy action from eq.~\eqref{eq:secondEFTfinal}, setting $\kappa^{-1}=M_{pl}=1$:
\beq
    S=\int d^4 x \sqrt{-g}\left[\frac{R}{2}-\frac{1}{2}g^{\mu \nu}\partial_\mu \phi \partial_\nu \phi-V(\phi)\right]
    \quad ,
\eeq
and focus on the case of a single exponential potential:
\beq
    V(\phi)=\frac{T}{2}e^{\sqrt{6} \, \gamma \, \phi}
    \quad .
\eeq

This describes a single scalar field minimally coupled to gravity. We restrict to flat FRW metrics:
\beq
    ds^2 = -e^{2 B (\xi)}d\xi^2 + e^{2 A(\xi)} \delta_{ij}dx^i dx^j
    \quad , 
    \label{eq:classofmetrics}
\eeq
with homogeneous field configuration: $\phi(\xi \, , \vec{x})=\phi (\xi)$ , 
where $\xi$ is a dimensionful parametric time. Identifying the dilaton with the inflaton, this setup mirrors standard single-field inflation, giving a string-theoretic interpretation to the inflaton. Cosmic time can be recovered via:
\beq
    dt_c=e^{B}d\xi
    \label{eq:simplechange}
    \quad .
\eeq

Specializing the action to this  and redefining variables as
\beq
    \tau \equiv M \, \sqrt{\frac{3}{2}} \, \xi \qquad , \qquad \varphi \equiv \sqrt{\frac{3}{2}} \, \phi \qquad , \qquad a \equiv 3A
    \quad ,
    \label{eq:substitutions}
\eeq
the final Lagrangian becomes:
\beq
    \mathcal{L}=M^2 \, e^{a-B}\left[\frac{1}{2}\dot{\varphi}^2-\frac{1}{2}\dot{a}^2-\frac{e^{2B}}{M^2} \, V(\varphi)\right]
    \quad ,
    \label{eq:singlelag}
\eeq
with potential
\beq
    V(\varphi)=\frac{T}{2}e^{2 \, \gamma \, \varphi}
    \quad .
    \label{eq:singleexp}
\eeq
$\tau$ is a dimensionless parametric time and $M$ is a mass scale to be interpreted later.
Varying the action with respect to $(B, a, \varphi)$ gives the following system of independent equations (the second depends on the other two consistently with hamiltonian formulation):
\beq
    \begin{cases}
        \dot{a}^2-\dot{\varphi}^2=\frac{2}{M^2} \, e^{2B} \, V(\varphi) \\
        \ddot{\varphi}+(\dot{a}-\dot{B})\dot{\varphi}+\frac{e^{2B}}{M^2}\frac{d V}{d \varphi}=0
    \end{cases}
    \quad .
\eeq

The system has gauge freedom since time slicing is not fixed. We choose the gauge:
\beq
    e^{2B} \, V(\varphi)=\frac{M^2}{2}
    \quad ,
    \label{eq:gauge}
\eeq
which defines $M$, and simplifies the system, that for an expanding universe reads:  
\beq
    \begin{cases}
        \dot{a}=\sqrt{1+\dot{\varphi}^2} \\
        \ddot{\varphi}+\sqrt{1+\dot{\varphi}^2}\, \dot{\varphi}+\frac{1}{2V}\frac{dV}{d \varphi}\left(1+\dot{\varphi}^2\right)=0
    \end{cases}
    \quad ,
    \label{eq:finalsystem}
\eeq
For the exponential potential~\eqref{eq:singleexp}, this gives:
\beq
    \ddot{\varphi}+\sqrt{1+\dot{\varphi}^2} \, \dot{\varphi}+\gamma \, \left(1+\dot{\varphi}^2\right)=0
    \quad .
    \label{eq:equazione}
\eeq
Since $\gamma$ can be taken positive (up to a redefinition), we study the range $0<\gamma<1$. Due to the $\mathbb{Z}_2$ symmetry of the Lagrangian~\eqref{eq:singlelag}, there are two solutions: $\varphi \to -\varphi$, $\gamma \to -\gamma$.

We define:
\beq
    \dot{a}=\cosh f(\tau) \qquad \dot{\varphi}=\sinh f(\tau)
    \quad ,
\eeq
which solves the Hamiltonian constraint. The equation becomes:
\beq
    \dot{f}+\sinh f + \gamma \cosh f = 0
    \label{eq:difff}
    \quad ,
\eeq
Separating variables and redefining the sign of $f$, we write:
\beq
    \int \frac{df}{ \gamma \cosh f-\sinh f }=\int d \tau=\tau + \text{const}
    \quad .
\eeq
Using $e^f = u$ and changing variables:
\bea
    \int \frac{2 \, e^f}{\gamma \left(e^{2 f}+1\right)- \left(e^{2 f}-1\right)}df&=&2 \int \frac{du}{\gamma \left(u^2+1\right)-(u^2-1)}\nonumber \\&=&2\int \frac{du}{(\gamma - 1)u^2+(1+\gamma)}=\frac{2}{1+\gamma}\int \frac{du}{1-\frac{1-\gamma}{1+\gamma}u^2}
    \quad ,
\eea
Rescaling with $\sqrt{\frac{1-\gamma}{1+\gamma}}u=x$, we get:
\beq
    \frac{2}{1+\gamma}\sqrt{\frac{1+\gamma}{1-\gamma}}\int \frac{dx}{1-x^2}
    \quad ,
\eeq
\bea
    \int \frac{dx}{1-x^2}&=&\frac{1}{2}\int dx \left(\frac{1}{1-x}+\frac{1}{1+x}\right)=\frac{1}{2}\log \left|\frac{1+x}{1-x}\right| + \text{const} \nonumber \\ 
    &=&\frac{1}{2}\log \left|\frac{1+\sqrt{\frac{1-\gamma}{1+\gamma}} \, u}{1-\sqrt{\frac{1-\gamma}{1+\gamma}} \, u}\right| =\frac{1}{2}\log \left|\frac{1+\sqrt{\frac{1-\gamma}{1+\gamma}} \, e^{-f}}{1-\sqrt{\frac{1-\gamma}{1+\gamma}} \, e^{-f}}\right|
    \quad ,
\eea

where the constant is eliminated setting initial time to zero.
From the modulus, we again get two branches. Choosing the one with positive argument and using $\frac{1}{2}\log \left(\frac{1+x}{1-x}\right)=\text{artanh} \, x$ , the first solution, known as the \textit{climbing solution}, is:
\bea
    \dot{\varphi}(\tau)&=\frac{1}{2}\left[\sqrt{\frac{1-\gamma}{1+\gamma}}\coth \left(\frac{\tau}{2}\sqrt{1-\gamma^2}\right)-\sqrt{\frac{1+\gamma}{1-\gamma}}\tanh \left(\frac{\tau}{2}\sqrt{1-\gamma^2}\right)\right] \quad ,\\
    \dot{a}(\tau)&=\frac{1}{2}\left[\sqrt{\frac{1-\gamma}{1+\gamma}}\coth \left(\frac{\tau}{2}\sqrt{1-\gamma^2}\right)+\sqrt{\frac{1+\gamma}{1-\gamma}}\tanh \left(\frac{\tau}{2}\sqrt{1-\gamma^2}\right)\right]
    \quad ,
    \label{eq:climb1}
\eea
Integrating with
\beq
    \int \coth x \, dx = \log \sinh x + \text{const} \qquad \int \tanh x \, dx = \log \cosh x + \text{const}
    \quad ,
\eeq
gives:
\bea
    \varphi(\tau) &=& \varphi_0+\frac{1}{1+\gamma}\log \sinh \left(\frac{\tau}{2}\sqrt{1-\gamma^2}\right)-\frac{1}{1-\gamma}\log \cosh \left(\frac{\tau}{2}\sqrt{1-\gamma^2}\right) \quad ,\\
    a(\tau)&=&\frac{1}{1+\gamma}\log \sinh \left(\frac{\tau}{2}\sqrt{1-\gamma^2}\right)+\frac{1}{1-\gamma}\log \cosh \left(\frac{\tau}{2}\sqrt{1-\gamma^2}\right) \quad ,
    \label{eq:climb2}
\eea

The additive constant in $a$ can be set to zero by rescaling spatial coordinates; $\varphi_0$ remains as an important integration constant.

The second solution, obtained via $\varphi \to -\varphi$, $\gamma \to -\gamma$, is:
\bea
    \dot{\varphi}(\tau)&=&\frac{1}{2}\left[\sqrt{\frac{1-\gamma}{1+\gamma}}\tanh \left(\frac{\tau}{2}\sqrt{1-\gamma^2}\right)-\sqrt{\frac{1+\gamma}{1-\gamma}}\coth \left(\frac{\tau}{2}\sqrt{1-\gamma^2}\right)\right] \quad ,\\
    \dot{a}(\tau)&=&\frac{1}{2}\left[\sqrt{\frac{1-\gamma}{1+\gamma}}\tanh \left(\frac{\tau}{2}\sqrt{1-\gamma^2}\right)+\sqrt{\frac{1+\gamma}{1-\gamma}}\coth \left(\frac{\tau}{2}\sqrt{1-\gamma^2}\right)\right] \quad ,
    \label{eq:des1}
\eea
\bea
    \varphi(\tau) &=& \varphi_0+\frac{1}{1+\gamma}\log \cosh \left(\frac{\tau}{2}\sqrt{1-\gamma^2}\right)-\frac{1}{1-\gamma}\log \sinh \left(\frac{\tau}{2}\sqrt{1-\gamma^2}\right) \quad ,\\
    a(\tau)&=&\frac{1}{1+\gamma}\log \cosh \left(\frac{\tau}{2}\sqrt{1-\gamma^2}\right)+\frac{1}{1-\gamma}\log \sinh \left(\frac{\tau}{2}\sqrt{1-\gamma^2}\right) \quad .
    \label{eq:des2}
\eea
the \textit{descending solution}.

For $0<\gamma<1$, both climbing and descending solutions exist. However, in the limit $\gamma \to 1$, only the climbing solution survives. Indeed, from~\eqref{eq:climb1}:
\bea
    \dot{\varphi}  \to \frac{1}{2}\left[\frac{1}{\tau}-\tau\right]
    \quad,
\eea
which is regular. In contrast, from~\eqref{eq:des1}:
\bea
    \dot{\varphi} \sim -\frac{1}{1-\gamma}\frac{1}{\tau}
    \quad ,
\eea
which diverges as $\gamma \to 1$. Hence, the descending solution becomes unphysical and disappears~\cite{Dudas:2010climbing,DudasKitazawaPatilSagnotti:2012,KitazawaSagnotti:2014}.

For $\gamma > 1$, only the climbing solution persists. Repeating the integration steps results in:
\bea
    \dot{\varphi}(\tau)&=&\frac{1}{2}\left[\sqrt{\frac{\gamma -1}{\gamma +1}}\cot \left(\frac{\sqrt{\gamma^2-1}}{2}\tau\right)-\sqrt{\frac{\gamma +1}{\gamma -1}}\tan \left(\frac{\sqrt{\gamma^2-1}}{2}\tau\right)\right] \quad ,\\
    \dot{a}(\tau)&=&\frac{1}{2}\left[\sqrt{\frac{\gamma -1}{\gamma +1}}\cot \left(\frac{\sqrt{\gamma^2-1}}{2}\tau\right)+\sqrt{\frac{\gamma +1}{\gamma -1}}\tan \left(\frac{\sqrt{\gamma^2-1}}{2}\tau\right)\right]
    \quad .
    \label{eq:trig1}
\eea
It is clear that the sign-flipped case cannot be considered since it leads to imaginary, unphysical solutions. Integrating for \(\gamma > 1\) yields the \textit{supercritical} solution:
\bea
    \varphi(\tau) &=& \varphi_0 + \frac{1}{\gamma+1} \log \sin \left( \frac{\sqrt{\gamma^2 - 1}}{2} \tau \right) - \frac{1}{\gamma-1} \log \cos \left( \frac{\sqrt{\gamma^2 - 1}}{2} \tau \right), \\
    a(\tau) &=& \frac{1}{\gamma+1} \log \sin \left( \frac{\sqrt{\gamma^2 - 1}}{2} \tau \right) + \frac{1}{\gamma-1} \log \cos \left( \frac{\sqrt{\gamma^2 - 1}}{2} \tau \right),
\eea
with \(\tau \in \left[0, \frac{\pi}{\sqrt{\gamma^2 - 1}}\right]\).

This solution has two singularities at the ends of the finite \(\tau\)-interval. The scalar emerges from the Big Bang while climbing the potential, reaches a turning point, and then descends, approaching infinite speed in parametric time. For \(\gamma > 1\), the \(\mathbb{Z}_2\) transformations \(\varphi \to -\varphi\) and \(\gamma \to -\gamma\), combined with finite \(\tau\)-translations, map solutions into themselves. This sharp change resembles a phase transition: for \(0 < \gamma < 1\) two branches (climbing and descending) exist, while for \(\gamma > 1\) only the climbing solution survives.

At the critical value \(\gamma = 1\), the solution simplifies to
\beq
    \dot{\varphi}(\tau) = \frac{1}{2} \left(\frac{1}{\tau} - \tau\right), \quad \dot{a}(\tau) = \frac{1}{2} \left(\frac{1}{\tau} + \tau\right),
\eeq
\beq
    \varphi(\tau) = \varphi_0 + \frac{1}{2} \log \tau - \frac{\tau^2}{4}, \quad a(\tau) = \frac{1}{2} \log \tau + \frac{\tau^2}{4}.
\eeq
Here the scalar can only emerge climbing the potential, then reverses motion at \(\tau=1\), producing a brief accelerated expansion phase before descending.

In summary, a scalar field can be forced to emerge from the Big Bang climbing an exponential potential when the logarithmic slope $\left|\frac{V'}{V}\right|$ 
reaches a critical value. Interestingly, in orientifold models of brane supersymmetry breaking \(\gamma\) is exactly critical, while in heterotic non-supersymmetric models \(\gamma > 1\). This suggests a deeper, yet unclear connection between orientifolds and early-universe scalar dynamics. This that we have described is the \textit{climbing scalar} phenomenon.

An instructive analogy is the motion of a Newtonian particle in a viscous medium under a constant force \(F\):
\beq
    m \dot{v}(t) + b v(t) = F, \quad v(t) = (v_0 - v_l) e^{-\frac{b t}{m}} + v_l,
\eeq
where $v_l = \frac{F}{b}$
is the limiting velocity. For finite \(b\), two solution branches exist depending on whether the initial velocity \(v_0\) lies above or below \(v_l\), but as \(b \to 0\) the upper branch disappears. Similarly, for \(\gamma < 1\) two cosmological solution branches exist (climbing and descending), but for \(\gamma \geq 1\) only the climbing solution remains, which admits a perturbative string realization. Descending solutions inevitably probe strong coupling at early times and are thus less physically viable.
\begin{figure}
    \centering    \includegraphics[width=0.35\linewidth]{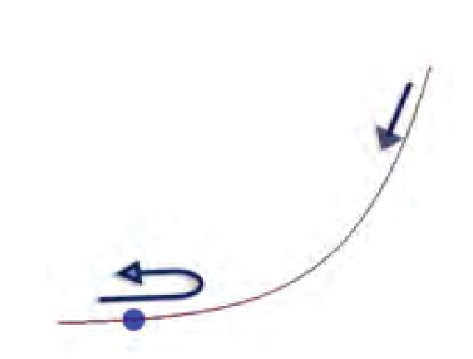}
    \caption{The two widely distinct scenarios of a climbing and
    a descending scalar. This picture is taken from~\cite{MouradSagnotti2021}.}
    \label{fig:climbinganddescendi}
\end{figure}
\subsection{Lucchin-Matarrese Attractor}
For both climbing and descending solutions (eqs.~\eqref{eq:climb1}, \eqref{eq:des1}), the scalar field asymptotically approaches the “limiting speed” in parametric time $\tau$:
\beq
    \dot{\varphi}(\tau) = -\frac{\gamma}{\sqrt{1-\gamma^2}} \quad ,
    \label{eq:limitspeed}
\eeq
defining the Lucchin-Matarrese (LM) attractor for $0 < \gamma < 1$. The exact solution reads
\beq
    \varphi(\tau) = \varphi_0 - \frac{\gamma}{\sqrt{1-\gamma^2}}\tau, \quad a(\tau) = \frac{\tau}{\sqrt{1-\gamma^2}} \quad .
    \label{eq:LM1}
\eeq

Using the gauge choice eq.~\eqref{eq:gauge}, cosmic and conformal times relate to $\tau$ as
\beq
    dt_c = e^B d\xi = \frac{1}{\sqrt{3 V(\varphi)}} d\tau = \sqrt{\frac{2}{3}} e^{-\gamma \varphi} \frac{d\tau}{M} \quad , 
    \label{eq:cosmictime}
\eeq
\beq
    d\eta = s^{-1} dt_c = \sqrt{\frac{2}{3}} e^{-\gamma \varphi - \frac{a}{3}} \frac{d\tau}{M} \quad , \quad s = e^{\frac{a}{3}} \quad .
    \label{eq:conformal time}
\eeq

This equations can be integrated exactly for the LM attractor, thus, in cosmic and conformal times:
\beq
    \varphi(t_c) = - \frac{1}{\gamma} \log\left(\sqrt{\frac{3}{2}} \frac{\gamma^2 M t_c}{\sqrt{1-\gamma^2}}\right), \quad
    a(t_c) = \frac{\varphi_0}{\gamma} + \frac{1}{\gamma^2} \log\left(\sqrt{\frac{3}{2}} \frac{\gamma^2 M t_c}{\sqrt{1-\gamma^2}}\right),
    \label{eq:LM2}
\eeq
\bea
    \varphi(\eta) &=& \frac{\varphi_0}{1 - 3 \gamma^2} + \frac{3 \gamma}{1 - 3 \gamma^2} \log\left[\frac{M(1 - 3 \gamma^2)}{\sqrt{6(1-\gamma^2)}}(-\eta)\right] \\
    a(\eta) &=& -\frac{3 \gamma \varphi_0}{1 - 3 \gamma^2} - \frac{3}{1 - 3 \gamma^2} \log\left[\frac{M(1 - 3 \gamma^2)}{\sqrt{6(1-\gamma^2)}}(-\eta)\right] \quad .
    \label{eq:LM3}
\eea

The scale factor exhibits power-law behavior:
\bea
    s(t_c) &=& \left(\sqrt{\frac{3}{2}} \frac{\gamma^2 M t_c}{\sqrt{1-\gamma^2}} e^{\gamma \varphi_0}\right)^{\frac{1}{3 \gamma^2}}, \\
    s(\eta) &=& \left(\frac{\sqrt{6(1-\gamma^2)}}{M (1 - 3 \gamma^2)} e^{-\gamma \varphi_0}\right)^{\frac{1}{1 - 3 \gamma^2}} (-\eta)^{-\frac{1}{1 - 3 \gamma^2}} \quad .
    \label{eq:LMscalefactor}
\eea
For $\gamma \to 0$, $s(\eta) \propto (-\eta)^{-1}$ reproduces the standard de Sitter solution.

The relations can be integrated exactly also close to the initial singularity. Climbing and descending solutions behave as
\beq
    \dot{\varphi}(\tau) \sim \dot{a}(\tau) \sim \pm \frac{1}{(1 \pm \gamma) \tau} \quad ,
\eeq
\beq
    \varphi(\tau) \sim \varphi_0 \pm \frac{1}{1 \pm \gamma} \log \tau, \quad a(\tau) \sim \pm \frac{1}{1 \pm \gamma} \log \tau \quad .
\eeq

So that
\beq
    \varphi(t_c) \sim a(t_c) \sim \pm \log t_c, \quad \varphi(\eta) \sim a(\eta) \sim \pm \frac{3}{3 \mp 1} \log \Delta \eta,
    \label{eq:nearsingular}
\eeq
and the scale factor behaves as
\beq
    s(t_c) \sim t_c^{\pm \frac{1}{3}}, \quad s(\eta) \sim \Delta \eta^{\pm \frac{1}{3 \mp 1}} \quad,
    \label{eq:scalefactornearsingular}
\eeq
where $\Delta \eta = -(\eta + \eta_s)$ with $\eta_s$ the conformal time of the initial singularity.

The relations between $\tau$ and $t_c$ are generally complicated; for example, for $0 < \gamma < 1$ and the climbing solution:
\beq
    t_c = \int d\tau \left[\sinh\left(\frac{\tau}{2} \sqrt{1-\gamma^2}\right)\right]^{\frac{\gamma}{1-\gamma}} \left[\cosh\left(\frac{\tau}{2} \sqrt{1-\gamma^2}\right)\right]^{-\frac{\gamma}{1+\gamma}} \quad .
\eeq
Substituting $x = \sinh^2\left(\frac{\tau}{2} \sqrt{1-\gamma^2}\right)$ gives
\beq
    t_c = \frac{1}{\sqrt{1-\gamma^2}} \int dx \, x^{\frac{\gamma - 1/2}{1-\gamma}} (1 + x)^{\frac{-\gamma - 1/2}{1+\gamma}} \quad ,
\eeq
which is generally expressed via hypergeometric functions:
\beq
    t_c \sim \left[\sinh^2\left(\frac{\tau}{2} \sqrt{1-\gamma^2}\right)\right]^{\frac{1/2}{1-\gamma}} 
    {}_2F_1\left(\frac{\gamma + 1/2}{1+\gamma}, \frac{1/2}{1-\gamma}; \frac{3/2}{1-\gamma}; -\sinh^2\left(\frac{\tau}{2} \sqrt{1-\gamma^2}\right)\right).
\eeq

The conformal time $\eta(\tau)$ relation is more complicated but can be numerically integrated, showing an asymptotic plateau that constrain $\eta$ in a finite domain $[-\eta_s, 0]$. This indicates a finite conformal time singularity characteristic of Kasner-like universes that will paly a crucial role for the approximation used in non-Gaussianity computations.

\section{Double Exponential Barrier}

The more general potential that we considere now is a double exponential:
\beq
    V(\varphi)=\frac{M^2}{2}\left(e^{2 \varphi}+e^{2 \gamma \varphi}\right), \quad 0 < \gamma < 1
    \quad .
    \label{eq:doubleexp}
\eeq
The steep term ($e^{2\varphi}$) drives an initial \textit{fast-roll} climbing phase dominated by kinetic energy, where slow-roll conditions are violated. The milder term ($e^{2 \gamma \varphi}$) is crucial to enable a slow-roll descent phase with accelerated expansion compatible with observations. 

This dual structure allows three dynamical phases: fast-roll climb, bounce, and slow-roll descent, producing a cosmologically viable scenario. 

Though no exact solution exists for the double exponential, single-exponential solutions approximate well near the singularity and at late times where one term dominates. For large positive $\varphi$, the steep term dominates forcing climbing, while descent is dominated by the mild term, approaching the limiting speed~\eqref{eq:limitspeed} from above.

The climbing forces initial conditions, leaving one free parameter $\varphi_0$ controlling the turning point and the extent of the “hard” exponential influence. This climbing mechanism naturally arises in string theory models such as the Sugimoto model~\cite{Sugimoto1999}, with $\varphi$ related to the dilaton controlling the string coupling $g_s = e^{\langle \phi_{10} \rangle}$. The climbing phenomenon persists in lower dimensions under certain stabilization assumptions for the other scalar field arising from compactification.

The value of $\varphi_0$ influences subsequent analysis of perturbations.
\subsection{Slow-Roll Restrictions on \texorpdfstring{$\gamma$} \ \ }
During the previous discussion, we emphasized the need to restrict the allowed range of \(\gamma\) to ensure the onset of a slow-roll phase after the scalar field’s motion reversal. We now analyze this aspect in detail, following strictly~\cite{Dudas:2010climbing}.

For the class of metrics in eq.~\eqref{eq:classofmetrics}, the Universe accelerates with respect to the cosmological time \(t_c\) when
\beq
    \frac{d^2 s}{dt_c^2}>0
    \quad .
\eeq
Expanding the scale factor and using the change of variables eq.~\eqref{eq:simplechange}, we derive that the quantity
\beq
    \mathcal{I} = \frac{d^2 A}{d\xi^2} + \frac{dA}{d\xi} \left( \frac{dA}{d\xi} - \frac{dB}{d\xi} \right)
    \quad ,
\eeq
must be positive.

Using equation of motion \eqref{eq:finalsystem}, redefinition of variables \eqref{eq:substitutions} and the gauge condition \eqref{eq:gauge}, this inequality translates into 
\beq
    |\dot{\varphi}| < \frac{1}{\sqrt{2}},
    \quad ,
\eeq
and, for example, using the Lucchin–Matarrese attractor solution eq.~\eqref{eq:LM1},
\beq
    \frac{\gamma}{\sqrt{1-\gamma^2}}<\frac{1}{\sqrt{2}} \quad \Rightarrow \quad 1-\gamma^2 >2 \gamma^2 \quad \Rightarrow \quad \gamma < \frac{1}{\sqrt{3}}
    \quad .
\eeq

Hence, to sustain the slow-roll inflationary phase from the LM attractor, the \(\gamma\)-domain condition is
\beq
    0 < \gamma < \frac{1}{\sqrt{3}}
    \quad .
    \label{eq:gammadomain}
\eeq
This can be generalized to \(d\)-dimensions as
\beq
    0 < \gamma < \frac{1}{\sqrt{d-1}}
    \quad .
\eeq

The upper bound \(\frac{1}{\sqrt{d-1}}\) lies well below the critical value \(\gamma=1\) associated with climbing, so the climbing phenomenon remains valid. As mentioned in the string theory section, typical \(\gamma\) values are \(O(0.1)\), comfortably within this domain.

\subsection{Slow-roll Parameters}
At this point, it is crucial to introduce the \textit{slow-roll parameters}, fundamental in characterizing inflationary dynamics and detecting deviations from slow-roll, particularly relevant in models with fast-roll phases like the one at hand~\cite{Baumann:Cosmology2022,Baumann2009TASI,Wald:GeneralRelativity,Peebles:PhysicalCosmology1993}.

Two families of definitions are typically used:
\begin{itemize}
    \item Hubble slow-roll parameters:
    \beq
        \varepsilon \equiv -\frac{1}{H^2}\frac{dH}{dt_c}, \quad \eta \equiv \frac{1}{H \, \varepsilon}\frac{d \varepsilon}{dt_c}
        \quad ;
    \eeq
    \item Potential slow-roll parameters:
    \beq
        \varepsilon_V \equiv \frac{1}{2} \left(\frac{V'}{V}\right)^2, \quad \eta_V \equiv \frac{V''}{V}
        \quad .
    \eeq
\end{itemize}

In the slow-roll limit (\(\varepsilon, \eta \ll 1\)), these definitions converge. However, in fast-roll phases (as in the climbing scenario), their dynamics differ significantly. In our model, the Hubble parameter-based \(\varepsilon\) is most informative for identifying the transition between fast and slow-roll, so now we derive its expression in terms of scalar field dynamics. Starting from the system of equation of motion for the background we obtain this equivalent formula that will be used also in the next section \cite{KitazawaSagnotti:2014}:
\beq
    \varepsilon(t_c) = -\frac{1}{H^2}\frac{dH}{dt_c} = \frac{1}{2}\frac{\left(\frac{d \phi}{d t_c}\right)^2}{\left(\frac{dA}{d t_c}\right)^2}
    \quad .
\eeq
Switching back to $(a, \varphi)$ and $\tau$:
\beq
    \varepsilon(\tau) = 3 \, \frac{\dot{\varphi}^2}{\dot{a}^2} = 3 \, \frac{\dot{\varphi}^2}{1+\dot{\varphi}^2}
    \quad .
\eeq

In the "critical climbing case", from eqs.~\eqref{eq:climb1} and~\eqref{eq:climb2}:
\beq
    \varepsilon(\tau) = 3 \frac{(1-\tau^2)^2}{(1+\tau^2)^2}
    \quad ,
    \label{eq:criticaleps}
\eeq
which shows: $\varepsilon \to 3$ near the singularity (fast-roll); $\varepsilon = 0$ at $\tau = 1$, the scalar inversion point $\tau^*$; $\varepsilon \to 3$ again as $\tau \to \infty$ (fast roll).

\begin{figure}
    \centering
    \includegraphics[width=0.35\linewidth]{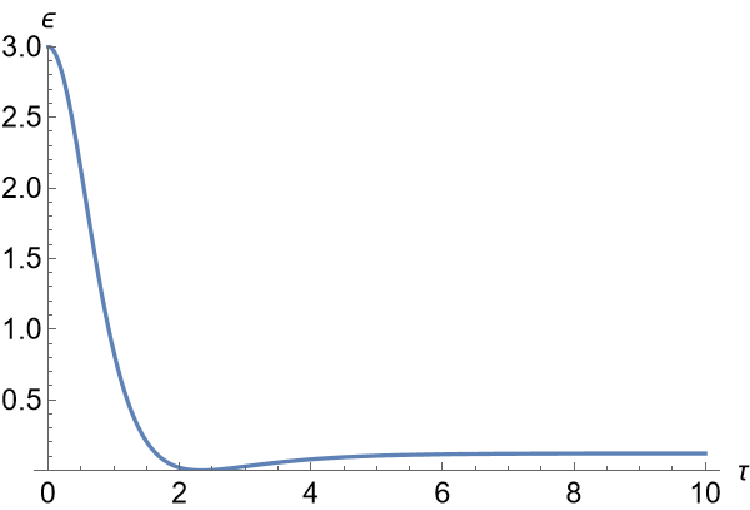}
    \caption{$\varepsilon(\tau)$ in a single exponential potential with $\gamma = 0.2$.}
    \label{fig:epspic}
\end{figure}

With the mild exponential, we get:
\beq
    \varepsilon (\tau) = 3 \frac{\left(-1+\gamma \cosh \left(\sqrt{1-\gamma^2} \, \tau\right)\right)^2}{\left(\gamma -\cosh \left(\sqrt{1-\gamma^2} \, \tau\right)\right)^2}
    \quad ,
\eeq
still yielding $\varepsilon(0) \to 3$, a zero at some $\tau^*$ (inversion), and now $\varepsilon \to 3 \gamma^2$
as $\tau \to \infty$, which lies in the slow-roll regime for $\gamma \sim O(0.1)$. A typical plot of this behavior is shown in Fig.~\ref{fig:epspic}.

\subsection{Some Numerical Results}
We summarize numerical results from~\cite{DudasKitazawaPatilSagnotti:2012, KitazawaSagnotti:2014} for the double exponential potential introduced in eq.~\eqref{eq:doubleexp}. Simulations performed with Runge–Kutta methods in \texttt{Maple}, within a single-scalar field minimally coupled to gravity, explore scenarios where the scalar \(\phi\) (e.g., the dilaton) starts with various initial values \(\phi_0 < 0\), consistent with weak string coupling and climbing dynamics.

\begin{figure}
    \centering
    \includegraphics[width=0.3\linewidth]{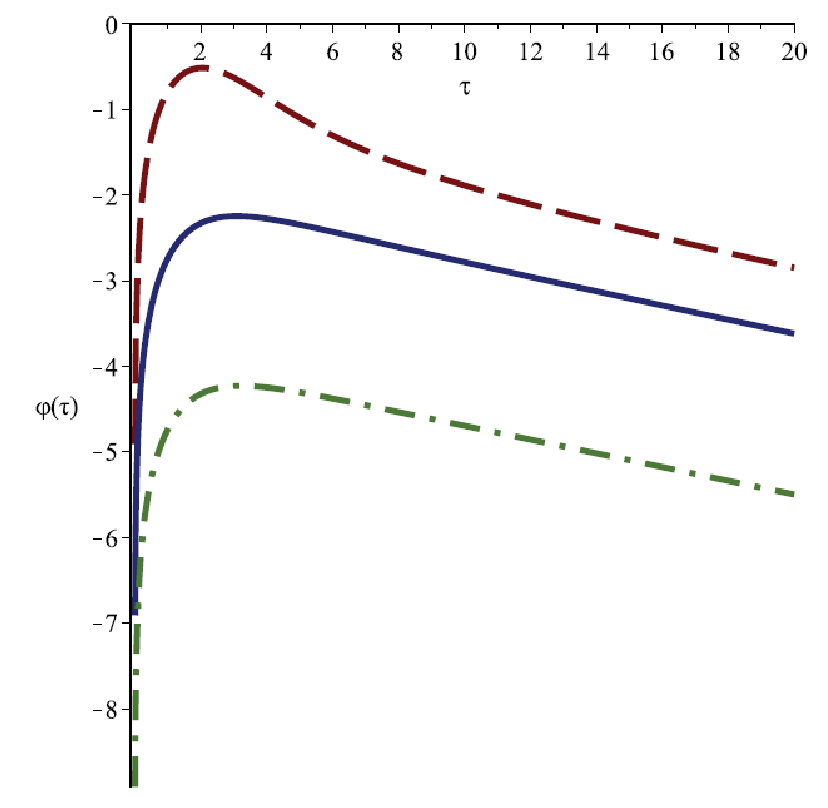}
    \caption{$\varphi(\tau)$ evolution for $\varphi_0=0$ (dashed), $-2$ (solid), and $-4$ (dash-dotted) with fixed $\gamma=0.007$. From~\cite{KitazawaSagnotti:2014}.}
    \label{fig:tauscalar}
\end{figure}

Figure~\ref{fig:tauscalar} shows a sharp reversal near the steep exponential for \(\varphi_0=0\), while \(\varphi_0=-2\) and \(-4\) produce similar trajectories. 

\begin{figure}[ht]
\centering
\begin{tabular}{ccc}
%\mbox{graphic} & \mbox{table} \\
\includegraphics[width=60mm]{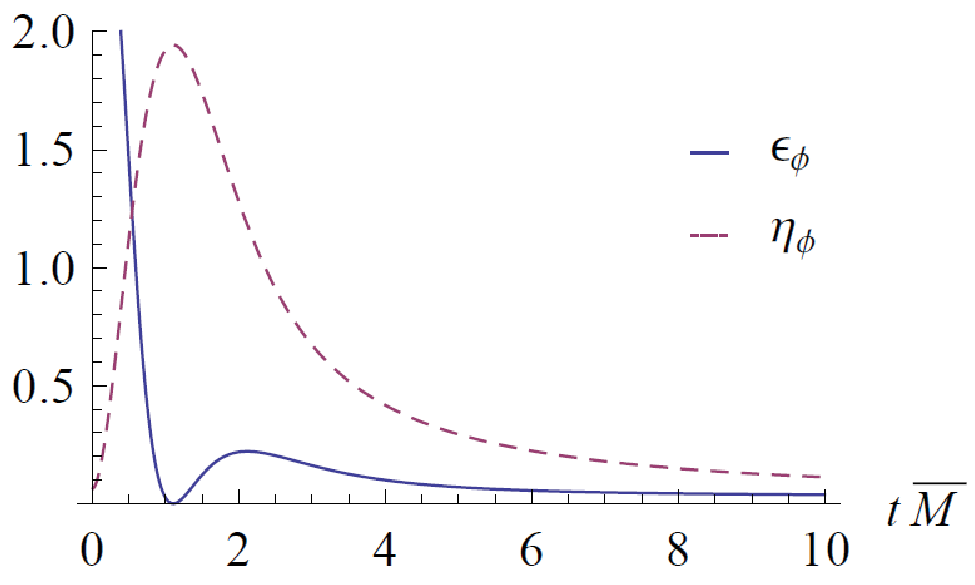} \quad  &
\includegraphics[width=60mm]{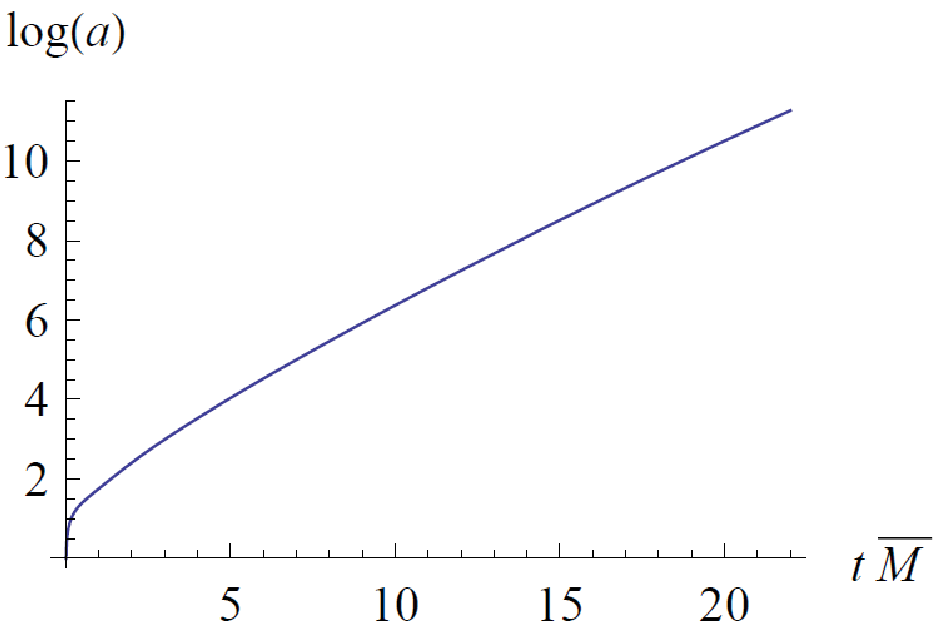}  \\
\end{tabular}
 \caption{\small Background solutions for the potential \eqref{eq:secondpotential} with 
 $\gamma =\frac{1}{4 \sqrt{6}}$ and $\phi_0=-4$. Left panel: $\varepsilon(t_c)$; Right panel: $\log s(t_c)$. Those are typical values given to these parameters for simulations. This picture is taken from \cite{DudasKitazawaPatilSagnotti:2012}.}
\label{fig:numerics}
\end{figure}

The scale factor evolution, plotted in Fig.~\ref{fig:numerics}, demonstrates attractor behavior emerges rapidly after the initial singularity. The Hubble parameter \(H\), analyzed in cosmic and parametric time, is given by \cite{KitazawaSagnotti:2014}
\beq
    H \equiv \frac{1}{s}\frac{ds}{dt_c} =  \sqrt{\frac{V}{3}(1+\dot{\varphi}^2)}
    \quad .
    \label{eq:Hubble}
\eeq

On the Lucchin–Matarrese attractor, cosmic-time asymptotics yield
$H(t_c)^2 \, t_c^2 \to \frac{1}{9 \gamma^4}$
but this regime occurs at late times \(t_c \bar{M} \sim O(10^4)\), while \(\log s(t_c)\) already shows near–de Sitter expansion soon after the singularity, influencing slow-roll parameter dynamics as was shown in \cite{DudasKitazawaPatilSagnotti:2012}.

Figure~\ref{fig:numerics} displays the slow-roll parameters \(\varepsilon\) and \(\eta\) as functions of cosmic time. Accelerated expansion (\(\varepsilon < 1\)) begins shortly after the climbing phase (\(t_c \bar{M} \sim 1\)). Although \(\eta\) converges more slowly to its attractor, temporarily large \(\eta\) does not hinder inflation, provided \(\varepsilon\) remains small. A short near–de Sitter phase with \(\varepsilon \approx 0\) appears near the inflection point. While attractor convergence occurs late, off-attractor inflation starts early and can yield significant e-folds despite lower efficiency.

Tau-dependent analyses~\cite{KitazawaSagnotti:2014} reveal that for \(\varphi_0=0\), the Hubble parameter \(H\) forms a quasi-exponential plateau near the inversion point, but \(\varepsilon\) grows quickly, indicating a fast bounce; simultaneously, \(\eta\) remains large, suppressing perturbation growth. 

Further discussions on the time evolution of \(\varepsilon\) will be provided in the context of non-Gaussianity computations.
\begin{center}
    ***
\end{center}
To summarize we adopt the double exponential potential with negative scalar values to control the string coupling and suppress corrections. Negative values of $\varphi$ ensures dominance of the mild exponential except near the inversion point, where the critical term enforces the climbing behavior tied to supersymmetry breaking.

Alternative potentials have been studied extensively~\cite{FreSagnottiSorin:2013_integrable}, but the double exponential remains preferable for its string theory origins and analytic tractability. Flat-space vacua are excluded by~\cite{Dudas:2000ff} due to dilaton tadpoles in broken supersymmetry, favoring cosmological solutions with spontaneous compactification.

We will next analyze perturbations on the climbing scalar background, assuming inflation occurs near the string and supersymmetry-breaking scales, reflecting the model’s structure without large hierarchies. Although limited by the low-energy effective theory this remains the best available framework to study these cosmologies. 
\chapter{Cosmological Perturbation Theory}
In this chapter, we develop a systematic framework for cosmological perturbation theory, focusing on primordial fluctuations arising from quantum effects during inflation. While linear perturbations are well understood, we extend the analysis to the cubic order to capture non-Gaussian features, crucially encoded in the three-point correlation function (bispectrum).

Before turning to cubic interactions, we compare the two-point correlation functions in the standard single-field slow-roll model and in the climbing scalar scenario, emphasizing the distinctive infrared suppression that characterizes the latter.

Starting from the single-field background dynamics, we introduce scalar perturbations in a gauge-invariant formalism and derive the action expanded up to third order. This leads to the Maldacena cubic action~\cite{maldacena2003non}, which forms the basis for computing the bispectrum and studying non-Gaussianities. The entire derivation is based on \cite{collins2011primordial}.
\section{The Background}
We begin from the canonical single-field inflationary action introduced in \eqref{eq:secondEFTfinal},
\beq
    S=\int d^{4}x \sqrt{-g} \left[\frac{1}{2}M^2_{pl}R - \frac{1}{2}g^{\mu \nu}\partial_{\mu}\phi \partial_{\nu}\phi - V(\phi)\right],
\eeq
where \(M_{pl}=(8\pi G)^{-1/2}\) is the reduced Planck mass. Setting \(M_{pl}=1\) for simplicity, variation with respect to the metric leads to Einstein's equations,
\beq
    R_{\mu \nu} - \frac{1}{2}g_{\mu \nu} R = T_{\mu \nu},
\eeq
with the scalar field energy-momentum tensor
\beq
    T_{\mu \nu} = \partial_{\mu}\phi \partial_{\nu}\phi - \frac{1}{2}g_{\mu \nu} g^{\lambda \sigma}\partial_{\lambda}\phi \partial_{\sigma}\phi + g_{\mu \nu} V(\phi).
\eeq
Variation with respect to \(\phi\) yields the curved-space Klein-Gordon equation,
\beq
    \frac{1}{\sqrt{-g}}\partial_{\mu} \left( \sqrt{-g} g^{\mu \nu} \partial_{\nu} \phi \right) - \frac{d V}{d \phi} = 0.
\eeq

Assuming a flat, homogeneous, and isotropic Friedmann–Robertson–Walker (FRW) metric,
\beq
    ds^2 = -dt^2 + e^{2\rho(t)} \delta_{ij} dx^i dx^j,
\eeq
the scale factor is \(a(t) = e^{\rho(t)}\) and the Hubble parameter is
\beq
    H(t) = \dot{\rho}(t) = e^{-\rho(t)} \frac{d}{dt} e^{\rho(t)}.
\eeq
Homogeneity also implies \(\phi = \phi(t)\). In this section dots denote derivatives in cosmic time, which here is $t$ for simplicity.

Developing the equation of motion we arrive to the system of Friedmann equations (with \(M_{pl} = 1\)):
\beq
    \begin{cases}
        \dot{\rho}^2 = \frac{1}{3} \left( \frac{1}{2} \dot{\phi}^2 + V(\phi) \right), \\
        -2 \ddot{\rho} - 3 \dot{\rho}^2 = \frac{1}{2} \dot{\phi}^2 - V(\phi).
    \end{cases}
    \label{eq:Friedmann}
\eeq
and the scalar field obeys
\beq
    \ddot{\phi} + 3 \dot{\rho} \dot{\phi} + \frac{d V}{d \phi} = 0.
\eeq
These are equivalent to the set \eqref{eq:finalsystem}
To characterize inflationary dynamics, the slow-roll parameters are introduced:
\beq
    \varepsilon \equiv -\frac{\dot{H}}{H^2} = -\frac{\ddot{\rho}}{\dot{\rho}^2}, \qquad \delta \equiv \frac{\ddot{\phi}}{H \dot{\phi}} = \frac{\ddot{\phi}}{\dot{\rho} \dot{\phi}}.
\eeq
From the Friedmann equations, one derives
\beq
    \ddot{\rho} = -\frac{1}{2} \dot{\phi}^2,
    \label{eq:back}
\eeq
which allows rewriting
\beq
    \varepsilon = \frac{1}{2} \frac{\dot{\phi}^2}{\dot{\rho}^2}.
\eeq
that was used also in chapter 2. Reintroducing \(M_{pl}\), this becomes \(\varepsilon = \frac{1}{2 M_{pl}^2} \frac{\dot{\phi}^2}{\dot{\rho}^2}\).

The slow-roll regime corresponds to \(\varepsilon, \delta \ll 1\), where the scalar evolves slowly and inflation is sustained. The parameter \(\varepsilon\) also quantifies deviations from exact de Sitter expansion, where \(\varepsilon = 0\).

To match notation used in the \textit{Climbing scalar} framework with that of~\cite{maldacena2003non,collins2011primordial}, one replaces \(\rho\) with \(a/3\), where \(a = \log ( \text{scale factor} )\). We will adopt notation of \cite{maldacena2003non,collins2011primordial} in this section.
\section{Perturbations}
Inflation provides a natural mechanism for primordial density fluctuations seeding structure formation. The inflaton \(\phi(t)\) acts as a “clock” for inflationary expansion, but quantum uncertainty induces spatial fluctuations \(\delta \phi(t,\vec{x})\), causing inflation to end at varying times and generating local density perturbations. Regions ending inflation earlier dilute more, becoming less dense; those inflating longer become denser.

Quantum corrections around the homogeneous background are:
\beq
    g_{\mu \nu}(t,\vec x)=g^{(0)}_{\mu \nu}(t)+\delta g_{\mu \nu}(t,\vec x) \qquad \phi(t,\vec x)=\phi^{(0)}(t)+\delta \phi(t,\vec x)
    \quad ,
\eeq
where \(\delta g_{\mu \nu}, \delta \phi\) fluctuate quantum mechanically, introducing spatial dependence. Inflation stretches these fluctuations to cosmic scales, freezing them beyond causal contact and seeding primordial perturbations.

Focusing on scalar perturbations (scalars under unbroken spatial rotations and translations; time translation symmetry is broken), the scalar degrees of freedom count as: \(\delta g_{00}\) (one scalar), \(\delta g_{0i}\) (derivable from scalar \(\partial_i \beta\)), \(\delta g_{ij}\) (two scalars: trace and \(\partial_i \partial_j \xi\)), and \(\delta \phi\) (one scalar), totaling five scalar fields. Gauge transformations \(x^\mu \to x^\mu + \delta x^\mu\) remove two scalars (one from \(\delta x^0\), one from \(\delta x^i \sim \delta^{ij}\partial_j f\)), and two more are non-propagating fixed by constraints, leaving one physical scalar degree of freedom.

Without matter, linear perturbations yield two tensorial degrees of freedom (gravitational wave polarizations). Inflation adds one scalar due to broken time translation invariance, totaling three physical degrees of freedom: two tensorial and one scalar. This work focuses on the scalar mode.

Using the ADM formalism (see \ref{appendix:ADM}), write the metric as:
\beq
    ds^2 = - N^2 dt^2 + h_{ij}(N^i dt + dx^i)(N^j dt + dx^j)= -(N^2-N_iN^i)dt^2+2N_idtdx^i+h_{ij}dx^idx^j
    \quad ,
\eeq
with lapse \(N(t,\vec{x})\) and shift \(N^i(t,\vec{x})\). The inverse metric components are:
\beq
    g^{00}=-\frac{1}{N^2} \quad , \quad g^{0i}=\frac{N^i}{N^2} \quad , \quad g^{ij}=h^{ij}-\frac{N^i N^j}{N^2}
    \quad .
\eeq
Curvature and covariant derivatives calculated with \(h_{ij}\) are denoted with a caret, e.g. \(\hat{R}\).

The action is (see \ref{appendix:ADM}):
\beq
    S=\frac{1}{2}\int d^4 x \sqrt{h} \Big[N\hat{R}+\frac{1}{N}(E_{ij}E^{ij}-E^2)+\frac{1}{N}(\dot{\phi}-N^i\partial_i \phi)^2-Nh^{ij}\partial_i\phi\partial_j\phi-2NV(\phi)\Big]
    \label{eq:ADM action}
\eeq
where
\beq
    E_{ij} = \frac{1}{2}(\dot{h}_{ij} - \hat{\nabla}_i N_j - \hat{\nabla}_j N_i)
    \quad .
\eeq
Adopting the same notation ~\eqref{eq:E formula}
\beq
    E_{ij}=\frac{1}{2}\big[\dot{h}_{ij}-\hat{\nabla}_iN_j-\hat{\nabla}_jN_i\big]=\frac{1}{2}\big[\dot{h}_{ij}-h_{ik}\partial_j N^k - h_{jk}\partial_i N^k - N^k\partial_k h_{ij}\big]
    \quad ,
\eeq
\beq
    E_{ij}E^{ij}-E^2=[h^{ik}h^{jl} - h^{ij}h^{kl}] E_{ij} E_{kl}
    \quad .
\eeq
Since \(N\) and \(N^i\) have no time derivatives, they are non-dynamical fields whose equations of motion produce two constraints, reducing by two the number of independent scalar degrees of freedom. Variation with respect to \(N \to N + \delta N\) gives:
\beq
    \hat{R} - \frac{1}{N^2}(E_{ij}E^{ij} - E^2) - \frac{1}{N^2}(\dot{\phi} - N^i \partial_i \phi)^2 - h^{ij} \partial_i \phi \partial_j \phi - 2V = 0
    \quad ,
\eeq
while for \(N^i \to N^i + \delta N^i\) (Euler-Lagrange):
\beq
    \hat{\nabla}_j \left[ N^{-1}(E^j_i - \delta^j_i E) \right] = \frac{1}{N} (\dot{\phi} - N^j \partial_j \phi) \partial_i \phi
    \quad ,
\eeq Fluctuations are studied by expanding the action to the required order: quadratic terms suffice for the two-point function, cubic terms are needed for the three-point function. The general parametrization of scalar fluctuations in the metric is:
\beq
    N=1+2\Phi(t,\vec{x}), \quad N^i = \delta^{ij} \partial_j B(t,\vec{x}), \quad h_{ij} = e^{2\rho(t)}\big[(1+2\zeta(t,\vec{x})) \delta_{ij} + \partial_i \partial_j \xi\big]
    \quad ,
\eeq
and for the scalar field fluctuations:
\beq
    \phi = \phi_0(t) + \delta \phi(t,\vec{x})
    \quad .
\eeq
Using coordinate freedom, two scalar functions can be removed by redefining time and spatial coordinates, setting \(\delta \phi=0\) and \(\xi=0\) (uniform density gauge), leaving:
\beq
    N=1+2\Phi(t,\vec{x}), \quad N^i = \delta^{ij} \partial_j B(t,\vec{x}), \quad h_{ij} = e^{2\rho(t)+2\zeta(t,\vec{x})} \delta_{ij}, \quad \phi = \phi(t)
    \quad .
\eeq
Note \(\zeta\) is redefined appearing in the exponent. Using this gauge in the constraint equations:
\beq
    \hat{R} - \frac{1}{N^2}(E_{ij}E^{ij} - E^2) - \frac{\dot{\phi}^2}{N^2} - 2V = 0
\eeq
\beq
    \hat{\nabla}_j \left[ N^{-1}(E^j_i - \delta^j_i E) \right] = 0
\eeq
Removing the potential using the first Friedmann equation \eqref{eq:Friedmann} yields:
\beq
    \hat{R} - \frac{1}{N^2}(E_{ij}E^{ij} - E^2) - 6 \dot{\rho}^2 + \left(1 - \frac{1}{N^2}\right) \dot{\phi}^2 = 0
    \quad .
\eeq
Using the A.D.M. formula for \(\hat{R}\) with this gauge choice we expand:
\beq
    \hat{R} = e^{-2\rho - 2\zeta}[-4\partial_k \partial^k \zeta - 2 \partial_k \zeta \partial^k \zeta] = -4 e^{-2\rho} \partial_k \partial^k \zeta + O(\zeta^2)
    \quad .
\eeq
With the definition of \(E_{ij}\):
\beq
    E_{ij} = \frac{1}{2} \big[ \dot{h}_{ij} - h_{ik} \partial_j N^k - h_{jk} \partial_i N^k - N^k \partial_k h_{ij} \big] = e^{2\rho} \big[ \dot{\rho} (1+2\zeta) \delta_{ij} + \dot{\zeta} \delta_{ij} - \partial_i \partial_j B \big] + O(\zeta^2)
    \quad .
\eeq
Thus,
\beq
    E_{ij} E^{ij} - E^2 = [h^{ik} h^{jl} - h^{ij} h^{kl}] E_{ij} E_{kl} = -6 \dot{\rho}^2 - 12 \dot{\rho} \dot{\zeta} + 4 \dot{\rho} \partial_k \partial^k B + O(\zeta^2)
    \quad .
\eeq
The constraint for \(N\) to first order reads:
\beq
    -3 \dot{\rho} [2 \dot{\rho} \Phi - \dot{\zeta}] - \partial_k \partial^k [\dot{\rho} B + e^{-2\rho} \zeta] + \dot{\phi}^2 \Phi = 0
    \quad .
\eeq
Similarly, the \(N^i\) constraint expanded to first order gives:
\beq
    2 \partial_i [2 \dot{\rho} \Phi - \dot{\zeta}] = 0 \quad \Rightarrow \quad \Phi = \frac{\dot{\zeta}}{2 \dot{\rho}}
    \quad .
    \label{eq:first constr}
\eeq
This fixes \(\Phi\), removing one scalar degree of freedom; the integration constant is fixed to zero to recover the background metric. Inserting this into the \(N\) constraint fixes \(B\):
\beq
    B = -\frac{e^{-2\rho}}{\dot{\rho}} \zeta + \chi, \quad \partial_k \partial^k \chi = \frac{1}{2} \frac{\dot{\phi}^2}{\dot{\rho}^2} \dot{\zeta}
    \quad .
\eeq
Having solved constraints eliminating \(\Phi\) and \(B\), the quadratic action for \(\zeta\) remains. Using the first Friedmann equation \(\eqref{eq:Friedmann}\) to remove \(V\), the action is
\beq
    S = \frac{1}{2} \int d^4 x \sqrt{h} \Big[ N \hat{R} + \frac{1}{N} (E_{ij} E^{ij} - E^2) - 6 N \dot{\rho}^2 + \left( N + \frac{1}{N} \right) \dot{\phi}^2 \Big]
    \quad .
\eeq
Expanding to second order in fluctuations,
\beq
    \hat{R} = e^{-2\rho} \big[-4 \partial_k \partial^k \zeta + 8 \zeta \partial_k \partial^k \zeta - 2 \partial_k \zeta \partial^k \zeta \big] + O(\zeta^3)
    \quad ,
\eeq
\bea
    && E_{ij} E^{ij} - E^2 = -6 \dot{\rho}^2 - 12 \dot{\rho} \dot{\zeta} + 4 \dot{\rho} \partial_k \partial^k B - 6 \dot{\zeta}^2 + 12 \dot{\rho} \partial_k \zeta \partial^k B + 4 \dot{\zeta} \partial_k \partial^k B + \nonumber \\ 
    && (\partial_i \partial_j B)(\partial^i \partial^j B) - (\partial_k \partial^k B)^2 + O(\zeta^3)
    \quad ,
\eea
\beq
    N + \frac{1}{N} = 2 + \frac{\dot{\zeta}^2}{\dot{\rho}^2} + O(\zeta^3)
    \quad ,
\eeq
\bea
    && \sqrt{h} N \hat{R} = 2 e^{\rho} \frac{\ddot{\rho}}{\dot{\rho}^2} \partial_k \zeta \partial^k \zeta - \partial_0 \left[ 2 e^{\rho} \frac{1}{\dot{\rho}} \zeta \partial_k \partial^k \zeta \right] + \nonumber \\ 
    && - 2 e^{\rho} \partial_k \left[ \left( 2 + \zeta + \frac{\ddot{\rho}}{\dot{\rho}^2} \zeta + \frac{\dot{\zeta}}{\dot{\rho}} \right) \partial^k \zeta - \frac{1}{\dot{\rho}} \zeta \partial^k \dot{\zeta} \right] + O(\zeta^3)
    \quad .
\eea
Using background equation \(\eqref{eq:back}\), only one term remains up to total derivatives:
\beq
    \sqrt{h} N \hat{R} = - e^{\rho} \frac{\dot{\phi}^2}{\dot{\rho}^2} \partial_k \zeta \partial^k \zeta + O(\zeta^3)
    \quad .
\eeq
The remaining second order terms give:
\bea
    && \sqrt{h} \left[ \frac{1}{N} (E_{ij} E^{ij} - E^2) - 6 N \dot{\rho}^2 + \left( N + \frac{1}{N} \right) \dot{\phi}^2 \right] = \nonumber \\
    &=& e^{3\rho} \frac{\dot{\phi}^2}{\dot{\rho}^2} \dot{\zeta}^2 + e^{3\rho} [2 \ddot{\rho} + \dot{\phi}^2] [2 + 6 \zeta + 9 \zeta^2] - 2 \partial_0 \left[ e^{3\rho} \dot{\rho} (2 + 6 \zeta + 9 \zeta^2) \right] + \nonumber \\
    && + e^{3\rho} \partial_k \left[ (4 \dot{\rho} + 12 \dot{\rho} \zeta - \partial_j \partial^j B) \partial^k B + (\partial_j B)(\partial^j \partial^k B) \right] + O(\zeta^3)
    \quad .
\eea
Again, using \(\eqref{eq:back}\), only one non-total derivative term remains:
\beq
    \sqrt{h} \left[ \frac{1}{N} (E_{ij} E^{ij} - E^2) - 6 N \dot{\rho}^2 + \left( N + \frac{1}{N} \right) \dot{\phi}^2 \right] = e^{3\rho} \frac{\dot{\phi}^2}{\dot{\rho}^2} \dot{\zeta}^2 + O(\zeta^3)
    \quad .
\eeq
Thus, after applying background equations and ignoring total derivatives, the quadratic action is:
\beq
    S^{(2)} = \int dt \frac{\dot{\phi}^2}{\dot{\rho}^2} \int d^3 \vec{x} e^{3\rho} \left[ \frac{1}{2} \dot{\zeta}^2 - \frac{1}{2} e^{-2\rho} \partial_k \zeta \partial^k \zeta \right]
    \label{eq:S2}
    \quad .
\eeq
Noting that referring to the background metric by a bar, the quadratic action can be expressed as:
\beq
    S^{(2)} = - \int d^4 x \sqrt{-\bar{g}} \, \varepsilon \left( \bar{g}^{\mu \nu} \partial_{\mu} \zeta \partial_{\nu} \zeta \right)
    \quad .
\eeq
This is the standard massless scalar action in an FRW background multiplied by \(2\varepsilon\). In exact de Sitter, \(\varepsilon = 0\), so the theory is infinitely strongly coupled, the second-order action vanishes, and no primordial fluctuations are generated. Hence, we always consider quasi-de Sitter backgrounds, e.g., slow-roll regime where \(\varepsilon \ll 1\) but nonzero, allowing perturbation theory.

Importantly, all results so far are independent of slow-roll assumptions.

\paragraph{Physical Meaning of $\zeta$}
\leavevmode\\
The scalar variable $\zeta$, also called the \textit{curvature perturbation}, is a gauge-invariant quantity describing scalar perturbations in cosmology. It arises from the scalar sector of the metric in a generic SVT decomposition and transforms non-trivially under diffeomorphisms, but combinations like
\beq
    \zeta = -C + \frac{1}{3}\nabla^2 A + 2\rho' \frac{\delta \sigma}{\sigma'}
\eeq
remain invariant (see \cite{Baumann:Cosmology2022} for notation). In the uniform density gauge, where $\delta \sigma = 0$, this reduces to $\zeta = -C + \frac{1}{3}\nabla^2 A$, and the spatial metric becomes
\beq
    h_{ij} = e^{2\rho}(1 + 2\zeta)\delta_{ij}
    \quad ,
\eeq
justifying the name "curvature perturbation". Indeed, the intrinsic curvature of spatial slices is
\beq
    \hat R = -4e^{-2\rho}\partial_k \partial^k \zeta + O(\zeta^2) \quad \Rightarrow \quad e^{2\rho} \hat R = -4\nabla^2 \zeta
    \quad ,
\eeq
so $\zeta$ directly encodes spatial curvature fluctuations. On superhorizon scales, if matter perturbations are adiabatic, $\zeta$ is conserved, making it a natural variable to describe the initial conditions for structure formation.

\subsection{Mukhanov-Sasaki Equation}
The fluctuations of interest originate from quantum mechanics, due to the Heisenberg uncertainty principle. To analyze them, it is convenient to recast the action in a form analogous to flat-space quantum field theory. This allows applying standard results, with care due to the curved, expanding background and high-energy effects near the Planck scale.

We define:
\beq
    v(t,\vec x)=e^{\rho}\frac{\dot \phi}{\dot \rho}\zeta(t, \vec x)
    \quad .
\eeq

This variable, known as the \textit{Mukhanov-Sasaki variable}, leads to: the action becomes :
\beq
    S^{(2)}=\frac{1}{2}\int d \eta\, d^3 x\left[(v')^2-\partial_k v\, \partial^k v+\frac{z''}{z}v^2\right]
    \quad ,
\eeq
where we defined
\beq
    z \equiv e^{\rho} \, \sqrt{2 \varepsilon}
\eeq
Most importantly to derive this form we use equation of motion and the relation that will prove useful later:
\beq
    \frac{\dot \varepsilon}{\varepsilon \dot \rho}=\frac{d}{dt}\left(\frac{\dot \phi^2}{\dot \rho^2}\right)\frac{\dot \rho}{\dot \phi^2}=2\frac{\ddot \phi}{\dot \phi \dot \rho}-2\frac{\ddot \rho}{\dot \rho^2}=2\frac{\ddot \phi}{\dot \phi \dot \rho}+\frac{\dot \phi^2}{\dot \rho^2}=2(\delta+\varepsilon)
    \quad ,
\eeq

From this point on, we assume that $\eta \in (-\infty, 0]$, so that time runs in the usual forward direction. Note that the final action resembles the Klein-Gordon action, with $\frac{z''}{z}$ playing the role of a time-dependent mass. We therefore define this term as the \textit{Mukhanov-Sasaki potential}:
\beq
    W_S(\eta)  \equiv \frac{z''}{z}
    \quad ,
    \label{eq:scalarpotential}
\eeq

The equation of motion is obtained by varying the action:
\beq
    v''-\nabla^2v-W_S(\eta)v=0
    \quad .
\eeq

Although formally similar to the equation of a free field in flat space, this is not invariant under Poincaré transformations, since the effective mass depends on the conformal time $\eta$. Nevertheless, the field can still be expanded in plane waves:
\beq
    v(t, \vec{x}) = \int \frac{d^3 \vec{k}}{(2\pi)^3} \left[ v_k(t)\, e^{i \vec{k} \cdot \vec{x}}\, a_{\vec{k}} + v_k^*(t)\, e^{-i \vec{k} \cdot \vec{x}}\, a_{\vec{k}}^\dagger \right]
    \quad .
\eeq

The time-dependent components then satisfy:
\beq
    v_k''(\eta)+\left(k^2-W_S(\eta)\right)v_k(\eta)=0
    \quad ,
    \label{eq:MS-equation}
\eeq

This is the well-known \textit{Mukhanov-Sasaki equation}.

\subsubsection{Solution for (quasi)dS-space}
The standard approach to solving the \textit{Mukhanov-Sasaki equation} is to specialize to (quasi) de Sitter (dS) space.

For pure dS we have
\beq
    \varepsilon = -\frac{\dot{H}}{H^2} = 0 \quad \rightarrow \quad H = \rho = \text{const}
    \quad,
\eeq
and the scale factor in cosmic time is $e^{Ht}$. Using conformal time with the negative branch,
\beq
    \eta(t) = \int dt\, e^{-\rho(t)} = \int dt\, e^{-Ht} = -\frac{e^{-Ht}}{H} \quad \rightarrow \quad e^{Ht} = -\frac{1}{H\eta}
    \quad,
\eeq
so the scale factor becomes
\beq
    e^{\rho} = -\frac{1}{H \eta}
    \quad.
    \label{eq:de-Sitter}
\eeq

The \textit{Mukhanov-Sasaki potential} reads
\beq
    W_S^{(\text{dS})}(\eta) = \frac{z''}{z} = e^{-\rho} \frac{d^2}{d\eta^2} \left(e^\rho\right) = -H\eta \frac{d^2}{d\eta^2} \left(-\frac{1}{H\eta}\right) = \frac{2}{\eta^2}
    \quad,
\eeq
thus the mode equation is
\beq
    v_k''(\eta) + \left(k^2 - \frac{2}{\eta^2}\right) v_k(\eta) = 0
    \quad.
\eeq

Setting $x = -k\eta$ and defining $v(x) = \sqrt{x} u(x)$, we find
\beq
    \frac{d^2}{dx^2} v(x) = \sqrt{x} \frac{d^2}{dx^2} u(x) + \frac{1}{\sqrt{x}} \frac{d}{dx} u(x) - \frac{u(x)}{4 x^{3/2}}
    \quad,
\eeq
which reduces to the Bessel equation for $u(x)$,
\beq
    u''(x) + \frac{1}{x} u'(x) + \left(1 - \frac{9}{4 x^2}\right) u(x) = 0
    \quad,
\eeq
matching the general Bessel form
\beq
    y'' + \frac{1}{x} y' + \left(1 - \frac{\nu^2}{x^2}\right) y = 0
    \quad,
\eeq
with $\nu = \frac{3}{2}$. The general solution is a linear combination of Hankel functions:
\beq
    u(x) = c_1 H^{(1)}_{\frac{3}{2}}(x) + c_2 H^{(2)}_{\frac{3}{2}}(x)
    \quad.
\eeq
For $\nu=\frac{3}{2}$, these simplify to
\beq
    H^{(1,2)}_{\frac{3}{2}}(x) = \mp \sqrt{\frac{2}{\pi}} \frac{1}{\sqrt{x}} \left(1 \pm \frac{i}{x}\right) e^{\pm i x}
    \quad,
\eeq
yielding the original $v$ modes for pure dS:
\beq
    v_k^{(1,2)}(\eta) = \mp \sqrt{\frac{2}{\pi}} \left(1 \mp \frac{i}{k\eta}\right) e^{\mp i k \eta}
    \quad.
    \label{eq:v1v2}
\eeq

Boundary conditions select the integration constants. The usual assumption is that spacetime is locally flat at arbitrarily small scales, consistent with general relativity’s equivalence principle. However, the notion of "short distance" depends on the time of evaluation due to cosmic expansion, and the inflationary Hubble scale $H$ may approach the Planck scale, where a consistent theory combining quantum field theory and gravity is lacking.

The standard prescription defines modes so that in the infinite past, $t \rightarrow -\infty$, they match the positive-energy modes of flat space, defining the \textit{Bunch-Davies vacuum} associated with the free part of the action.

This condition fixes one integration constant by requiring that modes at very short distances ($k \rightarrow \infty$) resemble flat-space quantum modes. The vacuum state satisfies
\beq
    a_{\vec{k}} |0\rangle = 0
    \quad.
\eeq
This choice corresponds to the lowest-energy state in flat space and is consistent because at early times the Mukhanov-Sasaki equation reduces to a standard harmonic oscillator. In terms of the conformal time the unique solution for the ground state of the harmonic oscillator that gives the Bunch-Davies vaccum is 
\beq
    v_k(\eta) = \frac{1}{\sqrt{2k}} e^{-ik\eta}
    \quad .
    \label{eq:Bunch-Davies}
\eeq
To apply the Bunch-Davies condition, we also impose canonical quantization via the commutation relation between the field \( v \) and its conjugate momentum \( \pi \):
\beq
    \left[ v(t, \vec{x}), \pi(t, \vec{y}) \right] = i \delta^{(3)}(\vec{x} - \vec{y})
    \quad ,
\eeq
with
\beq
    \pi = \frac{\partial \mathcal{L}}{\partial v'} = v'
    \quad .
\eeq
In modes \( v_k \), this leads to the Wronskian condition
\beq
    v_k {v'_k}^* - v_k^* v'_k = i
    \quad ,
\eeq
which fixes the normalization in ~\eqref{eq:Bunch-Davies}. The unique normalized solution is
\beq
    v_k(\eta)=-\frac{\sqrt{\pi}}{2}\sqrt{- \eta}H_{\frac{3}{2}}^{(1)}(-k \eta)=\frac{1}{\sqrt{2k}}\left(1-\frac{i}{k \eta}\right)e^{-ik \eta}
    \quad .
    \label{eq:standardmode}
\eeq

For quasi-de Sitter (quasi-dS) with slow-roll parameters \( \varepsilon, \delta \ll 1 \), considered constant, the Mukhanov-Sasaki potential is
\bea
    W_S(\eta)&=&\frac{z''}{z} \sim (\rho ')^2(2+2 \varepsilon+3 \delta)
    \quad .
\eea
The scale factor evolves as
\beq
    \rho ' = -\frac{1+\varepsilon}{\eta} + O(\varepsilon^2)
    \quad ,
\eeq
and the Mukhanov-Sasaki equation reads:
\beq
    v''_k(\eta)+\left(k^2-\frac{2+3(2\varepsilon+\delta)}{\eta^2}\right)v_k(\eta)=0
    \quad .
\eeq
Its solutions involve Hankel functions with order
\bea
    \nu &=& \frac{3}{2} + 2\varepsilon + \delta
    \quad ,
\eea
recovering \(\nu = \frac{3}{2}\) for pure dS. Imposing Bunch-Davies selects the Hankel function of the first kind:
\beq
    v_k(\eta)=e^{i \frac{\pi}{2}\left(\nu + \frac{1}{2}\right)}\frac{\sqrt{\pi}}{2}\sqrt{- \eta}H_{\nu}^{(1)}(-k \eta)
    \quad .
    \label{eq:modef1}
\eeq

At late times \((-k \eta \ll 1)\), the Mukhanov-Sasaki equation simplifies to
\beq
    v''_k(\eta)-\frac{2+3(2 \varepsilon+\delta)}{\eta^2}v_k(\eta)=0
    \quad ,
\eeq
with solutions
\beq
    v_k(\eta) = c_1(k) \eta^{\frac{1+\sqrt{1+4(2+3(2\varepsilon+\delta))}}{2}} + c_2(k) \eta^{\frac{1-\sqrt{1+4(2+3(2\varepsilon+\delta))}}{2}}
    \quad .
\eeq
Only the second solution dominates late times:
\beq
    v_k (\eta) \propto \eta^{-1}
    \quad .
\eeq

\begin{figure}
    \centering
    \includegraphics[width=0.35\linewidth]{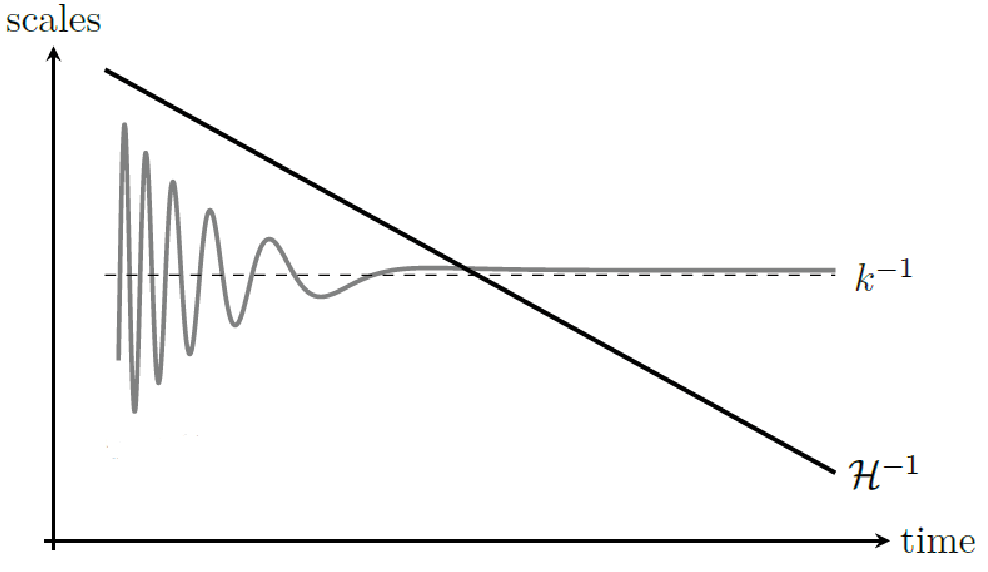}
    \caption{$\zeta$ freezing on superhorizon scales. Notice that the horizon is denoted with $\mathcal{H}$. This picture is taken from \cite{Baumann:Cosmology2022}.}
    \label{fig:freezing}
\end{figure}

Since \(\zeta = e^{-\rho}(2\varepsilon)^{-1/2} v\) and \(e^{-\rho} \propto \eta\), \(\zeta\) approaches a constant at late times, meaning scalar perturbations \(\zeta\) are frozen (Fig.~\ref{fig:freezing}). This occurs in the superhorizon limit \((-k \eta \ll 1)\), which physically means modes have stretched beyond the Hubble horizon \(1/\dot{\rho}\) during inflation.

Writing the physical momentum
\beq
    k_{\text{phys}} = e^{-\rho}k
    \quad ,
\eeq
the superhorizon condition is
\beq
    e^{-\rho}k \ll H \quad \Rightarrow \quad -k \eta \ll 1
    \quad .
\eeq

\subsection{Chibisov-Mukhanov Power Spectrum}
Using the exact mode functions, we can evaluate the two-point correlation function for the fluctuations:
\beq
    \langle 0(\eta) | \zeta(\eta, \vec{x}) \zeta(\eta, \vec{y}) | 0(\eta) \rangle
    \quad .
\eeq
This is often expressed via the power spectrum, its Fourier transform with conventional factors,
\beq
    \langle 0(\eta) | \zeta(\eta, \vec{x}) \zeta(\eta, \vec{y}) | 0(\eta) \rangle = \int \frac{d^3\vec{k}}{(2\pi)^3} e^{i \vec{k} \cdot (\vec{x} - \vec{y})} \frac{2\pi^2}{k^3} \mathcal{P}_k(\eta)
    \quad .
\eeq
Using the relation between \(\zeta\) and \(v\), the power spectrum is
\beq
    \mathcal{P}_k(\eta) = e^{-2\rho} \frac{\dot{\rho}^2}{\dot{\phi}^2} \frac{k^3}{2\pi^2} |v_k|^2
    \quad .
\eeq
Replacing \(\dot{\rho}^2 / \dot{\phi}^2\) with \(1/(2 M_{\text{pl}}^2 \varepsilon)\) and \(e^{-2\rho}\) by
\beq
    e^{-2\rho} = \frac{(-H\eta)^2}{(1 + \varepsilon)^2} + \cdots
    \quad ,
\eeq
the leading power spectrum reads (using \(H^{(1)*} = H^{(2)}\))
\beq
    \mathcal{P}_k(\eta) = \frac{H^2}{M_{\text{pl}}^2} \frac{(-k\eta)^3}{16\pi} H^{(1)}_\nu(-k\eta) H^{(2)}_\nu(-k\eta) \frac{1}{\varepsilon (1 + \varepsilon)^2}
    \quad .
\eeq

In the physical limit \(-k\eta \ll 1\), using Hankel asymptotics:
\bea
    H^{(1)}_\nu(x) &\sim& -\frac{i}{\pi}\Gamma(\nu)\left(\frac{x}{2}\right)^{-\nu} \\
    H^{(2)}_\nu(x) &\sim& +\frac{i}{\pi}\Gamma(\nu)\left(\frac{x}{2}\right)^{-\nu}
    \quad ,
    \label{eq:hankellimit}
\eea
and approximating \(\Gamma(\nu) \sim \frac{3}{2}\), \(2^{-\nu} \sim 2^{-\frac{3}{2}}\), we find
\beq
    \mathcal{P}_k = \frac{1}{8\pi^2} \frac{1}{\varepsilon} \frac{H^2}{M_{\text{pl}}^2} (-k\eta^*)^{3 - 2\nu} \sim \frac{1}{8\pi^2} \frac{1}{\varepsilon} \frac{H^2}{M_{\text{pl}}^2} (-k\eta^*)^{-4\varepsilon - 2\delta}
    \quad .
\eeq
Introducing a reference scale \(k_\ast\), we write
\beq
    \mathcal{P}_k = \frac{1}{8\pi^2} \frac{1}{\varepsilon} \frac{H^2}{M_{\text{pl}}^2} (-k_\ast \eta^*)^{-4\varepsilon - 2\delta} \left(\frac{k}{k_\ast}\right)^{-4\varepsilon - 2\delta}
    \quad .
    \label{eq:power spectrum}
\eeq
This form is used in data analysis, fitting a power-law spectrum,
\beq
    \mathcal{P}_k \equiv \mathcal{C}_s(k_\ast) \left(\frac{k}{k_\ast}\right)^{n_s - 1}
    \quad ,
\eeq
where the amplitude and tilt are
\beq
    \mathcal{C}_s(k_\ast) \approx \frac{1}{8\pi^2} \frac{1}{\varepsilon} \frac{H^2}{M_{\text{pl}}^2},
    \quad
    n_s = 1 - 4\varepsilon - 2\delta
    \quad .
\eeq
This \textit{Chibisov-Mukhanov power spectrum} \cite{Mukhanov:1981xt,Mukhanov:1990me} shows that scalar fluctuations in quasi-dS are nearly scale-invariant (\(n_s \sim 1\)) with small corrections from slow-roll parameters. Observations measure \(n_s \sim 0.96\). In pure dS, slow-roll parameters vanish, and the power spectrum is exactly scale-invariant. This near scale-invariance at late times is crucial for conformal treatments of dS on the boundary.
\section{Mukhanov-Sasaki Equation for the Climbing Scalar}
The climbing scalar model naturally triggers inflation through a pre-inflationary climbing phase (fast roll) that transitions to slow-roll, allowing inflation to start even with the inflaton off its attractor. This section studies general implications of such dynamics on cosmological perturbations and their signatures in the CMB power spectrum.

Restricting to two-exponential potentials simplifies the analysis while capturing key features, particularly the scalar field’s reversal. This setup can induce suppression of power at large angular scales (low multipoles), more pronounced for larger initial scalar values \(\varphi_0\), due to interaction with the potential’s “hard wall”~\cite{DudasKitazawaPatilSagnotti:2012}.

A suppressed quadrupole and power deficit at large scales are generic in cosmologies from an initial singularity. Climbing dynamics also produce residual peaks and oscillations at low \(\ell\), consistent with hints in WMAP9 and Planck data~\cite{Gruppuso2013}.

Studying these effects constrains \(\varphi_0\) and links climbing to string theory, where it appears in orientifold models~\cite{Sagnotti:1987Cargese,Pradisi:1989zz,Horava:1989fv,Horava:1989bg,BianchiSagnotti:1990,BianchiSagnotti:1991a,Bianchi1992,Sagnotti:1992note,Dudas:2000review,Angelantonj:2002ct,AngelantonjFlorakis:2024} with SUSY breaking tied to the inflationary scale~\cite{Dudas:2010climbing}. However, since climbing occurs near the initial singularity, curvature corrections and string approximations may break down, requiring caution~\cite{MouradSagnotti2021}.

As recalled earlier, the Mukhanov-Sasaki variable \(v\) reduces the problem to a time-dependent effective potential:
\beq
    W_S = \frac{z''}{z}
    \quad .
\eeq
A closed form in conformal time \(\eta\) is generally unavailable for climbing scalar, as \(W_S\) depends on both the scalar field and scale factor, and the relation \(\eta(\tau)\) is not integrable even with a single exponential \(V(\varphi)\).

Of particular interest is the LM attractor solution~\eqref{eq:LM3}
with scale factor~\eqref{eq:LMscalefactor} (denoted here by \(s\) since \(a\) is reserved for the logarithm):
\beq
    s(\eta) = \left(\frac{\sqrt{6(1-\gamma^2)}}{M (1 - 3 \gamma^2)} e^{-\gamma \varphi_0}\right)^{\frac{1}{1 - 3 \gamma^2}} (-\eta)^{-\frac{1}{1 - 3 \gamma^2}},
\eeq
and the slow-roll parameter is constant $\varepsilon = 3 \gamma^2$.
    
The Mukhanov-Sasaki potential simplifies to:
\beq
    \frac{z''}{z} = \frac{s''}{s} = \frac{2 - 3 \gamma^2}{(1 - 3 \gamma^2)^2} \frac{1}{\eta^2}
    \quad .
\eeq
Note that the Mukhanov-Sasaki potential for tensor perturbations has the same form. In the limit $\gamma \ll 1$, it reduces to the standard de Sitter case:
\beq
    W_S(\eta)=\frac{2}{\eta^2}
    \quad .
\eeq
The Mukhanov-Sasaki equation is exactly solvable in the LM case for arbitrary $\gamma$, introducing:
\beq
    \nu^2-\frac{1}{4}=\frac{2-3\gamma^2}{(1-3\gamma^2)^2} \qquad \to \qquad \nu = \frac{3}{2}\frac{1-\gamma^2}{1-3\gamma^2}
    \quad .
    \label{eq:nuvalue}
\eeq
It then takes the standard Bessel form:
\beq
    v''_k (\eta)+\left(k^2-\frac{\nu^2-\frac{1}{4}}{\eta^2}\right)v_k(\eta)=0
    \quad ,
    \label{eq:MSeqgeneral}
\eeq
with Bunch-Davies solution again given by~\eqref{eq:modef1}:
\beq
     v_k(\eta)=e^{i \frac{\pi}{2}\left(\nu + \frac{1}{2}\right)}\frac{\sqrt{\pi}}{2}\sqrt{- \eta}H_{\nu}^{(1)}(-k \eta)
     \label{eq:LMmode}
\eeq
For small $\gamma$, as in the mild exponential case, we recover $\nu = \frac{3}{2}$ and thus the standard de Sitter result.

However, the LM approximation misses key aspects of the dynamics. To capture the full physical picture—especially its impact on the CMB angular power spectrum—more general methods are needed. We now turn to these.
\subsection{Schr\"odinger Analogy}

\begin{figure}
    \centering    \includegraphics[width=0.35\linewidth]{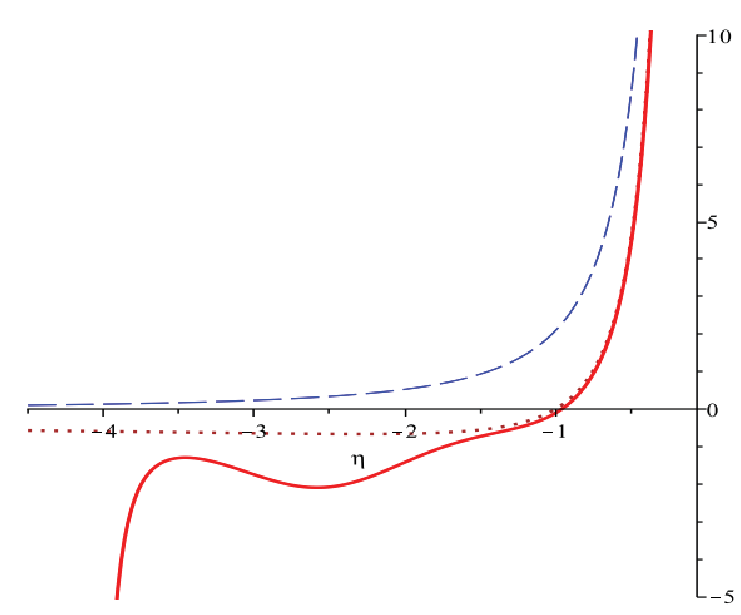}
    \caption{Comparison between attractor $W_S$ (dashed blue), two-exponential $W_S$ (continuous red)
    and the approximate form of ~\eqref{eq:deltaMS} (dotted brown). This picture is taken from~\cite{DudasKitazawaPatilSagnotti:2012}.}
    \label{fig:MSpotential2}.
\end{figure}

To analyze CMB power spectrum modifications in climbing scalar models, one can exploit the analogy between the Mukhanov–Sasaki (MS) equation and a one-dimensional Schrödinger equation using WKB methods~\cite{landau1977quantum}.

The MS equation from eq.~\eqref{eq:MS-equation} becomes:
\beq
    \frac{d^2}{d\eta^2} v_k(\eta) + p^2(k, \eta) v_k(\eta) = 0 \quad,
\eeq
with $p^2(k, \eta) = k^2 - W_S(\eta)$, 
where $W_S(\eta)$ is given in eq.~\eqref{eq:scalarpotential}. Its asymptotics are \cite{DudasKitazawaPatilSagnotti:2012,KitazawaSagnotti:2014}:
\beq
    W_S(\eta) \xrightarrow{\eta \to -\eta_s} -\frac{1}{4(\eta + \eta_s)^2} \quad ,
    \label{eq:singular-potential}
\eeq
\beq
    W_S(\eta) \xrightarrow{\eta \to 0} \frac{\nu^2 - \frac{1}{4}}{\eta^2} \quad .
    \label{eq:attractorpotential}
\eeq

Due to these limits, $W_S(\eta)$ crosses the real axis for $\eta \in (-\eta_s, 0)$, causing infrared suppression in the CMB power spectrum. Note the finite time singularity which presence will become crucial later.

The Schrödinger analogy includes a potential barrier at early times that, classically, would imply total reflection. However, to recover a Bunch-Davies-like vacuum near the singularity, solutions behave like plane waves in regions where $W_S$ is small, transitioning to growing and decaying modes. This leads to perturbation growth after horizon crossing.

In the classically forbidden region, the WKB approximation~\cite{landau1977quantum} gives:
\beq
    v_k(-\epsilon) \sim \frac{1}{\sqrt{2|p(-\epsilon)|}} \exp\left( \int_{-\eta^*}^{-\epsilon} |p(y)| \, dy \right) \quad,
\eeq
where $\epsilon \ll 1$ marks the end of inflation, and $\eta_*$ satisfies $W_S(-\eta_*) = k^2$.

For $\eta \to -\epsilon$, the LM attractor solution~\eqref{eq:LMmode} and Hankel limit~\eqref{eq:hankellimit} yield:
\beq
    v_k(\eta) \approx \frac{1}{\sqrt{2\pi}} e^{i\frac{\pi}{2}(\nu - \frac{1}{2})} \Gamma(\nu) k^{-\nu} \left( \frac{\epsilon}{2} \right)^{\frac{1}{2} - \nu} \quad,
\eeq
and the resulting power spectrum is:
\beq
    \mathcal{P}_k \propto k^{3 - 2\nu} \quad.
\eeq
The leading WKB approximation is:
\beq
    v_k(\eta) \sim \frac{1}{\sqrt{2}} \epsilon^{\frac{1}{2}-\tilde{\nu}}\left( \frac{2}{k} \right)^{\tilde{\nu}} \tilde{\nu}^{\tilde{\nu}-\frac{1}{2}} e^{-\tilde{\nu}} \quad,
\eeq
with $\tilde{\nu} = \sqrt{\nu^2 - \frac{1}{4}}$.

This approximation explains the qualitative spectrum behavior. Infrared suppression arises from the limited amplification due to negative $W_S$ at early times, where the $k^3$ factor dominates $\mathcal{P}_k$.

The dip in $W_S(\eta)$ near the scalar reversal induces oscillations, similar to quantum tunneling. Fig.~\ref{fig:MSpotential2} shows $W_S(\eta)$ for both the climbing solution and the LM attractor from eq.~\eqref{eq:doubleexp}. The climbing solution starts from a singularity, has a vertical asymptote, and approaches the attractor late.

Since $W_S(\eta)$ for the climbing case stays below the attractor curve, the WKB integral is smaller due to both a delayed turning point and reduced area under $|p(\eta)|$. Thus, the amplification is weaker, especially at small $k$. At large $k$, $W_S(\eta)$ aligns with the attractor, and spectra converge.

Though not analyzed in full detail, tensor spectra behave similarly: a mild infrared enhancement, followed by suppression as $W_T(\eta)$ crosses the real axis. This is expected from comparing scalar and tensor potentials $W_S \sim z''/z$ and $W_T \sim s''/s$, both depending on the inflaton velocity and $V$.

The ratio $W_S / W_T$ equals one at the limiting speed~\eqref{eq:limitspeed}, and drops below one for larger inflaton velocities in two-term potentials. Near the attractor, this ratio is symmetric, and deviations grow for $\gamma < 1/\sqrt{3}$, enabling slow-roll phases.
\subsection{Analytic Models for \texorpdfstring{$W_S(\eta)$}{WS}}
To understand qualitatively the power spectra from different cosmological evolutions, we study analytically tractable forms for the MS scalar and tensor potentials \(W_S\) and \(W_T\). Ignoring the Big Bang singularity, a key class involves deformations of the attractor potential \eqref{eq:attractorpotential}\eqref{eq:nuvalue}
which describes the leading behavior of MS potentials in the two-exponential model.

We focus on potentials of the form:
\beq
    W_S = \frac{\nu^2 - \frac{1}{4}}{\eta^2} \left[ c \left(1 + \frac{\eta}{\eta^*} \right) + (1 - c) \left(1 + \frac{\eta}{\eta^*} \right)^2 \right]
    \quad ,
    \label{eq:cpot}
\eeq
where \(\eta^*\) locates the zero and \(c\) controls physical behavior. For \(c > 1\), \(W_S(\eta)\) asymptotically approaches zero from below, modeling the two-exponential potential’s flat plateau. A transition at \(c=2\) separates potentials below (for \(1 < c < 2\)) or above (for \(c > 2\)) the LM attractor late-time behavior.

The MS equation becomes
\beq
    \frac{d^2 v_k}{d\eta^2} + \left[ k^2 - (1 - c)\frac{\nu^2 - \frac{1}{4}}{(\eta^*)^2} - (2 - c)\frac{\nu^2 - \frac{1}{4}}{\eta^* \eta} - \frac{\nu^2 - \frac{1}{4}}{\eta^2} \right] v_k = 0.
\eeq

Defining
\beq
    \alpha = \sqrt{(k\eta^*)^2 + (c - 1)\left(\nu^2 - \frac{1}{4}\right)}
    \quad , 
\eeq
the equation can be recast as
\beq
    \frac{d^2 v_k}{d\eta^2} + \left[ \frac{\alpha^2}{(\eta^*)^2} - (2 - c)\frac{\nu^2 - \frac{1}{4}}{\eta^* \eta} - \frac{\nu^2 - \frac{1}{4}}{\eta^2} \right] v_k = 0
    \quad .
\eeq

Introducing
\beq
    \rho = -\frac{\eta \alpha}{\eta^*}
    \quad ,
\eeq
the MS equation reads
\beq
    \frac{d^2 v_k}{d\rho^2} + \left[ 1 + \frac{(2 - c)(\nu^2 - \frac{1}{4})}{\alpha \rho} - \frac{\nu^2 - \frac{1}{4}}{\rho^2} \right] v_k = 0
    \quad .
\eeq

This is the Coulomb wave equation with
\beq
    L(L + 1) = \nu^2 - \frac{1}{4}, \quad L = \nu - \frac{1}{2}, \quad \beta = -\frac{(2 - c)(\nu^2 - \frac{1}{4})}{2\alpha} \quad .
\eeq

Thus,
\beq
    \frac{d^2 v_k}{d\rho^2} + \left[1 - \frac{2\beta}{\rho} - \frac{L(L + 1)}{\rho^2} \right] v_k = 0
    \quad ,
\eeq
is the standard Coulomb form ~\cite{abramowitz1964handbook}.

The normalized solution is
\beq
    v_k \sim \frac{1}{\sqrt{\alpha}} \left(F_L(\beta, \eta) + i G_L(\beta, \eta)\right)
    \quad ,
    \label{eq:coulombsol}
\eeq
where \(F_L\) and \(G_L\) are Coulomb wavefunctions. For \(\rho \to 0\), \(G_L\) dominates, yielding
\beq
    \mathcal{P}_k\sim \frac{k^3}{\alpha} \left| \epsilon^{\frac{1}{1-3\gamma^2}} G_L(-\epsilon) \right|^2
    \quad .
\eeq

Using \(G_L\) asymptotics,
\beq
    \mathcal{P}_k \sim k^3 e^{\pi \beta} \frac{|\Gamma(L + 1 + i\beta)|^{-2}}{\alpha^{2L + 1}}
    \quad ,
\eeq
which in original variables becomes
\beq
    \mathcal{P}_k \sim \frac{(k\eta^*)^3 \exp\left[ \frac{\pi(c/2 - 1)(\nu^2 - \frac{1}{4})}{\sqrt{(k\eta^*)^2 + (c - 1)(\nu^2 - \frac{1}{4})}} \right]}{ \left| \Gamma\left[ \nu + \frac{1}{2} + \frac{i(c/2 - 1)(\nu^2 - \frac{1}{4})}{\sqrt{(k\eta^*)^2 + (c - 1)(\nu^2 - \frac{1}{4})}} \right] \right|^{2} \, \left[(k\eta^*)^2 + (c - 1)(\nu^2 - \frac{1}{4})\right]^\nu } \quad .
\eeq

\begin{figure}
    \centering    \includegraphics[width=0.35\linewidth]{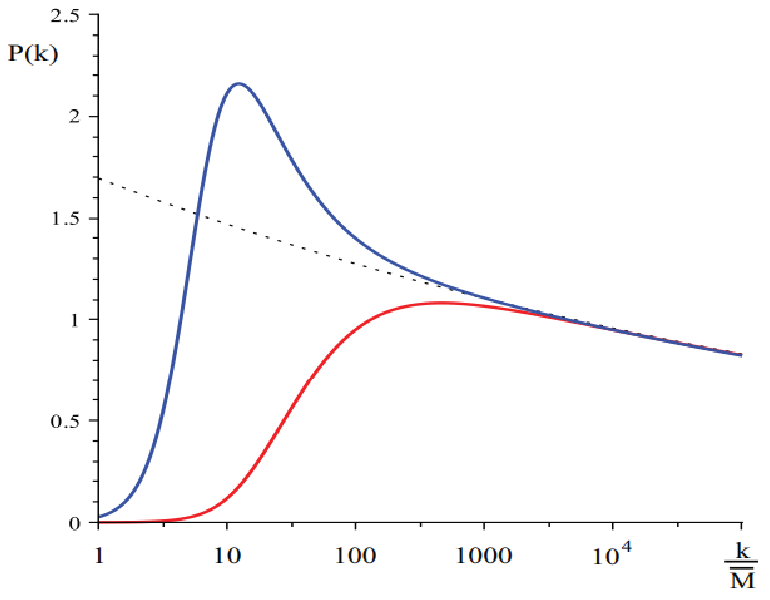}
    \caption{Analytic scalar (red) and tensor (blue) spectra vs attractor spectrum (dotted). This picture is taken from~\cite{DudasKitazawaPatilSagnotti:2012}.}
    \label{fig:Power1}
\end{figure}

This expression combines the attractor behavior \( \mathcal{P}(k) \sim k^{3 - 2\nu} \) at high \(k\) with infrared modifications controlled by \(\eta_1\) and \(c\). Fig.\ref{fig:Power1} shows scalar (red) and tensor (blue) spectra, displaying mild infrared suppression followed by enhancement, matching predicted features.

The leading WKB approximation reproduces high-frequency behavior well for large \(\nu\).

A simpler model corresponds to \(c=2\), with
\beq
    W_S(\eta)=\frac{\nu^2 - \frac{1}{4}}{\eta^2}-\Delta^2
    \label{eq:deltaMS}
\eeq
where \(\Delta\) introduces a new scale, shifting the MS potential down. This modification yields the same mode functions as quasi-dS but with
\beq
    k^2 \to \sqrt{k^2+\Delta^2}
    \label{eq:k-shift}
\eeq
so that
\beq
    v_k(\eta)=e^{i \frac{\pi}{2}\left(\nu + \frac{1}{2}\right)}\frac{\sqrt{\pi}}{2}\sqrt{- \eta}H_{\nu}^{(1)}(-\omega \eta)
     \label{eq:deltamode1}
     \qquad \text{with} \qquad \omega \equiv \sqrt{k^2+\Delta^2}
\eeq

The power spectrum becomes
\beq
    \mathcal{P}_k = C \frac{k^3}{\left(k^2+\Delta^2\right)^\nu}
    \label{eq:deltapowerspectrum}
\eeq
exhibiting the anticipated low-frequency suppression. This simple model will be key for studying non-Gaussian features.

Varying \(c\) changes the growth rate of the spectrum, allowing lowering, enhancement, or overshoot compared to the standard Chibisov-Mukhanov result. Additionally, local departures of \(W_S\) from its attractor shape induce oscillations superposed on these effects. First-order Schwinger-Keldysh perturbation theory (discussed next chapter) explains this. Coulomb-like potentials deviate over an infinite domain, so oscillations tend to be suppressed in eq.~\eqref{eq:deltapowerspectrum}.

\subsection{Power Spectrum}
This section reports results about numerical simulations of the power spectrum from~\cite{DudasKitazawaPatilSagnotti:2012,KitazawaSagnotti:2014}.

Simulations reveal scalar power suppression at long wavelengths and slight tensor enhancement, with superposed oscillations. Short-wavelength spectra match attractor solutions. The plots (e.g.,~\cite{DudasKitazawaPatilSagnotti:2012}) cover about 12 e-folds; WMAP observes only $\sim 6$, yet traces of the climbing phase may appear if the largest observable modes exited the horizon within $6$–$7$ e-folds after the climbing ended. The approach to the attractor takes longer than the onset of inflation itself (about one string time).

Key signatures of the initial singularity are:

\begin{itemize}
    \item[(a)] Inevitable low-frequency power suppression, varying from $\mathcal{O}(k^3)$ to milder or stronger~\cite{DudasKitazawaPatilSagnotti:2012,Destri2010,Piao2004,Piao2005,Jain2009,Jain2010,Liu2013}.
    \item[(b)] Possible overshoot if $W_S$ emerges more steeply than the attractor.
    \item[(c)] Localized Mukhanov-Sasaki potential perturbations induce localized oscillations in $k$-space, akin to those during fast-roll to slow-roll transitions~\cite{Destri2010,Piao2004,Piao2005,Jain2009,Jain2010,Liu2013}.
\end{itemize}

These oscillations have been numerically modeled using square-well perturbations to a Coulomb-like $W_S$, producing localized features in $k$-space as shown in~\cite{KitazawaSagnotti:2014}. These arise from equal-length perturbations centered on progressively smaller $\eta$ values.

Two-exponential potentials (eq.~\eqref{eq:doubleexp}) display incomplete transitions to slow roll, a phenomenon not emphasized in~\cite{DudasKitazawaPatilSagnotti:2012} but made manifest in~\cite{KitazawaSagnotti:2014}. In this case, the scalar field first climbs a mild exponential and then hits a hard exponential wall, often bouncing unless $\varphi_0$ is very negative. In that limit, the first term becomes negligible and the spectra coincide with the single-exponential case.

\begin{figure}
    \centering
    \includegraphics[width=0.5\linewidth]{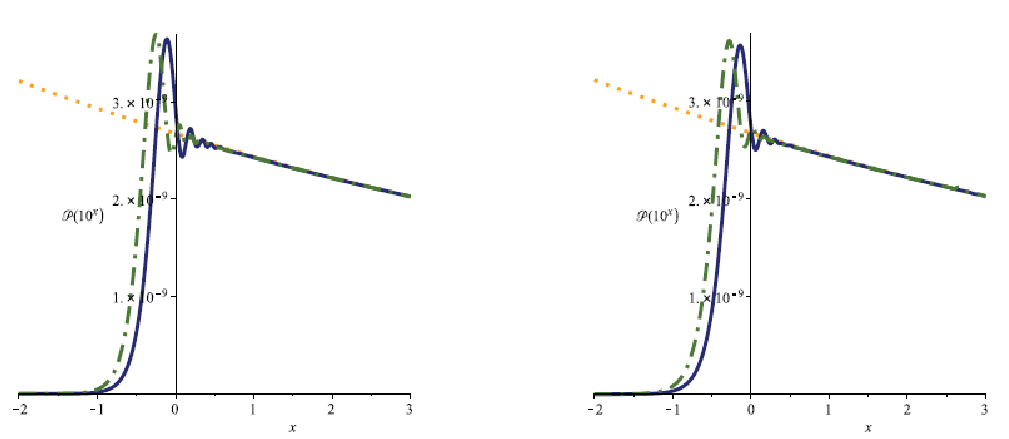}
    \caption{Scalar perturbation spectra for single-exponential potential (eq.~\eqref{eq:singleexp}): descending scalar $\varphi_0=0,-4$ (left) and climbing scalar $\varphi_0=0,-4$ (right). Similar imprints, typical of fast-roll to slow-roll transitions~\cite{Destri2010,Piao2004,Piao2005,Jain2009,Jain2010,Liu2013}. Increasing $\varphi_0$ shifts transitions to higher $k$. COBE-normalized at $k=38$, horizontal axis $k=10^x$. From~\cite{KitazawaSagnotti:2014}.}
    \label{fig:Power6}
\end{figure}

Figure~\ref{fig:Power6} (from~\cite{KitazawaSagnotti:2014}) shows the scalar perturbation spectra for single-exponential potentials. Both descending and climbing initial conditions (with $\varphi_0 = 0, -4$) result in similar features—oscillations associated with fast-roll to slow-roll transitions—almost independent of $\varphi_0$. This confirms the universality of these features, as also discussed in~\cite{Destri2010,Piao2004,Piao2005,Jain2009,Jain2010,Liu2013}.

For double-exponential potentials,~\cite{KitazawaSagnotti:2014} shows that for large $\varphi_0$, the scalar bounces against the “hard wall,” delaying slow-roll onset and suppressing the power spectrum over several decades in $k$. At intermediate $\varphi_0$, after the bounce, a brief slow-roll phase occurs, producing a pre-inflationary peak superimposed on a rising baseline. This peak coexists with the main slow-roll transition and depends sensitively on the initial condition.

Tensor spectra in~\cite{KitazawaSagnotti:2014} exhibit only a minor overshoot at $\varphi_0 = 0$, fading rapidly for smaller $\varphi_0$, as expected from their dependence on the scale factor evolution alone.

\subsection{Comparison with CMB Data}
Although this work does not focus on observational tests, it is important to highlight how the climbing-scalar scenario connects with CMB observations, especially at large angular scales where anomalies have been detected. The climbing phase, motivated by string theory, predicts a suppression of power in the primordial spectrum at low \(k\), which translates into a reduced CMB power \(C_\ell\) at low multipoles \(\ell\).

This suppression is described by the angular power spectrum
\beq
C_{\ell} \simeq \frac{2}{9\pi} \int \frac{dk}{k} \, \mathcal{P}_k \, j_\ell^2 \bigl[ k(\eta_0 - \eta_r) \bigr] \quad,
\eeq
where \(\mathcal{P}_k\) encodes the primordial fluctuations shaped by the climbing phase. The effect becomes significant if inflation starts close (within a few e-folds) to the horizon exit of the largest observable modes, and can exceed cosmic variance,
\beq
\frac{\Delta C_{\ell}}{C_{\ell}} = \sqrt{\frac{2}{2\ell + 1}} \quad,
\eeq
making it potentially observable.

This suppression is effectively modeled by introducing a low-\(k\) cutoff \(\Delta\) in the primordial spectrum, fitted to Planck data as
\beq
\Delta = (0.351 \pm 0.114) \times 10^{-3} \, \mathrm{Mpc}^{-1},
\eeq
which matches the expected energy scale of the climbing-to-inflation transition in string theory. This cutoff naturally explains the long-known large-scale power deficit seen in CMB data from COBE, WMAP, and Planck.

Besides suppression, the climbing model may produce oscillations in the spectrum, potentially improving fits to the full CMB data. While a comprehensive statistical analysis is still pending, the qualitative agreement between the climbing scalar predictions and observed anomalies suggests this model as a compelling link between fundamental theory and cosmological observations.

In short, the climbing scalar scenario provides a natural, non-fine-tuned explanation for large-scale CMB anomalies, making it a strong candidate for pre-inflationary physics rooted in string theory.

\section{Derivation of the Cubic Action \texorpdfstring{$S^{(3)}$}{S(3)}}
The three-point function comes from cubic interactions, so the action is expanded to third order in \(\zeta\). A direct expansion yields terms that obscure the true \(\varepsilon^2\) suppression. Alternative coordinates make this suppression clearer but are less effective for super-horizon fluctuations. Thus, \(\zeta(t, \vec{x})\), which remains nearly constant outside the horizon and relates to post-inflation spatial curvature fluctuations, is preferred.

The goal is to rewrite the cubic action in a simpler form containing only terms suppressed by \(\varepsilon^2\). This involves extensive integration by parts and careful reorganization into three groups: total derivatives and terms vanishing under the background equation
\beq
    \ddot{\rho} = - \frac{1}{2} \dot{\phi}^2,
\eeq
terms proportional to the linear equation of motion for \(\zeta\),
\beq
    \frac{d}{dt} \left[ e^{3\rho} \frac{\dot{\phi}^2}{\dot{\rho}^2} \dot{\zeta} \right] - e^{\rho} \frac{\dot{\phi}^2}{\dot{\rho}^2} \partial_k \partial^k \zeta=0,
\eeq
and finally the genuinely dynamical terms.
The second set does contribute in part to the three-point function, but it will be treated separately when computing the three-point function by means of a nonlinear shift in the field. The final set of terms are those that do not fall into either of these two classes. This set will be seen to be clearly order \( \varepsilon^2 \) and are therefore small in the slow-roll limit.
To begin, start once more with the coordinates that were introduced during the derivation of the quadratic action,
\beq
    N = 1 + \frac{\dot{\zeta}}{\dot{\rho}}, \quad N_i = \delta_{ij} \partial^j B, \quad h_{ij} = e^{2\rho + 2\zeta} \delta_{ij}, \quad \phi= \phi(t)
    \quad ,
\eeq
where one of the Lagrange constraints has already been used to define \( N \) in terms of \( \zeta \). As before, the action for this system is
\beq
    S = \frac{1}{2} \int d^4 x \sqrt{h} \left[ N \hat{R} + N^{-1} \left( E_{ij} E^{ij} - E^2 \right) - 6 N \dot{\rho}^2 + \left( N + N^{-1} \right) \dot{\phi}^2 \right]
    \quad ,
\eeq
once the background equation has been imposed to remove the explicit appearance of the potential \( V(\phi) \).
The exact expressions for \( \hat{R} \) and \( E_{ij} \) in these coordinates are
\beq
    \hat{R} = -4 e^{-2\rho} e^{-2\zeta} \partial_k \partial^k \zeta - 2 e^{-2\rho} e^{-2\zeta} \partial_k \zeta \partial^k \zeta
    \quad ,
\eeq
and
\beq
    E_{ij} = e^{2\rho + 2\zeta} \left[ \left( \dot{\rho} + \dot{\zeta} - \partial_k \zeta \partial^k B \right) \delta_{ij} - \partial_i \partial_j B \right]
    \quad ,
\eeq
so that
\bea
    E_{ij} E^{ij} - E^2 &=& -6 \left( \dot{\rho} + \dot{\zeta} - \partial_k \zeta \partial^k B \right)^2 + 4 \left( \dot{\rho} + \dot{\zeta} - \partial_k \zeta \partial^k B \right) \partial_l \partial^l B + \nonumber \\ &+& \left( \partial_k \partial_l B \partial^k \partial^l B - (\partial_k \partial^k B)^2 \right)
    \quad .
\eea
The rest of the factors need also to be expanded to third order,
\beq
    \sqrt{h} = e^{3\rho} e^{3\zeta} = e^{3\rho} \left( 1 + 3 \zeta + \frac{9}{2} \zeta^2 + \frac{9}{2} \zeta^3 + \cdots \right)
    \quad ,
\eeq
and
\beq
    \frac{1}{N} = 1 - \frac{\dot{\zeta}}{\dot{\rho}} + \frac{\dot{\zeta}^2}{\dot{\rho}^2} - \frac{\dot{\zeta}^3}{\dot{\rho}^3} + \cdots
    \quad .
\eeq
Expanding, for example, the first term in the Lagrangian produces the following set of third-order terms,
\beq
    \left. \sqrt{h} N \hat{R} \right|^{(3)} = -2 e^{\rho} \zeta^2 \partial_k \partial^k \zeta - 2 e^{\rho} \zeta \partial_k \zeta \partial^k \zeta - 4 e^{\rho} \frac{1}{\dot{\rho}} \dot{\zeta} \zeta \partial_k \partial^k \zeta - 2 e^{\rho} \frac{1}{\dot{\rho}} \dot{\zeta} \partial_k \zeta \partial^k \zeta
    \quad .
\eeq
The goal is to convert the action into a form where many of the leading contributions vanish because they are total derivatives. To start this process, convert this set of four terms into a single (nonderivative) term by noticing that the last two terms resemble the first two, once they have been integrated by parts to remove the time derivatives from the \( \zeta \) fields,
\beq
    -4 e^{\rho} \frac{1}{\dot{\rho}} \dot{\zeta} \zeta \partial_k \partial^k \zeta - 2 e^{\rho} \frac{1}{\dot{\rho}} \dot{\zeta} \partial_k \zeta \partial^k \zeta = 2 e^{\rho} \zeta^2 \partial_k \partial^k \zeta + 2 e^{\rho} \zeta \partial_k \zeta \partial^k \zeta - 2 e^{\rho} \frac{\ddot{\rho}}{\dot{\rho}^2} \zeta^2 \partial_k \partial^k \zeta \nonumber
\eeq
\beq
    - 2 e^{\rho} \frac{\ddot{\rho}}{\dot{\rho}^2} \zeta \partial_k \zeta \partial^k \zeta- \frac{d}{dt} \left( 2 e^{\rho} \frac{1}{\dot{\rho}} \zeta^2 \partial_k \partial^k \zeta + e^{\rho} \frac{1}{\dot{\rho}} \zeta \partial_k \zeta \partial^k \zeta \right) + \partial_k \left( 2 e^{\rho} \frac{1}{\dot{\rho}} \zeta^2 \partial^k \dot{\zeta} \right)
    \quad .
\eeq
This can be easily check expanding directly.
Inserted back into the previous equation, and integrating it by parts once more, yields a single term that is not a total derivative,
\beq
    \left. \sqrt{h} N \hat{R} \right|^{(3)} = 2 e^{\rho} \frac{\ddot{\rho}}{\dot{\rho}^2} \zeta \partial_k \zeta \partial^k \zeta - \frac{d}{dt} \left( 2 e^{\rho} \frac{1}{\dot{\rho}} \zeta \partial_k (\zeta \partial^k \zeta) \right) + \partial_k \left( 2 e^{\rho} \frac{1}{\dot{\rho}^2} \zeta^2 (\dot{\rho} \partial^k \dot{\zeta} - \ddot{\rho} \partial^k \zeta) \right)
    \quad .
\eeq
The background equation, \( \ddot{\rho} = -\frac{1}{2} \dot{\phi}^2 \), will be assumed frequently in this calculation; for example, the first term in this expression can be rewritten as
\bea
    &&\sqrt{h} N \hat{R} \bigg|^{(3)} = - e^{\rho} \frac{\dot{\phi}^2}{\dot{\rho}^2} \zeta \partial_k \zeta \partial^k \zeta + e^{\rho} \frac{1}{\dot{\rho}^2} \left( 2 \ddot{\rho} + \dot{\phi}^2 \right) \zeta \partial_k \zeta \partial^k \zeta - \frac{d}{dt} \left[ 2 e^{\rho} \frac{1}{\dot{\rho}} \zeta \partial_k(\zeta \partial^k \zeta)\right] \nonumber \\ && +\partial_k\left [2 e^{\rho}\frac{1}{\dot \rho ^2}\zeta^2(\dot \rho \partial^k \dot \zeta - \ddot \rho \partial^k \zeta)\right]
    \quad .
\eea
This step is useful; it makes the scaling of the initial term with the slow-roll parameter a little clearer, since the ratio \( \dot{\phi}^2/\dot{\rho}^2 \) is directly proportional to \( \varepsilon \).
The remaining terms in the Lagrangian contain the following cubic contributions:
\beq
    \sqrt{h} \left[ \frac{1}{N} \left( E_{ij} E^{ij} - E^2 \right) - 6N \dot{\rho}^2 + \left( N + \frac{1}{N} \right) \dot{\phi}^2 \right]^{(3)}
    \quad .
\eeq
Expanding this, we get:
\beq
    e^{3\rho} \frac{\dot{\phi}^2}{\dot{\rho}^2} \left( 3\zeta - \frac{\dot{\zeta}}{\dot{\rho}} \right) \dot{\zeta}^2
    + e^{3\rho} \left( 3\zeta - \frac{\dot{\zeta}}{\dot{\rho}} \right) \left( \partial_k \partial_l B \partial^k \partial^l B - (\partial_k \partial^k B)^2 \right)
    \quad \nonumber
\eeq
\beq
    - 4 e^{3 \rho}(\partial_k \zeta \partial^k B) (\partial_l \partial^l B)- \frac{d}{dt} \left( 18 e^{3\rho} \dot{\rho} \zeta^3 \right)
    + \partial_k \left( 18 e^{3\rho} \dot{\rho} \zeta^2 \partial^k B \right)
    + 9 e^{3\rho} \left( 2 \ddot{\rho} + \dot{\phi}^2 \right) \zeta^3
    \quad .
\eeq
Let us now define the full cubic Lagrangian for the fluctuations as:
\beq
    2\mathcal{L}^{(3)} = \sqrt{h} \left[ N \hat{R} + \frac{1}{N} (E_{ij}E^{ij} - E^2) - 6N\dot{\rho}^2 + \left( N + \frac{1}{N} \right) \dot{\phi}^2 \right]^{(3)}
    \quad ,
\eeq
which simplifies to:
\bea
    2\mathcal{L}^{(3)} &=& e^{3\rho} \frac{\dot{\phi}^2}{\dot{\rho}^2} \left( 3\zeta - \frac{\dot{\zeta}}{\dot{\rho}} \right) \dot{\zeta}^2
    - e^{\rho} \frac{\dot{\phi}^2}{\dot{\rho}^2} \zeta \partial_k \zeta \partial^k \zeta + \nonumber\\
    &+& e^{3\rho} \left( 3\zeta - \frac{\dot{\zeta}}{\dot{\rho}} \right)
    \left[ \partial_k \partial_l B \partial^k \partial^l B - (\partial_k \partial^k B)^2 \right]- 4 e^{3 \rho}\partial_k \zeta \partial^k B \partial_l \partial^l B
    + D_0
    \quad ,
\eea
where all total derivative and non-dynamical contributions are grouped into $D_0$.
Although this form of \( \mathcal{L}^{(3)} \) is correct, the \( \varepsilon^2 \) suppression of the interactions is not immediately obvious and requires further analysis to be made explicit.
To proceed, we eliminate the \( B \) field using the second constraint derived earlier:
\beq
    B = -\frac{e^{-2\rho}}{\dot{\rho}} \zeta + \chi \quad ,
    \quad \text{with} \quad \partial_k \partial^k \chi = \frac{1}{2} \frac{\dot{\phi}^2}{\dot{\rho}^2} \dot{\zeta}
    \quad .
\eeq
After substituting and organizing terms according to powers of \( e^\rho \), the Lagrangian becomes:
\beq
    2\mathcal{L}^{(3)} =
    e^{3\rho} \frac{\dot{\phi}^2}{\dot{\rho}^2} \left( 3\zeta - \frac{\dot{\zeta}}{\dot{\rho}} \right) \dot{\zeta}^2
    - \frac{1}{4} e^{3\rho} \frac{\dot{\phi}^4}{\dot{\rho}^4} \left( 3\zeta - \frac{\dot{\zeta}}{\dot{\rho}} \right) \dot{\zeta}^2
    - 2 e^{3\rho} \frac{\dot{\phi}^2}{\dot{\rho}^2} \dot{\zeta} \partial_k \zeta \partial^k \chi + \nonumber 
\eeq
\beq
     + e^{3\rho} \left( 3\zeta - \frac{\dot{\zeta}}{\dot{\rho}} \right) \partial_k \partial_l \chi \, \partial^k \partial^l \chi- e^{\rho} \frac{\dot{\phi}^2}{\dot{\rho}^2} \left( \zeta - 2\frac{\dot{\zeta}}{\dot{\rho}} \right) \partial_k \zeta \partial^k \zeta
    + e^{\rho} \frac{\dot{\phi}^2}{\dot{\rho}^3} \left( 3\zeta - \frac{\dot{\zeta}}{\dot{\rho}} \right) \dot{\zeta} \, \partial_k \partial^k \zeta + \nonumber
\eeq
\beq
    - 2e^{\rho} \frac{1}{\dot{\rho}} \left( 3\zeta - \frac{\dot{\zeta}}{\dot{\rho}} \right) \partial_k \partial_l \zeta \, \partial^k \partial^l \chi
    + 4e^{\rho} \frac{1}{\dot{\rho}} \partial_k \zeta \partial^k \chi \, \partial_l \partial^l \zeta + \nonumber 
\eeq
\beq
    + e^{-\rho} \frac{1}{\dot{\rho}^2} \left( 3\zeta - \frac{\dot{\zeta}}{\dot{\rho}} \right)
    \left( \partial_k \partial_l \zeta \, \partial^k \partial^l \zeta - (\partial_k \partial^k \zeta)^2 \right)- 4e^{-\rho} \frac{1}{\dot{\rho}^2} \partial_k \zeta \partial^k \zeta \partial_l \partial^l \zeta + D_0
    \quad .
\eeq
The cubic action is now written entirely in terms of \( \zeta \), having applied both Lagrange constraints. To make the \( \varepsilon^2 \) suppression manifest, the remaining calculation is structured according to the powers of \( e^\rho \) in each term.
\paragraph{The \texorpdfstring{$e^{-\rho}$}{e(-rho)}-terms}
\leavevmode\\
We begin by analyzing the terms proportional to $e^{-\rho}$, as they are the fewest in number. The contribution takes the form:
\bea
    2\mathcal{L}^{(3)}|_{e^{-\rho}} &=&\;
    3e^{-\rho} \frac{1}{\dot{\rho}^2} \zeta\, \partial_k \partial_l \zeta\, \partial^k \partial^l \zeta
    - 3e^{-\rho} \frac{1}{\dot{\rho}^2} \zeta \left( \partial_k \partial^k \zeta \right)^2  - 4e^{-\rho} \frac{1}{\dot{\rho}^2} \partial_k \zeta\, \partial^k \zeta\, \partial_l \partial^l \zeta \nonumber \\
    &-& e^{-\rho} \frac{1}{\dot{\rho}^3} \dot{\zeta}\, \partial_k \partial_l \zeta\, \partial^k \partial^l \zeta +
    + e^{-\rho} \frac{1}{\dot{\rho}^3} \dot{\zeta} \left( \partial_k \partial^k \zeta \right)^2
    \quad .
\eea
The last two terms differ structurally from the others due to the presence of time derivatives on $\zeta$. By integrating by parts sufficiently to eliminate these time derivatives, we obtain:
\bea
    &&- e^{-\rho} \frac{1}{\dot{\rho}^3} \dot{\zeta}\, \partial_k \partial_l \zeta\, \partial^k \partial^l \zeta
    + e^{-\rho} \frac{1}{\dot{\rho}^3} \dot{\zeta} \left( \partial_k \partial^k \zeta \right)^2 = \nonumber \\ &&
    - \frac{1}{3} e^{-\rho} \frac{1}{\dot{\rho}^2} \zeta \left[ \partial_k \partial_l \zeta\, \partial^k \partial^l \zeta - \left( \partial_k \partial^k \zeta \right)^2 \right]
    + \frac{1}{2} e^{-\rho} \frac{\dot{\phi}^2}{\dot{\rho}^4} \zeta \left[ \partial_k \partial_l \zeta\, \partial^k \partial^l \zeta - \left( \partial_k \partial^k \zeta \right)^2 \right] \nonumber \\
    &&+ \frac{1}{3} \frac{d}{dt} \left[ e^{-\rho} \frac{1}{\dot{\rho}^3} \zeta \left( \left( \partial_k \partial^k \zeta \right)^2 - \partial_k \partial_l \zeta\, \partial^k \partial^l \zeta \right) \right] + \frac{2}{3} \partial_k \left[ e^{-\rho} \frac{1}{\dot{\rho}^3} \dot{\zeta} \left( \partial^k \zeta\, \partial_l \partial^l \zeta - \partial^l \zeta\, \partial_l \partial^k \zeta \right) \right. \nonumber \\
    && \qquad\;\; \left. - e^{-\rho} \frac{1}{\dot{\rho}^3} \zeta \left( \partial^k \dot{\zeta}\, \partial_l \partial^l \zeta - \partial^l \dot{\zeta}\, \partial_l \partial^k \zeta \right) \right] \nonumber \\ &&+ e^{-\rho} \frac{1}{\dot{\rho}^4} \left( \ddot{\rho} + \frac{1}{2} \dot{\phi}^2 \right) \zeta \left[ \left( \partial_k \partial^k \zeta \right)^2 - \partial_k \partial_l \zeta\, \partial^k \partial^l \zeta \right]
    \quad .
\eea
Although this term appears dominant at order \( e^{-\rho} \), it combines with others into total derivatives, leaving a single relevant contribution:
\beq
    2\mathcal{L}^{(3)}_{e^{-\rho}} = \frac{1}{2} e^{-\rho} \frac{\dot{\phi}^2}{\dot{\rho}^4} \zeta \left( \partial_k \partial_l \zeta\, \partial^k \partial^l \zeta - \left( \partial_k \partial^k \zeta \right)^2 \right) + \cdots
    \quad ,
\eeq
which, despite appearing only \( \varepsilon \)-suppressed, gives a negligible contribution to the three-point function unless it is of order \( \varepsilon^2 \) or higher.
We can anticipate this by defining:
\beq
    F_A = \frac{1}{2} e^{-\rho} \frac{\dot{\phi}^2}{\dot{\rho}^4} \left( \partial_k \zeta\, \partial^k \zeta\, \partial_l \partial^l \zeta - \zeta\, \partial_k \partial_l (\partial^k \zeta\, \partial^l \zeta)\right)
    \quad ,
\eeq
after further spatial integrations by parts.
Thus, the total contribution from $e^{-\rho}$ terms to the third-order action is:
\beq
    2\mathcal{L}^{(3)}|_{e^{-\rho}} = F_A + D_A
    \quad ,
\eeq
where $D_A$ contains all total derivative contributions.
\paragraph{The \texorpdfstring{$e^{\rho}$}{e(rho)}-terms}
\leavevmode\\
The next set of terms to analyze is the group proportional to \( e_\rho \). This part of the action includes a total of seven interaction terms:
\bea
    2\mathcal{L}^{(3)}|_{e^\rho} &=& -e^\rho \frac{\dot{\phi}^2}{\dot{\rho}^2} \zeta \partial_k \zeta \partial^k \zeta 
    + 2e^\rho \frac{\dot{\phi}^2}{\dot{\rho}^3} \dot{\zeta} \partial_k \zeta \partial^k \zeta 
    + 3e^\rho \frac{\dot{\phi}^2}{\dot{\rho}^3} \dot{\zeta} \zeta \partial_k \partial^k \zeta - e^\rho \frac{\dot{\phi}^2}{\dot{\rho}^4} \dot{\zeta}^2 \partial_k \partial^k \zeta 
    \nonumber \\ &-& 6e^\rho \frac{1}{\dot{\rho}} \partial_k \partial_l \zeta \partial^k \partial^l \chi + 2e^\rho \frac{1}{\dot{\rho}^2} \dot{\zeta} \partial_k \partial_l \zeta \partial^k \partial^l \chi+ 4e^\rho \frac{1}{\dot{\rho}} \partial_k \zeta \partial^k \chi \partial_l \partial^l \zeta
    \quad .
\eea
This group, though not much larger than the \( e^{-\rho} \) terms, is structurally more complex. We begin by simplifying the terms involving the \( \chi \) field. In particular, the one with a time derivative on \( \zeta \) is integrated by parts, followed by selective integration by parts of spatial derivatives. Only after fully reducing the \( \chi \)-terms do we turn to the remaining terms in \( \mathcal{L}^{(3)}|_{e^\rho} \).
Let’s begin with the \( \chi \)-term involving \( \dot{\zeta} \). We integrate it by parts until no terms with four spatial derivatives contain \( \dot{\zeta} \):
\bea
    2e^\rho \frac{1}{\dot{\rho}^2} \dot{\zeta} \partial_k \partial_l \zeta \partial^k \partial^l \chi 
    &=& - e^\rho \frac{1}{\dot{\rho}^2} \zeta \partial_k \partial_l \zeta \partial^k \partial^l \dot{\chi}
    - e^\rho \frac{1}{\dot{\rho}} \zeta \partial_k \partial_l \zeta \partial^k \partial^l \chi \nonumber \\
    &&\quad - e^\rho \frac{\dot{\phi}^2}{\dot{\rho}^3} \zeta \partial_k \partial_l \zeta \partial^k \partial^l \chi
    + \frac{1}{2} e^\rho \frac{\dot{\phi}^2}{\dot{\rho}^4} \zeta\partial_k \dot{\zeta} \partial^k \dot{\zeta}  \nonumber \\
    &&\quad - \frac{1}{2} e^\rho \frac{\dot{\phi}^2}{\dot{\rho}^4} \dot{\zeta} \partial_k \zeta \partial^k \dot{\zeta}
    + \frac{d}{dt} \left( e^\rho \frac{1}{\dot{\rho}^2} \zeta \partial_k \partial_l \zeta \partial^k \partial^l \chi \right) \nonumber \\
    &&\quad + 2e^\rho \frac{1}{\dot{\rho}^3} \left( \ddot{\rho} + \frac{1}{2} \dot{\phi}^2 \right) \zeta \partial_k \partial_l \zeta \partial^k \partial^l \chi \nonumber \\
    &&\quad + \partial_k \left( e^\rho \frac{1}{\dot{\rho}^2} \left( \dot{\zeta} \partial_l \zeta - \zeta \partial_l \dot{\zeta} \right) \partial^k \partial^l \chi \right)
    \quad .
\eea
In this derivation, we have extensively used the constraint:
$\partial_k \partial^k \chi = \frac{1}{2} \frac{\dot{\phi}^2}{\dot{\rho}^2} \dot{\zeta}$.
The first term from the second line above can be rewritten as:
\bea
    - e^\rho \frac{\dot{\phi}^2}{\dot{\rho}^3} \zeta \partial_k \partial_l \zeta \partial^k \partial^l \chi 
    &=& -e^\rho \frac{\dot{\phi}^2}{\dot{\rho}^3} \left( \partial_k \zeta \partial^k \chi \partial_l \partial^l \zeta 
    - \zeta \partial_k \partial_l (\partial^k \zeta \partial^l \chi )\right) - \frac{e^\rho}{2} \frac{\dot{\phi}^4}{\dot{\rho}^5} \dot{\zeta} \zeta \partial_k \partial^k \zeta + \nonumber\\
    &&\quad - \partial_k \left[ e^\rho \frac{\dot{\phi}^2}{\dot{\rho}^3} \left( \zeta \partial_l (\partial^k \zeta \partial^l \chi )
    + \zeta \partial^k\partial^l \zeta \partial_l \chi 
    - \partial^l (\zeta \partial_l \zeta) \partial^k \chi \right) \right]
    \quad .
\eea
Finally, returning to the original \( e^\rho \) Lagrangian, its last term can be rewritten to match the format of the others, by reordering spatial derivatives:
\bea
    4e^\rho \frac{1}{\dot{\rho}} \partial_k \zeta \partial^k \chi \partial_l \partial^l \zeta 
    &=& 4e^\rho \frac{1}{\dot{\rho}} \zeta \partial_k \partial_l \zeta \partial^k \partial^l \chi 
    + 2e^\rho \frac{\dot{\phi}^2}{\dot{\rho}^3} \zeta \partial_k \zeta \partial^k \dot{\zeta} + e^\rho \frac{\dot{\phi}^2}{\dot{\rho}^3} \dot{\zeta} \partial_k \zeta \partial^k \zeta  \nonumber \\
    &&\quad + \partial_k \left( 2e^\rho \frac{1}{\dot{\rho}} \left( 2 \partial^k \zeta \partial_l \zeta \partial^l \chi 
    - 2\zeta \partial_l \zeta \partial^k \partial^l \chi 
    - \partial_l \zeta \partial^l \zeta \partial^k \chi \right) \right)
    \quad .
\eea
Combining all previously expanded terms involving the $\chi$ field yields a rather involved expression:
\bea
    && -6e^\rho \frac{1}{\dot{\rho}} \zeta\partial_k \partial_l \zeta\, \partial^k \partial^l \chi 
    + 2e^\rho \frac{1}{\dot{\rho}^2} \dot{\zeta}\, \partial_k \partial_l \zeta\, \partial^k \partial^l \chi 
    + 4e^\rho \frac{1}{\dot{\rho}} \partial_k \zeta\, \partial^k \chi\, \partial_l \partial^l \zeta = \nonumber \\
    &&= -e^\rho \frac{1}{\dot{\rho}^2} \zeta\, \partial_k \partial_l \zeta\, \partial^k \partial^l \dot{\chi} 
    - 3e^\rho \frac{1}{\dot{\rho}} \zeta\, \partial_k \partial_l \zeta\, \partial^k \partial^l \chi \nonumber \\ 
    &&- e^\rho \frac{\dot{\phi}^2}{\dot{\rho}^3} \left( \partial_k \zeta\, \partial^k \chi\, \partial_l \partial^l \zeta 
    - \zeta\, \partial_k \partial_l( \partial^k \zeta\, \partial^l \chi )\right) \nonumber \\
    && \quad + 2e^\rho \frac{\dot{\phi}^2}{\dot{\rho}^3} \zeta\, \partial_k \zeta\, \partial^k \dot{\zeta} 
    + e^\rho \frac{\dot{\phi}^2}{\dot{\rho}^3} \dot{\zeta}\, \partial_k \zeta\, \partial^k \zeta 
    - \frac{1}{2} e^\rho \frac{\dot{\phi}^4}{\dot{\rho}^5} \dot{\zeta}\zeta \, \partial_k \partial^k \zeta \nonumber \\
    && \quad + \frac{1}{2} e^\rho \frac{\dot{\phi}^2}{\dot{\rho}^4} \zeta \partial_k \dot{\zeta}\, \partial^k \dot{\zeta} 
    - \frac{1}{2} e^\rho \frac{\dot{\phi}^2}{\dot{\rho}^4} \dot{\zeta}\, \partial_k \zeta\, \partial^k \dot{\zeta} \nonumber \\
    && \quad + \frac{d}{dt} \left( e^\rho \frac{1}{\dot{\rho}^2} \zeta\, \partial_k \partial_l \zeta\, \partial^k \partial^l \chi \right)
    + 2e^\rho \frac{1}{\dot{\rho}^3} \left( \ddot{\rho} + \frac{1}{2} \dot{\phi}^2 \right) \zeta\, \partial_k \partial_l \zeta\, \partial^k \partial^l \chi \nonumber \\
    && \quad - \partial_k \left( e^\rho \frac{\dot{\phi}^2}{\dot{\rho}^3} \left( \zeta\, \partial_l (\partial^k \zeta\, \partial^l \chi) 
    + \zeta\,\partial^k \partial^l \zeta\, \partial_l \chi 
    - \partial^l (\zeta\, \partial_l \zeta)\, \partial^k \chi \right) \right) \nonumber \\
    && \quad + \partial_k \left( 2e^\rho \frac{1}{\dot{\rho}} \left( 2\partial^k \zeta\, \partial_l \zeta\, \partial^l \chi 
    - 2\zeta\, \partial_l \zeta\, \partial^k \partial^l \chi 
    - \partial_l \zeta\, \partial^l \zeta\, \partial^k \chi \right) \right) \nonumber \\
    && \quad + \partial_k \left( e^\rho \frac{1}{\dot{\rho}^2} \left( \dot{\zeta}\, \partial_l \zeta - \zeta\, \partial_l \dot{\zeta} \right) \partial^k \partial^l \chi \right)
    \quad .
\eea
Before incorporating the rest of the $e^\rho$-order Lagrangian, we integrate the first line by parts to simplify it:
\bea
    &&- e^\rho \frac{1}{\dot{\rho}^2} \zeta \, \partial_k \partial_{l}\zeta \partial^k \partial^{l} \dot \chi
    - 3 e^\rho \frac{1}{\dot{\rho}} \zeta \, \partial_k \partial_{l}\zeta \partial^k \partial^{l}\chi
    = - e^\rho \frac{1}{\dot{\rho}^2} \, \partial_k \zeta \, \partial^k \zeta \, \partial_l \partial^{l}\big[\dot \chi
    + 3 \dot{\rho} \chi\big] \nonumber \\ &&+ e^\rho \frac{1}{\dot{\rho}^2} [\dot{\chi} + 3 \dot{\rho} \chi] \, \partial_k \partial_l ( \partial^k \zeta \, \partial^{l}\zeta) -e^\rho \frac{1}{\dot \rho^2}\zeta \partial_k \partial^k \zeta \partial_l \partial^l \dot \chi - \frac{3}{2} e^\rho \frac{\dot{\phi}^2}{\dot{\rho}^3} \, \dot{\zeta} \, \zeta \, \partial_k\partial^k \zeta +
    \nonumber \\ &&- \partial_k \left[ e^\rho \frac{1}{\dot{\rho}^2} \left( \zeta \partial_{l}\zeta \partial^k \partial^{l}\dot\chi - \partial^k \zeta \, \partial_{l}\zeta \partial^{l}\dot \chi + \dot{\chi} \, \partial_l ( \partial^k \zeta \, \partial^{l}\zeta \right) \right]+ \nonumber \\ &&- \partial_k \left[3 e^\rho \frac{1}{\dot{\rho}} \left( \zeta \partial_{l}\zeta \partial^k \partial^{l}\chi - \partial^k \zeta \, \partial_{l}\zeta \partial^{l} \chi + {\chi} \, \partial_l ( \partial^k \zeta \, \partial^{l}\zeta \right) \right] \nonumber\\ &&+ \partial_k \left[ e^\rho \frac{1}{\dot{\rho}^2}  \zeta \partial^k\zeta \partial_l \partial^{l}\dot \chi\right] + \partial_k\left[\frac{3}{2}e^{\rho}\frac{\dot \phi^2}{\dot \rho^3}\dot \zeta \zeta \partial^k \zeta\right]
    \quad .
\eea
We apply another time the constraint inverting the Laplacian formally as:
\beq
    \chi = \frac{1}{2} \frac{\dot{\phi}^2}{\dot{\rho}^2} \partial^{-2} \dot{\zeta}
    \quad .
\eeq
This allows us to write the first line of the equation before the last to be written as
\bea
    &&- e^\rho \frac{1}{\dot{\rho}^2} \partial_k \zeta\, \partial^k \zeta\, \partial_l \partial^l \big[\dot{\chi} + 3\dot{\rho} \chi\big]+e^\rho \frac{1}{\dot \rho^2}\big[\dot \chi + 3 \dot \rho \chi\big]\partial_k\partial_l(\partial^k\zeta \partial^l\zeta)
    \nonumber \\ &&= -\frac{1}{2} e^{-2\rho} \frac{1}{\dot{\rho}^2} \frac{d}{dt} \left( e^{3\rho} \frac{\dot{\phi}^2}{\dot{\rho}^2} \dot{\zeta} \right) 
    \left[ \partial_k \zeta\, \partial^k \zeta - \partial^{-2} \partial_k \partial_l ( \partial^k \zeta\, \partial^l \zeta ) \right]
    \quad ,
\eea
where we have used the identity
\beq
    \left( \frac{d}{dt} + 3\dot{\rho} \right) \left( \frac{\dot{\phi}^2}{\dot{\rho}^2} \dot{\zeta} \right)
    = e^{-3\rho} \frac{d}{dt} \left( e^{3\rho} \frac{\dot{\phi}^2}{\dot{\rho}^2} \dot{\zeta} \right)
    \quad .
\eeq
Here, the inverse Laplacian operator $\partial^{-2}$ has been "formally integrated by parts", which is best understood by expressing two functions $f(\vec{x})$ and $g(\vec{x})$ in terms of their Fourier components. Then it's easy to derive
\beq
    f(\vec x)\partial^{-2}g(\vec x)=(\partial^{-2}f(\vec x))g(\vec x)
    \quad ,
\eeq
that we have applied before.
At last it is possible to assemble all the rest of the order $e^{\rho}$ terms (those that did not
contain the $\chi$ field) together with what was derived so far to find:
\bea
    2\mathcal{L}^{(3)}|_{e^\rho} &=&\;  
    -\frac{1}{2} \frac{e^{-2\rho}}{\dot{\rho}^2} \frac{d}{dt} \left( e^{3\rho} \frac{\dot{\phi}^2}{\dot{\rho}^2} \dot \zeta \right) \big[\partial_k \zeta \partial^k \zeta - \partial^{-2}\partial_k \partial_{l}(\partial^k \partial^l\zeta)\big] + \nonumber \\ &-& e^\rho \frac{\dot{\phi}^2}{\dot{\rho}^3} \left[ \partial^k \zeta\, \partial_k \chi\, \partial^l \partial_l \zeta - \zeta\, \partial^k \partial^l (\partial^k \zeta\, \partial^l \chi )\right]+ \nonumber \\ &-& e^\rho \frac{\dot{\phi}^2}{\dot{\rho}^2} \zeta\, \partial_k \zeta\, \partial^k \zeta 
    + 3e^\rho \frac{\dot{\phi}^2}{\dot{\rho}^3} \dot{\zeta}\, \partial_k \zeta\, \partial^k \zeta + \frac{3}{2} e^\rho \frac{\dot{\phi}^2}{\dot{\rho}^3} \dot{\zeta}\, \zeta\, \partial_k \partial^k \zeta 
    + 2e^\rho \frac{\dot{\phi}^2}{\dot{\rho}^3} \zeta\, \partial_k \zeta\, \partial^k \dot{\zeta} \nonumber \\
    &-& \frac{1}{2} e^\rho \frac{\dot{\phi}^4}{\dot{\rho}^5} \dot{\zeta}\, \zeta\, \partial_k \partial^k \zeta
    - e^\rho \frac{1}{\dot{\rho}^2} \zeta\, \partial_k \partial^k \zeta\, \partial_l \partial^l \dot \chi
    - e^\rho \frac{\dot{\phi}^2}{\dot{\rho}^4} \dot{\zeta}^2 \partial_k \partial^k \zeta 
    \nonumber \\ &+& \frac{1}{2} e^\rho \frac{\dot{\phi}^2}{\dot{\rho}^4} \zeta\, \partial_k\dot{\zeta}\, \partial^k \dot{\zeta} 
    - \frac{1}{2} e^\rho \frac{\dot{\phi}^2}{\dot{\rho}^4} \dot{\zeta}\, \partial_k \zeta\, \partial^k \dot{\zeta} 
    + \cdots
    \quad ,
\eea
plus a great many total derivative terms which will be shown explicitly at the end. After integrating everything on the last three lines by parts, one learns that the order $e^\rho$ part of the Lagrangian contains one dynamical term that is manifestly of order $\varepsilon^2$,
\beq
    2\mathcal{L}^{(3)}|_{e^\rho}=\frac{1}{2}e^\rho \frac{\dot \phi^4}{\dot \rho ^4}\zeta \partial_k \zeta \partial^k \zeta + F_B+D_B
    \quad .
\eeq
another dynamical contribution that will appear in the terms proportional to the linear equation of motion,
\bea
    F_B &=& 
    - \frac{1}{2}  \frac{e^{-2\rho}}{\dot{\rho}^{2}}\,
    \frac{d}{dt}\left( e^{3\rho}\,\frac{\dot{\phi}^{2}}{\dot{\rho}^{2}}\,\dot{\zeta}\right)\big[\partial_k \zeta \partial^k \zeta - \partial^{-2} \partial_k \partial_l(\partial^k\zeta \partial^l\zeta)\big]+ 
    \nonumber \\ &-& e^{\rho}\,\frac{\dot{\phi}^{2}}{\dot{\rho}^{2}}\left[\frac{\ddot{\phi}}{\dot \phi \dot \rho}\zeta^2 +\frac{1}{2}\frac{\dot \phi^2}{\dot \rho^2}\zeta^2 \frac{\dot{\phi}\,\dot{\rho}}{\zeta^{2}} + \frac{2}{\dot{\rho}}\,\dot{\zeta}\zeta\,\right]\partial_{k}\partial^{k}\zeta \nonumber \\ &-& e^{\rho}\,\frac{\dot{\phi}^{2}}{\dot{\rho}^{3}}\left[ \partial_{k}\zeta\,\partial^{k}\chi\,\partial_{l}\partial^{l}\zeta - \zeta\,\partial_{k}\partial_{l}(\partial^{k}\zeta\,\partial^{l}\chi) \right]
    \quad ,
\eea
plus total derivatives encoded in $D_B$.
\paragraph{The first \texorpdfstring{$e^{3\rho}$}{e(3rho)}-term}
\leavevmode\\
There remain only the \( e^{3\rho} \) terms. This set will be broken into two parts, with one of the terms treated separately:
\beq
    e^{3\rho} \frac{\dot{\phi}^2}{\dot{\rho}^2} \left( 3\zeta - \frac{\dot{\zeta}}{\dot{\rho}} \right) \dot{\zeta}^2
    \quad .
\eeq
This term produces the leading contribution to the three-point function that does not involve any spatial derivatives. Superficially, it appears proportional to \( \varepsilon \), but this too is misleading.
Start this time by integrating one of the \( \dot{\zeta} \) factors in the second term by parts:
\bea
    e^{3\rho} \frac{\dot{\phi}^2}{\dot{\rho}^2} \left( 3\zeta - \frac{\dot{\zeta}}{\dot{\rho}} \right) \dot{\zeta}^2
    &=& 3e^{3\rho} \frac{\dot{\phi}^2}{\dot{\rho}^2} \dot{\zeta}^2 \zeta 
    - e^{3\rho} \frac{\dot{\phi}^2}{\dot{\rho}^3} \dot{\zeta}^2 \frac{d}{dt} \zeta=  \nonumber \\
    &=& e^{3\rho} \frac{\dot{\phi}^2}{\dot{\rho}^2} \left[ \ddot{\zeta} + \frac{\ddot{\phi}}{\dot{\phi}} \dot{\zeta} + \frac{3}{4} \frac{\dot{\phi}^2}{\dot{\rho}} \dot{\zeta} + 3 \dot{\rho} \dot{\zeta} \right] \frac{2}{\dot{\rho}} \dot{\zeta} \zeta - \frac{d}{dt} \left[ e^{3\rho} \frac{\dot{\phi}^2}{\dot{\rho}^3} \dot{\zeta}^2 \zeta \right]
    \nonumber \\ &-& 3 e^{3\rho} \frac{\dot{\phi}^2}{\dot{\rho}^4} \left[ \ddot{\rho} + \frac{1}{2} \dot{\phi}^2 \right] \dot{\zeta}^2 \zeta
    \quad .
\eea
The dynamical term resembles the time-derivative part of the \( \zeta \) equation of motion:
\beq
    e^{3\rho} \frac{\dot{\phi}^2}{\dot{\rho}^2} \left( \ddot{\zeta} + 2\frac{\ddot{\phi}}{\dot{\phi}} \dot{\zeta} + \frac{\dot{\phi}^2}{\dot{\rho}} \dot{\zeta} + 3 \dot{\rho} \dot{\zeta} \right)
    = \frac{d}{dt} \left( e^{3\rho} \frac{\dot{\phi}^2}{\dot{\rho}^2} \dot{\zeta} \right)
    + 2e^{3\rho} \frac{\dot{\phi}^2}{\dot{\rho}^3} \left( \ddot{\rho} + \frac{1}{2} \dot{\phi}^2 \right) \dot{\zeta} 
    \quad ,
\eeq
though with slight differences in the coefficients. Adding and subtracting the necessary pieces yields:
\bea
    &&e^{3\rho} \frac{\dot{\phi}^2}{\dot{\rho}^2} \left[ 3\zeta - \frac{\dot{\zeta}}{\dot{\rho}} \right] \dot{\zeta}^2 
    = \frac{d}{dt} \left[ e^{3\rho} \frac{\dot{\phi}^2}{\dot{\rho}^2} \dot{\zeta} \right] \frac{2}{\dot{\rho}} \dot{\zeta} \zeta 
    - 2 e^{3\rho} \frac{\dot{\phi}^2}{\dot{\rho}^2} \left[ \frac{\ddot{\phi}}{\dot{\phi} \dot{\rho}} + \frac{1}{2} \frac{\dot{\phi}^2}{\dot{\rho}^2} \right] \dot{\zeta}^2 \zeta \nonumber \\ &+&\frac{1}{2} e^{3\rho} \frac{\dot{\phi}^4}{\dot{\rho}^4} \dot{\zeta}^2 \zeta - \frac{d}{dt} \left[ e^{3\rho} \frac{\dot{\phi}^2}{\dot{\rho}^3} \dot{\zeta}^2 \zeta \right] 
    + e^{3\rho} \frac{\dot{\phi}^2}{\dot{\rho}^4} \left[ \ddot{\rho} + \frac{1}{2} \dot{\phi}^2 \right] \dot{\zeta}^2 \zeta
    \quad .
\eea
It may not be obvious why the last term on the first line was separated, since it overlaps with part of the second term. However, it matches a term in the second line of the $F_B$ expression, which encodes the spatial part of the $\zeta$ equation of motion. Also, while the second dynamical term appears to be second order in slow-roll parameters, it is actually of even higher order. To make this fact apparent,
integrate a factor $\dot \zeta \zeta$ by parts, so that
\bea
    &&-2e^{3\rho} \frac{\dot{\phi}^2}{\dot{\rho}^2} \left[\frac{\ddot{\phi}}{\dot{\phi} \dot{\rho}} + \frac{1}{2} \frac{\dot{\phi}^2}{\dot{\rho}^2} \right] \dot{\zeta}^2 \zeta
    = -e^{3\rho} \frac{\dot{\phi}^2}{\dot{\rho}^2} \dot{\zeta} \left[ \frac{\ddot{\phi}}{\dot{\phi} \dot{\rho}} + \frac{1}{2} \frac{\dot{\phi}^2}{\dot{\rho}^2} \right] \frac{d}{dt} \zeta^2 = \nonumber \\
    &=& e^{3\rho} \frac{\dot{\phi}^2}{\dot{\rho}^2} \dot{\zeta} \zeta^2 \frac{d}{dt} \left[ \frac{\ddot{\phi}}{\dot{\phi} \dot{\rho}} + \frac{1}{2} \frac{\dot{\phi}^2}{\dot{\rho}^2} \right] +  \frac{d}{dt} \left[ e^{3\rho} \frac{\dot{\phi}^2}{\dot{\rho}^2} \dot{\zeta} \right]\left[ \frac{\ddot{\phi}}{\dot{\phi} \dot{\rho}} + \frac{1}{2} \frac{\dot{\phi}^2}{\dot{\rho}^2} \right] \zeta^2 + \nonumber \\
    &-& \frac{d}{dt} \left[ e^{3\rho} \frac{\dot{\phi}^2}{\dot{\rho}^2} \left( \frac{\ddot{\phi}}{\dot{\phi} \dot{\rho}} + \frac{1}{2} \frac{\dot{\phi}^2}{\dot{\rho}^2} \right) \dot{\zeta} \zeta^2 \right]
    \quad .
\eea
Thus, the first term in the cubic action can be rewritten as:
\beq
    e^{3\rho} \frac{\dot{\phi}^2}{\dot{\rho}^2} \left( 3\zeta - \frac{\dot{\zeta}}{\dot{\rho}} \right) \dot{\zeta}^2 
    = \frac{1}{2} e^{3\rho} \frac{\dot{\phi}^4}{\dot{\rho}^4} \dot{\zeta}^2 \zeta +e^{3 \rho}\frac{\dot \phi^2}{\dot \rho^2}\dot \zeta \zeta^2 \frac{d}{dt}\left[\frac{\ddot \phi}{\dot \phi \dot \rho}+\frac{1}{2}\frac{\dot \phi^2}{\dot \rho^2}\right]+F_C+D_C
    \quad ,
\eeq
where $D_C$ is something that has no dynamical effect, and where
\beq
    F_C = \frac{d}{dt}\left[e^{3 \rho} \frac{\dot \phi^2}{\dot \rho^2}\dot \zeta\right]\left[\frac{\ddot \phi}{\dot \phi \dot \rho}\zeta^2+\frac{1}{2}\frac{\dot \phi^2}{\dot \rho^2}\zeta^2+\frac{2}{\dot \rho}\dot \zeta \zeta\right]
    \quad ,
\eeq
will contribute yet another few terms proportional to the $\zeta$ equation of motion.
\paragraph{The remaining \texorpdfstring{$e^{3\rho}$}{e(3rho)}-terms}
\leavevmode\\
There is left just one final set to examine,
\bea
    2\mathcal{L}^{(3)}|_{e^{3\rho}} &=& 3e^{3\rho} \zeta\, \partial_k \partial_l \chi\, \partial^k \partial^l \chi 
    - e^{3\rho} \frac{1}{\dot{\rho}} \dot{\zeta}\, \partial_k \partial_l \chi\, \partial^k \partial^l \chi - \frac{3}{4} e^{3\rho} \frac{\dot{\phi}^4}{\dot{\rho}^4} \dot{\zeta}^2 \zeta 
    + \frac{1}{4} e^{3\rho} \frac{\dot{\phi}^4}{\dot{\rho}^5} \dot{\zeta}^3 
    \nonumber \\ &-& 2e^{3\rho} \frac{\dot{\phi}^2}{\dot{\rho}^2} \dot{\zeta} \partial_k \zeta\, \partial^k \chi
    \quad ,
\eea
those proportional to $e^{3\rho}$, aside from the one that was just evaluated.
Begin by integrating the second term by parts to remove the time derivative from the $\dot{\zeta}$,
and then integrate further some of the spatial derivatives to remove them from the term that
contains a $\dot{\chi}$ factor,
\bea
    &&- e^{3\rho} \frac{1}{\dot{\rho}} \dot{\zeta} \partial_k \partial_l \chi\, \partial^k \partial^l \chi 
    = \frac{1}{2} e^{3\rho} \frac{\dot{\phi}^2}{\dot{\rho}^{2}} \zeta\, \partial_k \partial_l \chi\, \partial^k \partial^l \chi 
    + 3e^{3\rho} \zeta\, \partial_k \partial_l \chi\, \partial^k \partial^l \chi+ \nonumber \\
    &&\quad - 2e^{3\rho} \frac{1}{\dot{\rho}} \dot{\chi} \partial_k \partial_l ( \partial^k \zeta\, \partial^l \chi )
    - 2e^{3\rho} \frac{1}{\dot{\rho}} \zeta\, \partial_k \chi\, \partial^k \partial_l \partial^l \dot{\chi}+  \nonumber \\
    &&\quad - \frac{d}{dt} \left( e^{3\rho} \frac{1}{\dot{\rho}} \zeta\, \partial_k \partial_l \chi\, \partial^k \partial^l \chi \right) 
    - \frac{e^{3\rho}}{\dot{\rho}^{2}} \left( \ddot{\rho} + \frac{1}{2} \dot{\phi}^2 \right) \zeta\, \partial_k \partial_l \chi\, \partial^k \partial^l \chi + \nonumber \\
    &&\quad + \partial_k \left[ 2e^{3\rho} \frac{1}{\dot{\rho}} 
    \left( \zeta\, \partial_l \chi\, \partial^k \partial^l \dot{\chi} 
    - \partial^k \zeta\, \partial_l \chi\, \partial^l \dot{\chi} 
    + \dot{\chi} \partial_l ( \partial^l \zeta\, \partial^k \chi ) \right) \right]
    \quad .
\eea
The second term in this expression is the same as the first term in $\mathcal{L}^{(3)}|_{ e^{3\rho}}$. Consider the
first two terms of the order $e^{3\rho}$ Lagrangian together, and integrate the spatial derivatives on
one of the $\chi$ fields by parts, to produce:
\bea
    &&3e^{3\rho} \zeta\, \partial_k \partial_l \chi\, \partial^k \partial^l \chi 
    - e^{3\rho} \frac{1}{\dot{\rho}} \dot{\zeta} \partial_k \partial_l \chi\, \partial^k \partial^l \chi 
    = \frac{1}{2} e^{3\rho} \frac{\dot{\phi}^2}{\dot{\rho}^{2}} \zeta\, \partial_k \partial_l \chi\, \partial^k \partial^l \chi 
    \nonumber \\ &+& 2e^{3\rho} \frac{1}{\dot{\rho}} \zeta\, \partial_k \partial^k \chi\, \partial_l \partial^l (\dot{\chi} + 3\dot{\rho} \chi) \nonumber \\
    &+& 2e^{3\rho} \frac{1}{\dot{\rho}} \partial_k \zeta\, \partial^k \chi\, \partial_l \partial^l (\dot{\chi} + 3\dot{\rho} \chi)
    - 2e^{3\rho} \frac{1}{\dot{\rho}} (\dot{\chi} + 3\dot{\rho} \chi)\, \partial_k \partial_l (\partial^k \zeta\, \partial^l \chi) \nonumber \\
    &-& \frac{d}{dt} \left( e^{3\rho} \frac{1}{\dot{\rho}} \zeta\, \partial_k \partial_l \chi\, \partial^k \partial^l \chi \right)
    - e^{3\rho} \frac{1}{\dot{\rho}^{2}} \left( \ddot{\rho} + \frac{1}{2} \dot{\phi}^2 \right) \zeta\, \partial_k \partial_l \chi\, \partial^k \partial^l \chi \nonumber \\
    &+& \partial_k \left[ 2e^{3\rho} \frac{1}{\dot{\rho}} 
    \left( \zeta\, \partial_l \chi\, \partial^k \partial^l \dot{\chi} 
    - \partial^k \zeta\, \partial_l \chi\, \partial^l \dot{\chi} 
    + \dot{\chi} \partial_l ( \partial^l \zeta\, \partial^k \chi ) \right) \right] \nonumber \\
    &+& \partial_k \left[ 6e^{3\rho} \left( \zeta\, \partial_l \chi\, \partial^k \partial^l \chi 
    - \partial^k \zeta\, \partial_l \chi\, \partial^l \chi 
    + \chi\, \partial_l ( \partial^l \zeta\, \partial^k \chi ) \right) \right] \nonumber \\
    &-& \partial_k \left[ 2e^{3\rho} \frac{1}{\dot{\rho}} \zeta\, \partial^k \chi\, \partial_l \partial^l (\dot{\chi} + 3\dot{\rho} \chi) \right]
    \quad .
\eea
Now apply the same trick used for the order $e^{\rho}$ terms, where some of the $\chi$ fields
were replaced with
\beq
    \chi = \frac{1}{2} \frac{\dot{\phi}^2}{\dot{\rho}^2} \partial^{-2} \dot{\zeta}
    \quad ,
\eeq
and then integrate the $\partial^{-2}$ operators by parts to remove them from the time derivative of the
$\zeta$ field, to arrive at
\bea
    &&3e^{3\rho} \zeta\, \partial_k \partial_l \chi\, \partial^k \partial^l \chi 
    - e^{3\rho} \frac{1}{\dot{\rho}} \dot{\zeta}\, \partial_k \partial_l \chi\, \partial^k \partial^l \chi 
    = \frac{1}{2} e^{3\rho} \frac{\dot{\phi}^2}{ \dot{\rho}^{2}} \zeta\, \partial_k \partial_l \chi\, \partial^k \partial^l \chi 
    + \nonumber \\&+&\frac{1}{2} \frac{\dot{\phi}^2}{\dot{\rho}^3} \dot{\zeta} \zeta \frac{d}{dt} \left( e^{3\rho} \frac{\dot{\phi}^2}{\dot{\rho}^2} \dot{\zeta} \right) +\frac{1}{\dot \rho}\frac{d}{dt}\left(e^{3\rho}\frac{\dot \phi^2}{\dot \rho^2}\dot \zeta\right)\left[\partial_k \zeta \partial^k \chi -\partial^{-2}\partial_k \partial_l (\partial^k \zeta \partial^l \chi)\right]+ \cdots
    \quad .
\eea
If the the third and the fourth terms of \( \mathcal{L}^{(3)}|_{e^{3\rho}} \) are combined with the second term of this last expression, together they yield:
\bea
    && -\frac{3}{4} e^{3\rho} \frac{\dot{\phi}^4}{ \dot{\rho}^4} \dot{\zeta}^2 \zeta 
    + \frac{1}{4} e^{3\rho} \frac{\dot{\phi}^4}{ \dot{\rho}^5}\dot{\zeta}^3 
    + \frac{1}{2} \frac{\dot{\phi}^2}{ \dot{\rho}^3} \dot{\zeta} \zeta \frac{d}{dt} \left( e^{3\rho} \frac{\dot{\phi}^2}{\dot{\rho}^2} \dot{\zeta} \right) \\
    &=& -\frac{1}{8} e^{3\rho} \frac{\dot{\phi}^6}{ \dot{\rho}^6} \dot{\zeta}^2 \zeta 
    + \frac{d}{dt} \left( \frac{1}{4} e^{3\rho} \frac{\dot{\phi}^4}{ \dot{\rho}^5} \dot{\zeta}^2 \zeta \right) 
    + \frac{1}{4} e^{3\rho} \frac{\dot{\phi}^4}{ \dot{\rho}^6} \left( \ddot{\rho} + \frac{1}{2} \dot{\phi}^2 \right) \dot{\zeta}^2 \zeta.
\eea
Putting all terms together, one obtains:
\beq
    2\mathcal{L}^{(3)}|_{e^{3\rho}} = -2e^{3\rho} \frac{\dot{\phi}^2}{\dot{\rho}^2} \dot{\zeta} \partial_k \zeta \, \partial^k \chi 
    + \frac{1}{2} e^{3\rho} \frac{\dot{\phi}^2}{\dot{\rho}^2} \zeta \, \partial_k \partial_l \chi \, \partial^k \partial^l \chi 
    - \frac{1}{8} e^{3\rho} \frac{\dot{\phi}^6}{\dot{\rho}^6} \dot{\zeta}^2 \zeta + F_D + D_D
    \quad ,
\eeq
where
\bea
    F_D &=& \frac{1}{\dot{\rho}} \frac{d}{dt} \left( e^{3\rho} \frac{\dot{\phi}^2}{\dot{\rho}^2} \dot{\zeta} \right) \left[\partial_k \zeta \, \partial^k \chi 
    + \partial^{-2} \partial_k \partial_l \left( \partial^k \zeta \, \partial^l \chi \right)\right]
    \quad ,
\eea
an $D_D$ encodes total derivatives.
\paragraph{Reassembling the Cubic Action}
\leavevmode\\
Finally, all contributions to the cubic action must be combined to determine their overall scaling with the slow-roll parameters. Crucially, the sum of the $F$-terms must be shown to be proportional to the variation of the quadratic action. Adding together the results of the four groupings of the cubic terms yields the following set of interactions:
\bea
    S^{(3)} &=& \int d^4x\, \Bigg\{ 
     \frac{1}{2} e^{3\rho} \frac{\dot{\phi}^4}{\dot{\rho}^4} \dot{\zeta}^2 \zeta
    + \frac{1}{2} e^{\rho} \frac{\dot{\phi}^4}{\dot{\rho}^4} \zeta\, \partial_k \zeta\, \partial^k \zeta
    - 2 e^{3\rho} \frac{\dot{\phi}^2}{\dot{\rho}^2} \dot{\zeta}\, \partial_k \zeta\, \partial^k \chi \notag + \nonumber \\
    && + e^{3\rho} \frac{\dot{\phi}^2}{\dot{\rho}^2} \dot{\zeta}\, \zeta^2\, \frac{d}{dt} \left[ \frac{\ddot{\phi}}{\dot{\phi} \dot{\rho}} + \frac{1}{2} \frac{\dot{\phi}^2}{\dot{\rho}^2} \right]
    - \frac{1}{8} e^{3\rho} \frac{\dot{\phi}^6}{\dot{\rho}^6} \dot{\zeta}^2 \zeta
    + \frac{1}{2} e^{3\rho} \frac{\dot{\phi}^2}{\dot{\rho}^2} \zeta\, \partial_k \partial_l \chi\, \partial^k \partial^l \chi \notag + \nonumber \\
    && + F + D \Bigg\}
    \quad ,
\eea
where $D$ represents the accumulated total derivative terms,
\beq
    D = D_0 + D_A + D_B + D_C + D_D
    \quad ,
\eeq
together with the terms that vanish when $\ddot{\rho} = -\frac{1}{2} \dot{\phi}^2$.
Analogously, $F$ represents the sum of terms that are proportional to
\beq
    \frac{d}{dt} \left[ e^{3\rho} \frac{\dot{\phi}^2}{\dot{\rho}^2} \dot{\zeta} \right] 
    - e^\rho \frac{\dot{\phi}^2}{\dot{\rho}^2} \partial_k \partial^k \zeta
    \quad .
\eeq
Together, these terms are:
\bea
    F &=& \left( \frac{d}{dt} \left[ e^{3\rho} \frac{\dot{\phi}^2}{\dot{\rho}^2} \dot{\zeta} \right] 
    - e^\rho \frac{\dot{\phi}^2}{\dot{\rho}^2} \partial_k \partial^k \zeta \right)
    \left( \frac{\ddot{\phi}}{\dot{\phi} \dot{\rho}} \zeta^2 
    + \frac{1}{2} \frac{\dot{\phi}^2}{\dot{\rho}^2} \zeta^2 
    + 2 \frac{1}{\dot{\rho}} \dot{\zeta} \zeta \right) \notag + \nonumber \\
    &&\quad + \frac{d}{dt} \left[ e^{3\rho} \frac{\dot{\phi}^2}{\dot{\rho}^2}    \dot{\zeta} \right] 
    \left( \frac{1}{\dot{\rho}} \left[\partial_k \zeta\, \partial^k \chi 
    - \partial^{-2} \partial^k \partial^l (\partial_k \zeta\, \partial_l \chi)\right] \right) \notag + \nonumber \\
    &&\quad - \frac{d}{dt} \left[ e^{3\rho} \frac{\dot{\phi}^2}{\dot{\rho}^2} \dot{\zeta} \right] 
    \frac{1}{2} e^{-2\rho} \frac{1}{\dot{\rho}^2} 
    \left( \partial_k \zeta\, \partial^k \zeta 
    - \partial^{-2} \partial^k \partial^l (\partial_k \zeta\, \partial_l \zeta) \right)+ \nonumber \\
    &&\quad - e^\rho \frac{\dot{\phi}^2}{\dot{\rho}^3} 
    \left( \partial_k \zeta\, \partial^k \chi\, \partial_l \partial^l \zeta 
    - \zeta\, \partial_k \partial_l (\partial^k \zeta\, \partial^l \chi) \right) \notag + \nonumber \\
    &&\quad + \frac{1}{2} e^{-\rho} \frac{\dot{\phi}^2}{\dot{\rho}^4} 
    \left( \partial_k \zeta\, \partial^k \zeta\, \partial_l \partial^l \zeta 
    - \zeta\, \partial_k \partial_l (\partial^k \zeta\, \partial^l \zeta) \right)
    \quad ,
\eea
The first line is already proportional to the $\zeta$ equation of motion, but the last two lines do not yet quite match with the two preceding them; but they can be converted by taking two
of the $\zeta$ ’s in them and inserting a factor of the identity operator in the form, $\partial^{-2}\partial_k\partial^k \zeta$ , and
then integrating by parts one last time so that these lines become,
\bea
    && \cdots + \Bigg( 
     - e^{\rho} \frac{\dot{\phi}^2}{\dot{\rho}^2} \partial_k \partial^k \zeta \Bigg)
    \frac{1}{\dot{\rho}} \left( \partial_l \zeta\, \partial^l \chi 
    - \partial^{-2} \partial^j \partial^l (\partial_j \zeta\, \partial_l \chi) \right) \nonumber \\
    && - \Bigg(- e^{\rho} \frac{\dot{\phi}^2}{\dot{\rho}^2} \partial_k \partial^k \zeta \Bigg)
    \frac{1}{2} \frac{e^{-2\rho}}{\dot{\rho}^2} 
    \left( \partial_l \zeta\, \partial^l \zeta 
    - \partial^{-2} \partial^j \partial^l (\partial_j \zeta\, \partial_l \zeta) \right)
    \quad .
\eea
The expression for $F$ then becomes,
\bea
    F &=& \left( \frac{d}{dt} \left[ e^{3\rho} \frac{\dot{\phi}^2}{\dot{\rho}^2} \dot{\zeta} \right] 
    - e^\rho \frac{\dot{\phi}^2}{\dot{\rho}^2} \partial_j \partial^j \zeta \right)
    \left( \frac{\ddot{\phi}}{\dot{\phi} \dot{\rho}} \zeta^2 + \frac{1}{2} \frac{\dot{\phi}^2}{\dot{\rho}^2} \zeta^2 + 2 \frac{1}{\dot{\rho}} \dot{\zeta} \zeta \right) \notag \\
    &&\quad + \left( \frac{d}{dt} \left[ e^{3\rho} \frac{\dot{\phi}^2}{\dot{\rho}^2} \dot{\zeta} \right] 
    - e^\rho \frac{\dot{\phi}^2}{\dot{\rho}^2} \partial_j \partial^j \zeta \right)
    \left( \frac{1}{\dot{\rho}} \left[\partial_k \zeta\, \partial^k \chi - \partial^{-2} \partial_k \partial_l (\partial^k \zeta\, \partial^l \chi)\right] \right) \notag \\
    &&\quad - \left( \frac{d}{dt} \left[ e^{3\rho} \frac{\dot{\phi}^2}{\dot{\rho}^2} \dot{\zeta} \right] 
    - e^\rho \frac{\dot{\phi}^2}{\dot{\rho}^2} \partial_j \partial^j \zeta \right)
    \Bigg( \frac{1}{2} e^{-2\rho} \frac{1}{\dot{\rho}^2} \Big[\partial_k \zeta\, \partial^k \zeta 
    \nonumber \\ &-& \partial^{-2} \partial_k \partial_l (\partial^k \zeta\, \partial^l \zeta)\Big] \Bigg) \quad .
\eea
Collecting all the terms yields
\bea
    F &=& \left( \frac{d}{dt} \left[ e^{3\rho} \frac{\dot{\phi}^2}{\dot{\rho}^2} \dot{\zeta} \right] 
    - e^\rho \frac{\dot{\phi}^2}{\dot{\rho}^2} \partial_j \partial^j \zeta \right)
    \Bigg( 
    \frac{\ddot{\phi}}{\dot{\phi} \dot{\rho}} \zeta^2 
    + \frac{1}{2} \frac{\dot{\phi}^2}{\dot{\rho}^2} \zeta^2 
    + 2 \frac{1}{\dot{\rho}} \dot{\zeta} \zeta \notag + \nonumber\\
    &-& \frac{1}{2} \frac{e^{-2\rho}}{\dot{\rho}^2} 
    \left( \partial_k \zeta\, \partial^k \zeta 
    - \partial^{-2} \partial_k \partial_l ( \partial^k \zeta\, \partial^l \zeta ) \right)
    + \frac{1}{\dot{\rho}} 
    \left( \partial_k \zeta\, \partial^k \chi 
    - \partial^{-2} \partial_k \partial_l ( \partial^k \zeta\, \partial^l \chi ) \right)
    \Bigg) \nonumber \\ &.&
\eea
So, up to total derivatives, the cubic part of the action for an inflationary theory with a single inflaton field, written in terms of coordinates where the inflaton has no fluctuations and the independent scalar field $\zeta(t,\vec{x})$ corresponds to the fluctuation in the scale factor that multiplies a flat spatial metric, is:
\bea
    S^{(3)} &=& \int d^4x\, \Bigg\{ 
     \frac{1}{4} e^{3\rho} \frac{1}{M_{pl}^2} \frac{\dot{\phi}^4}{\dot{\rho}^4} \dot{\zeta}^2 \zeta 
    + \frac{1}{4} e^{\rho} \frac{1}{M_{pl}^2} \frac{\dot{\phi}^4}{\dot{\rho}^4} \zeta\, \partial_k \zeta\, \partial^k \zeta +
    \nonumber \\ &-& \frac{1}{2} e^{3\rho} \frac{1}{M_{pl}^2} \frac{\dot{\phi}^4}{\dot{\rho}^4} \dot{\zeta}\, \partial_k \zeta\, \partial^k \left( \partial^{-2} \dot{\zeta} \right) \notag + \nonumber \\
    &+& \frac{1}{2} e^{3\rho}   \frac{\dot{\phi}^2}{\dot{\rho}^2} \dot{\zeta} \zeta^2 \frac{d}{dt} \left[ \frac{\ddot{\phi}}{\dot{\phi} \dot{\rho}} + \frac{1}{2 M_{pl}^2} \frac{\dot{\phi}^2}{\dot{\rho}^2} \right]+ \nonumber \\ &-& \frac{1}{16} e^{3\rho} \frac{1}{M_{pl}^4} \frac{\dot{\phi}^6}{\dot{\rho}^6} 
    \left( \dot{\zeta}^2 \zeta - \zeta\, \partial_k \partial_l \left( \partial^{-2} \dot{\zeta} \right) 
    \partial^k \partial^l \left( \partial^{-2} \dot{\zeta} \right)  \right) +\nonumber \\
    &+& \left( \frac{d}{dt} \left[ e^{3\rho} \frac{\dot{\phi}^2}{\dot{\rho}^2} \dot{\zeta} \right] 
    - e^\rho \frac{\dot{\phi}^2}{\dot{\rho}^2} \partial_j \partial^j \zeta \right) \Bigg[
    \frac{1}{2} \left( \frac{\ddot{\phi}}{\dot{\phi} \dot{\rho}} + \frac{1}{2 M_{pl}^2} \frac{\dot{\phi}^2}{\dot{\rho}^2} \right) \zeta^2 
    + \frac{1}{\dot{\rho}} \dot{\zeta} \zeta +\nonumber \\
     &-& \frac{1}{4} \frac{e^{-2\rho}}{\dot{\rho}^2} 
    \left( \partial_k \zeta\, \partial^k \zeta 
    - \partial^{-2} \partial_k \partial_l (\partial^k \zeta\, \partial^l \zeta) \right)+ \nonumber \\
    &+& \frac{1}{4 M_{pl}^2} \frac{\dot{\phi}^2}{\dot{\rho}^3} 
    \left( \partial_k \zeta\, \partial^k (\partial^{-2} \dot{\zeta}) 
    - \partial^{-2} \partial_k \partial_l (\partial^k \zeta\, \partial^l (\partial^{-2} \dot{\zeta}) ) \right) \Bigg] 
    \Bigg\} \quad .
\eea
Here, all of the proper factors of $M_{pl}$
have once again been restored.
To better understand the scaling with the slow-roll parameters, this set of cubic interactions can be written schematically as:
\bea
    S^{(3)} &=& M_{pl}^2 \int d^4x\, \Bigg\{ \varepsilon^2 e^{3\rho} \dot{\zeta}^2 \zeta 
    + \varepsilon^2 e^\rho \zeta\, \partial_k \zeta\, \partial^k \zeta 
    - 2 \varepsilon^2 e^{3\rho} \dot{\zeta}\, \partial_k \zeta\, \partial^k \left( \partial^{-2} \dot{\zeta} \right) \nonumber \\
    &+& \varepsilon (\dot{\delta} + \dot{\varepsilon}) e^{3\rho} \dot{\zeta} \zeta^2 
    - \frac{1}{2} \varepsilon^3 e^{3\rho} \left( \dot{\zeta}^2 \zeta 
    - \zeta\, \partial_k \partial_l (\partial^{-2} \dot{\zeta})\, \partial^k \partial^l (\partial^{-2} \dot{\zeta}) \right) \nonumber +\\
    &+& \left[ \frac{d}{dt} ( \varepsilon e^{3\rho} \dot{\zeta} ) 
    - \varepsilon e^{\rho} \partial_j \partial^j \zeta \right]
    \Big[ (\delta + \varepsilon) \zeta^2 
    + \frac{2}{\dot{\rho}} \dot{\zeta} \zeta \nonumber +\\
    &-& \frac{1}{2} \frac{e^{-2\rho}}{\dot{\rho}^2} \left( \partial_k \zeta\, \partial^k \zeta 
    - \partial^{-2} \partial_k \partial_l (\partial^k \zeta\, \partial^l \zeta) \right)+\nonumber\\
    &+& \frac{1}{\dot{\rho}} \varepsilon \left( \partial_k \zeta\, \partial^k (\partial^{-2} \dot{\zeta}) 
    - \partial^{-2} \partial_k \partial_l (\partial^k \zeta\, \partial^l (\partial^{-2} \dot{\zeta}) ) \right)
    \Big]
    \Bigg\} \quad .
\eea
The first-line operators are clearly $\varepsilon^2$-suppressed, while those below are even more negligible. Some terms proportional to the $\zeta$ equation of motion carry only a single $\varepsilon$ factor, but these do not contribute significantly to the three-point function in the observational limit.

Up to this point, the derivation closely follows Maldacena’s original treatment of non-Gaussianities~\cite{maldacena2003non}, where the size of each term in the Lagrangian was estimated in terms of slow-roll parameters. This is key to understanding why single-field slow-roll inflation predicts small non-Gaussianities. When applied to the \textit{climbing scalar model}, slow-roll is explicitly violated, yet the same cubic action form is used. It is thus important to stress that the structure of the cubic action does not rely on the slow-roll approximation. In the climbing scalar notation, where the scale factor is written as $e^{a/3}$, we identify $\rho = a/3$. We also rewrite the piece with the second slow roll coefficient using the previously proved relation: 
\beq
    \frac{\dot \varepsilon}{\varepsilon \dot \rho}=2(\delta+\varepsilon) 
\eeq Once again this is completely slow-roll independent.
\bea
    S^{(3)} &=& M_{pl}^2 \int d^4x\, \Bigg\{
     \varepsilon^2 e^{a} \dot{\zeta}^2 \zeta 
    + \varepsilon^2 e^{\frac{a}{3}} \zeta\, \partial_k \zeta\, \partial^k \zeta 
    - 2 \varepsilon^2 e^{a} \dot{\zeta}\, \partial_k \zeta\, \partial^k \left( \partial^{-2} \dot{\zeta} \right) \nonumber \\ &+& \frac{3}{2}\varepsilon \frac{d}{dt}\left(\frac{\dot \varepsilon}{\varepsilon \dot a}\right) e^{a} \dot{\zeta} \zeta^2 - \frac{1}{2} \varepsilon^3 e^{a} \left( \dot{\zeta}^2 \zeta 
    - \zeta\, \partial_k \partial_l (\partial^{-2} \dot{\zeta})\, \partial^k \partial^l (\partial^{-2} \dot{\zeta}) \right)  +f(\zeta)\frac{\delta \mathcal{L}}{\delta \zeta}
    \Bigg\} \quad .
    \label{eq:cubic action}
\eea
where we have used the notation $\frac{\delta \mathcal{L}}{\delta \zeta}$ to refer to the equation of motion for $\zeta$ and 
\bea
    f(\zeta)&=&\bigg[ \frac{3}{2}\frac{\dot \varepsilon}{\varepsilon a} \zeta^2 
    + \frac{6}{\dot{a}} \dot{\zeta} \zeta   - \frac{9}{2} \frac{e^{-\frac{2}{3}a}}{\dot{a}^2} \left( \partial_k \zeta\, \partial^k \zeta 
    - \partial^{-2} \partial_k \partial_l (\partial^k \zeta\, \partial^l \zeta) \right)+ \nonumber \\ 
    &+& \frac{3}{\dot{a}} \varepsilon \left( \partial_k \zeta\, \partial^k (\partial^{-2} \dot{\zeta}) 
    - \partial^{-2} \partial_k \partial_l (\partial^k \zeta\, \partial^l (\partial^{-2} \dot{\zeta}) ) \right)
    \bigg] \quad .
    \label{eq: field redefinition}
\eea
\begin{center}
    ***
\end{center}
In this chapter, we analyzed the climbing scalar’s impact on the two-point function and derived the cubic action for scalar perturbations, forming the basis for studying primordial non-Gaussianities.

Three key motivations emerge for deeper study of the climbing scalar model:
\begin{itemize}
    \item It explains the suppression of low multipoles in the CMB ($\ell \lesssim 40$).
    \item It provides a dynamical pre-inflationary mechanism triggering slow-roll inflation.
    \item It offers a string-theoretic interpretation of the inflaton, notably the dilaton’s climbing behavior.
\end{itemize}

The next step is to compute the three-point function using the \textit{Schwinger--Keldysh} formalism, focusing on perturbative expansion and time integrals during the climbing phase.

The final chapter will present original results on the amplitude and shape of non-Gaussianities, their dependence on model parameters, scenario comparisons, and a benchmark against Maldacena’s standard slow-roll prediction~\cite{maldacena2003non}.

\chapter{Non-Gaussian Amplitudes}
In this final chapter, we summarize the key steps leading to the main result: the computation of the full three-point scalar amplitude for cosmological perturbations in a climbing scalar background.

We first introduce the computational framework, focusing on the interplay between the slow-roll parameter $\varepsilon$, the scale factor $s$, and the mode functions $v$, which are central to the perturbative analysis.

Two alternative strategies for resolving the model’s singularity are examined, each with its motivations and drawbacks. We justify the choice ultimately adopted in the study.

We then outline the computational techniques used to evaluate the three-point correlator. The final sections present the general structure of the amplitude and compare it to the Maldacena single-field slow-roll result.

The chapter concludes with a discussion on the physical implications of the findings, emphasizing the specific predictions for scalar amplitude behavior in the climbing scalar scenario.

The material discussed in this section was previously presented by the authors in \cite{meo2025preinflationarynongaussianities}.
\section{Setup}
We begin this section by outlining the methodology for computing three-point functions in cosmological backgrounds, where the lack of time translation invariance requires expressing correlators as time integrals over the entire time domain.

These integrals involve interaction vertices derived from the interaction Lagrangian (eqs.~\eqref{eq:cubic action} and \eqref{eq: field redefinition}), contributing three mode functions per integral, often accompanied by time derivatives and combinations of the slow-roll parameter and scale factor (see next section).

The computation is performed in conformal time, with the relevant domain for a climbing scalar scenario given by $[-\eta_s, 0]$, where $-\eta_s$ marks the singularity.

Despite the applicability of the standard \textit{in-in} formalism (see next section), the structure of the Mukhanov–Sasaki potential near the singularity causes ultraviolet divergences in the three-point function. This indicates a breakdown of the two-derivative effective field theory in the singular regime. To examine this issue more closely, we analyze the behavior near the singularity.

The MS potential becomes \eqref{eq:singular-potential}
\beq
    W_S(\eta)=-\frac{1}{4(\eta + \eta_s)^2} \quad ,
\eeq
which means that the MS-equation becomes
\beq
    v''_k(\eta)+\left(k^2+\frac{1}{4(\eta + \eta_s)^2}\right)v_k(\eta)=0 \quad,
\eeq
that is the same shape of \eqref{eq:MSeqgeneral} with $\nu=0$, so the solution is \eqref{eq:modef1}:
\beq
    v_k(\eta)=e^{i \frac{\pi}{4}}\frac{\sqrt{\pi}}{2}\sqrt{- (\eta+\eta_s)}H_{0}^{(1)}\left(-k( \eta+\eta_s)\right) \quad.
\eeq
Near the initial singularity, so in the limit $\eta \to - \eta_s$, which means that the Hankel0 argument goes to zero, the mode function goes like (see \cite{abramowitz1964handbook,DLMF})
\beq
    v_k(\eta) \sim \frac{e^{i \frac{3}{4}\pi}}{\sqrt{\pi}}\sqrt{- (\eta+\eta_s)}\log \left(-k (\eta + \eta_s)\right) \quad .
\eeq
Now we anticipate that the integrals to be computed are of the type:
\beq
    \int^{0}_{- \eta_s} d \eta s^2(\eta) \varepsilon^2 \frac{d}{d \eta} \left(\frac{v_{k_1}(\eta)}{s (\eta) \sqrt{\varepsilon}}\right)\frac{d}{d \eta} \left(\frac{v_{k_2}(\eta)}{s (\eta) \sqrt{\varepsilon}}\right)\left(\frac{v_{k_3}(\eta)}{s (\eta) \sqrt{\varepsilon}}\right) \quad .
    \label{eq:O1tent}
\eeq
\beq
    \int^{0}_{- \eta_s} d \eta s^2(\eta) \varepsilon^2 \left(\frac{v_{k_1}(\eta)}{s (\eta) \sqrt{\varepsilon}}\right) \left(\frac{v_{k_2}(\eta)}{s (\eta) \sqrt{\varepsilon}}\right)\left(\frac{v_{k_3}(\eta)}{s (\eta) \sqrt{\varepsilon}}\right) \quad .
    \label{eq:O2tent}
\eeq
From chapter 2 we know that near the initial singularity $\varepsilon$ goes to a constant value $3$, so we can ignore it in determing the behavior of the integral. On the other hand the scale factor is \eqref{eq:scalefactornearsingular} (we choose of course the climbing solution)
\beq
    s(\eta) \sim \left(-(\eta + \eta_s)\right)^{1/2}
    \quad .
\eeq
This is also a confirmation of the shape of the $W_S$ which demostration was omitted in \eqref{eq:singular-potential}, infact if we derive twice, since $\varepsilon$ is constant, we have
\beq
    \frac{z''}{z}=\frac{s''}{s}=-\frac{1}{4 (\eta + \eta_s)^2}
    \quad .
\eeq
Plugging inside the integral
\beq
    \frac{v_k(\eta)}{s(\eta)}\sim \log (-k (\eta + \eta_s))
    \quad ,
    \quad
    \frac{d}{d \eta} \left(\frac{v_k (\eta)}{s (\eta)}\right)=\frac{1}{(\eta + \eta_s)}
    \quad ,
\eeq
\beq
    s^2 (\eta)\frac{d}{d \eta} \left(\frac{v_{k_1} (\eta)}{s (\eta)}\right)\frac{d}{d \eta} \left(\frac{v_{k_2} (\eta)}{s (\eta)}\right)\left(\frac{v_{k_3} (\eta)}{s (\eta)}\right) \sim -\frac{\log (-k_3 (\eta + \eta_s))}{\eta + \eta_s}
    \quad ,
\eeq
and once we integrate in $\eta$
\beq
    -\frac{1}{2}\left[\log \left(-k(\eta + \eta_s)\right)\right]^2
    \quad .
\eeq
That diverges near the initial singularity. 
The second integral is not divergent:
\bea
    &&s^2 (\eta) \left(\frac{v_{k_1} (\eta)}{s (\eta)}\right)\left(\frac{v_{k_2} (\eta)}{s (\eta)}\right)\left(\frac{v_{k_3} (\eta)}{s (\eta)}\right) \sim \nonumber \\ &&\sim -(\eta + \eta_s) \, \log \left(-k_1 (\eta + \eta_s)\right) \, \log \left(-k_2 (\eta + \eta_s)\right) \, \log \left(-k_3 (\eta + \eta_s)\right)
    \quad ,
\eea
This integral is more difficult, but still duable. We can solve it using the trick $\log a = \frac{d}{dx}a^x \Big|_{x=0}$
so,
\bea
    &&-\int^{0}_{-\eta_s}d \eta (\eta + \eta_s) \, \log \left(-k_1 (\eta + \eta_s)\right) \, \log \left(-k_2 (\eta + \eta_s)\right) \, \log \left(-k_3 (\eta + \eta_s)\right)= \nonumber \\&& =-\frac{d}{d \alpha_1} \Bigg|_{\alpha_1 = 0}\frac{d}{d \alpha_2} \Bigg|_{\alpha_2 = 0}\frac{d}{d \alpha_3} \Bigg|_{\alpha_3 = 0} (-k_1)^{\alpha_1} \, (-k_2)^{\alpha_2} \, (-k_3)^{\alpha_3} \, \int^{0}_{- \eta_s}d \eta \, (\eta + \eta_s)^{\alpha_1 + \alpha_2 + \alpha_3 + 1} = \nonumber \\ &&=\left[-\frac{d}{d \alpha_1} \Bigg|_{\alpha_1 = 0}\frac{d}{d \alpha_2} \Bigg|_{\alpha_2 = 0}\frac{d}{d \alpha_3} \Bigg|_{\alpha_3 = 0} (-k_1)^{\alpha_1} \, (-k_2)^{\alpha_2} \, (-k_3)^{\alpha_3} \, \frac{(\eta + \eta_s)^{\alpha_1 + \alpha_2 + \alpha_3+2}}{\alpha_1 + \alpha_2 + \alpha_3+2}\right]^{0}_{-\eta_s}
\eea
Doing derivatives we obtain a finite contribution as anticipated.

To address the divergence issue, we propose the central hypothesis of this work: higher-curvature corrections resolve the initial singularity, effectively regularizing the cosmological background. Concretely, the conformal time singularity is not physical but an artifact of the two-derivative approximation. \textit{Once higher-order corrections are properly accounted for, the singularity is effectively "pushed" to infinite past time, thus removing the divergent behavior associated with the three-point function}.

This idea is supported by modified gravity models and string theory, where higher-curvature terms become relevant in high-curvature regimes. While the exact form of these corrections is unknown, they are expected to smooth out the singularity.

Since the full corrected dynamics is analytically intractable, we introduce a physically motivated ansatz that captures the expected features of a regularized background. This allows us to perform the perturbative analysis within a smooth geometry, without full knowledge of the UV-complete theory.

A key assumption is that, within this regularized background, perturbations remain governed by the standard two-derivative effective theory. Although higher-order corrections modify the background, their influence on perturbations is assumed to be encoded entirely in the modified geometry. This separation is common in cosmological contexts.

We acknowledge that this assumption is both central and delicate, representing the main limitation of our approach. Nevertheless, it offers a tractable framework for obtaining well-defined predictions for the three-point function in climbing scalar models.

Importantly, our hypothesis involves only gravitational higher-curvature corrections, excluding string loop or quantum effects. Hence, the perturbative expansion remains valid within the standard cosmological framework.

To implement this idea, we introduce two ansätze reflecting features expected from higher-curvature regularization. Ultimately, only one resolves the divergence without introducing pathologies. However, both are discussed in the following sections, as each contributes to constructing a consistent and physically meaningful background.
\subsection{Persistent de Sitter Scale Factor}
The singularity resolution ansatz is built around three interrelated functions: $\varepsilon(\eta)$, $s(\eta)$, and $v_k(\eta)$, or equivalently $\varepsilon(\eta)$, $s(\eta)$, and $W_S(\eta)$.

The first ansatz assumes that higher-curvature corrections extend the LM attractor behavior—observed just after the singularity (see Fig.~\ref{fig:numerics})—to all times. For $\gamma^2 \ll 1$, corresponding to $\gamma \sim \mathcal{O}(0.1)$, this implies a de Sitter background, which is nonsingular and free from divergence issues in standard non-Gaussianity computations \cite{maldacena2003non,BaumannJoyceCosmoCorrelators2023}.

This leads to the conformal time domain $(-\infty, 0)$ and the de Sitter scale factor:
\beq
    s(\eta) = -\frac{1}{\mathcal{H}\eta} \quad .
    \label{eq:persistentds}
\eeq
Fixing this form for $s(\eta)$, we determine the required behavior of $\varepsilon(\eta)$ and $W_S(\eta)$ to maintain consistency. (Note the change in Hubble constant notation, discussed later.)

To maintain physical accuracy, the MS potential near $\eta = 0$ should recover the de Sitter behavior $\frac{2}{\eta^2}$ for $\gamma \ll 1$. However, oversimplifying would reduce the setup to standard slow-roll inflation. Our goal is to adapt the computation in \cite{maldacena2003non} to include key features of the climbing scalar background while preserving analytic control.

\begin{figure}
    \centering
    \includegraphics[width=0.3\linewidth]{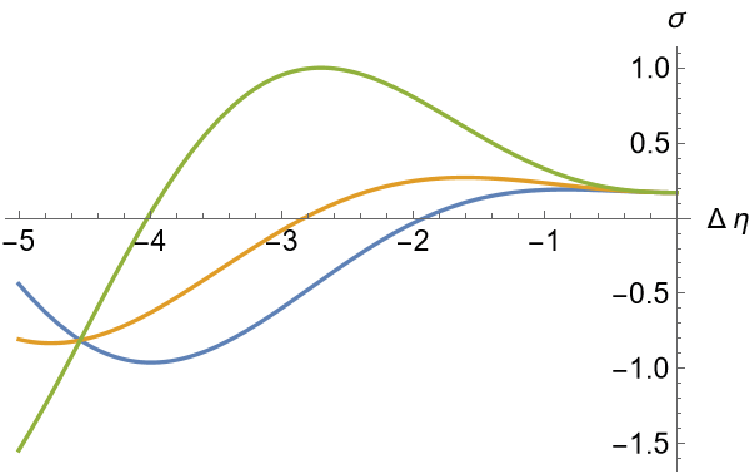}
    \caption{$\sigma (\Delta \eta )$ with the non-trivial choice of $\nu = 1.53$ that correspond to the usual value of $n_s \sim 0.96$, and with different values of $c_1$: the blue curve corresponds to $c_1 = - 0.2$, the orange curve corresponds to $c_1=0$, the green curve corresponds to $c_1 = 0.4$. Note that for $c_1 > 0$ $\sigma$ develops the undesired bump. The first zero is in the region $[-2/\Delta \, , \, 0]$.}
    \label{fig:sigma}
\end{figure}

Therefore, we adopt a modified MS potential, motivated both theoretically \cite{DudasKitazawaPatilSagnotti:2012,KitazawaSagnotti:2014} and observationally \cite{GruppusoKitazawaMandolesiNatoliSagnotti2016}, given by:
\beq
    W_S(\eta) = \frac{2}{\eta^2} - \Delta^2 \quad .
\eeq
With $s(\eta)$ and $W_S(\eta)$ specified, we can analyze the differential equation governing $\varepsilon(\eta)$:
\beq
    W_S(\eta) = \frac{z''}{z} \qquad z = s(\eta) \sqrt{2 \varepsilon(\eta)}
\eeq
So under the discussed assumptions, defining $\sigma (\eta) \equiv \sqrt{\varepsilon (\eta)}$, we have (at this stage we can use just attractor expressions without requiring $\gamma \ll 1$ so $s(\eta) \sim (-\eta)^{ \frac{1}{2}-\nu}$ and $W_S(\eta) \sim \frac{\nu^2 - \frac{1}{4}}{\eta^2}-\Delta^2$)
\bea
    \frac{z''}{z}&=&\frac{1}{s \, \sigma} \frac{d^2}{d \eta^2}(s \, \sigma)= \frac{\sigma '' \, s \, + \, 2 \, \sigma ' \, s' \, + \, \sigma \, s''}{s \, \sigma}=\frac{\sigma ''}{\sigma} \, + \, 2 \, \frac{\sigma '}{\sigma}\frac{s'}{s} \, + \, \frac{s''}{s}= \nonumber \\&=& \frac{\sigma ''}{\sigma} \, + \, \frac{(2 \nu - 1)}{-\eta} \, \frac{\sigma'}{\sigma} \, + \, \frac{\nu^2-\frac{1}{4}}{\eta^2} \, = \, \frac{\nu^2-\frac{1}{4}}{\eta^2} \, - \, \Delta^2 \quad ,
\eea
\beq
    \sigma '' \, + \, \frac{1-2 \nu}{\eta} \sigma ' \, + \,  \Delta^2 \sigma \, = \, 0 \quad .
\eeq
The solution of this equation is Bessel-like (following the same procedure done for mode functions in Chapter 3): 
\beq
    \sigma(\eta)=c_1 (-\eta)^{\nu}J_{\nu}(-\Delta \eta) + c_2 (-\eta)^{\nu} Y_{\nu}(-\Delta \eta) \quad ,
\eeq
\beq
    \varepsilon(\eta)=[c_1 (-\eta)^{\nu} J_{\nu} (-\Delta \eta) + c_2 (-\eta)^{\nu} Y_{\nu}(-\Delta \eta)]^2
\eeq
Now we have to impose boundary conditions. The first obvious one is the slow-roll value obtained by $\varepsilon$ once $\eta \to 0$, which means $\sigma \to \sqrt{3} \gamma$: using the known limit expressions for the Bessell functions:
\bea
    J_{\nu}(x) &\to& 2^{- \nu}\left(\Gamma(\nu + 1)\right)^{-1} \, x^\nu \\
     Y_{\nu}(x) &\to& -\frac{1}{\pi} 2^{\nu}\Gamma(\nu) \, x^{-\nu}
\eea
\bea
    \sigma(\eta)&\to& c_1 \, (-\eta)^{\nu} \, 2^{-\nu} \, \Delta^{\nu} \, (-\eta)^{\nu} \,\left(\Gamma(\nu + 1)\right)^{-1} \, + \,  c_2  \,(- \eta)^\nu \, (-1/\pi) \, \Gamma(\nu) \, 2^{\nu} \, \Delta^{-\nu} \,(-\eta)^{-\nu} \nonumber \\ &\to& \, c_2 \, (-1/(\pi \, \Delta^\nu)) \, \Gamma (\nu) 2^\nu  = \sqrt{\frac{2 \nu-3}{2 \nu-1}} \quad .
\eea
Throughout this derivation we have used the relation \eqref{eq:nuvalue} to express $\gamma^2$. From the matching we have in the end \beq
    \varepsilon(\eta)=\left[c_1 (-\eta)^{\nu} J_{\nu}(-\Delta \eta) - \frac{\pi}{2^{\nu}\Gamma(\nu)}\sqrt{\frac{2\nu-3}{2\nu-1}}(-\Delta \eta)^{\nu} Y_{\nu}(-\Delta \eta)\right]^2 \quad ,
    \label{eq:epsilon}
\eeq
The parameter \( c_1 \) controls the bump in the profile of \( \sigma \), or equivalently \( \varepsilon \) (see Fig.~\ref{fig:sigma}). For slightly negative \( c_1 \), the bump is negligible and the first zero \( \eta_t \) approaches \( -\,\frac{2}{\Delta} \). For positive \( c_1 \), \( \eta_t \) shifts toward \( -\,\frac{4}{\Delta} \), accompanied by a pronounced bump, indicating a fast-roll phase during descent. This would imprint a pre-inflationary peak on the power spectrum, distinct from the Chibisov–Mukhanov regime analyzed in~\cite{KitazawaSagnotti:2014} and shown in Fig.~\ref{fig:Power6}. We focus on small \( c_1 \) values, corresponding to \( \eta_t \geq -\,\frac{2}{\Delta} \), thus avoiding such non-universal features that do not significantly affect \( \eta_t \).

Here, deviations from the de Sitter case are significant: assuming \( \gamma \ll 1 \Rightarrow \nu \sim \frac{3}{2} \) and enforcing an exact zero bump leads to exact de Sitter, i.e. \( \varepsilon = 0 \), which would yield a featureless result. This approximation is not suitable at this stage. Instead, we retain the turning point of the scalar—corresponding to the zero of \( \varepsilon \)—as the first nontrivial feature, localized in time and deviating from the slow-roll value \( 3\gamma^2 \).

This analysis is valid up to the first zero of \( \sigma \) or \( \varepsilon \), denoted by \( \eta_t \), which we link to the scale \( \Delta \).

Dimensional analysis suggests \( \eta_t \sim \Delta^{-1} \), though the exact coefficient depends on \( c_1 \). Within the allowed range avoiding a bump in \( \sigma \), this coefficient lies between 0 and 2. We define:
\beq
    \eta_t =  \frac{\alpha}{\Delta} \, \, \, \, \, \text{with} \, \, \, \alpha \in [-2,0) \quad .
\label{eq:etat}
\eeq
The reference value is \( \eta_t \sim - \frac{1}{\Delta} \), which will be important in constructing the three-point function.

 The presence of a turning point for the scalar where the slow--roll parameter $\varepsilon$ vanishes, makes the three--point amplitude based on the Mukhanov--Sasaki variable apparently singular. Or, if you will, making curvature and scalar perturbations somehow inequivalent. If one insists on working with curvature perturbations, which are closer in spirit to the actual observations, the singularity is resolved by the Schwinger--Keldysh contour, and the contributions from the turning point have peculiar and potentially interesting features.

However, extending the same MS potential form to all conformal times leads to inconsistency: the resulting \( \varepsilon \) oscillates indefinitely as \( \eta \to -\infty \), contrary to expected scalar dynamics, where \( \varepsilon \to 3 \) under strict climbing behavior (see Chapter 2).

This also explains why no consistent boundary condition in the infinite past can uniquely fix \( c_1 \).

From the differential equation for \( \sigma \), we see a constant solution requires the \( \Delta \)-term to vanish—i.e., \( \Delta \to 0 \). Thus, maintaining the attractor form of the scale factor necessitates an MS potential that rises after crossing the \( \eta \)-axis.

In this view, \( \Delta \) represents a localized distortion in the MS potential that fades in the far past. This aligns with the actual potential profile shown in Fig.~\ref{fig:MSpotential2}, where the MS potential increases before the singularity. Resolving the singularity by pushing it to \( \eta \to -\infty \) naturally eliminates the effect of \( \Delta \).

A key issue is the asymptotic behavior of \( \varepsilon \). While climbing dynamics suggest \( \varepsilon = 3 \), this contradicts the near–de Sitter scale factor assumed in our framework. Since
\beq
    \varepsilon = - \frac{1}{\mathcal{H}^2}\frac{d \mathcal{H}}{dt_c} \quad ,
\eeq
this parameter must remain small in quasi-de Sitter evolution.

\begin{figure}[ht]
\begin{center}
    \includegraphics[width=3in]{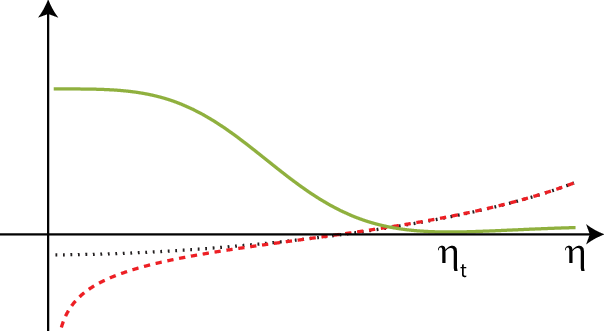}
\end{center}
\caption{Typically the scale factor (red, dashed) approaches the attractor (black, dotted) while the scalar field (green, solid) is still in the climbing phase with large $\varepsilon$. Tracing backwards, the region with $\varepsilon \simeq 3$ signals the initial singularity. The dip in $\varepsilon$ marks the turning point \cite{meo2025preinflationarynongaussianities}.}
\label{fig:attractor}
\end{figure}

We explored whether higher-curvature corrections could alter this relation, allowing large \( \varepsilon \) alongside a de Sitter scale factor. Numerical results (see Figs.~\ref{fig:numerics}, \ref{fig:attractor}) appear to support this: \( \varepsilon \) remains large even as the scale factor reaches the attractor. However, relying on uncontrolled corrections would contradict our hypothesis, which assumes the background is resolved without modifying the two-derivative theory.

This led us to reject toy models for \( \varepsilon \) (e.g., square wells between 3 and \( 3\gamma^2 \)) despite their numerical appeal. Instead, we concluded that resolving the singularity must restore \( \varepsilon \to 3\gamma^2 \) in the infinite past. This avoids fast-roll asymptotics while preserving the key feature: the localized dip at the turning point.

\begin{figure}[ht]
\begin{center}
    \includegraphics[width=3in]{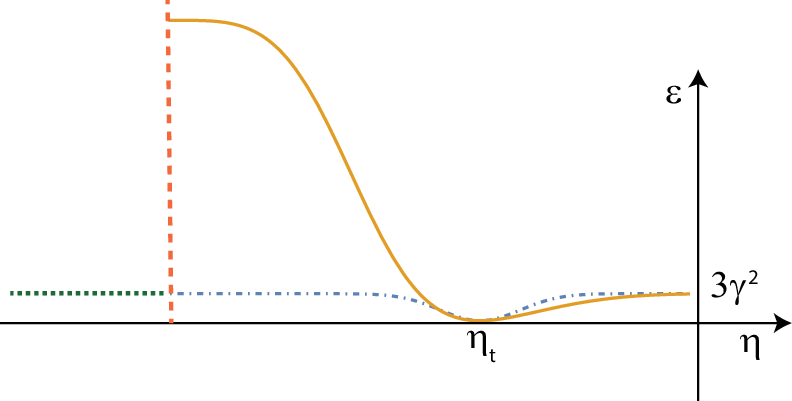}
\end{center}
\caption{Simplified model for $\varepsilon(\eta)$ (blue, dash-dotted), featuring a dip at the turning point $\eta_t$. The actual $\varepsilon(\eta)$ (orange, solid) grows rapidly as the Universe collapses. We assume singularity resolution allows a past-extending de Sitter–like phase with $\varepsilon = 3\gamma^2$ outside the dip \cite{meo2025preinflationarynongaussianities}.}
\label{fig:model_eps}
\end{figure}

Having fixed \( s(\eta) \) and \( \varepsilon(\eta) \), we now determine the shape of \( W_S(\eta) \). The initial segment must correspond to the modified MS potential containing \( \Delta \). As discussed, \( \Delta \) must eventually vanish, implying a rising potential at early times.

\begin{figure}[ht]
\centering
\begin{tabular}{ccc}
\includegraphics[width=50mm]{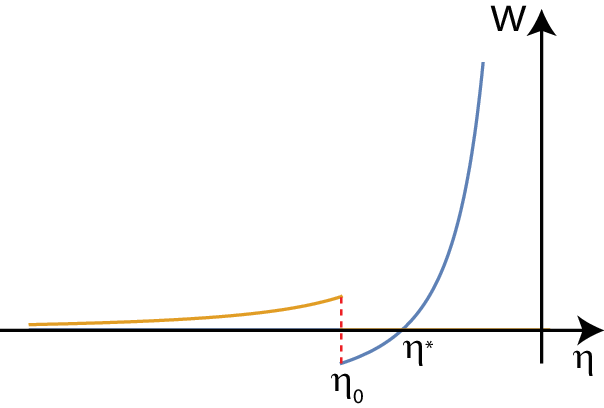} \quad  &
\includegraphics[width=50mm]{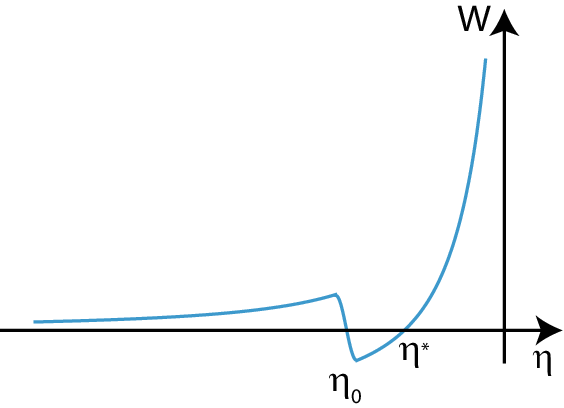}  \\
\end{tabular}
 \caption{\small Left: simple model where $W$ jumps at $\eta_0$ from the perturbed to attractor form. Right: more realistic smooth transition \cite{meo2025preinflationarynongaussianities}.}
\label{fig:Wsm}
\end{figure}

Enforcing the quasi-de Sitter attractor for \( s(\eta) \) and constant \( \varepsilon(\eta) \) implies that the MS potential must asymptotically behave as:
\begin{equation}
    W_S(\eta) = \frac{z''}{z} = \frac{s''}{s} \sim \frac{2}{\eta^2} \quad.
\end{equation}
This behavior requires \( W_S \) to cross the \( \eta \)-axis again before reaching early-time asymptotics. Thus, \( \Delta \) must vanish at some \( \eta_0 \), after which \( W_S \) returns to its attractor form, as shown in Fig.~\ref{fig:Wsm} (left: sharp jump, right: smooth transition).

The abrupt-jump model is analytically solvable using standard matching conditions~\cite{landau1977quantum}, since \( W_S(\eta) \) is:
\begin{align}
    W_S(\eta) &= \frac{2}{\eta^2} - \Delta^2 \quad \text{for} \quad -\eta_0 < \eta < 0 \nonumber \\
    W_S(\eta) &= \frac{2}{\eta^2} \quad \text{for} \quad \eta < -\eta_0 \quad.
\end{align}
The corresponding mode functions are:
\bea
    v_k (\eta) &=& \frac{1}{\sqrt{2k}}\left(1 - \frac{i}{k \eta}\right)e^{-ik \eta} \quad \text{for} \quad \eta < - \eta_0 \\
    v_k (\eta) &=& \frac{1}{\sqrt{2\omega}}\left[
        c_1\left(1 - \frac{i}{\omega \eta}\right)e^{-i\omega \eta} +
        c_2\left(1 + \frac{i}{\omega \eta}\right)e^{+i\omega \eta}
    \right] \quad \text{for} \quad - \eta_0 < \eta < 0 \quad,
\eea
where \( \omega = \sqrt{k^2 + \Delta^2} \), and the Bunch–Davies condition is imposed. Matching at \( \eta = -\eta_0 \) gives:
\bea
    c_1 &=& \frac{(k + \omega) \, e^{-i \eta_0 (k - \omega)} \left[k(\eta_0 \omega + i) - i \omega \right]}{2 \eta_0 k^{3/2} \omega^{3/2}} \quad, \\
    c_2 &=& -\frac{(k - \omega) \, e^{-i \eta_0 (k + \omega)} \left[k(\eta_0 \omega - i) - i \omega \right]}{2 \eta_0 k^{3/2} \omega^{3/2}} \quad.
\eea
When \( \omega = k \), \( c_2 = 0 \) and only the standard outgoing wave survives. But for \( \omega \neq k \), the reflected component \( H^{(2)}_{\frac{3}{2}}(-\omega \eta) \) is unavoidable.

This leads to a critical issue: the reflected wave introduces an infrared (IR) divergence. In the non-Gaussianity contributions like \eqref{eq:O2tent}, the outgoing mode yields:
\beq
    \frac{\sin x - x \cos x}{x^2} \quad,
\eeq
which is finite near the origin. But the reflected wave gives:
\beq
    \frac{\cos x - x \sin x}{x^2} \quad,
\eeq
which diverges as \( x \to 0 \) like \( \frac{1}{x^2} \), signaling an IR divergence.

Neglecting this term would induce errors up to 60\% in the relevant \( k \)-range. Therefore, eliminating the reflected wave became a priority, requiring a more refined design of the MS potential. The next section explores possible strategies to achieve this.

\subsubsection{Attempts to Eliminate the Reflected Wave}
We first attempted a standard Quantum Mechanics approach, writing the MS potential as a de Sitter background plus a localized perturbation:
\beq
    W_S^{(0)}(\eta) = \frac{2}{\eta^2} \quad \text{and} \quad W_S(\eta) = W_S^{(0)}(\eta) + V(\eta) \quad,
\eeq
with \( V(\eta) \) small and confined in a narrow region. Assuming \( \Delta \ll k \), we expanded the mode function as:
\beq
    v(\eta) = v_{(0)}(\eta) + v_{(1)}(\eta) \quad,
\eeq
where the unperturbed solution is:
\beq
    v_{(0)} (\eta) = - \frac{\sqrt{\pi}}{2}\sqrt{- \eta} \, H^{(1)}_{\frac{3}{2}}(-k \eta) \quad.
\eeq
The correction satisfies:
\beq
    v_{(1)}'' + \left(k^2 - \frac{2}{\eta^2}\right)v_{(1)} = V(\eta)\, v_{(0)} \quad,
\eeq
and can be solved using the Green function:
\beq
    v_{(1)}(\eta) \sim \int d \eta' \, G(\eta , \eta') \, V(\eta') \, v_{(0)}(\eta') \quad.
\eeq
With Bunch–Davies conditions, the Green function is:
\beq
    G(\eta,\eta')=\frac{v_{(0)}^2(\eta) \, v_{(0)}^1(\eta')}{W(\eta')} \quad, \quad W(\eta') \sim \frac{1}{\eta'} \quad,
\eeq
which leads to:
\beq
    v_{(1)}(\eta) \sim \sqrt{- \eta} \, H_{\frac{3}{2}}^{(2)}(-k\eta)\int d \eta'  \, (\eta')^2\left(H_{\frac{3}{2}}^{(1)}(-k \eta')\right)^2 V(\eta') \quad.
\eeq
This again generates the reflected wave component, \( H^{(2)}_{\frac{3}{2}} \), which persists regardless of how localized \( V(\eta) \) is. Even restricting the upper limit of the integral to the end of inflation (\( \eta \sim e^{-60} \)) doesn't sufficiently suppress it—unless \( V(\eta) \) is so small as to render the correction negligible.

Hence, this first perturbative strategy fails to resolve the infrared divergence without trivializing the problem.
\begin{center}
    ***
\end{center}
Secondly, we considered that removing the reflected wave must arise from a non-perturbative effect, since any discontinuity inevitably produces it. Thus, we tried to build a smooth, exactly solvable potential avoiding the second intersection with the \(\eta\)-axis, yet maintaining the correct asymptotic behavior toward zero.

A relevant example is the potential from \cite{DudasKitazawaPatilSagnotti:2012}, choosing \( c = 1 \) in \eqref{eq:cpot}, which tends to zero in the infinite past as required. The solutions here are Coulomb functions, and imposing Bunch–Davies conditions yields:
\beq
    v_k \sim \frac{1}{\sqrt{\alpha}} \left(F_L(\beta, \eta) + i G_L(\beta , \eta)\right) \quad,
\eeq
with \( L=1 \) to mimic quasi-de Sitter, and parameters
$\alpha \sim \frac{k}{\Delta} \quad, \quad \beta = -\frac{1}{\alpha} \sim -\frac{\Delta}{k}$, where \(\Delta\) relates to the previous intersection \(\eta^*\).

Comparing real and imaginary parts of this solution with the Bunch–Davies mode in standard de Sitter space for a trial \( k \) shows that while the real parts can be roughly matched by fine-tuning \(\Delta\), the imaginary parts are completely out of phase. This clearly indicates the unavoidable presence of a reflected wave, confirming that this second approach also fails.
\begin{center}
    ***
\end{center}
In the end, we made a final attempt based on the following observation: in supersymmetric quantum mechanics, some models feature reflectionless potentials. For example, the potential
\beq
    V(x) = -\frac{2 \kappa^2}{\cosh^2 (\kappa x)} \quad,
\eeq
is known to be reflectionless~\cite{sakurai2020modern, schwabl1995quantum, landau1977quantum}.

Although this potential cannot exactly mimic the Mukhanov–Sasaki potential, it can be viewed as a deformation that smooths the transition without trivializing the problem. The main question is whether adding the centrifugal barrier spoils the reflectionless property. The equation under study is
\beq
    \frac{d^2 v}{d \eta^2} + \left(k^2 - \frac{2}{\eta^2} + \frac{\Delta^2}{\cosh^2(\alpha \eta)}\right) v = 0 \quad,
\eeq
where the coefficient \(\Delta^2\) matches the simplified model near the origin.

Using approximation techniques from low-energy physics~\cite{you2013solutions}, we approximate the centrifugal barrier by a hyperbolic function, rewriting the equation as a Pöschl–Teller type:
\beq
    \frac{d^2 v}{d \eta^2} + \left(k^2 - \frac{2 \alpha^2}{\sinh^2 (\alpha \eta)} + \frac{\Delta^2 - \frac{2}{3}\alpha^2}{\cosh^2(\alpha \eta)}\right) v = 0 \quad.
\eeq
Defining \(\rho = \alpha \eta\) and \(z = \tanh^2 \rho\), the equation becomes
\bea
    &&\frac{d^2 v}{dz^2} + \frac{1 - 3z}{2 z (1 - z)} \frac{dv}{dz} + \left[ \frac{\kappa^2}{4 z (1 - z)^2} + \frac{\gamma^2}{4 z (1 + z)} - \frac{1}{2 z (1 + z)^2} \right] v = 0 \nonumber \\
    &&\quad \text{with} \quad \gamma^2 = \left(\frac{\Delta}{\alpha}\right)^2 - \frac{2}{3}, \quad \kappa = \frac{k}{\alpha} \quad.
\eea
Using the ansatz \(v(z) = z (1 - z)^{-\frac{i \kappa}{2}} w(z)\), this reduces to a hypergeometric equation:
\beq
    (1 - z) z w''(z) + \frac{1}{2} \left[ 5 + (-7 + 2 i \kappa) z \right] w'(z) + \frac{1}{4} \left[\gamma^2 + \kappa (\kappa + 5 i) - 6 \right] w(z) = 0 \quad.
\eeq
The general solution reads
\bea
    v_k(\eta) &=& \tanh^2(\alpha \eta) \left(1 - \tanh^2(\alpha \eta)\right)^{-\frac{i \kappa}{2}} \times \nonumber \\
    &\times& \Bigg[ \frac{i c_1 \, _2F_1\left(-\frac{i \kappa}{2} - \frac{1}{4} \sqrt{4 \gamma^2 + 1} - \frac{1}{4}, -\frac{i \kappa}{2} + \frac{1}{4} \sqrt{4 \gamma^2 + 1} - \frac{1}{4}; -\frac{1}{2}; \tanh^2(\alpha \eta) \right)}{\tanh^{3}(\alpha \eta)} \nonumber \\
    &+& c_2 \, _2F_1\left(-\frac{i \kappa}{2} - \frac{1}{4} \sqrt{4 \gamma^2 + 1} + \frac{5}{4}, -\frac{i \kappa}{2} + \frac{1}{4} \sqrt{4 \gamma^2 + 1} + \frac{5}{4}; \frac{5}{2}; \tanh^2(\alpha \eta) \right) \Bigg] \quad.
\eea
Imposing that near the origin only the incoming wave \(H^{(1)}_{\frac{3}{2}}(-\omega \eta)\) is present, fixes the constants:
\beq
    c_1 = - \frac{\alpha}{\omega} \quad, \quad c_2 = -\frac{1}{3} \left(\frac{\omega}{\alpha}\right)^2 \quad,
\eeq
so that
\bea
    v_k(\eta) &=& \tanh^2(\alpha \eta) \left(1 - \tanh^2(\alpha \eta)\right)^{-\frac{i \kappa}{2}} \times \nonumber \\
    &\times& \Bigg[ i \left(-\frac{\alpha}{\omega}\right) \frac{\,_2F_1\left(-\frac{i \kappa}{2} - \frac{1}{4} \sqrt{4 \gamma^2 + 1} - \frac{1}{4}, -\frac{i \kappa}{2} + \frac{1}{4} \sqrt{4 \gamma^2 + 1} - \frac{1}{4}; -\frac{1}{2}; \tanh^2(\alpha \eta) \right)}{\tanh^{3}(\alpha \eta)} \nonumber \\
    &-& \frac{1}{3} \left(\frac{\omega}{\alpha}\right)^2 \, _2F_1\left(-\frac{i \kappa}{2} - \frac{1}{4} \sqrt{4 \gamma^2 + 1} + \frac{5}{4}, -\frac{i \kappa}{2} + \frac{1}{4} \sqrt{4 \gamma^2 + 1} + \frac{5}{4}; \frac{5}{2}; \tanh^2(\alpha \eta) \right) \Bigg] \quad.
\eea
Examining the asymptotic behavior for \(\eta \to \infty\), one finds a combination of two waves:
\beq
    v(\eta) \to A e^{-i k \eta} + B e^{i k \eta} \quad,
\eeq
with explicit expressions for \(A\) and \(B\) in terms of Gamma functions and parameters \(\alpha, \omega, \gamma, k\) (see~\cite{DLMF, abramowitz1964handbook}). 

Eliminating the reflected wave, i.e., imposing \(B = 0\), requires a very restrictive condition on \(\omega\) or on allowed \(k\) values, which is not physically reasonable.

Therefore, this third approach also failed, confirming the inevitability of two waves and motivating a reconsideration of the initial assumptions on the scale factor and the quasi-de Sitter attractor form.
\subsection{Bounce Cosmology}
\begin{figure}[ht]
\centering
\begin{tabular}{ccc}
%\mbox{graphic} & \mbox{table} \\
\includegraphics[width=50mm]{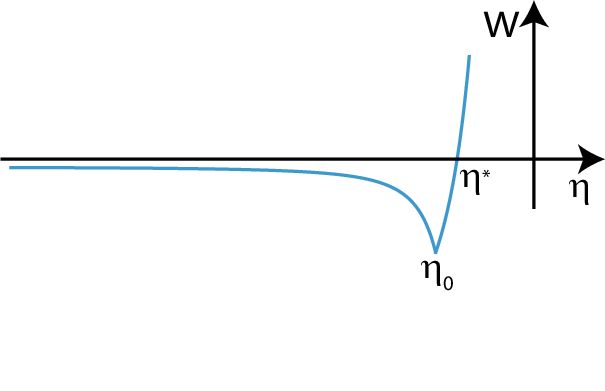} \quad  &
\includegraphics[width=50mm]{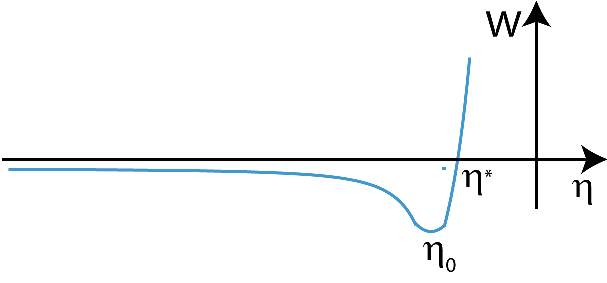}  \\
\end{tabular}
 \caption{\small Left panel: a sharp transition between an earlier epoch of compression and the expansion that evolves into the inflationary scenario. Right panel: a smooth model of the transition \cite{meo2025preinflationarynongaussianities}.}
\label{fig:Wsm2}
\end{figure}
In the end, we formulated an alternative picture that allows us to avoid the difficulties encountered in the previous approaches. As shown in Fig.~\ref{fig:MSpotential2}, the MS potential $W_S$ initially increases as $\eta$ decreases, before being dominated by the singularity at early times. But what if this growth persisted, as illustrated in Fig.~\ref{fig:Wsm2}? If $\varepsilon$ were to recover its slow-roll value $3\,\gamma^2$, then \textit{the Universe would be compelled to expand again for decreasing values of $\eta$}, and the initial singularity would be replaced by a bounce occurring at a particular conformal time, denoted by $\eta_0$.

In more detail, once we assume an early growth of $W_S$ as shown, the second branch can be modeled using the standard form:
\beq
    W_S(\eta)=\frac{\xi^2 - \frac{1}{4}}{\eta^2}
    \quad ,
    \label{eq:bouncing MS}
\eeq
with the requirement that $\xi^2 - \frac{1}{4} < 0$. This structure resembles the behavior of the actual MS potential in the early branch, while avoiding the appearance of a singularity in finite conformal time. As anticipated, this scenario is consistent with the ``pushing'' effect expected from higher-curvature string corrections.

Unlike before, the scale factor in this branch is not quasi-de Sitter. However, since $\varepsilon$ is constant, we still have:
\beq
    W_S(\eta)=\frac{s''}{s}=\frac{\xi^2-\frac{1}{4}}{\eta^2} \quad \to \quad s(\eta) \sim (-\eta)^{\frac{1}{2}\pm \xi} \quad .
\eeq
Because we have assumed $\xi < \frac{1}{2}$, both solutions describe a contracting Universe that undergoes a bounce precisely at $-\eta_0$, where a Big Crunch is followed by a new de Sitter expansion.

A typical value of $\eta_0$ can be estimated by matching the two branches of the MS potential given in \eqref{eq:bouncing MS} and \eqref{eq:deltaMS}. Choosing $\xi = 0$—which reproduces the growth observed in the actual MS potential \eqref{eq:singular-potential}—yields:
\beq
    \eta_0 = - \frac{1}{\Delta}
\eeq
This value is slightly smaller than $\eta_t$, but since both are of order $O\left(-\, \frac{1}{\Delta}\right)$, we can simply identify them. If one assumes that the bounce is induced by higher-derivative string corrections, then the Bunch--Davies condition is naturally imposed at the moment of maximal compression. This cuts off all integrals involved in the computation of the three-point function at $\eta_0$, thus avoiding any singularity and preventing the appearance of reflected waves, since only the second branch of the mode functions from \eqref{eq:deltaMS} is involved.

While we cannot derive this bounce from first principles, the idea that higher-curvature string corrections could resolve the singularity is widely accepted. Indeed, this assumption is not foreign to other approaches to non-Gaussianities—starting from the original work of Maldacena~\cite{maldacena2003non,collins2011primordial}. A bounce scenario of this type was also explored long ago in the context of String Theory by Gasperini and Veneziano~\cite{gasperini1993,gasperini2003}, relying on $T$-duality. In our case, it provides a consistent framework that resolves the main issues discussed so far, and therefore forms the basis for the remainder of our analysis.

To summarize, our setup assumes a \textit{bouncing universe}, with a Big Crunch at $\eta_0$, which marks the beginning of conformal time. Thus, the relevant domain becomes $(-\eta_0, 0)$. Within this interval, the MS potential is:
\beq
    W_S(\eta) = 
    \begin{cases}
        \frac{\xi^2-\frac{1}{4}}{\eta^2} \quad \text{for} \quad \eta < \eta_0 \\
        \frac{2}{\eta^2}-\Delta^2 \quad \text{for} \quad \eta_0 < \eta < 0
    \end{cases} \quad .
\eeq
However, since the domain is restricted to the second branch, we can focus only on:
\beq
    W_S(\eta) = \frac{2}{\eta^2}-\Delta^2 \quad ,
\eeq
which corresponds to the following Bunch--Davies solution for the mode function:
\beq
    v_k (\eta) = \frac{1}{\sqrt{2\omega}}\left(1 - \frac{i}{\omega \eta}\right)e^{-i \omega \eta}
     \label{eq:deltamode2}
     \qquad \text{with} \qquad \omega \equiv \sqrt{k^2+\Delta^2} \quad .
\eeq
In this branch, we can safely adopt the standard de Sitter form for the scale factor, which remains:
\beq
    s(\eta) = -\frac{1}{\mathcal{H} \eta} \quad .
\eeq
Finally, we assume a slow-roll parameter $\varepsilon(\eta)$ as shown in Fig.~\ref{fig:model_eps}, where it remains equal to $3 \gamma^2$ except for a localized dip around the turning point $\eta_t$, caused by the climbing scalar. This dip will give rise to a distinct contribution in the three-point function. 
\subsection{Energy Scales}
 To conclude this section—and more generally the setup for the forthcoming computation—we now address how to minimize the number of parameters by expressing all relevant energy scales in terms of the only one we ultimately care about: $\Delta$.

In order to complete the analysis, we must still consider $\mathcal{H}$, the Hubble scale during inflation, and the Planck mass $M_{pl}$. The latter can be effectively absorbed into $\mathcal{H}$, since if $M_{pl}$ were too large, all contributions to the three-point function would be suppressed. As we shall see, these contributions are weighted by the factor $\frac{\mathcal{H}^4}{M_{pl}^4}$.

The key point is to relate $\Delta$ to $\mathcal{H}$. This can be done by tracing $\Delta$ back from its observed present-day value to its value at the beginning of inflation, following the evolution of the scale factor. The procedure relies on standard estimates of the number of e-folds~\cite{Baumann:Cosmology2022} and the cosmological expansion history, as outlined for example in~\cite{GruppusoKitazawaMandolesiNatoliSagnotti2016}.

By combining the expansion during inflation with the subsequent radiation-, matter-, and dark energy–dominated eras, one can estimate that the comoving scale $\Delta$ at the start of inflation is enhanced by a factor:
\beq
    \left( \frac{s_5}{s_1} \right) \sim 3 \times 10^{56} \cdot e^{N - 60} \cdot \left( \frac{\mathcal{H}}{M_{pl}} \right)^{1/2} \quad ,
\eeq
where $N$ is the total number of e-folds of inflation, and we have assumed constant $\mathcal{H}$ during the process.

Using the observed value $\Delta_{\text{now}} = 3.5 \times 10^{-4}\, \text{Mpc}^{-1}$ and converting it back to the beginning of inflation via this scaling, one finds:
\beq
    \Delta_{\text{beg:infl}} \sim 7 \times 10^{14} \cdot e^{N - 60} \cdot \left( \frac{\mathcal{H}}{M_{pl}} \right)^{1/2} \, \text{GeV} \quad .
\eeq
Switching to the reduced Planck mass $\mu_{pl} = M_{pl} / \sqrt{8\pi} \approx 2.4 \times 10^{18}\, \text{GeV}$, we rewrite the expression as:
\beq
    \Delta_{\text{beg:infl}} = 3.2 \times 10^{14} \cdot e^{N - 60} \cdot \left( \frac{\mathcal{H}}{\mu_{pl}} \right)^{1/2} \, \text{GeV} \quad .
\eeq
The upper bound on the inflationary scale from PLANCK observations gives:
\beq
    \frac{\mathcal{H}}{\mu_{pl}} < 3.6 \times 10^{-5} \quad ,
\eeq
which implies that:
\beq
    \Delta < 2 \times 10^{12} \cdot e^{N - 60} \, \text{GeV} \quad .
\eeq
Putting everything together, we arrive at the key estimate:
\beq
    \Delta = 2 \times 10^{12} \cdot e^{N - 60} \, \text{GeV} \quad ,
\eeq
or, equivalently, expressing $\Delta$ in terms of $\mathcal{H}$:
\beq
    \Delta = \frac{\mathcal{H}}{50} e^{x} \quad \to \quad \mathcal{H} = 50 \cdot 10^{-x} \cdot \Delta \quad .
    \label{eq:Hdelta}
\eeq
where $x \equiv N - 60$. This reveals an important feature: the relation between $\Delta$ and $\mathcal{H}$ is highly sensitive to the total number of e-folds $N$ of the inflationary process.
This will be crucial in the analysis of the resulting non-Gaussianity, as it determines the size of all terms suppressed by $\mathcal{H} / M_{pl}$.

Secondly, we aim to eliminate the only remaining parameter from the scalar dynamics: $\varphi_0$. This parameter plays a role only near the inversion point, where the dynamics of the scalar field influence the behavior of the Hubble parameter. Assuming a nearly constant value of ${\cal H}$ in that region allows us to establish a connection between ${\cal H}$, $\varphi_0$, and the microscopic parameter $M$ introduced in Chapter~2.

Near the turning point, the Hubble parameter evolves as:
\beq
    H (\tau)= \sqrt{\frac{V}{3}(1+\dot{\varphi}^2)} =\frac{M}{\sqrt{6}} \, e^{\varphi} \, \dot{a}
    \quad ,
\eeq
where we used the exponential potential \eqref{eq:singleexp} and the Hamiltonian constraint (first equation of \eqref{eq:finalsystem}), identifying the scale $M$ with the string tension $T$.

Using the explicit solutions for the climbing scalar \eqref{eq:climb1} and \eqref{eq:climb2}, one finds:
\beq
    H(\tau)=\frac{M}{2 \sqrt{6}}\left(\frac{1}{\tau}+\tau\right) \, \tau^{1/2} \, e^{- \tau/4} \, e^{\varphi_0} \quad .
\eeq
Evaluating this expression at the turning point $\tau_t = 1$—identified with the moment where $H = {\cal H}$—gives:
\beq
    {\cal H} \ = \ \frac{M}{\sqrt{6}}\ e^{-\frac{1}{4}}e^{{\varphi}_0}  \quad .
    \label{eq:H matching}
\eeq
This equation motivates the use of the notation ${\cal H}$ for the Hubble constant in our setup, since it encapsulates both $M$ and $\varphi_0$.

In the actual computation of non-Gaussianities, the scalar dynamics enters only when converting between different time variables (cosmic, conformal, and parametric), as already shown in \eqref{eq:cosmictime} and \eqref{eq:conformal time}. For the critical case $\gamma = 1$, these conversions involve the factor $e^{-\varphi}/M$. 

Using the relation \eqref{eq:H matching} to express $M$, one finds that all occurrences of $\varphi_0$ cancel out. Thus, this parameter has no effect on the final observables and can be safely ignored in the forthcoming computation.

These considerations complete the setup. We are now ready to review the foundational formalism used to compute such correlators: the IN-IN formalism.
\section{Schwinger Keldysh Formalism}
The standard computation of cosmological correlation functions relies on the IN-IN, or Schwinger–Keldysh, formalism \cite{Schwinger1961,BakshiMahanthappa1963a,Keldysh1965,Jordan1986,CalzettaHu1987}, which will be central to the next section. Its application to cosmology was revived by Maldacena for computing the three-point function in slow-roll inflation \cite{maldacena2003non}, and later systematized by Weinberg for studying loop corrections \cite{Weinberg2005}. Useful reviews with cosmological focus are \cite{chen2010non,Wang2014}, though we closely follow \cite{BaumannJoyceCosmoCorrelators2023}.

In flat-space quantum field theory, the main object of interest is the $S$-matrix, encoding transition amplitudes between asymptotic ``in'' and ``out'' states \cite{landau1977quantum,Weinberg_QTF1_1995,Maggiore_QFT_2005}, requiring boundary conditions in both the past and future.

In cosmology, however, these conditions fail in expanding spacetimes such as de Sitter: wave packets cannot be asymptotically localized, gravitational particle production cannot be turned off, stable asymptotic states are ambiguous, a single observer cannot access a full Cauchy surface, and the Universe is not prepared by an experimentalist, rendering transition amplitudes less meaningful.

Instead, cosmology focuses on quantum expectation values of operators at fixed time with respect to a chosen initial state, usually the Bunch--Davies vacuum. These are known as \emph{in-in correlators} and are given by:
\beq
    \langle O(t) \rangle =
    \langle 0 | \bar{T} \exp\!\left(i \int_{-\infty^+}^{t} dt' \, H_{\text{int}}(t')\right)
    \, O(t) \,
    T \exp\!\left(-i \int_{-\infty^-}^{t} dt' \, H_{\text{int}}(t')\right) |0 \rangle \quad ,
    \label{eq:inin}
\eeq
where $T$ and $\bar{T}$ denote time and anti-time ordering, respectively. Unlike the $S$-matrix, the same initial state appears on both sides of the operator insertion, hence the name \emph{IN-IN formalism}. Full derivations of this master formula can be found in \cite{BaumannJoyceCosmoCorrelators2023,collins2011primordial}. 

The in-in formula includes an $i \epsilon$ prescription, implemented via $t \to t(1-i\epsilon)$, and corresponds to a closed time path contour in the complex plane: operators on $C^+$ are time-ordered, while those on $C^-$ are anti-time-ordered. Although exact, the formula is typically evaluated perturbatively. At first order in the interaction Hamiltonian—sufficient for our purposes—it becomes:
\beq
    \langle O(t) \rangle
    = 2 \, \text{Im} \left[
        \int_{-\infty}^{t} dt' \;
        \langle 0 | O(t)\, H_{\rm int}(t') | 0 \rangle
    \right] .
\eeq
assuming $\langle 0 |O(t)|0\rangle = 0$. This can be evaluated using diagrammatic rules developed in~\cite{Weinberg2005,Weinberg_QTF1_1995,Peskin:1995ev,Giddings:2010pp,Baumann:2019dut}. These rules, which we now outline and illustrate with examples, are:

\begin{enumerate}
    \item Draw the final time hypersurface at $t = t_c$.
    
    \item Sketch all relevant diagrams, allowing vertices both above and below the final time surface.
    
    \item Lines crossing or ending on the surface use the Wightman propagator:
    \begin{equation}
        W_k(t, t') = f_k^*(t)\, f_k(t') \, \quad ,
    \end{equation}
    with the left time argument for the higher vertex and the right for the lower one.

    Lines entirely \emph{below} use the Feynman propagator:
    \begin{equation}
        G_F(k; t, t') = W_k(t, t')\, \theta(t - t') + W_k(t', t)\, \theta(t' - t) \, \quad,
    \end{equation}
    Lines entirely \emph{above} use the time-reversed Feynman propagator:
    \begin{equation}
        G_F(k; t', t) = G_F^*(k; t, t')  \quad.
    \end{equation}
    \item Assign a factor $-iV$ for vertices below the surface (with $V$ from the interaction Hamiltonian), and $iV^\dagger$ for vertices above. Impose momentum conservation at each vertex, and a global delta function for total momentum conservation.
    \item Integrate each vertex over time using the measure:$\mathrm{d}t\, a^3(t) \,.$
    For diagrams with loops, also integrate over loop momenta.
    \item Account for symmetry factors by dividing appropriately. If diagrams are quotiented by reflections across the final time surface (step 2), the result is multiplied by $2 \, \text{Re}$.
\end{enumerate}

We now apply the diagrammatic rules to the three-point function of a scalar field minimally coupled to gravity. The first contribution comes from the diagram with the interaction vertex below the final time surface:
\bea
    \vcenter{\hbox{\includegraphics[height=2.5em]{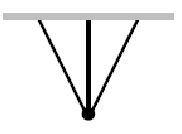}}}
    \quad &=& -iV \int_{-\infty}^{t_c} dt\, a^3(t)\, W_{k_1}(t_c, t)\, W_{k_2}(t_c, t)\, W_{k_3}(t_c, t) \nonumber \\ &=& 
    -iV\, f_{k_1}^*(t_c)\, f_{k_2}^*(t_c)\, f_{k_3}^*(t_c)
    \int_{-\infty}^{t_c} dt\, a^3(t)\, f_{k_1}(t)\, f_{k_2}(t)\, f_{k_3}(t) \quad .
\eea
Similarly, the second diagram, with the vertex above the surface, gives:
\bea
    \vcenter{\hbox{\includegraphics[height=2.5em]{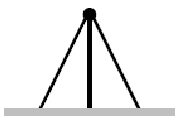}}}
    \quad &=& +iV \int_{-\infty}^{t_c} dt\, a^3(t)\, W_{k_1}(t, t_c)\, W_{k_2}(t, t_c)\, W_{k_3}(t, t_c) \nonumber \\ &=&
    +iV\, f_{k_1}(t_c)\, f_{k_2}(t_c)\, f_{k_3}(t_c)
    \int_{-\infty}^{t_c} dt\, a^3(t)\, f_{k_1}^*(t)\, f_{k_2}^*(t)\, f_{k_3}^*(t) \quad .
\eea
The sum of the two diagrams therefore indeed reproduces the result that would have been obtained fully expanding the perturbative formula. Surely this diagramatic works also for interesting exchange diagrams, however in this work we shall focus only on three point contact interactions coming from the derived interaction piece of the action for $\zeta$ \eqref{eq:cubic action}. 

Before the end of the present section we want to use the IN-IN formalism to make a useful example related to what we discussed for the two-point function, or equivalently, the power spectrum. We can think about the MS potential once again as
\beq
    \frac{z''}{z}=\frac{2}{\eta^2}+\delta_S(\eta) \quad .
\eeq
In particular, if $\left|\frac{\delta_S(\eta)}{\left(\frac{2}{\eta^2}\right)}\right|\ll 1$ this can be done efficiently working with IN-IN formalism. One can compute, for instance, the correction to the two-point correlation function in terms of the Weightman functions for the LM attractor background, that are the product of two \eqref{eq:LMmode} at two different times.

Within  this setup we can perturbatively derive the power spectrum \eqref{eq:deltapowerspectrum}. We consider $\delta_S=-\Delta^2$ and extend the conformal time domain as $[- \infty, \eta_{\text{end}}]$, where $\eta_{\text{end}}$ denotes the conformal time of the end of inflation (we know that $\eta_{\text{end}}\sim 0$). We can perform the computation with generic $\nu$. The perturbation acts in the following way
\beq
    \mathcal{P}_k = \mathcal{P}_k^{(0)}+ \delta \mathcal{P}_k
    \quad ,    
\eeq
\beq
    \delta \mathcal{P}_k =  \frac{k^3}{2 \pi^2} \, \delta \langle \zeta(k) \zeta(k) \rangle \quad ,
\eeq
\beq
    \delta \langle \zeta(k) \zeta(k) \rangle= i \frac{\mathcal{H}^2}{2\varepsilon} \, \eta_{\text{end}}^2 \,\left[-\Delta^2 v_k^2(\eta_{\text{end}})\int_{- \infty}^{\eta_{\text{end}}}d \eta' v_k^2(\eta') - \text{c.c}  \right] \quad .
\eeq
All the external factors cospire to recostruct the Power spectrum of the unperturbed form, that is standard Chibisov-Mukhanov $\mathcal{P}_k = C \, k^{3-2 \nu}$. The correction to the power spectrum now is
\bea
    \delta \mathcal{P}_k &=& C \, k^{3 - 2 \nu} \left[-i\Delta^2 \, \int_{- \infty}^{\eta_{\text{end}}}d \eta' v_k^2(\eta') - \text{c.c}\right] \nonumber \\
    \delta \mathcal{P}_k &=& C \, k^{3 - 2 \nu} \left[ \frac{\pi}{4} \, i \,\Delta^2 \, \int_{- \infty}^{0}d \eta' \, \eta ' \, \left(H^{(1)}_{\nu}(-k \eta')\right)^2 - \text{c.c}\right] \nonumber \\
    \delta \mathcal{P}_k &=& -C \, k^{3 - 2 \nu} \left[ \pi  \,\Delta^2 \, \int_{- \infty}^{0}d \eta' \, \eta ' \, J_{\nu}(-k \eta') \, Y_{\nu}(-k \eta')\right] \quad ,
\eea
where we have used the fact that the complex conjugated of the $H_\nu^{(1)}$ is $H_\nu^{(2)}$ and their expressions in terms of the Bessell functions. Changing variable with the definition $x \equiv - k \eta '$, we obtain
\beq
    \delta \mathcal{P}_k = C \, k^{3 - 2 \nu} \left[ \pi  \,\left(\frac{\Delta}{k}\right)^2 \, \int_{0}^{+ \infty}d x \, x \, J_{\nu}(x) \, Y_{\nu}(x)\right] \quad .
\eeq
This integral can be solved with standard results of\cite{DLMF,abramowitz1964handbook} in terms of hypergeometric functions and employing the $i \epsilon$ prescription we simply obtain
\beq
    \delta \mathcal{P}_k =- C \, k^{3 - 2 \nu} \,\left(\frac{\Delta}{k}\right)^2 \, \nu \quad .
\eeq
So finally
\beq
    \mathcal{P}_k = C \, k^{3-2\nu} \, \left(1-\nu \, \left(\frac{\Delta}{k}\right)^2\right) \quad,
\eeq
that is precisely the result of \eqref{eq:deltapowerspectrum} once we consider $\Delta \ll k^2$.

We are now ready to discuss non-gaussianites in full generality.

\section{Computation of the Amplitude}
The cubic action from \eqref{eq:cubic action} determines the standard inflationary prediction for the three-point function, which encodes non-Gaussian features beyond the two-point function.

A momentum-dependent amplitude \( \mathcal{A}_{k_1, k_2, k_3}(t) \) is defined as:
\beq
\begin{aligned}
    \langle 0(t) |\, \zeta(t, \vec{x})\, \zeta(t, \vec{y})\, \zeta(t, \vec{z})\, | 0(t) \rangle &=
    \int \frac{d^3 \vec{k}_1}{(2\pi)^3}
    \frac{d^3 \vec{k}_2}{(2\pi)^3}
    \frac{d^3 \vec{k}_3}{(2\pi)^3} \,
    e^{i \vec{k}_1 \cdot \vec{x}} e^{i \vec{k}_2 \cdot \vec{y}} e^{i \vec{k}_3 \cdot \vec{z}} \,
    (2\pi)^3 \delta^3(\vec{k}_1 + \vec{k}_2 + \vec{k}_3) \\
    &\quad \times \frac{1}{32 \, \varepsilon^2} \frac{\mathcal{H}^4}{M_{pl}^4} \frac{1}{k_1^3 k_2^3 k_3^3} \,
    \mathcal{A}_{k_1, k_2, k_3}(t)
\end{aligned}
\eeq
Momentum conservation is enforced by the delta function, which is usually left explicit for symmetry. The calculation simplifies by removing terms in the cubic action proportional to the linear equation of motion via the nonlinear field redefinition:
\beq
    \zeta = \zeta_n + f(\zeta_n) \quad .
\eeq
Here, \( f(\zeta_n) \) matches the function in \eqref{eq: field redefinition}. Most generated terms are negligible in the late-time limit. The redefinition induces new cubic terms from the quadratic action:
\beq
    S^{(2)}(\zeta) = S^{(2)}(\zeta_n) + \tilde{S}^{(3)}(\zeta_n) + O(\zeta_n^4) \quad .
\eeq
Since the action is quadratic in \( \zeta \), the form of \( S^{(2)}(\zeta_n) \) is unchanged:
\beq
    S^{(2)}(\zeta_n) = \int d^4x\, e^{a}
    \, \varepsilon \, \left[\dot{\zeta}_n^2
    - e^{-2 a/3} (\partial_k \zeta_n)^2
    \right] \quad .
\eeq
Cubic terms arising from the shift are:
\beq
    \tilde{S}^{(3)} = \int d^4x\, e^{a} 
    2 \varepsilon \left[ \dot{\zeta}_n \dot{f}(\zeta_n)
    - e^{-2 a/3} \partial_k \zeta_n \partial_k f(\zeta_n)
    \right] \quad .
\eeq
Integration by parts yields:
\beq
    \tilde{S}^{(3)} = -\int d^4x\, f(\zeta_n)
    \left[
    \frac{d}{dt} \left( 9 e^{a} \frac{\dot{\phi}^2}{\dot{a}^2} \dot{\zeta}_n \right)
    - 9 e^{a/3} \frac{\dot{\phi}^2}{\dot{a}^2} \partial^2 \zeta_n
    \right] \quad ,
\eeq
cancelling the unwanted part of \( S^{(3)}(\zeta_n) \). Thus, the full action becomes:
\beq
    S^{(3)}(\zeta) = S^{(3)}(\zeta_n) + O(\zeta_n^4)
    \quad .
\eeq
The remaining interaction terms are:
\bea
    S^{(3)} &=& M_{pl}^2 \int d^4x\, \Bigg\{
     \varepsilon^2 e^{a} \dot{\zeta}^2 \zeta 
    + \varepsilon^2 e^{\frac{a}{3}} \zeta\, \partial_k \zeta\, \partial^k \zeta 
    - 2 \varepsilon^2 e^{a} \dot{\zeta}\, \partial_k \zeta\, \partial^k \left( \partial^{-2} \dot{\zeta} \right) \nonumber \\ &+& \frac{3}{2}\varepsilon \frac{d}{dt}\left(\frac{\dot \varepsilon}{\varepsilon \dot a}\right) e^{a} \dot{\zeta} \zeta^2 - \frac{1}{2} \varepsilon^3 e^{a} \left( \dot{\zeta}^2 \zeta 
    - \zeta\, \partial_k \partial_l (\partial^{-2} \dot{\zeta})\, \partial^k \partial^l (\partial^{-2} \dot{\zeta}) \right) 
    \Bigg\} \quad .
    \label{eq:finalcubic}
\eea
This action drives the evolution of \( \zeta \)-states and defines the interaction Hamiltonian used in the IN-IN formalism, via \( H_{\text{int}} = -L_{\text{int}} \), with \( S^{(3)}_n = \int dt\, L_{\text{int}}(t) \). This identity holds for cubic interactions, but fails beyond that order due to modifications in the canonical momentum.

After the field redefinition, the three-point function splits into separate contributions:
\beq
\begin{aligned}
\langle 0(t) | \zeta(t, \vec{x}) \zeta(t, \vec{y}) \zeta(t, \vec{z}) | 0(t) \rangle =\ & \langle 0(t) | \zeta_n(t, \vec{x}) \zeta_n(t, \vec{y}) \zeta_n(t, \vec{z}) | 0(t) \rangle \\
&+ \langle 0(t) | \zeta_n(t, \vec{x}) \zeta_n(t, \vec{y}) f[\zeta_n(t, \vec{z})] | 0(t) \rangle \\
&+ \langle 0(t) | \zeta_n(t, \vec{x}) f[\zeta_n(t, \vec{y})] \zeta_n(t, \vec{z}) | 0(t) \rangle \\
&+ \langle 0(t) | f[\zeta_n(t, \vec{x})] \zeta_n(t, \vec{y}) \zeta_n(t, \vec{z}) | 0(t) \rangle \\
&+ \text{higher order terms}.
\end{aligned}
\eeq
Higher-order terms (involving five or more \( \zeta_n \)) are suppressed or vanish at late times. The non-Gaussian part of \( \zeta_n \)'s three-point function arises dynamically:
\beq
    \langle 0(t) | \zeta_n(t, \vec{x}) \zeta_n(t, \vec{y}) \zeta_n(t, \vec{z}) | 0(t) \rangle
\eeq
This is computed using the IN-IN formalism.
\begin{figure}[ht]
\begin{center}
    \includegraphics[width=3.6in]{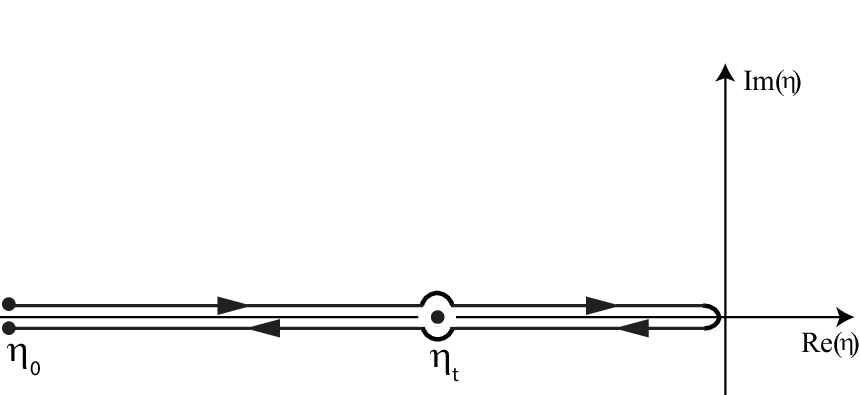}
\end{center}
\caption{The Schwinger--Keldysh contour starts slightly above the real axis of the $\eta$ plane at $\eta_0$, proceeds along the semicircle above $\eta_t$, turns back at the origin, proceeds along the semicircle below $\eta_t$ and finally ends at $\eta_0$ slightly below the real axis \cite{meo2025preinflationarynongaussianities}.}
\label{fig:contour}
\end{figure}

Compared to~\cite{maldacena2003non,collins2011primordial}, this setup uses a different contour (Fig.~\ref{fig:contour}), truncated at the bounce \( \eta_0 \), with Bunch--Davies conditions imposed there. A distinctive contribution arises at the turning point \( \eta_t \), where the scalar field inverts. These will be addressed in the next section.

The three terms involving \( f(\zeta_n) \) are cyclically related and quartic in \( \zeta_n \). Their leading contribution arises in the free vacuum and can be computed as:
\beq
    \langle 0(t) | \zeta_n(t, \vec{x}) \zeta_n(t, \vec{y}) f[\zeta_n(t, \vec{z})] | 0(t) \rangle = \langle 0 | \zeta_n(t, \vec{x}) \zeta_n(t, \vec{y}) f[\zeta_n(t, \vec{z})] | 0 \rangle + \cdots
\eeq
This matches the result in~\cite{maldacena2003non,collins2011primordial}.

In summary, the three-point amplitude receives three contributions:
\begin{enumerate}
    \item Principal-part: from $\eta_0 \leq \eta < 0$, excluding \( \eta_t \), with $\varepsilon = 3\gamma^2$, along a Schwinger--Keldysh contour deformed away from the real axis;
    \item Local: around \( \eta_t \), where \( \varepsilon(\eta) = 0 \); although seemingly singular, these are regularized by the contour;
    \item Field redefinition: matching the original computation in~\cite{maldacena2003non,collins2011primordial}.
\end{enumerate}

\begin{figure}[ht]
\begin{center}
    \includegraphics[width=1.5in]{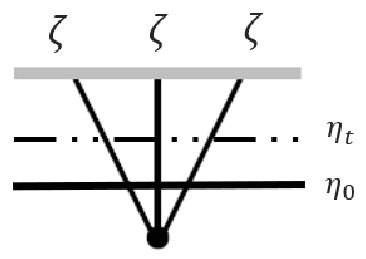}
    \end{center}
    \caption{IN-IN diagram of the present three point function.}
    \label{fig:diagram}
\end{figure}

The next section details each contribution. One final ingredient is needed for the IN-IN computation: the Wightman functions.

As introduced in the IN-IN formalism, Wightman functions are expectation values of two quantum fields at different spacetime points. There are two types, \( W_k^< \) and \( W_k^> \), related by:
\[
W^<_k(t, t') = W^>_k(t', t), \quad W^<_k(t, t') = \left[W^>_k(t, t')\right]^*,
\]
so only one needs to be computed.

In terms of the mode function \( v_k \), the Wightman function in momentum space reads:
\bea
    W^>_k(t, t') &=& \frac{e^{-a(t)/3}}{M_{pl}} \frac{e^{-a(t')/3}}{M_{pl}} \frac{1}{\sqrt{2 \varepsilon(t)}} \frac{1}{\sqrt{2 \varepsilon (t')}} \zeta_k(t) \zeta^*_k(t') \quad .
\eea

The calculation is usually performed in conformal time, and in our case we work in the domain \( [\eta_0, 0] \), where the MS potential is given by \eqref{eq:deltaMS} and the scale factor is de Sitter. Using \eqref{eq:deltamode2}, Wightman functions and their time derivatives can be computed.

For the principal-part region, where \( \varepsilon = 3\gamma^2 \) is constant, derivatives of \( \varepsilon \) can be neglected. However, near the localized contribution around \( \eta_t \), where \( \varepsilon \) varies, they must be included.

At late times \( \eta \to 0 \), the general expressions become:
\bea
      W^>_k(0, \eta) &\sim& \frac{1}{4} \frac{\mathcal{H}^2}{M_{pl}^2} \frac{1}{\omega^3} \frac{1}{\sqrt{3 \gamma^2}} \frac{(1 - i \omega\eta)}{\sqrt{\varepsilon(\eta)}}\, e^{i\omega\eta} \quad , \\
    \dot{W}^>_k(0, \eta) &\sim& -\frac{1}{4} \frac{\mathcal{H}^3}{M_{pl}^2} \frac{1}{\omega^3} \frac{1}{\sqrt{3\gamma^2}} \frac{\eta }{\sqrt{\varepsilon(\eta)}} \left[\omega^2 \eta - \frac{1}{2}(1-i\omega \eta)\frac{\varepsilon'(\eta)}{\varepsilon(\eta)}\right]e^{i\omega\eta} \quad ,
    \label{eq:generalW}
\eea

For the constant-\( \varepsilon \) case:
\bea
      W^>_k(0, \eta) &\sim& \frac{1}{12 \gamma^2} \frac{\mathcal{H}^2}{M_{pl}^2} \frac{1}{\omega^3} (1 - i\omega\eta) e^{i\omega\eta} \quad, \\
    \dot{W}^>_k(0, \eta') &\sim& -\frac{1}{12 \gamma^2} \frac{\mathcal{H}^3}{M_{pl}^2} \frac{\eta^2}{\omega} e^{i\omega\eta} \quad.
    \label{eq:simpleW}
\eea

Note that when the second argument tends to zero, derivatives vanish. Thus, operators involving time derivatives are suppressed in the late-time limit, unless compensated by factors like \( 1/\eta^2 \).

\subsection{Field Redefinition}
Of the three contributions to the three-point function, the quartic ones (from the field redefinition) are the most straightforward. Since they differ only by spatial permutations, we compute a representative term:
\beq
    \langle 0 | \zeta_n(t,\vec{x}) \zeta_n(t,\vec{y}) f(\zeta_n(t,\vec{z})) | 0 \rangle \quad.
\eeq
Expanding \( f(\zeta_n) \), the expectation value yields several terms:
\bea
    \langle 0 | \zeta_n(t,\vec{x}) \zeta_n(t,\vec{y}) f(\zeta_n(t,\vec{z})) | 0 \rangle 
    &=& \frac{\delta + \varepsilon}{2}
    \langle \zeta_n \zeta_n \zeta_n \zeta_n \rangle 
    + \frac{3}{\dot{a}} \langle \zeta_n \zeta_n \dot{\zeta}_n \zeta_n \rangle \notag \\
    &-& \frac{9}{4} \frac{e^{-2 a/3}}{\dot{a}^2} 
    \langle \zeta_n \zeta_n \partial_k \zeta_n \partial^k \zeta_n \rangle 
    + \frac{9}{4} \frac{e^{-2a/3}}{\dot{a}^2} 
    \langle \zeta_n \zeta_n \partial^{-2} \partial_k \partial_l ( \partial^k \zeta_n \partial^l \zeta_n ) \rangle \notag \\
    &+& \frac{3}{2\dot{a}} \varepsilon
    \langle \zeta_n \zeta_n \partial_k \zeta_n \partial^k \dot{\zeta}_n \rangle 
    - \frac{3}{2\dot{a}} \varepsilon
    \langle \zeta_n \zeta_n \partial^{-2} \partial_k \partial_l ( \partial^k \zeta_n \partial^l \dot{\zeta}_n ) \rangle \quad.
\eea
In the late-time limit:
- Terms with \( \dot{\zeta}_n \) vanish (\( \dot{G}_k^>(t,t) \to 0 \)),
- Spatial derivatives are suppressed by \( e^{-2a/3} \sim \eta^2 \to 0 \).

Thus, only the first term survives. Using Wick contractions and Fourier expansion:
\bea
    &&\langle \zeta_n(\vec{x}) \zeta_n(\vec{y}) f(\zeta_n(\vec{z})) \rangle 
    = \int \frac{d^3\vec{k}_1}{(2\pi)^3} \frac{d^3\vec{k}_2}{(2\pi)^3} 
    e^{i \vec{k}_1 \cdot (\vec{x} - \vec{z})} e^{i \vec{k}_2 \cdot (\vec{y} - \vec{z})} \nonumber \\
    &\times& \left(\delta + \varepsilon\right)
    W^>_{k_1}(t,t) W^>_{k_2}(t,t) + \cdots
\eea
Taking the late-time limit of the Wightman function:
\beq
    \lim_{t \to \infty} W^>_k(t,t) = \frac{1}{4\varepsilon} \frac{\mathcal{H}^2}{M_{pl}^2} \frac{1}{\omega^3} \quad ,
\eeq
\bea
    &&\langle \zeta_n(\vec{x}) \zeta_n(\vec{y}) f(\zeta_n(\vec{z})) \rangle
    = \int \frac{d^3 \vec{k}_1}{(2\pi)^3} \frac{d^3 \vec{k}_2}{(2\pi)^3}
    e^{i \vec{k}_1 \cdot \vec{x}} e^{i \vec{k}_2 \cdot \vec{y}} e^{-i (\vec{k}_1 + \vec{k}_2) \cdot \vec{z}} \,
    \frac{1}{16 \varepsilon^2} \frac{\mathcal{H}^4}{M_{pl}^4} \frac{1}{\omega_1^3 \omega_2^3}
    \left(\delta + \varepsilon \right) \nonumber \\ &.&
\eea
To symmetrize:
\bea
    &&\langle \zeta_n(\vec{x}) \zeta_n(\vec{y}) f(\zeta_n(\vec{z})) \rangle + \text{2 perms}
    = \int \frac{d^3 \vec{k}_1}{(2\pi)^3} \frac{d^3 \vec{k}_2}{(2\pi)^3} \frac{d^3 \vec{k}_3}{(2\pi)^3}
    e^{i \vec{k}_1 \cdot \vec{x}} e^{i \vec{k}_2 \cdot \vec{y}} e^{i \vec{k}_3 \cdot \vec{z}} \nonumber \\
    &&\times (2\pi)^3 \delta^{(3)}(\vec{k}_1 + \vec{k}_2 + \vec{k}_3)
    \frac{1}{16 \varepsilon^2} \frac{\mathcal{H}^4}{M_{pl}^4}
    \frac{1}{\omega_1^3 \omega_2^3 \omega_3^3}
    \left(\varepsilon + \delta \right)
    (\omega_1^3 + \omega_2^3 + \omega_3^3) 
\eea
Removing the delta-function prefactor, the result is:
\beq
    \langle \mathcal{O}_f \rangle = \frac{1}{16 \varepsilon^2} \frac{\mathcal{H}^4}{M_{\rm pl}^4}
    \left(\varepsilon + \delta \right)
    \frac{\sum_i \omega_i^3}{\prod_i \omega_i^3} \quad .
    \label{eq:Of}
\eeq
On the attractor solution of the single exponential model, we have \( \delta = -\varepsilon \) (from \eqref{eq:LM2}), so this contribution vanishes exactly.
\subsection{Principal Parts}
In this section we compute contributions coming from simple integration over the conformal time domain $[\eta_0,0]$ using constant slow roll parameter $\varepsilon = 3 \gamma^2$. This computation is really similar to the one of \cite{maldacena2003non,collins2011primordial} apart from the modified $k$s due to the presence of $\Delta$ and the limitation on the domain that will introduce one of the striking features of the computation: oscillations. We devide this computation in five pieces associated to the five interactions of \eqref{eq:finalcubic}. We follow the IN-IN formalism.
\newline
\newline
- \hspace{0.5em}$\langle \mathcal{O}_1\rangle_{\text{PP}}$:
\newline
\newline
The interaction vertex is (we use the notation $s$ for the scale factor)
\beq
    M_{pl}^2 \, s^3 \,\varepsilon^2 \, \dot{\zeta}^2 \, \zeta \quad ,
\eeq
so taking into account the symmetry factor, the integration becomes (we use the explicit formula for the scale factor and the slow-roll parameter)
\bea
    \langle O_1 \rangle_{\text{PP}} =  2i M_{pl}^2 \int_{ \eta_0}^{0} d \eta \frac{9 \, \gamma^4}{(\mathcal{H} \eta)^2}\left[\, W_>'(k_1)\, W_>'(k_2) \,W_>(k_3)\  - \mathrm{c.c. \, + \, 2 \, \text{perms}}\right] \quad,
    \label{eq:integral1}
\eea
where we changed a bit the notation for the Wightman functions to report this formula. The substitution of \eqref{eq:simpleW} gives:
\bea
    \langle O_1 \rangle_{\text{PP}} =  - \frac{1}{48 \, \gamma^2}\frac{\mathcal{H}^4}{M_{pl}^4} \, \int^0_{\eta_0}d \eta \left[\, \omega_1^2 \, \omega_2^2 \left(\sin \Omega \, \eta - \omega_3 \eta \cos \Omega \, \eta\right) \, + \, 2 \, \text{perms}\right] \quad ,
\eea
where $\Omega \equiv \omega_1 + \omega_2 + \omega_3$
The integral is solved by parts
\beq
    \int d \eta  (\sin \Omega \,\eta  - \omega_3 \eta \, \cos \Omega \, \eta)=-\frac{1}{\Omega}\left[
    \cos \Omega \, \eta \, + \eta \, \omega_3 \, \sin \Omega \, \eta + \frac{\omega_3}{\Omega} \, \cos \Omega \eta\right]  \quad .
\eeq
We obtain
\bea
    \langle \mathcal{O}_1\rangle_{\text{PP}} &=&\frac{\mathcal{H}^4}{48 \gamma ^2 \text{Mp}^4 \Omega ^2 \omega_1^3 \omega_2^3 \omega_3^3} \times \nonumber \\
    &\times& \left[\left(1- \Omega\eta_0 \, \sin \Omega \eta_0 - \cos \Omega \eta_0 \right) \prod_i \omega_i \sum_{j<l}\omega_j \omega_l + \left(1-\cos \Omega \eta_0\right)\Omega \sum_{i<j}\omega_i^2 \, \omega_j^2\right]
    \nonumber \\&&\quad .
    \label{eq:O1PP}
\eea
- \hspace{0.5em}$\langle \mathcal{O}_2\rangle_{\text{PP}}$:
\newline
\newline
The interaction vertex is (we use the notation $s$ for the scale factor)
\beq
    M_{pl}^2 \, s \,\varepsilon^2 \, \zeta \, \partial_k\zeta \, \partial^k \zeta \quad ,
\eeq
so taking into account the symmetry factor, the integration becomes (we use the explicit formula for the scale factor and the slow-roll parameter)
\bea
    \langle O_2 \rangle_{\text{PP}} =  -2i M_{pl}^2 \int_{ \eta_0}^{0} d \eta\frac{9 \, \gamma^4}{(\mathcal{H} \eta)^2}\left[\, \left(\vec{k}_1 \cdot \vec{k}_2\right)W_>(k_1)\, W_>(k_2) \,W_>(k_3)\  - \text{c.c.} + 2 \text{perms}\right] \, ,
    \label{eq:integral2}
\eea
where we changed a bit the notation for the Wightman functions to report this formula. The substitution of \eqref{eq:simpleW} gives:
\bea
    \langle O_2 \rangle_{\text{PP}} &=&  \frac{1}{48 \, \gamma^2} \, \frac{\mathcal{H}^4}{M_{pl}^4} \frac{\sum_{i<j} \vec{k}_i \cdot \vec{k}_j}{\omega_1^3 \, \omega_2^3 \, \omega_3^3} \int^{0}_{\eta_0} d \eta \Bigg[\frac{\sin \Omega \eta - \Omega \eta \cos \Omega \eta }{\eta^2}-\sum_{i<j}\omega_i \, \omega_j \, \sin \Omega \eta \nonumber \\ &+& \prod_i \omega_i \, \eta \, \cos \Omega \eta\Bigg] \quad ,
\eea
where $\Omega \equiv \omega_1 + \omega_2 + \omega_3$.
The two non-trivial integrals are solved by parts:
\beq
    \int d \eta \,  \frac{\sin \Omega \eta - \Omega \eta \cos \Omega \eta }{\eta^2}=-\frac{\sin \Omega \eta}{\eta}, \qquad
    \int d \eta \,\eta \, \cos \Omega \eta =\frac{\Omega \eta \sin \Omega \eta + \cos \Omega \eta}{\Omega^2} \quad .
\eeq
We can sum the $k$s using momentum conservation:
\beq
    \sum_{i<j} \vec{k}_i \cdot \vec{k}_j = \frac{1}{2}\left[k_3^2-k_1^2-k_2^2+k_1^2-k_2^2-k_3^2+k_2^2-k_3^2-k_1^2\right]=-\frac{1}{2}\sum_i k_i^2
\eeq
We obtain
\bea
    \langle O_2 \rangle_{\text{PP}} &=&  \frac{1}{96 \, \gamma^2} \, \frac{\mathcal{H}^4}{M_{pl}^4} \frac{\sum_{i} k_i^2}{\omega_1^3 \, \omega_2^3 \, \omega_3^3} \times \nonumber \\ &\times&  \left[ \Omega - \frac{\sin \Omega \eta_0}{\eta_0}+\frac{1}{\Omega}\left((\cos \Omega \eta_0-1)\sum_{i<j} \omega_i \, \omega_j - \frac{1}{\Omega}(1-\Omega \eta_0 \sin \Omega \eta_0 - \cos \Omega \eta_0)\prod_i \omega_i \right)\right] \nonumber \\ && \quad .
    \label{eq:O2PP}
\eea
- \hspace{0.5em}$\langle \mathcal{O}_3\rangle_{\text{PP}}$:
\newline
\newline
The interaction vertex is (we use the notation $s$ for the scale factor)
\beq
    -2M_{pl}^2 \, s^3 \,\varepsilon^2 \, \dot{\zeta} \, \partial_k \zeta \partial^k \zeta \left(\partial^{-2} \, \dot{\zeta}\right) \quad ,
    \label{eq:integral3}
\eeq
so the integration becomes (we use the explicit formula for the scale factor and the slow-roll parameter)
\bea
    \langle O_3 \rangle_{\text{PP}} = - 2i M_{pl}^2 \int_{ \eta_0}^{0} d \eta \frac{9 \, \gamma^4}{(\mathcal{H} \eta)^2}\left\{\left[\frac{\vec k_1 \cdot \vec k_3}{k_1^2} +  \frac{\vec k_2 \cdot \vec k_3}{k_2^2}\right]\, W_>'(k_1)\, W_>'(k_2) \,W_>(k_3)\  - \mathrm{c.c. + 2 \text{perms}}\right\} \nonumber \\ && \quad,
\eea
where we changed a bit the notation for the Wightman functions to report this formula. This is the same integral of the first contribution with a different structure in $k$s. The $k$s structure is developed using momentum conservation. The result is (we write it as a sum over permutations to make it simplier)
\bea
    \langle \mathcal{O}_3\rangle_{\text{PP}} &=&\frac{\mathcal{H}^4}{48 \gamma ^2 \text{Mp}^4 \Omega ^2 \omega_1^3 \omega_2^3 \omega_3^3} \times \nonumber \\
    &\times& \sum_{\text{perms}(i,j,l)}\Bigg\{\left[-2+\frac{k_j^2}{k_l^2}+\frac{k_l^2}{k_j^2}-\frac{k_i^2}{k_l^2}-\frac{k_i^2}{k_j^2}\right] \omega_i^2 \omega_l^2 ((\cos \Omega \eta_0-1) (\omega_i+2 \omega_j+\omega_l)
    \nonumber \\&+&\Omega \eta_0  \omega_j \sin  \Omega \eta_0 ))\Bigg\}\quad .
    \label{eq:O3PP}
\eea
- \hspace{0.5em}$\langle \mathcal{O}_4\rangle_{\text{PP}}$:
\newline
\newline
The interaction vertex is (we use the notation $s$ for the scale factor and $\frac{\dot{a}}{3}=\mathcal{H}$)
\beq
    \frac{M_{pl}^2}{2} \, \varepsilon \, \frac{d}{dt_c}\left(\frac{\dot{\varepsilon}}{\varepsilon \, \mathcal{H}}\right)\, s^3 \, \dot{\zeta} \, \zeta^2 \quad .
\eeq
Since we are considering a constant $\varepsilon$ contribution, this term is zero
\bea
    \langle \mathcal{O}_4\rangle_{\text{PP}}=0 \quad .
    \label{eq:O4PP}
\eea
- \hspace{0.5em}$\langle \mathcal{O}_5\rangle_{\text{PP}}$:
\newline
\newline
The interaction vertex is (we use the notation $s$ for the scale factor)
\beq
    -\frac{1}{2}M_{pl}^2 \, s^3 \,\varepsilon^3 \, \left[\dot{\zeta}^2 \, \zeta - \zeta \partial_k \partial_l \left(\partial^{-2}\dot{\zeta}\right)\partial^k \partial^l \left(\partial^{-2}\dot{\zeta}\right)\right] \quad ,
    \label{eq:integral5}
\eeq
so the integration becomes (we use the explicit formula for the scale factor and the slow-roll parameter)
\bea
    \langle O_5 \rangle_{\text{PP}} = - i M_{pl}^2 \int_{ \eta_0}^{0} d \eta \frac{27 \, \gamma^6}{(\mathcal{H} \eta)^2}\left\{\left[1 \, - \, \frac{\left(\vec{k}_2\cdot \vec{k}_3\right)^2}{k_2^2\,k_3^2}\right]\, W_>'(k_1)\, W_>'(k_2) \,W_>(k_3)\  - \mathrm{c.c. + 2 \text{perms}}\right\} \nonumber \\ && \quad,
\eea
where we changed a bit the notation for the Wightman functions to report this formula. This is the same integral of the first contribution with a different structure in $k$s up to another $\varepsilon$ and a numeric factor that makes this term subdominant with respect to the others. The $k$s structure is developed using momentum conservation. The result is (we write it as a sum over permutations to make it simplier)
\bea
    \langle \mathcal{O}_5\rangle_{\text{PP}} &=&\frac{\mathcal{H}^4}{128 \text{Mp}^4 \Omega ^2 \,\omega_1^3 \, \omega_2^3 \, \omega_3^3 \, k_1^3 \, k_2^3 \, k_3^3} \, \sum_i k_i \, \prod_{\text{perms}(i,j,l)} (k_i + k_j - k_3) \times \nonumber \\
    &\times& \Bigg\{\Omega \eta_0 \sin \Omega \eta_0 \prod_m \omega_m \sum_{i<j} k_i^2 \omega_i \omega_j  + (1-\cos \Omega \eta_0)\sum_{\text{perms}(i,j,l)} k_i^2 \omega_i^2 \omega_j^2 (\omega_i^2 + \omega_j^2 + 2 \omega_l)\Bigg\} \nonumber \\ &&\quad .
    \label{eq:O5PP}
\eea
\subsection{Localized Contributions}
In this section we compute for each of the previously listed operators the corresponding localized contribution. This contributions, as explained, arises because the scalar iverts its motion and this translates in a zero for $\varepsilon$. By the fact that $\varepsilon$ is present in the denominator of Wightman functions, that now have to be expressed as \eqref{eq:generalW}, this contribution could give a new divergence. However we have to remember that the IN-IN formalism is shaped on an $i \varepsilon$ prescription that dispalace the contour from the real a making a simple pole divergence, as the considered one, solvable. Let's get into mathematics. Near the turning point we can approx the potential as single critical exponential as discussed in chapter 2, this means that the $\varepsilon$ expression is \eqref{eq:criticaleps}
\beq
    \varepsilon(\tau)=3 \left(\frac{1-\tau^2}{1+\tau^2}\right)^2 \quad ,
\eeq
that has a zero in $\tau=1$. Since $\varepsilon$ is present in denominators of the integrand, this gives rise to poles. The integral that we are considering, going to parametric time, is
\beq
    \int_{1-\lambda}^{1+\lambda} d \tau f(\tau)
\eeq
If $f(\tau)$ has a simple pole in $\tau = 1$ we can use the well known formula from complex analysis and theory of distributions \cite{ahlfors_complex_1979} to get
\beq
    \lim_{\varepsilon \to 0^+} \int^{1+\lambda}_{1-\lambda} d \tau \frac{g(\tau)}{\tau-1 \pm i \varepsilon}= \text{P.P.}\int^{1+\lambda}_{1-\lambda} d \tau \frac{g(\tau)}{\tau-1} \mp i \pi g(1) \quad, 
    \label{eq:simplepoleformula}
\eeq
where $g(\tau)$ is the analytic part of the integrand. Since $\lambda$ is small the principal part contribution is always zero and we are left with 
\beq
    \lim_{\varepsilon \to 0^+} \int^{1+\lambda}_{1-\lambda} d \tau \frac{g(\tau)}{\tau-1 \pm i \varepsilon}= \mp i \pi g(1) \quad. 
\eeq
Note that both signs have to be taken into account since the $i \varepsilon$ prescription changes sign on the two branches of the contour. 

This is the simplest case, as we will see the singular part will always start from a simple pole. If $f(\tau)$ has a simple pole plus higher order poles in $\tau=1$, the previous formula generalizes to 
\beq
    \lim_{\varepsilon \to 0^+} \int_{1 - \lambda}^{1 + \lambda} \frac{g(\tau)}{(\tau - 1 \pm i\varepsilon)^n} \, d\tau
    = \mathrm{P.P.} \int_{1 - \lambda}^{1 + \lambda} \frac{g(\tau)}{(\tau - 1)^n} \, d\tau
    \mp i\pi \cdot \frac{(-1)^{n-1}}{(n-1)!} g^{(n-1)}(1) \quad .
\eeq
Principal parts are once again zero. If we call $n$ the heighest order of poles, the second term, once we sum on $n$, correspond to expand $g$ to up to order $n-1$ in Taylor. So we can think of this case as: expand as a Laurent series the singular part and recognize the value of $n$, then expand the analytic part $g(\tau)$ up to $n-1$ in Taylor, multiply the two series and take only the term with the simple pole, finally apply the formula \eqref{eq:simplepoleformula}.

Surely to use this analysis we have to express everything in terms of $\tau$. In this sense we have to remember formulas to change variables (\eqref{eq:cosmictime}\eqref{eq:conformal time}):
\beq
    dt_c = \sqrt{\frac{2}{3}}e^{-\varphi}\frac{d\tau}{M}=\frac{e^{-1/4}}{3 \mathcal{H}}\tau^{-1/2}e^{\tau^2/4}\, d\tau \quad ,
    \quad
    d\eta = \sqrt{\frac{2}{3}}e^{-\varphi}\frac{d\tau}{M}=\frac{e^{-1/4}}{3 \mathcal{H}}\tau^{-2/3}e^{\tau^2/6}\, d\tau \quad ,
\eeq
where we employed \eqref{eq:H matching}. This formulas and the fact that $\eta(1)=\eta_t$ by definition are sufficient to expand in Taylor once we need it. We are now ready to start computing this contributions:
\newline
\newline
- \hspace{0.5em}$\langle \mathcal{O}_1\rangle_{\text{t}}$:
\newline
\newline
The integral to compute is once again \eqref{eq:integral1} but with a time dependent $\varepsilon$ and localized around $\eta_t$.
We follow this steps:
\begin{itemize}
    \item we express $\varepsilon$ in terms of $\tau$ with the previously described formula. Its derivative with respect to conformal time is obtained with the chain rule:
    \beq
        \varepsilon ' = \frac{72 \,  e^{\frac{1}{4}} \, \mathcal{H} \, \tau ^{5/3}  \,  \left(\tau ^2-1\right)   \, e^{-\frac{\tau ^2}{6}}}{\left(\tau ^2+1\right)^3} \quad ;
    \eeq
    \item we can easily simplify square roots of squares without dealing with branch cuts because of the displacement of contours. The final expression (we still have to subtract c.c. ,to symmetrize and add a prefactor coming from $W$s) has only a simple pole:
    \beq
        -\frac{18 \, \sqrt{3} \,  M_{pl}^2 \, (\eta\omega_1+i) \, (\eta\omega_2+i) (\eta\omega_3+i) \, e^{\frac{1}{6}+i   \Omega \eta}}{\eta ^2 (\tau -1)}
    \eeq
    We note that the simple pole is symmetric
    \item Since the change of measure and the function at the denominator are analytic in $\tau$, we can use \eqref{eq:simplepoleformula} remembering that $\eta(1)=\eta_t$. We have to take the real part because the change of sign on the two branches select that. Employing also the change of measure we obtain
    \beq
        \frac{12 \, \sqrt{3} \, e^{1/12} \, M_{pl}^2 \, \pi}{\mathcal{H} \, \eta_t^2} \, \left[\left(1-\eta_t^2 \sum_{i<j} \omega_i \omega_j\right)\cos \Omega \eta_t + \eta_t \left(\Omega - \eta_t^2 \prod_i \omega_i\right)\sin \Omega \eta_t\right] \quad ;
    \eeq
    \item finally we symmetrize and add the prefactor, which means we multiply for 3 times the prefactor.
\end{itemize}
The final result is
\bea
    \langle \mathcal{O}_1\rangle_{\text{t}}=\frac{3  \, e^{1/12} \, \mathcal{H}^5 \, \pi}{16 \, M_{pl}^4 \, \gamma^3 \eta_t^2 \, \omega_1^3 \, \omega_2^3 \omega_3^3 } \, \left[\left(1-\eta_t^2 \sum_{i<j} \omega_i \omega_j\right)\cos \Omega \eta_t + \eta_t \left(\Omega - \eta_t^2 \prod_i \omega_i\right)\sin \Omega \eta_t\right] \quad .
    \label{eq:O1t}
\eea
- \hspace{0.5em}$\langle \mathcal{O}_2\rangle_{\text{t}}$:
\newline
\newline
The integral to compute is
once again \eqref{eq:integral2} but with a time dependent $\varepsilon$ and localized around $\eta_t$.
The absence of the derivatives makes manifest that the slow roll factor appears with positive power. This means that the integrand has no poles and this local contribution vanishes.
\bea
    \langle \mathcal{O}_2\rangle_{\text{t}}=0 \quad .
    \label{eq:O2t}
\eea
- \hspace{0.5em}$\langle \mathcal{O}_3\rangle_{\text{t}}$:
\newline
\newline
The integral to compute is once again \eqref{eq:integral3} but with a time dependent $\varepsilon$ and localized around $\eta_t$.
This integral is precisely the same of the first contribution except for the $k$s structure. Following the same steps the result is the same as for $\mathcal{O}_1$ since the residue is symmetric and the $k$s add to a constant.
\bea
    \langle \mathcal{O}_3\rangle_{\text{t}}=\frac{3  \, e^{1/12} \, \mathcal{H}^5 \, \pi}{16 \, M_{pl}^4 \, \gamma^3 \eta_t^2 \, \omega_1^3 \, \omega_2^3 \omega_3^3 } \, \left[\left(1-\eta_t^2 \sum_{i<j} \omega_i \omega_j\right)\cos \Omega \eta_t + \eta_t \left(\Omega - \eta_t^2 \prod_i \omega_i\right)\sin \Omega \eta_t\right] \quad .
    \label{eq:O3t}
\eea
- \hspace{0.5em}$\langle \mathcal{O}_4\rangle_{\text{t}}$:
\newline
\newline
The integral to compute is
\bea
    \langle O_4 \rangle_{\text{t}} =  i M_{pl}^2 \int_{ \eta_t} \frac{d \eta}{(\mathcal{H} \eta)^2} \, \frac{\varepsilon}{\mathcal{H}}\,\frac{d^2}{dt_c^2}\log \varepsilon\, \left[ \, W_>'(k_1)\, W_>(k_2) \,W_>(k_3)\  - \mathrm{c.c.}+ 2 \text{perms}\right] \quad .
\eea
The procedure to compute this integral is precisly the same as before. However the presence of that second derivative with respect to cosmic time makes the integrand more singular and the expression has a pole of order four. This requires to expand the analytic part up to order three in Taylor around $\tau=1$. This makes the computation heavier, but the algorithm doesn't change. The final result is
\bea
    \langle \mathcal{O}_4\rangle_{\text{t}}&=&\frac{\pi \mathcal{H}}{288 \, e^{\frac{1}{4}} \, \gamma ^3 \, \eta_t^6 \, M_{pl}^4 \, \omega_1^3 \, \omega_2^3 \, \omega_3^3} \times \nonumber \\ &\times&  \Bigg\{\cos \Omega \eta_t \Bigg[\eta_t^4 \Omega  \left(\sum_{i<j}\omega_i^2 \omega_j + 5 \prod_i \omega_i\right) -\eta_t^2 \left(7 \sum_i \omega_i^2 +20 \sum_{i<j} \omega_i \omega_j\right)\nonumber \\ &-&90 \,e^{\frac{1}{4}} \,\eta_t^3 \, \mathcal{H}^3 \left(\eta_t^2 \sum_{i<j}\omega_i \omega_j-1\right) +2 \, e^{\frac{1}{6}} \, \eta_t^2 \, \mathcal{H}^2 \left(-\eta_t^4 \Omega \prod_i \omega_i +\eta_t^2 \left(\sum_i \omega_i^2-5 \sum_{i<j} \omega_i \omega_j\right)+5\right)\ \nonumber \\ &+&e^{\frac{1}{12}} \,\eta_t \, \mathcal{H} \left(\eta_t^4 \Omega ^2 \sum_{i<j} \omega_i \omega_j-\eta_t^2 \left(5 \sum_i \omega_i^2 + 12 \sum_{i<j} \omega_i \omega_j\right) +12\right)+20\Bigg]\nonumber \\ &+&\eta_t \sin \Omega \eta_t \Bigg[\Omega \left(20+12e^{\frac{1}{12}} \mathcal{H} \eta_t+10e^{\frac{1}{6}} \mathcal{H}^2 \eta_t^2+90e^{\frac{1}{4}} \mathcal{H}^3 \eta_t^3\right)\nonumber \\ &-&\left(\sum_i \omega_i^3\right) \left(\eta_t^2+e^{\frac{1}{12}}\mathcal{H}\eta_t^3\right)\nonumber \\ &+&\left(\sum_{i<j}\omega_i^2 \omega_j\right)\left(2e^{\frac{1}{6}}\mathcal{H}^2 \eta_t^4 -5e^{\frac{1}{12}}\mathcal{H} \eta_t^3-7 \eta_t^2\right)\nonumber \\ &+& \left(\prod_i \omega_i\right)\left(-20 \eta_t^2 - 12 e^{\frac{1}{12}} \mathcal{H} \eta_t^3 -10 e^{\frac{1}{6}} \mathcal{H}^2 \eta_t^4 -90 e^{\frac{1}{4}}\mathcal{H}^3 \eta_t^5 + \eta_t^4 \Omega^3 + e^{\frac{1}{12}}\mathcal{H}\eta_t^5 \Omega^3\right)\Bigg]\Bigg\} \quad .
    \label{eq:O4t}
\eea
\newline
\newline
- \hspace{0.5em}$\langle \mathcal{O}_5\rangle_{\text{t}}$:
\newline
\newline
The integral to compute is
is once again \eqref{eq:integral5} but with a time dependent $\varepsilon$ and localized around $\eta_t$.
This looks like the first. However the presence of another $\varepsilon$ cancel poles, so this contribution vanishes.
\bea
    \langle \mathcal{O}_5\rangle_{\text{t}}=0 \quad .
    \label{eq:O5t}
\eea
An important feature of these amplitudes is that they scale differently with \( \mathcal{H} \) and \( M_{pl} \), implying a strong dependence on the number of e-folds via \eqref{eq:Hdelta}. This does not affect the principal part, where all Hubble factors cancel in \( f_{NL} \) (see next section).

\begin{figure}[ht]
\begin{center}
    \includegraphics[width=2in]{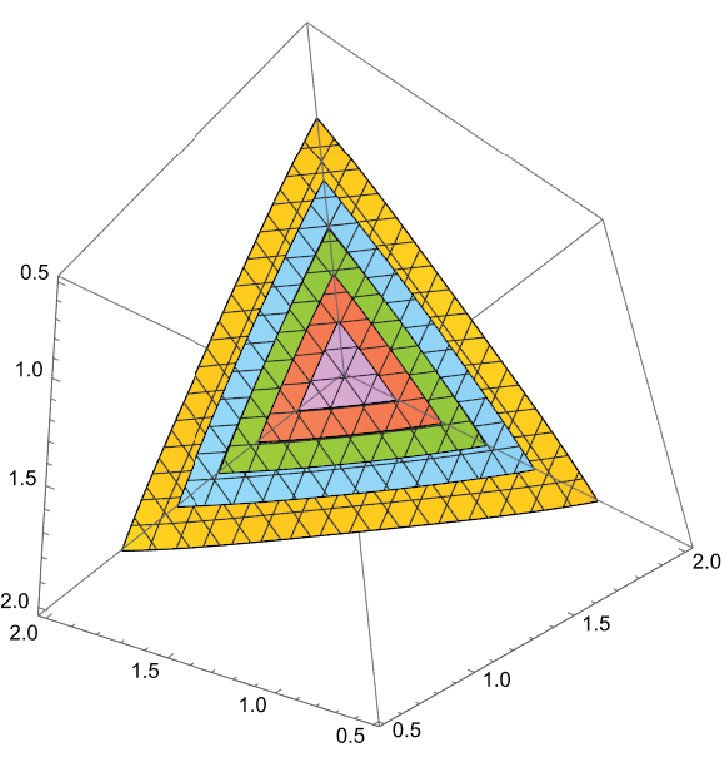}
\end{center}
\caption{3D plot of the bispectrum in the region \( k_i \in [\Delta/2, 2\Delta] \), showing a clear peak in the equilateral configuration.}
\label{fig:equil}
\end{figure}

The result for the three-point function is general and valid for arbitrary momentum configurations, as long as \( \sum_i \vec{k}_i = 0 \). However, it is known \cite{Baumann2009TASI,chen_huang_kachru_shiu_2007,babich_creminelli_zaldarriaga_2004_shape} that contact interactions like ours peak in the equilateral configuration \( k_1 = k_2 = k_3 \), as confirmed by the numerical plot in Fig.~\ref{fig:equil}.

Before specializing to equilateral shapes, we briefly recall the squeezed limit, where \( k_1 \ll k_2 \sim k_3 \). Here, the bispectrum is fixed by the consistency relation (Maldacena’s theorem) \cite{maldacena2003non}, which links the squeezed bispectrum to the tilt of the power spectrum. Physically, a long-wavelength mode \( \zeta_L \) acts as a local rescaling on short modes \( \zeta_S \).

This result has been extended using symmetry arguments \cite{creminelli_zaldarriaga_2004_consistency} and the EFT of inflation \cite{cheung2008consistency}, showing that the squeezed limit is universal in single-field inflation.

In the squeezed limit, the bispectrum takes the universal form \cite{maldacena2003non,creminelli_zaldarriaga_2004_consistency,cheung2008consistency}
\beq
    \lim_{k_1 \to 0} \langle \zeta_{\mathbf{k}_1} \zeta_{\mathbf{k}_2} \zeta_{\mathbf{k}_3} \rangle = -(2\pi)^3 \delta^3(\mathbf{k}_1 + \mathbf{k}_2 + \mathbf{k}_3) \, \mathcal{P}(k_1) \mathcal{P}(k_3) \frac{d \ln \left(k_3^3 \mathcal{P}(k_3)\right)}{d \ln k_3} \quad,
\eeq
in agreement with Maldacena’s consistency relation. This expression relates the squeezed bispectrum to the spectral tilt \( n_s - 1 \), and any deviation signals the breakdown of single-field inflation.

In our case, applying this relation to the power spectrum \eqref{eq:deltapowerspectrum} (even in the regime \( \Delta \ll k \)) leads to a corrected tilt:
\beq
    n_s - 1 \;\longrightarrow\; n_s - 1 - 2 \nu \left(\frac{\Delta}{k}\right)^2 \quad,
\eeq
suggesting enhanced non-Gaussianity. However, this result is not fully reliable, as it neglects the analyticity properties of the full correlator and therefore cannot capture the precise behavior of the bispectrum. Nonetheless, it provides a useful slow-roll independent motivation to further investigate this regime.

We now turn to the equilateral configuration and present the full expression for the resulting non-Gaussianity.

\bea
    \langle \, \zeta \, \zeta \, \zeta \, \rangle_{\text{equil}} &=& \int \frac{d^3 \vec k_1}{(2 \pi)^3} \frac{d^3 \vec k_2}{(2 \pi)^3} \frac{d^3 \vec k_3}{(2 \pi)^3}e^{i \vec k_1 \cdot \vec x} e^{i \vec k_2 \cdot \vec y} e^{i \vec k_3 \cdot \vec z}(2 \pi)^3 \delta^{(3)}(\vec k_1 + \vec k_2 + \vec k_3)\nonumber \\
    && \Big\{\langle \mathcal{O}_1 \rangle + \langle \mathcal{O}_2 \rangle + \langle \mathcal{O}_3 \rangle + \langle \mathcal{O}_4 \rangle + \langle \mathcal{O}_5 \rangle \Big\}_{\text{equil}} \quad  ,
    \label{eq:nongaussianity}
\eea
\bea
    - \hspace{0.5em}\langle \mathcal{O}_1 \rangle_{\text{PP}} &=&  \frac{\mathcal{H}^4}{144 \, \gamma^2 \, {M_{pl}}^4 \omega^6}\ \Big[4 \left(1 \ - \ \cos (3 \eta_0 \, \omega )\right) \ - \ 3 \,\eta_0 \, \omega \sin (3 \eta_0 \,\omega )  \Big] \ , \nonumber\\
    - \hspace{0.5em}\langle \mathcal{O}_2 \rangle_{\text{PP}} &=& \frac{\mathcal{H}^4\ k^2}{288 \,\gamma^2 \, \eta_0 \, {M_{pl}}^4 \omega^9}   \Big[3 \left(\eta_0^2 \,\omega^2\,-\,3\right) \sin (3 \eta_0 \, \omega )\ + \ \eta_0 \, \omega  \left(10 \cos (3 \eta_0 \,\omega )\,+\,17\right)\Big] \ ,\nonumber\\
    - \hspace{0.5em}\langle \mathcal{O}_3 \rangle_{\text{PP}} &=&  \frac{\mathcal{H}^4}{144 \, \gamma^2 \, {M_{pl}}^4 \omega^6}\ \Big[4 \left(1 \, - \, \cos (3 \eta_0 \, \omega )\right) \ - \ 3 \,\eta_0 \, \omega \sin (3 \eta_0 \,\omega )  \Big] \ , \nonumber\\
    - \hspace{0.5em}\langle \mathcal{O}_4 \rangle_{\text{PP}} &=&  0 \ , \nonumber\\
    - \hspace{0.5em}\langle \mathcal{O}_5 \rangle_{\text{PP}} &=&  - \ \frac{\mathcal{H}^4}{128 \, {M_{pl}}^4 \omega^6} \   \Big[4 \left(1 \, - \, \cos (3 \eta_0 \, \omega )\right) \ - \ 3 \,\eta_0 \, \omega \sin (3 \eta_0 \,\omega )  \Big] \ , \label{eq:PP_contributions}
\eea
\bea
    - \hspace{0.5em}\langle \mathcal{O}_1 \rangle_{\text{t}} &=& -\ \frac{3\,\pi\, \mathcal{H}^5 \,{e}^\frac{1}{12}}{16\, \gamma^3 \,\eta_t^2\, {M_{pl}}^4 \omega^9} \  \Big[\eta_t \omega  \left(\eta_t^2 \omega ^2-3\right) \sin (3 \eta_t \omega )\ + \ \left(3 \eta_t^2 \omega ^2-1\right) \cos (3 \eta_t \omega )\Big]  \ , \nonumber\\
    - \hspace{0.5em}\langle \mathcal{O}_2 \rangle_{\text{t}} &=&  0 \ ,\nonumber\\
    - \hspace{0.5em}\langle \mathcal{O}_3 \rangle_{\text{t}} &=&  -\ \frac{3\,\pi\, \mathcal{H}^5 \,{e}^\frac{1}{12}}{16\, \gamma^3 \,\eta_t^2\, {M_{pl}}^4 \omega^9} \  \Big[\eta_t \omega  \left(\eta_t^2 \omega ^2-3\right) \sin (3 \eta_t \omega )\ + \ \left(3 \eta_t^2 \omega ^2-1\right) \cos (3 \eta_t \omega )\Big] \ , \nonumber\\
    - \hspace{0.5em}\langle \mathcal{O}_4 \rangle_{\text{t}} &=&  \frac{\pi  \mathcal{H}}{288 \,{e}^\frac{1}{4} \gamma ^3 \eta_t^6 {M_{pl}}^4 \omega ^9} \Big\{\eta_t \omega  \sin (3 \eta_t \omega ) \Big[\eta_t \Big(9 \eta_t^3 \omega ^4-65 \eta_t \omega^2-90 {e}^\frac{1}{4} \eta_t^2 \mathcal{H}^3 \left(\eta_t^2 \omega ^2-3\right)\nonumber \\ &+& 2 {e}^\frac{1}{6} \eta_t \mathcal{H}^2 \left(\eta_t^2 \omega ^2+15\right) \ + \ 9 {e}^\frac{1}{12} \mathcal{H} \left(\eta_t^4 \omega ^4-5 \eta_t^2 \omega ^2+4\right)\Big)\ + \ 60\Big]\nonumber \\
    &+&\cos (3 \eta_t \omega ) \Big[\eta_t \Big(33 \eta_t^3 \omega^4-81 \eta_t \omega^2+90 {e}^\frac{1}{4} \eta_t^2 \mathcal{H}^3 \left(1-3 \eta_t^2 \omega ^2\right)\nonumber \\ &-& 2 {e}^\frac{1}{6} \eta_t \mathcal{H}^2 \left(3 \eta_t^4 \omega^4+12 \eta_t^2 \omega^2-5\right)\ + \ 3 {e}^\frac{1}{12} \mathcal{H} \left(9 \eta_t^4 \omega ^4-17 \eta_t^2 \omega ^2+4\right)\Big) \ +\ 20\Big]\Big\} \ , \nonumber\\
    - \hspace{0.5em}\langle \mathcal{O}_5 \rangle_{\text{t}} &=& 0 \ . \label{eq:t_contributions}
\eea
Before concluding this section, it is important to highlight a key point. If one averages trigonometric functions over a full period—determined by \(\eta_0\) for the principal part (PP) terms and by \(\eta_t\) for the transient (t) terms—the result precisely reproduces that of \cite{maldacena2003non,collins2011primordial}, up to the subleading term \(\langle O_5 \rangle\), which was neglected in those works.

 Furthermore, substituting \(\omega \to k\) (justified by the disappearence of the \(\Delta\)-dependence), we exactly recover the standard equilateral limit of Maldacena’s result:
\beq
    \frac{\mathcal{H}^4 \left(-3 \gamma ^4 + 11 \gamma ^2 \right)}{96 \gamma ^4 k^6 M_{pl}^4} \quad.
    \label{eq:maldacenaequil}
\eeq
Here we have used our definitions of the slow-roll parameters, \(\varepsilon = 3 \gamma^2\) and \(\delta = -3 \gamma^2\), making the connection between our generalized calculation and the standard result manifest.
\section{Comparisons and Plots}
In this section we visualize the derived result about of the previous section. Actually Non-gaussianity is usually visualized using the $f_{NL}$ parameter that wheights the magnitude of the three point function compared to the square of the two point function. 

The standard definition used to define $f_{NL}$ is through the field redefinition
\beq
    \zeta = \zeta_g + \frac{3}{5} f_{NL} \, \zeta_g^2 \quad,
\eeq
where $\zeta_g$ is a gaussian field. Using the modified power spectrum of \eqref{eq:deltapowerspectrum} written as
\beq
    \mathcal{P}_k = \frac{1}{8\pi^2}  \frac{1}{\varepsilon}  \frac{\mathcal{H}^2}{M_{pl}^2} \frac{\left(\frac{k}{k_*}\right)^3}{\left[\left(\frac{k}{k_*}\right)^2+\left(\frac{\Delta}{k_*}\right)^2\right]^{\nu}} \quad .
\eeq
And taking as reference value $\Delta$ the $f_{NL}$ formula is 
\beq
    f_{NL}(k)=40 \, \gamma^4 \, \frac{M_{pl}^4}{\mathcal{H}^4} \, \Delta^{6} \,\left[\left(\frac{k}{\Delta}\right)^2 + 1\right]^{2 \, \nu} \, \langle \zeta(k) \zeta(k) \zeta(k) \rangle ' \quad .
    \label{eq:fnl2}
\eeq
where:
\beq
    \langle \zeta(k) \zeta(k) \zeta(k) \rangle '= \langle \mathcal{O}_1 \rangle + \langle \mathcal{O}_2 \rangle + \langle \mathcal{O}_3 \rangle + \langle \mathcal{O}_4 \rangle + \langle \mathcal{O}_5 \rangle \quad,
\eeq
\begin{center}
    ***
\end{center}

\begin{figure}[ht]
\centering
\begin{tabular}{ccc}
%\mbox{graphic} & \mbox{table} \\
\includegraphics[width=65mm]{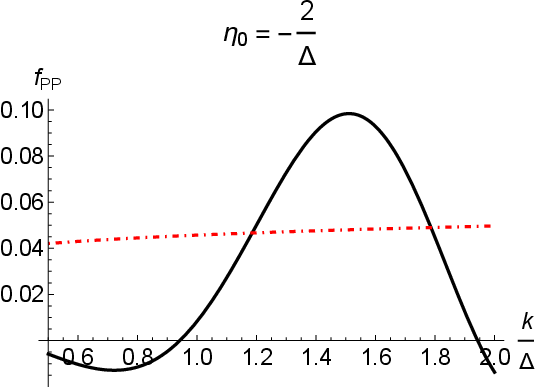} \quad  &
\includegraphics[width=65mm]{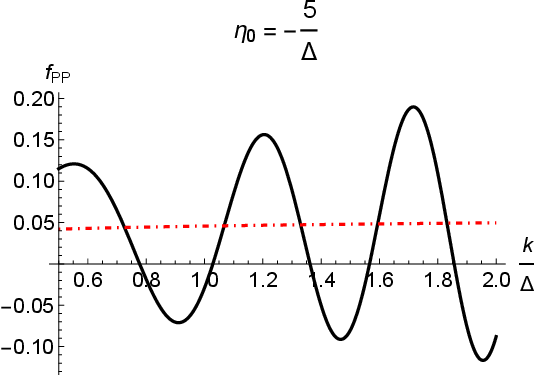}  \\
\end{tabular}
 \caption{\small Left panel: $f_{PP}$ (black, solid) compared with $f_{NL}^{(M)}$ (red, dot-dashed), for $\eta_0= -\,\frac{2}{\Delta}$. Right panel: $f_{PP}$ (black, solid) compared with $f_{NL}^{(M)}$ (red, dot-dashed), for $\eta_0= -\,\frac{5}{\Delta}$. The results displayed correspond to the choice $\varepsilon=0.03$, and the oscillations are determined by $\eta_0$ \cite{meo2025preinflationarynongaussianities}.}
\label{fig:fpp}
\end{figure}

We decompose $f_{NL}$ as
\beq
    f_{NL}(k) \ = \ f_{\text{PP}}(k) \ + \ f_{\text{t}}(k) \quad ,
\eeq
\bea
    f_{\text{PP}}(k) &=&  \frac{5 \gamma^2 \Delta ^6 \left(\frac{k^2}{\Delta ^2}+1\right)^{2 \nu }}{144\, \eta_0 \,\omega^9}  \Big\{3 \sin (3 \eta_0 \,\omega ) \Big[\left(9 \gamma ^2-16\right) \eta_0^2 \,\omega ^4+4 k^2 \left(\eta_0^2\, \omega ^2-3\right)\Big] \nonumber \\
    &+& 4 \,\eta_0\, \omega  \cos (3 \eta_0 \omega ) \Big[\left(9 \gamma^2-16\right) \omega ^2+10 k^2\Big]\ + \ 68 \,\eta_0 k^2 \omega \ + \ 4 \left(16-9 \gamma^2\right) \eta_0 \omega^3 \Big\}\, \nonumber \\ &&\quad,
    \label{eq:fNLPP}
\eea
and 
\bea
    f_{\text{t}}(k) &=& \frac{5 \pi  \gamma  \Delta ^6 \left(\frac{k^2}{\Delta ^2}+1\right)^{2 \nu }}{36 \, e^{\frac{1}{4}} \, \eta_t^6 \, \mathcal{H}^3 \, \omega ^9} \Bigg\{\eta_t \omega  \sin 3 \eta_t \omega  \bigg[\eta_t \Big(9 \eta_t^3 \omega ^4-65 \eta_t \omega ^2-108 e^{\frac{1}{3}} \eta_t^3 \mathcal{H}^4 \left(\eta_t^2 \omega ^2-3\right) \nonumber \\ &-&90 e^{\frac{1}{4}} \eta_t^2 \mathcal{H}^3 \left(\eta_t^2 \omega ^2-3\right)+2 e^{\frac{1}{6}} \eta_t \mathcal{H}^2 \left(\eta_t^2 \omega ^2+15\right)+9 e^{\frac{1}{12}} \mathcal{H} \left(\eta_t^4 \omega ^4-5 \eta_t^2 \omega ^2+4\right)\Big)+60\bigg] \nonumber \\ &+&\cos (3 \eta_t \omega ) \bigg[\eta_t \Big(33 \eta_t^3 \omega ^4-81 \eta_t \omega ^2+108 e^{\frac{1}{3}} \eta_t^3 \mathcal{H}^4 \left(1-3 \eta_t^2 \omega ^2\right)+90 e^{\frac{1}{4}} \eta_t^2 \mathcal{H}^3 \left(1-3 \eta_t^2 \omega ^2\right) \nonumber \\ &-&2 e^{\frac{1}{6}} \eta_t \mathcal{H}^2 \left(3 \eta_t^4 \omega ^4+12 \eta_t^2 \omega ^2-5\right)+3 e^{\frac{1}{12}} \mathcal{H} \left(9 \eta_t^4 \omega ^4-17 \eta_t^2 \omega ^2+4\right)\Big)+20\bigg]\Bigg\} \quad ,
    \label{eq:fNLt}
\eea
and in the following we shall consider separately the two contributions
of PP-terms and t-terms comparing them with the original result in~\cite{maldacena2003non,collins2011primordial}, which in the same conventions reads
\beq
    f_{NL}^{(M)}(k)\ = \ \frac{5}{12}\, \left(\frac{k}{\Delta}\right)^{4 \, \nu - 6} \left[-\,3\, \gamma^4\,+\,11\, \gamma^2\right] \ .
    \label{eq:fnlM}
\eeq
To plot figures we have made quanties adimensional mesuring them with $\Delta$ and, as explained, we have written the $\mathcal{H}$ dependence with \eqref{eq:Hdelta} that introduces dependace to the number of e-folds with $x$.
\begin{figure}[ht]
\centering
\begin{tabular}{ccc}
%\mbox{graphic} & \mbox{table} \\
\includegraphics[width=65mm]{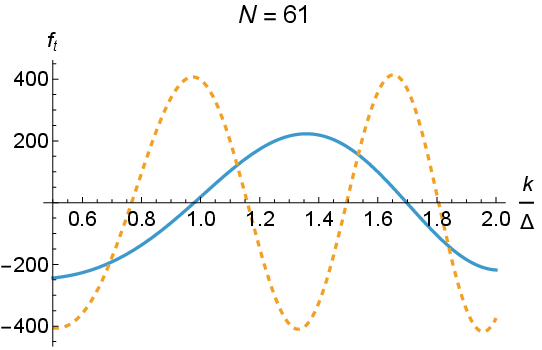} \quad  &
\includegraphics[width=65mm]{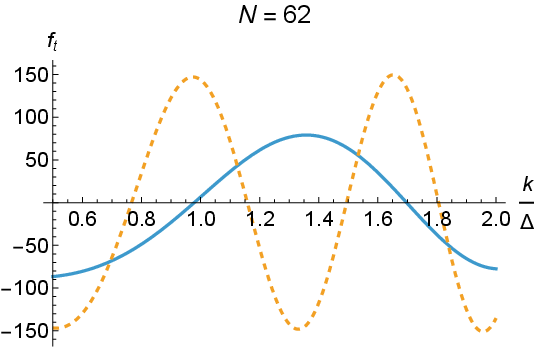}  \\
\end{tabular}
 \caption{\small Left panel: $f_{t}$ for $\eta_t= -\,\frac{2}{\Delta}$ (blue, solid) and for $\eta_t= -\,\frac{4}{\Delta}$ (orange, dashed) for $N=61$. Right panel: $f_{t}$ for $\eta_t= -\,\frac{2}{\Delta}$ (blue, solid) and for $\eta_t= -\,\frac{4}{\Delta}$ (orange, dashed) for $N=62$.  The results displayed correspond to the choice $\varepsilon=0.03$, and the oscillations are determined by $\eta_t$ \cite{meo2025preinflationarynongaussianities}.}
\label{fig:ft1}
\end{figure}
\begin{figure}[ht]
\centering
\begin{tabular}{ccc}
%\mbox{graphic} & \mbox{table} \\
\includegraphics[width=65mm]{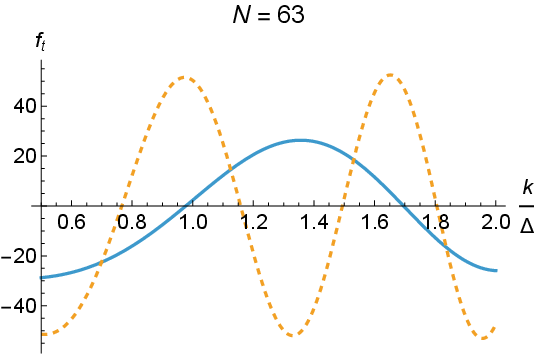} \quad  &
\includegraphics[width=65mm]{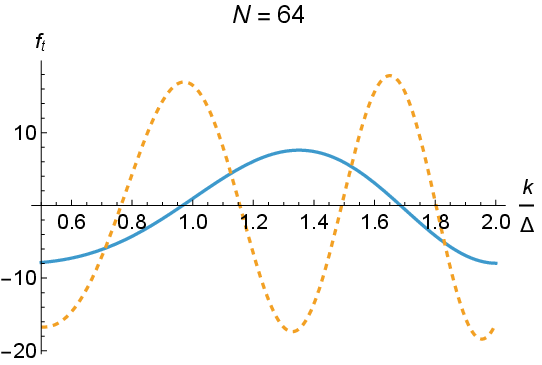}  \\
\end{tabular}
 \caption{\small Left panel: $f_{t}$ for $\eta_t= -\,\frac{2}{\Delta}$ (blue, solid) and for $\eta_t= -\,\frac{4}{\Delta}$ (orange, dashed) for $N=63$. Right panel: $f_{t}$ for $\eta_t= -\,\frac{2}{\Delta}$ (blue, solid) and for $\eta_t= -\,\frac{4}{\Delta}$ (orange, dashed) for $N=64$.  The results displayed correspond to the choice $\varepsilon=0.03$, and the oscillations are determined by $\eta_t$ \cite{meo2025preinflationarynongaussianities}.}
\label{fig:ft2}
\end{figure}
\begin{figure}[ht]
\centering
\begin{tabular}{ccc}
%\mbox{graphic} & \mbox{table} \\
\includegraphics[width=65mm]{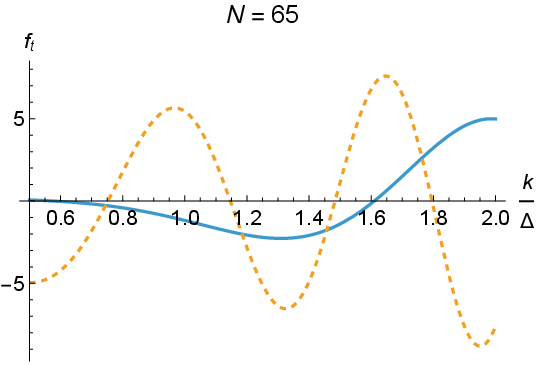} \quad  &
\includegraphics[width=65mm]{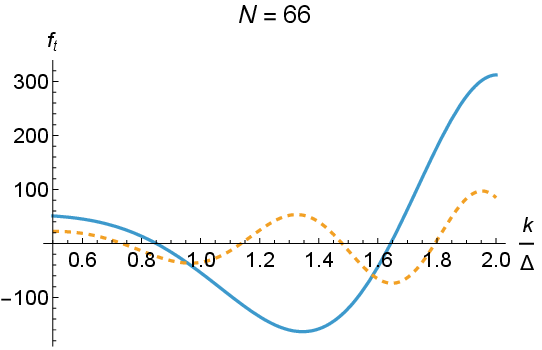}  \\
\end{tabular}
 \caption{\small Left panel: $f_{t}$ for $\eta_t= -\,\frac{2}{\Delta}$  (blue, solid) and for $\eta_t= -\,\frac{4}{\Delta}$ (orange, dashed) for $N=65$. Right panel: $f_{t}$ for $\eta_t= -\,\frac{2}{\Delta}$  (blue, solid) and for $\eta_t= -\,\frac{4}{\Delta}$ (orange, dashed) for $N=66$.  The results displayed correspond to the choice $\varepsilon=0.03$, and the oscillations are determined by $\eta_t$ \cite{meo2025preinflationarynongaussianities}.}
\label{fig:ft3}
\end{figure}

Fig.~\ref{fig:fpp} compares $f_{PP}$ to $f_{NL}^{(M)}$, within the relevant range $\frac{\Delta}{2}<k<2\,\Delta$, for two values of the conformal time $\eta_0$ that signal the beginning of the expanding phase. The two signals are comparable in size, as expected. Note how the choice of $\eta_0$ affects the frequency of the oscillations around the original result in~\cite{maldacena2003non}.

\begin{figure}[ht]
\begin{center}
    \includegraphics[width=3.4in]{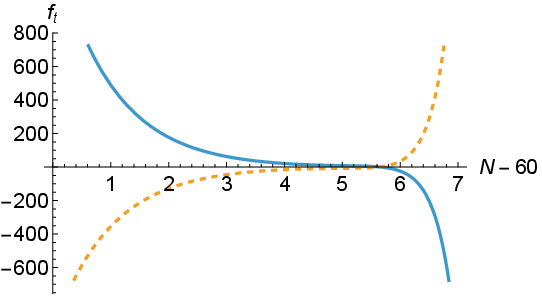}
\end{center}
\caption{The typical behavior of $f_t$ for $k=1.2\,\Delta$ (blue, solid) and for $k=2\,\Delta$ (orange, dashed) and $\varepsilon=0.03$, as the number of $e$-folds increases from $60$ to $67$. Note that a plateau is present in the region $63<N<65$. This is due to the combined effect of $\langle O_{1}\rangle$ and $\langle O_{3}\rangle$, which decrease in absolute value for increasing values of $N$, and to the opposite behavior of $\langle O_4 \rangle$ \cite{meo2025preinflationarynongaussianities}.}
\label{fig:fx}
\end{figure}

Figs.~\ref{fig:ft1}, ~\ref{fig:ft2} and~\ref{fig:ft3} compare the behavior of $f_t$, within the relevant range $\frac{\Delta}{2}<k<2\,\Delta$ for different numbers of $e$--folds in the interval $61 \leq N \leq 66$. The solid curves refer to $\eta_t= -\,\frac{2}{\Delta}$ and the dashed one to $\eta_t= -\,\frac{5}{\Delta}$. Note that for $63\leq N \leq 65$ the values of $f_t$ are appreciable and yet within ranges not excluded by the {\it Planck} collaboration and that the values of $\eta_t$ also affect in important ways the size of $f_t$. For $N<63$ or $N>65$ the peaks typically grow far beyond these values, so there is a clear preference for a small window in the number of $e$-folds. This can also be seen from fig.~\ref{fig:fx}, which displays the behavior of $f_t$ for $k= 1.2\,\Delta$ (solid line) and for $k=2\,\Delta$ (dashed line). 

In summary, $f_{PP}$ undergoes small oscillations around the original result in~\cite{maldacena2003non,collins2011primordial}, but $f_t$ can offer prospects for future detection and has the interesting feature of selecting a very small range of $e$-folds where our results are sizable and yet not too large.
In the present setup, as we have stressed, $\eta_0$ should be at least slightly below $\eta_t$. AS we have stressed, the original result in~\cite{maldacena2003non,collins2011primordial} is formally recovered averaging  over a period determined by $\omega$ and $\eta_0$, while the same operation eliminates $\eta_t$ altogether. Let us recall that, according to the analysis in~\cite{GruppusoKitazawaMandolesiNatoliSagnotti2016}, $k=\Delta$ translates into $\ell \simeq 11$ in the CMB angular power spectrum, which justifies our choice of confining the attention to the range $\frac{\Delta}{2}<k<2\,\Delta$.
All preceding results become unreliable for values of $k$ well beyond $\Delta$. The $k \to \infty$ limit is equivalent to letting $\Delta \to 0$, which would extend the integrals to the standard range $- \infty< \eta<0$. This would require the usual contour deformation that makes it move farther and farther away from the real axis for large negative values of $\eta$, which can be avoided within the limited range of interest for $k$.

One last comment might perhaps help in comparisons with data. The oscillations are actually determined by the combinations $3 \eta_0 \sqrt{k^2+ \Delta^2}$ and $3 \eta_t \sqrt{k^2+ \Delta^2}$, and thus become periodic in $k$ only when it grows beyond a few times $\Delta$. For typical values of $\eta_0\ \simeq \ \eta_t \ \simeq \ - \ \frac{2}{\Delta}$ or lower ones, beyond the first couple of intervals of order $\Delta$ for $k$, averaging the trigonometric functions completely eliminates the effects that we are highlighting.

\begin{appendices}

\chapter{The A.D.M. formalism}
\label{appendix:ADM}
This section provides a brief review of the Arnowitt–Deser–Misner (A.D.M.) formalism, which recasts General Relativity in Hamiltonian form. The focus will be primarily on deriving the key expressions that will be useful for the computations presented throughout the text. A concise summary of their physical interpretation will also be given. We will use the notation $f_{,x}$ for partial derivatives and $f_{;x}$ for covariant derivatives. This section is based on ~\cite{poisson2004relativist,landau1975classical} which review the famous work~\cite{Arnowitt:1962hi} .
\section{Toolkit of Differential Geometry}
In this section I define differential geometry tools useful to understand what follows.
\paragraph{Induced Metric and Extrinsic Curvature}
\leavevmode\\
In a four-dimensional spacetime, a hypersurface $\Sigma$ is a three-dimensional submanifold that may be timelike, spacelike, or null. It can be defined either by a coordinate condition
\beq
    \Phi(x^{\alpha})=0 \quad ,
\eeq
or parametrically as
\beq
    x^{\alpha}=x^{\alpha}(y^a) \quad ,
\eeq
with $y^a$ coordinates intrinsic to $\Sigma$.

The gradient $\Phi_{,\alpha}$ is normal to $\Sigma$. For non-null hypersurfaces, a unit normal vector $n_{\alpha}$ can be defined satisfying
\beq
    n^{\alpha}n_{\alpha}=\varepsilon\equiv 
    \begin{cases}
        -1 \qquad \text{if $\Sigma$ is spacelike}\\
        +1 \qquad \text{if $\Sigma$ is timelike}
    \end{cases} \quad ,
\eeq
and oriented so that $n^{\alpha}\Phi_{,\alpha} > 0$. The expression for $n_{\alpha}$ is
\beq
    n_{\alpha}=\frac{\varepsilon\Phi_{,\alpha}}{|g^{\mu \nu}\Phi_{,\mu}\Phi_{,\nu}|^{1/2}} \quad .
\eeq
The induced metric on $\Sigma$ comes from restricting the line element to the hypersurface. Using the parametrization, the tangent vectors
\beq
    e^{\alpha}_a=\frac{\partial x^{\alpha}}{\partial y^a} \quad ,
\eeq
are orthogonal to the normal. The line element becomes
\beq
    ds_{\Sigma}^2=g_{\alpha \beta}dx^{\alpha}dx^{\beta}=g_{\alpha \beta}\Bigg(\frac{\partial x^{\alpha}}{\partial y^a}dy^a\Bigg)\Bigg(\frac{\partial x^{\beta}}{\partial y^b}dy^b\Bigg)=h_{ab}dy^ady^b \quad ,
\eeq
with induced metric
\beq
    h_{ab}=g_{\alpha \beta}e^{\alpha}_a e^{\beta}_b \quad .
\eeq
In the non-null case, the completeness relation for the inverse metric is
\beq
    g^{\alpha \beta}=\varepsilon n^{\alpha} n^{\beta}+h^{ab}e^{\alpha}_a e^{\beta}_b \quad ,
\eeq
where $h^{ab}$ is the inverse of $h_{ab}$.

The extrinsic curvature is defined as
\beq
    E_{ab}\equiv n_{\alpha ; \beta}e_a^{\alpha}e_b^{\beta} \quad ,
\eeq
and its trace is $E = h^{ab}E_{ab}$. Using the completeness relation and the orthogonality of $n^{\alpha}$ to its covariant derivative, this becomes $E = n^{\alpha}_{;\alpha}$.

For non-null hypersurfaces, the oriented volume element is
\beq
    d \Sigma_{\alpha} \equiv \varepsilon |h|^{1/2} n_{\alpha} d^3y \quad .
\eeq
\paragraph{Extrinsic Curvature Properties}
\leavevmode\\
We consider the covariant derivative of a spacetime vector field \( A^\alpha \), and analyze its decomposition relative to a hypersurface with basis vectors \( e^\alpha_a \) and normal vector \( n^\alpha \). Projecting the derivative onto the hypersurface gives the tangential part:
\beq
    A^a{}_{|b} = A^\alpha{}_{;\beta} \, e^a_\alpha e^b_\beta
    \quad .
\eeq
To identify any normal component, we decompose the metric as:
\beq
    g^{\alpha\mu} = \varepsilon n^\alpha n^\mu + h^{am} e^\alpha_a e^\mu_m
    \quad ,
\eeq
which leads to:
\beq
    A^\alpha{}_{;\beta} e^\beta_b = \varepsilon (n^\mu A_{\mu;\beta} e^\beta_b) n^\alpha + A^a{}_{|b} e^\alpha_a
    \quad .
\eeq

Assuming the vector is tangential, \( A^\mu n_\mu = 0 \), and using the identity:$n^\mu A_{\mu;\beta} = -n^\mu{}_{;\beta} A_\mu$
we find the decomposition:
\beq
    A^\alpha{}_{;\beta} e^\beta_b = A^a{}_{|b} e^\alpha_a - \varepsilon A^a E_{ab} n^\alpha
    \quad .
\eeq
Choosing \( A^\alpha = e^\alpha_a \), one obtains the \emph{Gauss--Weingarten equation}:
\beq
    e^\alpha_{a;\beta} \, e^\beta_b = \Gamma^c_{ab} \, e^\alpha_c - \varepsilon E_{ab} \, n^\alpha
    \quad .
\eeq
The extrinsic curvature tensor \( E_{ab} \) is symmetric, and can be written in terms of the Lie derivative of the metric along the normal:
\beq
    E_{ab} = \frac{1}{2} \left( \mathcal{L}_n g_{\alpha\beta} \right) e^\alpha_a e^\beta_b
    \quad .
\eeq
Thus, the extrinsic curvature characterizes how the induced geometry on the hypersurface changes along the normal direction.
\paragraph{Ricci Scalar Decomposition}
\leavevmode\\
We define the intrinsic curvature on the hypersurface via the commutator of covariant derivatives:
\beq
    A^c{}_{|ab} - A^c{}_{|ba} = -\hat R^c{}_{dab} A^d
    \quad ,
\eeq
with the intrinsic Riemann tensor expressed as
\beq
    \hat R^c{}_{dab} = \Gamma^c_{db,a} - \Gamma^c_{da,b} + \Gamma^c_{ma} \Gamma^m_{db} - \Gamma^c_{mb} \Gamma^m_{da}
    \quad .
\eeq
To relate this to the full spacetime Riemann tensor \( R^\gamma{}_{\delta\alpha\beta} \), we start from the Gauss--Weingarten equation and compare second derivatives of the basis vectors. Antisymmetrizing and projecting leads to the Gauss equation:
\beq
    R_{\alpha\beta\gamma\delta} e^\alpha_a e^\beta_b e^\gamma_c e^\delta_d = \hat R_{abcd} + \varepsilon \left( E_{ad} E_{bc} - E_{ac} E_{bd} \right)
    \quad .
\eeq
Projecting one index along the normal vector gives the Codazzi equation:
\beq
    R_{\mu\alpha\beta\gamma} n^\mu e^\alpha_a e^\beta_b e^\gamma_c = E_{ab|c} - E_{ac|b}
    \quad ,
\eeq
These are the \textit{Gauss--Codazzi equations}, relating the spacetime curvature to intrinsic and extrinsic geometry.

Some components, such as
\beq
    R_{\mu\alpha\nu\beta} n^\mu e^\alpha_a n^\nu e^\beta_b
    \quad ,
\eeq
cannot be fully expressed in terms of \( h_{ab} \), \( E_{ab} \), or their derivatives, as they describe curvature in the normal-normal directions.

The Ricci tensor is obtained from contraction:
\beq
    R_{\alpha\beta} = g^{\mu\nu} R_{\mu\alpha\nu\beta}
    = \varepsilon R_{\mu\alpha\nu\beta} n^\mu n^\nu + h^{mn} R_{\mu\alpha\nu\beta} e^\mu_m e^\nu_n
    \quad ,
\eeq
and the Ricci scalar becomes:
\beq
    R = g^{\alpha\beta} R_{\alpha\beta}
    = 2 \varepsilon h^{ab} R_{\mu\alpha\nu\beta} n^\mu e^\alpha_a n^\nu e^\beta_b
    + h^{ab} h^{mn} R_{\mu\alpha\nu\beta} e^\mu_m e^\alpha_a e^\nu_n e^\beta_b
    \quad .
\eeq
Using the Gauss–Codazzi relations and symmetry of \( E_{ab} \), we find:
\beq
    h^{ab} h^{mn} R_{\mu\alpha\nu\beta} e^\mu_m e^\alpha_a e^\nu_n e^\beta_b = \hat R + \varepsilon \left( E_{ab} E^{ab} - E^2 \right)
    \quad ,
\eeq
and
\beq
    2 \varepsilon h^{ab} R_{\mu\alpha\nu\beta} n^\mu e^\alpha_a n^\nu e^\beta_b = 2 \varepsilon R_{\alpha \beta} n^\alpha n^\beta
    \quad .
\eeq
Evaluating \( R_{\alpha \beta} n^\alpha n^\beta \) and collecting terms, we arrive at the Ricci scalar decomposition:
\beq
    R = \hat R + \varepsilon \left( E^2 - E_{ab} E^{ab} \right)
    + 2 \varepsilon \left( n^\alpha{}_{;\beta} n^\beta - n^\alpha n^\beta{}_{;\beta} \right)_{;\alpha}
    \quad .
    \label{eq:Ricci}
\eeq
This expresses the four-dimensional Ricci scalar in terms of intrinsic curvature \( \hat R \), extrinsic curvature \( E_{ab} \), and a total divergence term.
\section{Complete Gravitational Action}
The total action in general relativity includes a gravitational part and a matter part. The gravitational part is
\beq
    S_G[g] = S_H[g] + S_B[g] - S_0
    \label{eq:complete gravitational action}
    \quad ,
\eeq
where
\beq
    S_H[g] = \frac{M_{pl}^2}{2} \int_{\mathcal{V}} d^4 x \sqrt{-g} R
    \quad , \quad
    S_B[g] = M_{pl}^2 \oint_{\partial \mathcal{V}} d^3 y \, \varepsilon E |h|^{1/2}
    \quad , \quad
    S_0 = M_{pl}^2 \oint_{\partial \mathcal{V}} d^3 y \, \varepsilon E_0 |h|^{1/2}.
\eeq
The second piece is known as \textit{Gibbons-Hawking-York term}.
Here, \( R \) is the Ricci scalar in the bulk, \( E \) is the trace of the extrinsic curvature of the boundary, and \( \varepsilon = +1 \) for timelike and \( -1 \) for spacelike boundaries. The induced metric on the boundary has determinant \( h \).

To obtain the field equations, we vary the action under Dirichlet boundary conditions:
\beq
    \delta g_{\alpha \beta} |_{\partial \mathcal{V}} = 0
    \quad ,
\eeq
which fixes the induced metric on \( \partial \mathcal{V} \).

Using standard identities:
\beq
    \delta g_{\alpha \beta} = -g_{\alpha \mu} g_{\beta \nu} \delta g^{\mu \nu}
    \quad , \qquad
    \delta \sqrt{-g} = -\frac{1}{2} \sqrt{-g} g_{\alpha \beta} \delta g^{\alpha \beta}
    \quad ,
\eeq
the variation of the Hilbert action becomes:
\bea
    \frac{2}{M_{pl}^2} \delta S_H &=&
    \int_{\mathcal{V}} d^4x \sqrt{-g} \left( R_{\alpha \beta} - \frac{1}{2} R g_{\alpha \beta} \right) \delta g^{\alpha \beta}
    + \int_{\mathcal{V}} d^4x \sqrt{-g} g^{\alpha \beta} \delta R_{\alpha \beta}
    \quad .
\eea
The second term is reduced using:
\beq
    \delta R_{\alpha \beta} = (\delta \Gamma^{\mu}_{\alpha \beta})_{;\mu} - (\delta \Gamma^{\mu}_{\mu \alpha})_{;\beta}
    \quad , \qquad
    g^{\alpha \beta} \delta R_{\alpha \beta} = \bar{\delta} v^\mu{}_{;\mu}
    \quad , \qquad
    \bar{\delta} v^\mu = g^{\alpha \beta} \delta \Gamma^\mu_{\alpha \beta} - g^{\alpha \mu} \delta \Gamma^\beta_{\alpha \beta}
    \quad .
\eeq

Applying Gauss's theorem and evaluating on the boundary:
\beq
    \int_{\mathcal{V}} d^4x \sqrt{-g} \bar{\delta} v^\mu{}_{;\mu}
    = \oint_{\partial \mathcal{V}} d^3 y \, \varepsilon |h|^{1/2} \bar{\delta} v^\mu n_\mu
    \quad .
\eeq

Given that \( \delta g^{\alpha \beta} = 0 \) on \( \partial \mathcal{V} \), we find:
\beq
    \bar{\delta} v^\mu = g^{\alpha \beta} ( \delta g_{\mu \beta, \alpha} - \delta g_{\alpha \beta, \mu} )
    \quad ,
\eeq
and tangential derivatives of \( \delta g_{\alpha \beta} \) vanish on \( \partial \mathcal{V} \), leading to:
\beq
    n^\mu \bar{\delta} v_\mu = -n^\mu h^{\alpha \beta} \delta g_{\alpha \beta, \mu}
    \quad .
\eeq
So the full variation of the Hilbert term becomes:
\beq
    \frac{2}{M_{pl}^2} \delta S_H
    = \int_{\mathcal{V}} d^4x \sqrt{-g} G_{\alpha \beta} \delta g^{\alpha \beta}
    - \oint_{\partial \mathcal{V}} d^3y \, \varepsilon |h|^{1/2} n^\mu h^{\alpha \beta} \delta g_{\alpha \beta, \mu}
    \quad .
    \label{eq:var_SH}
\eeq
Now, varying the boundary term:
\beq
    E = h^{\alpha \beta} n_{\alpha ; \beta}
    \quad , \qquad
    \delta E = \frac{1}{2} h^{\alpha \beta} \delta g_{\alpha \beta, \mu} n^\mu
    \quad ,
\eeq
\beq
    \frac{2}{M_{pl}^2} \delta S_B
    = \oint_{\partial \mathcal{V}} d^3y \, \varepsilon |h|^{1/2} n^\mu h^{\alpha \beta} \delta g_{\alpha \beta, \mu}
    \quad .
    \label{eq:var_SB}
\eeq
Adding \eqref{eq:var_SH} and \eqref{eq:var_SB}, the boundary terms cancel, and since \( \delta S_0 = 0 \), we obtain:
\beq
    \delta S_G = \frac{M_{pl}^2}{2} \int_{\mathcal{V}} d^4x \sqrt{-g} G_{\alpha \beta} \delta g^{\alpha \beta}
    \quad .
\eeq
This explains the presence of the boundary term. Varying the matter action yields:
\beq
    \delta S_M = -\frac{1}{2} \int_{\mathcal{V}} d^4x \sqrt{-g} T_{\alpha \beta} \delta g^{\alpha \beta}
    \quad ,
\eeq
leading to the Einstein field equations:
\beq
    \delta S = 0 \quad \Rightarrow \quad G_{\alpha \beta} = \frac{1}{M_{pl}^2} T_{\alpha \beta}
    \quad .
\eeq
Finally, the subtraction term \( S_0 \) is defined by:
\beq
    E_0 = \text{extrinsic curvature of } \partial \mathcal{V} \text{ embedded in flat spacetime}
    \quad ,
\eeq
ensuring that \( S_G = 0 \) in flat space and yielding a finite action in asymptotically flat spacetimes.

\section{3+1 Decomposition}
To express the action in Hamiltonian form, we foliate spacetime with spacelike hypersurfaces $\Sigma_t$ labeled by a time function \( t(x^\alpha) \). This defines a $3+1$ decomposition, where each hypersurface is described by coordinates \( y^a \), constant along a family of non-orthogonal curves with tangent vector \( t^\alpha \), satisfying $t^\alpha \partial_\alpha t = 1$.

Spatial coordinates remain constant along these curves:
$y^a(P) = y^a(P') = y^a(P'')$. This leads to the adapted coordinate system \( (t, y^a) \), related to the original coordinates via $x^\alpha = x^\alpha(t, y^a)$.

The tangent and spatial basis vectors are given by
\beq
    t^{\alpha}=\bigg(\frac{\partial x^{\alpha}}{\partial t}\bigg)_{y^a} \qquad e^{\alpha}_a=\bigg(\frac{\partial x^{\alpha}}{\partial y^a}\bigg)_t
    \quad .
\eeq
In this coordinate system, \( t^{\alpha}=\delta^{\alpha}_t \) and \( e^{\alpha}_a=\delta^{\alpha}_a \), and the basis vectors satisfy $\mathcal{L}_te^{\alpha}_a=0$.

\begin{figure}
    \centering
    \includegraphics[width=0.35\linewidth]{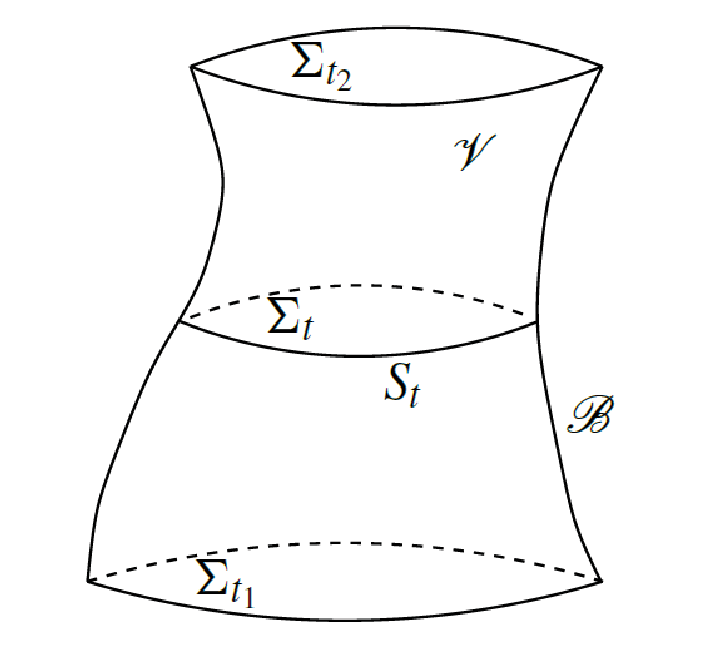}
    \caption{Foliation of the region $\partial \mathcal{V}$. This picture is taken from \cite{poisson2004relativist}.}
    \label{fig:foliation}
\end{figure}

The unit normal to the hypersurfaces is introduced as
\beq
    n_{\alpha}=-N\partial_{\alpha}t \qquad n_{\alpha}e^{\alpha}_a=0
    \quad ,
\eeq
with \( N \) being the lapse function. The non-orthogonal relation between $t^\alpha$ and $n^\alpha$ leads to the decomposition
\beq
    t^{\alpha}=Nn^{\alpha}+N^a e^{\alpha}_a
    \quad ,
\eeq
where \( N^a \) is the shift vector. Using this, the differential $dx^\alpha$ becomes
\beq
    dx^{\alpha}=t^{\alpha}dt+e^{\alpha}_ady^a=(N dt)n^{\alpha}+(dy^a+N^a dt)e^{\alpha}_a
    \quad .
\eeq
Substituting into the line element yields the metric in $3+1$ form:
\beq
    ds^2 = - N^2 dt^2 +h_{ab}(dy^a+N^a dt)(dy^b+N^b dt)
    \quad ,
    \label{eq:metric}
\eeq
with \( h_{ab}=g_{\alpha \beta}e^{\alpha}_a e^{\beta}_b \) being the induced metric on $\Sigma_t$. Finally, the determinant of the metric is expressed as
\beq
    \sqrt{-g}=N\sqrt{h}
    \quad ,
    \label{eq:det}
\eeq
which will be useful in rewriting the action in Hamiltonian form.
\paragraph{Foliation of the boundary}
\leavevmode\\
The discussion in this section refers to Fig.~\ref{fig:foliation}, which illustrates the foliation of the timelike boundary \( \mathcal{B} = \partial \mathcal{V} \) into closed two-surfaces \( \mathcal{S}_t \), each forming the boundary of a spatial hypersurface \( \Sigma_t \). These surfaces can be defined either implicitly, \( f(y^a) = 0 \), or parametrically via \( y^a(\theta^A) \), where \( \theta^A \) are coordinates on \( \mathcal{S}_t \).

Let \( r^a \) be the unit normal to \( \mathcal{S}_t \) within \( \Sigma_t \), and its corresponding spacetime vector is $r^\alpha = r^a e^\alpha_a$. This satisfies \( r^\alpha r_\alpha = 1 \) and \( r^\alpha n_\alpha = 0 \), where \( n^\alpha \) is the normal to \( \Sigma_t \). The tangent vectors to \( \mathcal{S}_t \) are defined by \( e^a_A = \partial y^a / \partial \theta^A \), and satisfy \( r_a e^a_A = 0 \). These induce the spacetime vectors
\beq
    e^\alpha_A \equiv e^\alpha_a e^a_A = \frac{\partial x^\alpha}{\partial \theta^A} \,
    \quad ,
\eeq
and the induced metric on \( \mathcal{S}_t \) becomes
\beq
    ds^2 = \sigma_{AB} \, d\theta^A d\theta^B \,
    \quad ,
    \quad
    \sigma_{AB} = h_{ab} \, e^a_A e^b_B = g_{\alpha\beta} \, e^\alpha_a e^\beta_b e^a_A e^b_B \,
    \quad ,
    \quad
    \sigma_{AB} = g_{\alpha\beta} \, e^\alpha_A e^\beta_B \,
    \quad .
\eeq
Let \( \sigma^{AB} \) be the inverse of \( \sigma_{AB} \). The 3-dimensional completeness relation reads
\beq
    h_{ab} = r_a r_b + \sigma_{AB} \, e^a_A e^b_B \,
    \quad ,
\eeq
and the full spacetime decomposition becomes
\beq
    g_{\alpha\beta} = -n_\alpha n_\beta + r_\alpha r_\beta + \sigma_{AB} \, e^\alpha_A e^\beta_B \,
    \quad .
\eeq
The extrinsic curvature of \( \mathcal{S}_t \subset \Sigma_t \) is defined by
\beq
    k_{AB} = r_{\alpha;\beta} \, e^\alpha_A e^\beta_B \,
    \quad ,
    \quad
    k = \sigma^{AB} k_{AB} \,
    \quad .
\eeq
To relate the coordinates on different \( \mathcal{S}_t \), we introduce a congruence of curves \( \beta \) on \( \mathcal{B} \), orthogonal to each \( \mathcal{S}_t \) and with tangent vector aligned with \( n^\alpha \). If \( \theta^A \) labels a point on \( \mathcal{S}_t \), the same coordinates label its image on \( \mathcal{S}_{t''} \). Since \( \theta^A \) is constant along each curve, we have:
\beq
    n^\alpha = \frac{1}{N} \left( \frac{\partial x^\alpha}{\partial t} \right)_{\theta^A}
    \quad ,
\eeq
ensuring proper normalization and orthogonality with \( e^\alpha_A \) on \( \mathcal{B} \). The hypersurface \( \mathcal{B} \) is described by coordinates \( z^i \), with tangent vectors \( e^\alpha_i = \partial x^\alpha / \partial z^i \), and induced metric
\beq
    \gamma_{ij} = g_{\alpha\beta} \, e^\alpha_i e^\beta_j
    \quad ,
\eeq
satisfying the decomposition
\beq
    g_{\alpha\beta} = r_\alpha r_\beta + \gamma_{ij} \, e^\alpha_i e^\beta_j
    \quad .
\eeq
Using the convenient choice \( z^i = (t, \theta^A) \), a displacement on \( \mathcal{B} \) becomes
\beq
    dx^\alpha = \left( \frac{\partial x^\alpha}{\partial t} \right)_{\theta^A} dt + \left( \frac{\partial x^\alpha}{\partial \theta^A} \right)_t d\theta^A = N n^\alpha dt + e^\alpha_A d\theta^A
    \quad .
\eeq
The line element on \( \mathcal{B} \) then reads:
\beq
    ds^2_\mathcal{B} = g_{\alpha\beta} dx^\alpha dx^\beta = -N^2 dt^2 + \sigma_{AB} d\theta^A d\theta^B
    \quad .
\eeq
Thus, the metric decomposition is
\beq
    \gamma_{ij} \, dz^i dz^j = -N^2 dt^2 + \sigma_{AB} \, d\theta^A d\theta^B
    \quad ,
\eeq
and the determinant relation is $\sqrt{-\gamma} = N \sqrt{\sigma}$,
where \( \gamma \) and \( \sigma \) are the determinants of \( \gamma_{ij} \) and \( \sigma_{AB} \). Finally, the extrinsic curvature of \( \mathcal{B} \) is defined by
\beq
    K_{ij} = r_{\alpha;\beta} \, e^\alpha_i e^\beta_j
    \quad , \quad
    K = \gamma^{ij} K_{ij}
    \quad .
\eeq
\paragraph{Decomposition of the Gravitational Action}
\leavevmode\\
We start from the Einstein–Hilbert action with the Gibbons–Hawking–York boundary term~\eqref{eq:complete gravitational action}, temporarily omitting the background subtraction term. The spacetime region \( \mathcal{V} \) is bounded by two spacelike hypersurfaces \( \Sigma_{t_1}, \Sigma_{t_2} \) and a timelike boundary \( \mathcal{B} \), such that
\beq
    \partial \mathcal{V} = \Sigma_{t_2} \cup (-\Sigma_{t_1}) \cup \mathcal{B}
    \quad .
\eeq
The minus sign indicates that the normal to \( \Sigma_{t_1} \) points inward, opposite to the required outward orientation. Geometric quantities like \( n^\alpha \), \( h_{ab} \), and \( E_{ab} \) are interpreted with respect to the spacelike slices \( \Sigma_t \). The gravitational action becomes:
\beq
    \frac{2}{M_{pl}^2} S_G = \int_{\mathcal{V}} R \sqrt{-g} \, d^4x 
    - 2 \int_{\Sigma_{t_2}} E \sqrt{h} \, d^3y 
    + 2 \int_{\Sigma_{t_1}} E \sqrt{h} \, d^3y 
    + 2 \int_{\mathcal{B}} K \sqrt{-\gamma} \, d^3z
    \quad .
\eeq
The bulk term is the integral of the Ricci scalar \( R \) over the spacetime region \( \mathcal{V} \), while the second and third terms are surface integrals over the spacelike hypersurfaces \( \Sigma_{t_1} \) and \( \Sigma_{t_2} \), involving the trace of the extrinsic curvature \( E \) and the determinant of the induced spatial metric \( h \). The minus sign in the \( \Sigma_{t_1} \) term accounts for orientation. The last term corresponds to the contribution from the timelike boundary \( \mathcal{B} \), with extrinsic curvature \( K \) and metric determinant \( \gamma \).

Using the foliation of \( \mathcal{V} \), the Ricci scalar from~\eqref{eq:Ricci} becomes:
\beq
    R = \hat R - \left( E^2 - E_{ab} E^{ab} \right) - 2 \left( n^\alpha{}_{;\beta} n^\beta - n^\alpha n^\beta{}_{;\beta} \right)_{;\alpha}
    \quad ,
\eeq
and the volume element from~\eqref{eq:det} is:
\beq
    \sqrt{-g} \, d^4x = N \sqrt{h} \, dt \, d^3y
    \quad ,
\eeq
so the bulk term becomes:
\bea
    &&\int_{\mathcal{V}} R \sqrt{-g} \, d^4x = \int_{t_1}^{t_2} dt \int_{\Sigma_t} \left( \hat R + E_{ab} E^{ab} - E^2 \right) N \sqrt{h} \, d^3y +\nonumber \\ &&- 2 \oint_{\partial \mathcal{V}} \left( n^\alpha{}_{;\beta} n^\beta - n^\alpha n^\beta{}_{;\beta} \right) d\Sigma_\alpha
    \quad .
\eea
The new boundary term can be broken down into integrals over \(\Sigma_{t_1}\), \(\Sigma_{t_2}\), and \( \mathcal{B} \). On \(\Sigma_{t_1}\), \(d\Sigma_\alpha = n_\alpha \sqrt{h} \, d^3y\) — this also incorporates an extra minus sign — and
\beq
    -2 \int_{\Sigma_{t_1}} \left( n^\alpha{}_{;\beta} n^\beta - n^\alpha n^\beta{}_{;\beta} \right) d\Sigma_\alpha = -2 \int_{\Sigma_{t_1}} n^\beta{}_{;\beta} \sqrt{h} \, d^3y = -2 \int_{\Sigma_{t_1}} E \sqrt{h} \, d^3y
    \quad .
\eeq
where we have used orthogonality.
We see that this term cancels out the other integral over \(\Sigma_{t_1}\) coming from the original boundary term in the gravitational action. The integrals over \(\Sigma_{t_2}\) cancel out also. There remains a contribution from \(\mathcal{B}\), on which \(d\Sigma_\alpha = r_\alpha \sqrt{-\gamma} \, d^3z\), giving
\beq
    -2 \int_{\mathcal{B}} \left( n^\alpha{}_{;\beta} n^\beta - n^\alpha n^\beta{}_{;\beta} \right) d\Sigma_\alpha = -2 \int_{\mathcal{B}} n^\alpha{}_{;\beta} n^\beta r_\alpha \sqrt{-\gamma} \, d^3z = 2 \int_{\mathcal{B}} r^\alpha{}_{;\beta} n_\alpha n^\beta \sqrt{-\gamma} \, d^3z
    \quad ,
\eeq
where we have used \(n^\alpha r_\alpha = 0\).
Collecting the results, we have
\beq
    \frac{2}{M_{pl}^2}S_G = \int_{t_1}^{t_2} dt \int_{\Sigma_t} \left( \hat R + E_{ab} E^{ab} - E^2 \right) N \sqrt{h} \, d^3y + 2 \int_{\mathcal{B}} \left( K + r^\alpha{}_{;\beta} n_\alpha n^\beta \right) \sqrt{-\gamma} \, d^3z
    \quad .
\eeq
We now use the fact that \({\mathcal{B}}\) is foliated by the closed two-surfaces \({\mathcal{S}}_t\). We substitute
\beq
    -\gamma \, d^3z = N \sqrt{\sigma} \, dt \, d^2\theta
    \quad ,
\eeq
and express \(K\) as
\beq
    K = \gamma^{ij} K_{ij} = \gamma^{ij} \left( r_\alpha{}_{;\beta} e^\alpha_i e^\beta_j \right) = r_\alpha{}_{;\beta} \left( \gamma^{ij} e^\alpha_i e^\beta_j \right) = r_\alpha{}_{;\beta} (g^{\alpha\beta} - r^\alpha r^\beta)
    \quad ,
\eeq
so that the integrand becomes
\bea
    K + r_\alpha{}_{;\beta} n^\alpha n^\beta &=& r_\alpha{}_{;\beta} \left( g^{\alpha\beta} - r^\alpha r^\beta + n^\alpha n^\beta \right) = r_\alpha{}_{;\beta} \left( \sigma^{AB} e^\alpha_A e^\beta_B \right)= \nonumber \\ &=& \sigma^{AB} \left( r_\alpha{}_{;\beta} e^\alpha_A e^\beta_B \right) = \sigma^{AB} k_{AB} = k
    \quad .
\eea
Substituting this into our previous expression for the gravitational action, we arrive at
\beq
    S_G = \frac{M_{pl}^2}{2} \int_{t_1}^{t_2} dt \Bigg[ \int_{\Sigma_t} \left( \hat R + E_{ab} E^{ab} - E^2 \right) N \sqrt{h} \, d^3y + 2 \oint_{{\mathcal{S}}_t} (k - k_0) N \sqrt{\sigma} \, d^2\theta\Bigg]
    \quad .
\eeq
We have re-instated the subtraction term by inserting \(k_0\) into the integral over \({\mathcal{S}}_t\).

Apart from the boundary term, which was not included for simplicity (does not affect any of previous results), this is exactly the form of the gravitational part of the action~\eqref{eq:ADM action}. We can make this more explicit using their notation $E_{ab} \rightarrow N^{-1}E_{ij}$, $M_{pl}=1$, $d^4x = dt d^3y$ and using the A.D.M. form of the metric to write the matter action:
\bea
    g^{\mu \nu}\partial_{\mu}\phi \partial_{\nu} \phi &=& -\frac{1}{N^2}\dot \phi ^2 +2\frac{N^i}{N^2}\partial_{i}\phi \dot \phi + \frac{1}{N^2}[N^2h^{ij}-N^iN^j]\partial_i\phi \partial_j \phi =\nonumber \\ &=& -\frac{1}{N^2}(\dot \phi - N^i \partial_i \phi )^2 + h^{ij} \partial_i \phi \partial_j \phi 
    \quad ,
\eea
\beq
    \sqrt{-g}\Bigg(-\frac{1}{2}g^{\mu \nu}\partial_{\mu}\phi \partial_{\nu} \phi-V(\phi)\Bigg) =\frac{1}{2}\sqrt{h}\Bigg(\frac{1}{N}(\dot{\phi}-N^i\partial_i \phi)^2-Nh^{ij}\partial_i\phi\partial_j\phi-2NV(\phi)\Bigg)
    \quad .
\eeq
This makes manifest the final shape of the action \eqref{eq:ADM action}.

To conclude we also derive the formula for the extrinsic curvature in terms of the spatial metric and shift vectors. Following the hamiltonian construction of the General Relativity
\beq
    \dot h_{ab}\equiv \mathcal{L}_t h_{ab}=\mathcal{L}_t(g_{\alpha \beta}e^{\alpha}_ae^{\beta}_b)= (\mathcal{L}_tg_{\alpha \beta})e^{\alpha}_ae^{\beta}_b 
    \quad ,
\eeq
where we have used that $\mathcal{L}_te^{\alpha}_a=0$. Now we can develop the Lie derivative using A.D.M. language:
\bea
    \mathcal{L}_t g_{\alpha\beta} &=& t_{\alpha;\beta} + t_{\beta;\alpha} \nonumber \\
    &=& (N n_\alpha + N_\alpha)_{;\beta} + (N n_\beta + N_\beta)_{;\alpha} \nonumber \\
    &=& n_\alpha N_{,\beta} + N_{,\alpha} n_\beta + N(n_{\alpha;\beta} + n_{\beta;\alpha}) + N_{\alpha;\beta} + N_{\beta;\alpha},
\eea

where \( N_\alpha = N_a e^a_\alpha \). Finally, projecting along \( e^a_\alpha e^b_\beta \) gives:

\beq
    \dot{h}_{ab} = 2N E_{ab} + N_{a|b} + N_{b|a}
    \quad ,
\eeq
where we have used the definition of $E_{ab}$ and the intrinsic differentiation defined above. We derive
\beq
    E_{ab} = \frac{1}{2N} \left( \dot{h}_{ab} - N_{a|b} - N_{b|a} \right)
    \quad .
\eeq
Coming back to the same notation used in the cosmology chapter (remember that $E_{ab} \rightarrow N^{-1}E_{ij}$) we obtain
\beq
     E_{ij}=\frac{1}{2}\big[\dot{h}_{ij}-\hat{\nabla}_iN_j-\hat{\nabla}_jN_i\big]
     \quad ,
     \label{eq:E formula}
\eeq
that is the desidered result.
\end{appendices}
\renewcommand{\bibname}{Bibliography}
\bibliographystyle{unsrturl}
\bibliography{bibliography}

\begin{thebibliography}{100}

\bibitem{Goroff:1985th}
M.H. Goroff and A.~Sagnotti.
\newblock Quantum gravity at two loops.
\newblock {\em Phys. Lett. B}, 160:81--86, 1985.

\bibitem{GoroffSagnotti1986}
M.H. Goroff and A.~Sagnotti.
\newblock The ultraviolet behavior of einstein gravity.
\newblock {\em Nuclear Physics B}, 266(3‑4):709‑736, 1986.

\bibitem{tHooftVeltman1974}
G.'t Hooft and M.J.G. Veltman.
\newblock One‑loop divergencies in the theory of gravitation.
\newblock {\em Annales de l’Institut Henri Poincaré. Section A, Physique
  Théorique}, 20(1):69‑94, 1974.

\bibitem{Veneziano1968}
G.~Veneziano.
\newblock Construction of a crossing-symmetric, regge-behaved amplitude for
  linearly rising trajectories.
\newblock {\em Il Nuovo Cimento A}, 57(1):190–197, 1968.

\bibitem{ShapiroVirasoro1969}
J.A. Shapiro and M.A. Virasoro.
\newblock Soft singularities of the dual resonance model.
\newblock {\em Physical Review D}, 1(10):2933–2936, 1970.

\bibitem{Ramond1971}
P.~Ramond.
\newblock Dual theory for free fermions.
\newblock {\em Physical Review D}, 3(10):2415--2418, 1971.

\bibitem{NeveuSchwarz1971}
A.~Neveu and J.H. Schwarz.
\newblock Factorizable dual model of pions.
\newblock {\em Nuclear Physics B}, 31:86--112, 1971.

\bibitem{ScherkSchwarz1974}
J.~Scherk and J.H. Schwarz.
\newblock Dual models for non-hadrons.
\newblock {\em Nuclear Physics B}, 81(1):118--144, 1974.

\bibitem{Yoneya1974}
T.~Yoneya.
\newblock Connection of dual models to electrodynamics and gravidynamics.
\newblock {\em Progress of Theoretical Physics}, 51(2):190--195, 1974.

\bibitem{Gliozzi:1977}
F.~Gliozzi, J.~Scherk, and D.I. Olive.
\newblock Supersymmetry, supergravity theories and the dual spinor model.
\newblock {\em Nucl. Phys. B}, 122:253, 1977.

\bibitem{GreenSchwarzWitten1987}
M.B. Green, J.H. Schwarz, and E.~Witten.
\newblock {\em Superstring Theory}, volume~1.
\newblock Cambridge University Press, Cambridge, 1987.
\newblock Volume 1: Introduction.

\bibitem{Gross1985}
D.J. Gross, J.A. Harvey, E.J. Martinec, and R.~Rohm.
\newblock Heterotic string theory. 1. the free heterotic string.
\newblock {\em Nucl. Phys. B}, 256:253--284, 1985.

\bibitem{Gross1986}
D.J. Gross, J.A. Harvey, E.J. Martinec, and R.Rohm.
\newblock Heterotic string theory. 2. the interacting heterotic string.
\newblock {\em Nucl. Phys. B}, 267:75--124, 1986.

\bibitem{GreenSchwarz1984}
M.B. Green and J.H. Schwarz.
\newblock Anomaly cancellation in supersymmetric d=10 gauge theory and
  superstring theory.
\newblock {\em Physics Letters B}, 149(1-3):117--122, 1984.

\bibitem{Candelas1985}
P.~Candelas, G.T. Horowitz, A.~Strominger, and E.~Witten.
\newblock Vacuum configurations for superstrings.
\newblock {\em Nuclear Physics B}, 258(1):46--74, 1985.

\bibitem{DixonHarveyVafaWitten1985}
L.J. Dixon, J.A. Harvey, C.~Vafa, and E.~Witten.
\newblock Strings on orbifolds.
\newblock {\em Nuclear Physics B}, 261(4):678--686, 1985.

\bibitem{Polchinski1995}
J.~Polchinski.
\newblock Dirichlet-branes and ramond-ramond charges.
\newblock {\em Phys. Rev. Lett.}, 75(26):4724--4727, 1995.

\bibitem{Witten1995}
E.~Witten.
\newblock String theory dynamics in various dimensions.
\newblock {\em Nuclear Physics B}, 443(1–2):85--126, 1995.
\newblock \href {https://arxiv.org/abs/hep-th/9503124}
  {\path{arXiv:hep-th/9503124}}.

\bibitem{HullTownsend1995}
C.M. Hull and P.K. Townsend.
\newblock Unity of superstring dualities.
\newblock {\em Nucl. Phys. B}, 438(1-2):109--137, 1995.
\newblock \href {https://arxiv.org/abs/hep-th/9410167}
  {\path{arXiv:hep-th/9410167}}.

\bibitem{BoussoPolchinski2000}
R.~Bousso and J.~Polchinski.
\newblock Quantization of four-form fluxes and dynamical neutralization of the
  cosmological constant.
\newblock {\em JHEP}, 06:006, 2000.
\newblock \href {https://arxiv.org/abs/hep-th/0004134}
  {\path{arXiv:hep-th/0004134}}.

\bibitem{DouglasKachru2007}
M.R. Douglas and S.~Kachru.
\newblock Flux compactification.
\newblock {\em Rev. Mod. Phys.}, 79(2):733--796, 2007.
\newblock \href {https://arxiv.org/abs/hep-th/0610102}
  {\path{arXiv:hep-th/0610102}}.

\bibitem{Linde1990}
A.D. Linde.
\newblock {\em Particle Physics and Inflationary Cosmology}.
\newblock Harwood Academic Publishers, Chur, Switzerland, 1990.

\bibitem{Guth2007}
A.H. Guth.
\newblock Eternal inflation and its implications.
\newblock {\em J. Phys. A: Math. Theor.}, 40(25):6811--6826, 2007.
\newblock \href {https://arxiv.org/abs/hep-th/0702178}
  {\path{arXiv:hep-th/0702178}}.

\bibitem{Planck2018}
N.A. Aghanim, Y.~Akrami, M.~Ashdown, J.~Aumont, C.~Baccigalupi, M.~Ballardini,
  A.J. Banday, R.~Barreiro, N.~Bartolo, S.~Basak, and et~al.
\newblock {Planck} 2018 results. vi. cosmological parameters.
\newblock {\em Astron. Astrophys.}, 641:A6, 2020.
\newblock \href {https://arxiv.org/abs/1807.06209} {\path{arXiv:1807.06209}}.

\bibitem{Contaldi2003}
C.R. Contaldi, M.~Peloso, L.~Kofman, and A.~Linde.
\newblock Suppressing the lower multipoles in the cosmic microwave background
  anisotropies.
\newblock {\em JCAP}, 07:002, 2003.
\newblock \href {https://arxiv.org/abs/astro-ph/0303636}
  {\path{arXiv:astro-ph/0303636}}.

\bibitem{Sugimoto1999}
S.~Sugimoto.
\newblock Anomaly cancellations in type i d9--d9-bar system and the usp(32)
  string theory.
\newblock {\em Progress of Theoretical Physics}, 102:685--699, 1999.
\newblock \href {https://arxiv.org/abs/hep-th/9905159}
  {\path{arXiv:hep-th/9905159}}.

\bibitem{Antoniadis1999}
I.~Antoniadis, E.~Dudas, and A.~Sagnotti.
\newblock Brane supersymmetry breaking.
\newblock {\em Physics Letters B}, 464:38--45, 1999.
\newblock \href {https://arxiv.org/abs/hep-th/9908023}
  {\path{arXiv:hep-th/9908023}}.

\bibitem{Angelantonj2000}
C.~Angelantonj.
\newblock Comments on open-string orbifolds with a nonvanishing b(ab).
\newblock {\em Nuclear Physics B}, 566:126--150, 2000.
\newblock \href {https://arxiv.org/abs/hep-th/9908064}
  {\path{arXiv:hep-th/9908064}}.

\bibitem{Dudas:2000ff}
E.~Dudas and J.~Mourad.
\newblock Brane solutions in strings with broken supersymmetry and dilaton
  tadpoles.
\newblock {\em Phys. Lett. B}, 486:172--178, 2000.
\newblock \href {https://arxiv.org/abs/hep-th/0004165}
  {\path{arXiv:hep-th/0004165}}.

\bibitem{DudasKitazawaPatilSagnotti:2012}
E.~Dudas, N.~Kitazawa, S.P. Patil, and A.~Sagnotti.
\newblock Cmb imprints of a pre‑inflationary climbing phase.
\newblock {\em Journal of Cosmology and Astroparticle Physics}, 2012(05):012,
  2012.
\newblock \href {https://arxiv.org/abs/1202.6630} {\path{arXiv:1202.6630}}.

\bibitem{KitazawaSagnotti:2014}
N.~Kitazawa and A.~Sagnotti.
\newblock Pre‑inflationary clues from string theory?
\newblock {\em Journal of Cosmology and Astroparticle Physics}, 2014(04):017,
  2014.
\newblock \href {https://arxiv.org/abs/1402.1418} {\path{arXiv:1402.1418}}.

\bibitem{Dudas:2010climbing}
E.~Dudas, N.~Kitazawa, and A.~Sagnotti.
\newblock On climbing scalars in string theory.
\newblock {\em Phys. Lett. B}, 694:80--83, 2011.
\newblock \href {https://arxiv.org/abs/1009.0874} {\path{arXiv:1009.0874}}.

\bibitem{Martin2014}
J.~Martin, C.~Ringeval, and V.~Vennin.
\newblock Encyclopædia inflationaris.
\newblock {\em Physics of the Dark Universe}, 5-6:75--235, 2014.
\newblock \href {https://arxiv.org/abs/1303.3787} {\path{arXiv:1303.3787}}.

\bibitem{GruppusoKitazawaMandolesiNatoliSagnotti2016}
A.~Gruppuso, N.~Kitazawa, N.~Mandolesi, P.~Natoli, and A.~Sagnotti.
\newblock Pre‑inflationary relics in the cmb?
\newblock {\em Physics of the Dark Universe}, 11:68, 2016.
\newblock \href {https://arxiv.org/abs/1508.00411} {\path{arXiv:1508.00411}}.

\bibitem{maldacena2003non}
J.M. Maldacena.
\newblock Non-gaussian features of primordial fluctuations in single field
  inflationary models.
\newblock {\em Journal of High Energy Physics}, 2003(05):013, 2003.

\bibitem{collins2011primordial}
H.~Collins and R.~Holman.
\newblock Primordial non-gaussianities from inflation.
\newblock {\em International Journal of Modern Physics D}, 20(05):1073--1083,
  2011.

\bibitem{meo2025preinflationarynongaussianities}
M.~Meo and A.~Sagnotti.
\newblock On pre-inflationary non-gaussianities.
\newblock 2025.
\newblock arXiv:2510.01360 [hep-th], to appear in JHEP.
\newblock \href {https://arxiv.org/abs/2510.01360} {\path{arXiv:2510.01360}}.

\bibitem{Freedman:1976xh}
D.Z. Freedman, P.~van Nieuwenhuizen, and S.~Ferrara.
\newblock Progress toward a theory of supergravity.
\newblock {\em Phys. Rev. D}, 13(12):3214--3218, 1976.

\bibitem{Deser:1976eh}
S.~Deser and B.~Zumino.
\newblock Consistent supergravity.
\newblock {\em Phys. Lett. B}, 62:335--337, 1976.

\bibitem{Freedman:2012zz}
D.Z. Freedman and A.~Van Proeyen.
\newblock {\em Supergravity}.
\newblock Cambridge University Press, Cambridge, 2012.

\bibitem{Sagnotti:1987Cargese}
A.~Sagnotti.
\newblock Open strings and their symmetry groups.
\newblock In G.~'t~Hooft, A.~Jaffe, G.~Mack, P.K. Mitter, and R.~Stora,
  editors, {\em Non-Perturbative Quantum Field Theory}, pages 521--528, New
  York, 1988. Plenum Press.
\newblock Based on lectures given at the Carg\`ese Summer Institute, Carg\`ese,
  France, July 16--30, 1987. Also available as ROM2F-87-025, and reprinted in
  \texttt{arXiv:hep-th/0208020}.
\newblock \href {https://arxiv.org/abs/hep-th/0208020}
  {\path{arXiv:hep-th/0208020}}.

\bibitem{Pradisi:1989zz}
G.~Pradisi and A.~Sagnotti.
\newblock Open string orbifolds.
\newblock {\em Phys. Lett. B}, 216:59--67, 1989.

\bibitem{Horava:1989fv}
P.~Hořava.
\newblock Strings on world sheet orbifolds.
\newblock {\em Nucl. Phys. B}, 327:461--484, 1989.

\bibitem{Horava:1989bg}
P.~Hořava.
\newblock Background duality of open string models.
\newblock {\em Phys. Lett. B}, 231:251--257, 1989.

\bibitem{BianchiSagnotti:1990}
M.~Bianchi and A.~Sagnotti.
\newblock On the systematics of open string theories.
\newblock {\em Phys. Lett. B}, 247:517--524, 1990.

\bibitem{BianchiSagnotti:1991a}
M.~Bianchi and A.~Sagnotti.
\newblock Twist symmetry and open string wilson lines.
\newblock {\em Nucl. Phys. B}, 361:519--538, 1991.

\bibitem{Bianchi1992}
M.~Bianchi, G.~Pradisi, and A.~Sagnotti.
\newblock Toroidal compactification and symmetry breaking in open string
  theories.
\newblock {\em Nuclear Physics B}, 376:365--386, 1992.

\bibitem{Sagnotti:1992note}
A.~Sagnotti.
\newblock A note on the green‑schwarz mechanism in open string theories.
\newblock {\em Phys. Lett. B}, 294:196--203, 1992.
\newblock \href {https://arxiv.org/abs/hep-th/9210127}
  {\path{arXiv:hep-th/9210127}}.

\bibitem{Dudas:2000review}
E.~Dudas.
\newblock Theory and phenomenology of type i strings and m theory.
\newblock {\em Class. Quant. Grav.}, 17:R41--R75, 2000.
\newblock \href {https://arxiv.org/abs/hep-ph/0006190}
  {\path{arXiv:hep-ph/0006190}}.

\bibitem{Angelantonj:2002ct}
C.~Angelantonj and A.~Sagnotti.
\newblock Open strings.
\newblock {\em Phys. Rept.}, 371:1--150, 2002.
\newblock \href {https://arxiv.org/abs/hep-th/0204089}
  {\path{arXiv:hep-th/0204089}}.

\bibitem{AngelantonjFlorakis:2024}
C.~Angelantonj and I.~Florakis.
\newblock A lightning introduction to string theory, 2024.
\newblock arXiv:2406.09508 [hep-th].

\bibitem{Aldazabal1999}
G.~Aldazabal and A.M. Uranga.
\newblock Tachyon-free non-supersymmetric type iib orientifolds via
  brane-antibrane systems.
\newblock {\em Journal of High Energy Physics}, 1999(10):024, 1999.
\newblock \href {https://arxiv.org/abs/hep-th/9908072}
  {\path{arXiv:hep-th/9908072}}.

\bibitem{Polchinski:1998rq}
J.~Polchinski.
\newblock {\em String Theory. Vol. 1: An Introduction to the Bosonic String}.
\newblock Cambridge University Press, 1998.

\bibitem{Polchinski:1998rr}
J.~Polchinski.
\newblock {\em String Theory. Vol. 2: Superstring Theory and Beyond}.
\newblock Cambridge University Press, 1998.

\bibitem{Angelantonj:2000ct}
C.~Angelantonj and A.~Armoni.
\newblock Nontachyonic type0b orientifolds, nonsupersymmetric gauge theories
  and cosmological rg flow.
\newblock {\em Nucl.\ Phys.\ B}, 578:239--255, 2000.
\newblock \href {https://arxiv.org/abs/hep-th/9912257}
  {\path{arXiv:hep-th/9912257}}.

\bibitem{Fay1973}
J.D. Fay.
\newblock Theta functions on riemann surfaces.
\newblock In {\em Lecture Notes in Mathematics}, volume 352. Springer, Berlin,
  1973.

\bibitem{Mumford1983}
D.~Mumford.
\newblock {\em Tata Lectures on Theta}.
\newblock Birkhäuser, Basel, 1983.

\bibitem{Erdelyi1953}
A.~Erdélyi, editor.
\newblock {\em Higher Transcendental Functions}, volume 1--3.
\newblock McGraw-Hill, New York, 1953.

\bibitem{WhittakerWatson1927}
E.T. Whittaker and G.N. Watson.
\newblock {\em A Course of Modern Analysis}.
\newblock Cambridge University Press, Cambridge, 1927.

\bibitem{Belavin:1984vu}
A.A. Belavin, A.M. Polyakov, and A.B. Zamolodchikov.
\newblock Infinite conformal symmetry in two-dimensional quantum field theory.
\newblock {\em Nucl. Phys. B}, 241:333--380, 1984.

\bibitem{Friedan:1985ge}
D.~Friedan, E.J. Martinec, and S.H. Shenker.
\newblock Conformal invariance, supersymmetry and string theory.
\newblock {\em Nucl. Phys. B}, 271:93--165, 1986.

\bibitem{Ginsparg:1988ui}
P.~Ginsparg.
\newblock Applied conformal field theory.
\newblock Technical Report HUTP-88-A054, Harvard University, 1988.
\newblock Lectures given at Les Houches Summer School, June-August 1988.

\bibitem{Cardy:1988}
J.L. Cardy.
\newblock Conformal invariance and statistical mechanics.
\newblock In {\em Les Houches Summer School in Theoretical Physics}, 1988.
\newblock Lectures given at Les Houches Summer School, June-August 1988.

\bibitem{ItzyksonDrouffe1989v1}
C.~Itzykson and J.‑M. Drouffe.
\newblock {\em Statistical Field Theory, Volume 1: From Brownian Motion to
  Renormalization and Lattice Gauge Theory}.
\newblock Cambridge University Press, Cambridge, 1989.

\bibitem{ItzyksonDrouffe1989v2}
C.~Itzykson and J.‑M. Drouffe.
\newblock {\em Statistical Field Theory, Volume 2: Strong Coupling, Monte Carlo
  Methods, Conformal Field Theory and Random Systems}.
\newblock Cambridge University Press, Cambridge, 1989.

\bibitem{Zwiebach2004}
B.~Zwiebach.
\newblock {\em A First Course in String Theory}.
\newblock Cambridge University Press, Cambridge, 1st edition, 2004.

\bibitem{Paton:1969je}
J.E. Paton and H.M. Chan.
\newblock Generalized veneziano model with isospin.
\newblock {\em Nucl. Phys. B}, 10:516--520, 1969.

\bibitem{cp2}
J.H. Schwarz.
\newblock Superstring theory.
\newblock {\em Physics Reports}, 89:223, 1982.

\bibitem{cp3}
J.H. Schwarz.
\newblock Gauge groups for type i superstrings, 1982.
\newblock CALT-68-906-REV.
\newblock \href {https://arxiv.org/abs/CALT-68-906-REV}
  {\path{arXiv:CALT-68-906-REV}}.

\bibitem{cp4}
N.~Marcus and A.~Sagnotti.
\newblock Tree level constraints on gauge groups for type i superstrings.
\newblock {\em Physics Letters B}, 119(B):97, 1982.

\bibitem{Polchinski:1988}
J.~Polchinski and Y.~Cai.
\newblock Consistency of open superstring theories.
\newblock {\em Nucl. Phys. B}, 296:91, 1988.

\bibitem{Aldazabal:1999}
G.~Aldazabal, D.~Badagnani, L.E. Ibáñez, and A.M. Uranga.
\newblock Tadpole versus anomaly cancellation in d = 4; 6 compact iib
  orientifolds.
\newblock {\em JHEP}, 9906:031, 1999.
\newblock \href {https://arxiv.org/abs/hep-th/9904071}
  {\path{arXiv:hep-th/9904071}}.

\bibitem{Scrucca:1999}
C.A. Scrucca and M.~Serone.
\newblock Gauge and gravitational anomalies in d = 4 n = 1 orientifolds.
\newblock {\em JHEP}, 9912:024, 1999.
\newblock \href {https://arxiv.org/abs/hep-th/9912108}
  {\path{arXiv:hep-th/9912108}}.

\bibitem{Scrucca:2000}
C.A. Scrucca and M.~Serone.
\newblock Anomaly cancellation in k3 orientifolds.
\newblock {\em Nucl. Phys. B}, 564:555, 2000.
\newblock \href {https://arxiv.org/abs/hep-th/9907112}
  {\path{arXiv:hep-th/9907112}}.

\bibitem{Bianchi:2000}
M.~Bianchi and J.F. Morales.
\newblock Anomalies and tadpoles.
\newblock {\em JHEP}, 0003:030, 2000.
\newblock \href {https://arxiv.org/abs/hep-th/0002149}
  {\path{arXiv:hep-th/0002149}}.

\bibitem{GreenSchwarzWitten1987Vol2}
M.B. Green, J.H. Schwarz, and E.~Witten.
\newblock {\em Superstring Theory}, volume~2.
\newblock Cambridge University Press, Cambridge, 1987.
\newblock Volume 2: Loop Amplitudes, Anomalies and Phenomenology.

\bibitem{Lust:1989}
D.~Lüst and S.~Theisen.
\newblock {\em Lectures on String Theory}, volume 346 of {\em Lecture Notes in
  Physics}.
\newblock Springer, 1989.

\bibitem{Kiritsis:1998}
E.~Kiritsis.
\newblock {\em Introduction to Superstring Theory}.
\newblock Leuven University Press, Leuven, 1998.
\newblock \href{https://arxiv.org/abs/hep-th/9709062}{arXiv:hep-th/9709062}.

\bibitem{Gliozzi:1976}
F.~Gliozzi, J.~Scherk, and D.I. Olive.
\newblock Supergravity and the spinor dual model.
\newblock {\em Phys. Lett. B}, 65:282, 1976.

\bibitem{Kachru1999}
S.~Kachru, J.~Kumar, and E.~Silverstein.
\newblock Vacuum energy cancellation in a non-supersymmetric string.
\newblock {\em Phys. Rev. D}, 59:106004, 1999.
\newblock \href {https://arxiv.org/abs/hep-th/9807076}
  {\path{arXiv:hep-th/9807076}}.

\bibitem{Harvey1999}
J.A. Harvey.
\newblock String duality and non-supersymmetric strings.
\newblock {\em Phys. Rev. D}, 59:026002, 1999.
\newblock \href {https://arxiv.org/abs/hep-th/9807213}
  {\path{arXiv:hep-th/9807213}}.

\bibitem{Angelantonj1999}
C.~Angelantonj, I.~Antoniadis, and K.~Forger.
\newblock Non-supersymmetric type i strings with zero vacuum energy.
\newblock {\em Nucl. Phys. B}, 555:116--134, 1999.
\newblock \href {https://arxiv.org/abs/hep-th/9904092}
  {\path{arXiv:hep-th/9904092}}.

\bibitem{BeckerBeckerSchwarz2006}
K.~Becker, M.~Becker, and J.H. Schwarz.
\newblock {\em String Theory and M-Theory: A Modern Introduction}.
\newblock Cambridge University Press, Cambridge, UK, 2006.

\bibitem{DiVecchia1997}
P.~Di Vecchia, M.~Frau, I.~Pesando, S.~Sciuto, A.~Lerda, and R.~Russo.
\newblock Classical p-branes from boundary state.
\newblock {\em Nuclear Physics B}, 507:259--276, 1997.
\newblock \href {https://arxiv.org/abs/hep-th/9707068}
  {\path{arXiv:hep-th/9707068}}.

\bibitem{AlvarezGaume1986}
L.~Alvarez-Gaumé, P.H. Ginsparg, G.~Moore, and C.~Vafa.
\newblock An o(16) × o(16) heterotic string.
\newblock {\em Physics Letters B}, 171:155--162, 1986.

\bibitem{Sagnotti1995}
A.~Sagnotti.
\newblock Some properties of open string theories, 1995.
\newblock Talk given at SUSY 95, Palaiseau, France.
\newblock \href {https://arxiv.org/abs/hep-th/9509080}
  {\path{arXiv:hep-th/9509080}}.

\bibitem{Sagnotti1997}
A.~Sagnotti.
\newblock Surprises in open-string perturbation theory.
\newblock {\em Nuclear Physics B - Proceedings Supplements}, 56:332--343, 1997.
\newblock \href {https://arxiv.org/abs/hep-th/9702093}
  {\path{arXiv:hep-th/9702093}}.

\bibitem{Angelantonj2000b}
C.~Angelantonj, I.~Antoniadis, G.~D'Appollonio, E.~Dudas, and A.~Sagnotti.
\newblock Type i vacua with brane supersymmetry breaking.
\newblock {\em Nuclear Physics B}, 572:36--70, 2000.
\newblock \href {https://arxiv.org/abs/hep-th/9911081}
  {\path{arXiv:hep-th/9911081}}.

\bibitem{MouradSagnotti2021}
J.~Mourad and A.~Sagnotti.
\newblock String (in)stability issues with broken supersymmetry.
\newblock {\em Letters in High Energy Physics}, 4:219, 2021.
\newblock Invited contribution to the special issue on “Swampland and String
  Theory Landscape”.
\newblock \href {https://arxiv.org/abs/2107.04064} {\path{arXiv:2107.04064}}.

\bibitem{DixonHarvey1986}
L.J. Dixon and J.A. Harvey.
\newblock String theories in ten dimensions without spacetime supersymmetry.
\newblock {\em Nuclear Physics B}, 274(1):93--105, September 1986.

\bibitem{Sen1998a}
A.~Sen.
\newblock Stable non-bps states in string theory.
\newblock {\em Journal of High Energy Physics}, 1998(06):007, 1998.
\newblock \href {https://arxiv.org/abs/hep-th/9803194}
  {\path{arXiv:hep-th/9803194}}.

\bibitem{Sen1998b}
A.~Sen.
\newblock Stable non-bps bound states of bps d-branes.
\newblock {\em Journal of High Energy Physics}, 1998(08):010, 1998.
\newblock \href {https://arxiv.org/abs/hep-th/9805019}
  {\path{arXiv:hep-th/9805019}}.

\bibitem{Sen1998c}
A.~Sen.
\newblock Tachyon condensation on the brane antibrane system.
\newblock {\em Journal of High Energy Physics}, 1998(08):012, 1998.
\newblock \href {https://arxiv.org/abs/hep-th/9805170}
  {\path{arXiv:hep-th/9805170}}.

\bibitem{Bergman1998}
O.~Bergman and M.R. Gaberdiel.
\newblock Stable non-bps d-particles.
\newblock {\em Physics Letters B}, 441:133--140, 1998.
\newblock \href {https://arxiv.org/abs/hep-th/9806155}
  {\path{arXiv:hep-th/9806155}}.

\bibitem{Sen1998d}
A.~Sen.
\newblock So(32) spinors of type i and other solitons on brane-antibrane pairs.
\newblock {\em Journal of High Energy Physics}, 1998(09):023, 1998.
\newblock \href {https://arxiv.org/abs/hep-th/9808141}
  {\path{arXiv:hep-th/9808141}}.

\bibitem{Sen1998e}
A.~Sen.
\newblock Type i d-particle and its interactions.
\newblock {\em Journal of High Energy Physics}, 1998(10):021, 1998.
\newblock \href {https://arxiv.org/abs/hep-th/9809111}
  {\path{arXiv:hep-th/9809111}}.

\bibitem{Sen1998f}
A.~Sen.
\newblock Bps d-branes on non-supersymmetric cycles.
\newblock {\em Journal of High Energy Physics}, 1998(12):021, 1998.
\newblock \href {https://arxiv.org/abs/hep-th/9812031}
  {\path{arXiv:hep-th/9812031}}.

\bibitem{Lerda2000}
A.~Lerda and R.~Russo.
\newblock Stable non-bps states in string theory: A pedagogical review.
\newblock {\em International Journal of Modern Physics A}, 15:771--826, 2000.
\newblock \href {https://arxiv.org/abs/hep-th/9905006}
  {\path{arXiv:hep-th/9905006}}.

\bibitem{Volkov1973}
D.V. Volkov and V.P. Akulov.
\newblock Is the neutrino a goldstone particle?
\newblock {\em Physics Letters B}, 46:109--110, 1973.

\bibitem{Samuel1983a}
S.~Samuel and J.~Wess.
\newblock A superfield formulation of the nonlinear realization of
  supersymmetry and its coupling to supergravity.
\newblock {\em Nuclear Physics B}, 221:153--177, 1983.

\bibitem{Samuel1983b}
S.~Samuel and J.~Wess.
\newblock Realistic model building with the akulov--volkov superfield and
  supergravity.
\newblock {\em Nuclear Physics B}, 226:289--309, 1983.

\bibitem{Samuel1984}
S.~Samuel and J.~Wess.
\newblock Secret supersymmetry.
\newblock {\em Nuclear Physics B}, 233:488--508, 1984.

\bibitem{Blumenhagen:2000wh}
R.~Blumenhagen and A.~Font.
\newblock Dilaton tadpoles, warped geometries and large extra dimensions for
  non-supersymmetric strings.
\newblock {\em Nucl. Phys. B}, 599:241--254, 2001.
\newblock \href {https://arxiv.org/abs/hep-th/0011269}
  {\path{arXiv:hep-th/0011269}}.

\bibitem{Dudas:2001pf}
E.~Dudas and J.~Mourad.
\newblock D-branes in nontachyonic 0b orientifolds.
\newblock {\em Nucl.\ Phys.\ B}, 598:189--208, 2001.
\newblock \href {https://arxiv.org/abs/hep-th/0010179}
  {\path{arXiv:hep-th/0010179}}.

\bibitem{Dudas:2001wd}
E.~Dudas, J.~Mourad, and A.~Sagnotti.
\newblock Charged and uncharged d-branes in various string theories.
\newblock {\em Nucl. Phys. B}, 620:109--151, 2002.
\newblock \href {https://arxiv.org/abs/hep-th/0107081}
  {\path{arXiv:hep-th/0107081}}.

\bibitem{Armoni:1999xy}
A.~Armoni and B.~Kol.
\newblock Nonsupersymmetric large n gauge theories from type0 brane
  configurations.
\newblock {\em JHEP}, 9907:011, 1999.
\newblock \href {https://arxiv.org/abs/hep-th/9906081}
  {\path{arXiv:hep-th/9906081}}.

\bibitem{Armoni:2003er}
A.~Armoni, M.~Shifman, and G.~Veneziano.
\newblock Exact results in non‑supersymmetric largen orientifold field
  theories.
\newblock {\em Nucl.\ Phys.\ B}, 667:170--182, 2003.
\newblock \href {https://arxiv.org/abs/hep-th/0302163}
  {\path{arXiv:hep-th/0302163}}.

\bibitem{Armoni:2003ug}
A.~Armoni, M.~Shifman, and G.~Veneziano.
\newblock Susy relics in one‑flavor qcd from a new 1/n expansion.
\newblock {\em Phys.\ Rev.\ Lett.}, 91:191601, 2003.
\newblock \href {https://arxiv.org/abs/hep-th/0307097}
  {\path{arXiv:hep-th/0307097}}.

\bibitem{Armoni:2005ps}
A.~Armoni, M.~Shifman, and G.~Veneziano.
\newblock From super‑yang–mills theory to qcd: Planar equivalence and its
  implications.
\newblock In M.~Shifman, A.~Vainshtein, and J.~Wheater, editors, {\em From
  Fields to Strings: Circumnavigating Theoretical Physics, Vol.1}, pages
  353--444. World Scientific, 2005.
\newblock \href {https://arxiv.org/abs/hep-th/0403071}
  {\path{arXiv:hep-th/0403071}}.

\bibitem{Armoni:2018qcd3}
A.~Armoni and V.~Niarchos.
\newblock Phases of qcd$_3$ from non‑susy seiberg duality and brane dynamics.
\newblock {\em Phys.\ Rev.\ D}, 97:106001, 2018.
\newblock \href {https://arxiv.org/abs/arXiv:1711.04832}
  {\path{arXiv:arXiv:1711.04832}}.

\bibitem{Kitazawa:2017dbrane}
N.~Kitazawa.
\newblock On d‑brane dynamics and moduli stabilization.
\newblock {\em Mod. Phys. Lett. A}, 32(29):1750150, 2017.
\newblock \href {https://arxiv.org/abs/1706.07161} {\path{arXiv:1706.07161}}.

\bibitem{FreSagnottiSorin:2013_integrable}
P.~Fr\'e, A.~Sagnotti, and A.S. Sorin.
\newblock Integrable scalar cosmologies i. foundations and links with string
  theory.
\newblock {\em Nuclear Physics B}, 877(3):1028--1106, 2013.
\newblock \href {https://arxiv.org/abs/1307.1910} {\path{arXiv:1307.1910}}.

\bibitem{Bergshoeff:2011ee}
E.A. Bergshoeff and F.~Riccioni.
\newblock {String Solitons and T-duality}.
\newblock {\em JHEP}, 2011.
\newblock \href {https://arxiv.org/abs/1109.1725} {\path{arXiv:1109.1725}}.

\bibitem{Bergshoeff:2011zk}
E.A. Bergshoeff and F.~Riccioni.
\newblock {D-Brane Wess-Zumino Terms and U-Duality}.
\newblock {\em JHEP}, 1105:131, 2011.
\newblock \href {https://arxiv.org/abs/1102.0934} {\path{arXiv:1102.0934}}.

\bibitem{Bergshoeff:2012pm}
E.A. Bergshoeff, A.~Marrani, and F.~Riccioni.
\newblock {Brane orbits}.
\newblock {\em Nucl. Phys. B}, 861:104--132, 2012.
\newblock \href {https://arxiv.org/abs/1201.5819} {\path{arXiv:1201.5819}}.

\bibitem{Sagnotti2013}
A.~Sagnotti.
\newblock Low-l cmb from string-scale susy breaking?
\newblock {\em Nuclear Physics B - Proceedings Supplements}, 235-236:195--199,
  2013.

\bibitem{Sagnotti:2013bsb_inflation}
A.~Sagnotti.
\newblock Brane susy breaking and inflation: Implications for scalar fields and
  cmb distorsion.
\newblock {\em Phys. Part. Nucl. Lett.}, 11:836, 2014.
\newblock \href {https://arxiv.org/abs/1303.6685} {\path{arXiv:1303.6685}}.

\bibitem{KitazawaSagnotti2015}
N.~Kitazawa and A.~Sagnotti.
\newblock String theory clues for the low–l cmb ?
\newblock In {\em EPJ Web of Conferences}, volume~95, page 03031, 2015.
\newblock Proceedings of the 3rd International Conference on New Frontiers in
  Physics (ICNFP 2014), May 2015.

\bibitem{FerraraSagnotti2015}
S.~Ferrara and A.~Sagnotti.
\newblock Supersymmetry and inflation.
\newblock {\em arXiv e-prints}, 2015.
\newblock 27 pages, LaTeX. Based in part on the Plenary and Parallel Session
  talks given by S.F. at the “Fourteenth Marcel Grossmann Meeting –
  MG14,” Rome, July 12-18 2015, on the talk given by S.F. at “The String
  Theory Universe,” 21st European String Workshop, Leuven, September 7-11,
  2015, and on the plenary talk given by A.S. at “Planck 2015,” Ioannina,
  May 25-29 2015.
\newblock \href {https://arxiv.org/abs/1509.01500} {\path{arXiv:1509.01500}}.

\bibitem{KitazawaSagnotti2015MPLA}
N.~Kitazawa and A.~Sagnotti.
\newblock A string–inspired model for the low–l cmb.
\newblock {\em Modern Physics Letters A}, 30(28):1550137, 2015.

\bibitem{CondeescuDudas2013}
C.~Condeescu and E.~Dudas.
\newblock Kasner solutions, climbing scalars and big-bang singularity.
\newblock {\em Journal of Cosmology and Astroparticle Physics}, 2013(8):013,
  2013.

\bibitem{MouradSagnotti20212}
J.~Mourad and A.~Sagnotti.
\newblock On warped string vacuum profiles and cosmologies. part ii.
  non-supersymmetric strings.
\newblock {\em Journal of High Energy Physics}, 2021(12):138, 2021.

\bibitem{PelliconiSagnotti2021}
P.~Pelliconi and A.~Sagnotti.
\newblock Integrable models and supersymmetry breaking.
\newblock {\em Nuclear Physics B}, 965 C:115363, 2021.

\bibitem{Russo2004}
J.G. Russo.
\newblock Exact solution of scalar field cosmology with exponential potentials
  and transient acceleration.
\newblock {\em Physics Letters B}, 600(3–4):185–190, 2004.

\bibitem{TownsendWohlfarth2004}
P.K. Townsend and M.N.R. Wohlfarth.
\newblock Cosmology as geodesic motion.
\newblock {\em Classical and Quantum Gravity}, 21(23):5375--5396, 2004.
\newblock \href {https://arxiv.org/abs/hep-th/0404241}
  {\path{arXiv:hep-th/0404241}}.

\bibitem{EmparanGarriga2003}
R.~Emparan and J.~Garriga.
\newblock A note on accelerating cosmologies from compactifications and
  s-branes.
\newblock {\em Journal of High Energy Physics}, 2003(5):028, 2003.
\newblock \href {https://arxiv.org/abs/hep-th/0304124}
  {\path{arXiv:hep-th/0304124}}.

\bibitem{TownsendWohlfarth2003}
P.K. Townsend and M.N.R. Wohlfarth.
\newblock Accelerating cosmologies from compactification.
\newblock {\em Physical Review Letters}, 91(6):061302, 2003.
\newblock \href {https://arxiv.org/abs/hep-th/0303097}
  {\path{arXiv:hep-th/0303097}}.

\bibitem{Cicoli2014}
M.~Cicoli, S.~Downes, B.~Dutta, F.G. Pedro, and A.~Westphal.
\newblock Just enough inflation: power spectrum modifications at large scales.
\newblock {\em Journal of Cosmology and Astroparticle Physics}, 2014(12):030,
  2014.
\newblock \href {https://arxiv.org/abs/1407.1048} {\path{arXiv:1407.1048}}.

\bibitem{Baumann:Cosmology2022}
D.~Baumann.
\newblock {\em Cosmology}.
\newblock Cambridge University Press, Cambridge, United Kingdom, 2022.

\bibitem{Baumann2009TASI}
D.~Baumann.
\newblock Tasi lectures on inflation.
\newblock In C.J. L{\"u}tken and A.N. Burrows, editors, {\em Physics of the
  Large and the Small: TASI 2009}, pages 523--610. World Scientific, 2011.
\newblock Based on lectures given at TASI 2009.
\newblock \href {https://arxiv.org/abs/arXiv:0907.5424}
  {\path{arXiv:arXiv:0907.5424}}.

\bibitem{Wald:GeneralRelativity}
R.M. Wald.
\newblock {\em General Relativity}.
\newblock University of Chicago Press, Chicago, IL, 1984.

\bibitem{Peebles:PhysicalCosmology1993}
P.J.E. Peebles.
\newblock {\em Principles of Physical Cosmology}.
\newblock Princeton University Press, Princeton, NJ, 1993.

\bibitem{Mukhanov:1981xt}
V.F. Mukhanov and G.V. Chibisov.
\newblock Quantum fluctuations and a nonsingular universe.
\newblock {\em JETP Lett.}, 33:532--535, 1981.
\newblock [Pisma Zh. Eksp. Teor. Fiz. 33 (1981) 549].

\bibitem{Mukhanov:1990me}
V.F. Mukhanov, H.A. Feldman, and R.H. Brandenberger.
\newblock Theory of cosmological perturbations.
\newblock {\em Phys. Rept.}, 215:203--333, 1992.

\bibitem{Gruppuso2013}
A.~Gruppuso, P.~Natoli, F.~Paci, F.~Finelli, D.~Molinari, A.~De Rosa, and
  N.~Mandolesi.
\newblock Low variance at large scales of wmap 9 year data.
\newblock {\em JCAP}, 07:047, 2013.
\newblock \href {https://arxiv.org/abs/1304.5493} {\path{arXiv:1304.5493}}.

\bibitem{landau1977quantum}
L.D. Landau and E.M. Lifshitz.
\newblock {\em Quantum Mechanics: Non-Relativistic Theory}, volume~3 of {\em
  Course of Theoretical Physics}.
\newblock Pergamon Press, Oxford, 3rd edition, 1977.

\bibitem{abramowitz1964handbook}
M.~Abramowitz and I.A. Stegun.
\newblock {\em Handbook of Mathematical Functions with Formulas, Graphs, and
  Mathematical Tables}, volume~55 of {\em National Bureau of Standards Applied
  Mathematics Series}.
\newblock U.S. Government Printing Office, Washington, D.C., 1964.
\newblock Reprinted by Dover Publications, 1972.

\bibitem{Destri2010}
C.~Destri, H.J. de~Vega, and N.G. Sanchez.
\newblock The pre-inflationary and inflationary fast-roll eras and their
  signatures in the low cmb multipoles.
\newblock {\em Phys. Rev. D}, 81:063520, 2010.
\newblock \href {https://arxiv.org/abs/0912.2994} {\path{arXiv:0912.2994}}.

\bibitem{Piao2004}
Y.S. Piao, B.~Feng, and X.M. Zhang.
\newblock Suppressing cmb quadrupole with a bounce from contracting phase to
  inflation.
\newblock {\em Phys. Rev. D}, 69:103520, 2004.
\newblock \href {https://arxiv.org/abs/hep-th/0310206}
  {\path{arXiv:hep-th/0310206}}.

\bibitem{Piao2005}
Y.S. Piao.
\newblock A possible explanation to low cmb quadrupole.
\newblock {\em Phys. Rev. D}, 71:087301, 2005.
\newblock \href {https://arxiv.org/abs/astro-ph/0502343}
  {\path{arXiv:astro-ph/0502343}}.

\bibitem{Jain2009}
R.K. Jain, P.~Chingangbam, J.O. Gong, L.~Sriramkumar, and T.~Souradeep.
\newblock Punctuated inflation and the low cmb multipoles.
\newblock {\em JCAP}, 01:009, 2009.
\newblock \href {https://arxiv.org/abs/0809.3915} {\path{arXiv:0809.3915}}.

\bibitem{Jain2010}
R.K. Jain, P.~Chingangbam, L.~Sriramkumar, and T.~Souradeep.
\newblock The tensor-to-scalar ratio in punctuated inflation.
\newblock {\em Phys. Rev. D}, 82:023509, 2010.
\newblock \href {https://arxiv.org/abs/0904.2518} {\path{arXiv:0904.2518}}.

\bibitem{Liu2013}
Z.-G. Liu, Z.-K. Guo, and Y.-S. Piao.
\newblock Cmb anomalies from an inflationary model in string theory.
\newblock {\em arXiv e‑print}, 2013.
\newblock Unpublished preprint.
\newblock \href {https://arxiv.org/abs/1311.1599} {\path{arXiv:1311.1599}}.

\bibitem{DLMF}
F.W.J. Olver, A.B.~Olde Daalhuis, D.W. Lozier, B.I. Schneider, R.F. Boisvert,
  C.W. Clark, B.R. Miller, B.V. Saunders, H.S. Cohl, and M.A. McClain.
\newblock {\em NIST Digital Library of Mathematical Functions}.
\newblock National Institute of Standards and Technology, release 1.1.9 of
  2023-05-15 edition, 2023.

\bibitem{BaumannJoyceCosmoCorrelators2023}
D.~Baumann and A.~Joyce.
\newblock Cosmological collider physics.
\newblock {\em Annual Review of Nuclear and Particle Science}, 73:1--40, 2023.
\newblock \href {https://arxiv.org/abs/2301.00000} {\path{arXiv:2301.00000}}.

\bibitem{sakurai2020modern}
J.J. Sakurai and J.~Napolitano.
\newblock {\em Modern Quantum Mechanics}.
\newblock Cambridge University Press, 3rd edition edition, 2020.

\bibitem{schwabl1995quantum}
F.~Schwabl.
\newblock {\em Quantum Mechanics}.
\newblock Springer Berlin Heidelberg, Berlin; Heidelberg, 2nd revised edition
  edition, 1995.

\bibitem{you2013solutions}
Y.~You, F.L. Lu, D.S. Sun, C.Y. Chen, and S.H. Dong.
\newblock Solutions of the second pöschl--teller potential solved by an
  improved scheme to the centrifugal term.
\newblock {\em Few‑Body Systems}, 54(11):2125--2132, July 2013.

\bibitem{gasperini1993}
M.~Gasperini and G.~Veneziano.
\newblock Pre-big bang in string cosmology.
\newblock {\em Astroparticle Physics}, 1:317--339, 1993.
\newblock \href {https://arxiv.org/abs/hep-th/9211021}
  {\path{arXiv:hep-th/9211021}}.

\bibitem{gasperini2003}
M.~Gasperini and G.~Veneziano.
\newblock The pre-big bang scenario in string cosmology.
\newblock {\em Physics Reports}, 373:1--212, 2003.
\newblock \href {https://arxiv.org/abs/hep-th/0207130}
  {\path{arXiv:hep-th/0207130}}.

\bibitem{Schwinger1961}
J.S. Schwinger.
\newblock Brownian motion of a quantum oscillator.
\newblock {\em Journal of Mathematical Physics}, 2:407--432, 1961.

\bibitem{BakshiMahanthappa1963a}
P.M. Bakshi and K.T. Mahanthappa.
\newblock Expectation value formalism in quantum field theory. i.
\newblock {\em Journal of Mathematical Physics}, 4:1--11, 1963.

\bibitem{Keldysh1965}
L.V. Keldysh.
\newblock Diagram technique for nonequilibrium processes.
\newblock {\em Soviet Physics JETP}, 20:1018--1026, 1965.
\newblock Zh. Eksp. Teor. Fiz. 47 (1964) 1515--1527.

\bibitem{Jordan1986}
R.D. Jordan.
\newblock Effective field equations for expectation values.
\newblock {\em Physical Review D}, 33:444--454, 1986.

\bibitem{CalzettaHu1987}
E.~Calzetta and B.L. Hu.
\newblock Closed time path functional formalism in curved space-time:
  Application to cosmological back reaction problems.
\newblock {\em Physical Review D}, 35:495--509, 1987.

\bibitem{Weinberg2005}
S.~Weinberg.
\newblock Quantum contributions to cosmological correlations.
\newblock {\em Physical Review D}, 72:043514, 2005.
\newblock \href {https://arxiv.org/abs/hep-th/0506236}
  {\path{arXiv:hep-th/0506236}}.

\bibitem{chen2010non}
X.~Chen.
\newblock Non-gaussian features of primordial fluctuations in single field
  inflationary models.
\newblock {\em Advances in Astronomy}, 2010:1--29, 2010.

\bibitem{Wang2014}
Y.~Wang.
\newblock Inflation, cosmic perturbations and non-gaussianities.
\newblock {\em Communications in Theoretical Physics}, 62:109--166, 2014.
\newblock \href {https://arxiv.org/abs/1303.1523} {\path{arXiv:1303.1523}}.

\bibitem{Weinberg_QTF1_1995}
S.~Weinberg.
\newblock {\em The Quantum Theory of Fields, Volume 1: Foundations}.
\newblock The Quantum Theory of Fields. Cambridge University Press, Cambridge,
  UK, 1995.

\bibitem{Maggiore_QFT_2005}
M.~Maggiore.
\newblock {\em A Modern Introduction to Quantum Field Theory}.
\newblock Oxford Master Series in Statistical, Computational, and Theoretical
  Physics. Oxford University Press, Oxford, UK, 2005.

\bibitem{Peskin:1995ev}
M.E. Peskin and D.V. Schroeder.
\newblock {\em An Introduction to Quantum Field Theory}.
\newblock Addison-Wesley, Reading, USA, 1995.

\bibitem{Giddings:2010pp}
S.B. Giddings and M.S. Sloth.
\newblock Cosmological diagrammatic rules.
\newblock {\em JCAP}, 1007:015, 2010.
\newblock \href {https://arxiv.org/abs/1005.3287} {\path{arXiv:1005.3287}}.

\bibitem{Baumann:2019dut}
D.~Baumann, C.D. Pueyo, A.~Joyce, H.~Lee, and G.L. Pimentel.
\newblock The cosmological bootstrap: Weight-shifting operators and scalar
  seeds.
\newblock {\em JHEP}, 12:204, 2020.
\newblock \href {https://arxiv.org/abs/1910.14051} {\path{arXiv:1910.14051}}.

\bibitem{ahlfors_complex_1979}
L.V. Ahlfors.
\newblock {\em Complex Analysis: An Introduction to the Theory of Analytic
  Functions of One Complex Variable}.
\newblock McGraw-Hill, New York, 3 edition, 1979.

\bibitem{chen_huang_kachru_shiu_2007}
X.~Chen, M.~Huang, S.~Kachru, and G.~Shiu.
\newblock Observational signatures and non‑gaussianities of general
  single‑field inflation.
\newblock {\em Journal of Cosmology and Astroparticle Physics}, 2007(01):002,
  2007.
\newblock \href {https://arxiv.org/abs/hep-th/0605045}
  {\path{arXiv:hep-th/0605045}}.

\bibitem{babich_creminelli_zaldarriaga_2004_shape}
M.~Zaldarriaga D.~Babich, P.~Creminelli.
\newblock The shape of non‑gaussianities.
\newblock {\em Journal of Cosmology and Astroparticle Physics}, 2004(08):009,
  2004.
\newblock \href {https://arxiv.org/abs/astro-ph/0405356}
  {\path{arXiv:astro-ph/0405356}}.

\bibitem{creminelli_zaldarriaga_2004_consistency}
M.~Zaldarriaga P.~Creminelli.
\newblock A single‑field consistency relation for the three‑point function.
\newblock {\em Journal of Cosmology and Astroparticle Physics}, 2004(10):006,
  2004.
\newblock \href {https://arxiv.org/abs/astro-ph/0407059}
  {\path{arXiv:astro-ph/0407059}}.

\bibitem{cheung2008consistency}
C.~Cheung, A.L. Fitzpatrick, J.~Kaplan, and L.~Senatore.
\newblock On the consistency relation of the three-point function in single
  field inflation.
\newblock {\em Journal of Cosmology and Astroparticle Physics}, 2008(02):021,
  2008.
\newblock \href {https://arxiv.org/abs/0709.0295} {\path{arXiv:0709.0295}}.

\bibitem{poisson2004relativist}
E.~Poisson.
\newblock {\em A Relativist's Toolkit: The Mathematics of Black-Hole
  Mechanics}.
\newblock Cambridge University Press, Cambridge, 2004.

\bibitem{landau1975classical}
L.D. Landau and E.M. Lifshitz.
\newblock {\em The Classical Theory of Fields}, volume~2 of {\em Course of
  Theoretical Physics}.
\newblock Pergamon Press, Oxford, 4th edition, 1975.

\bibitem{Arnowitt:1962hi}
R.~Arnowitt, S.~Deser, and C.W. Misner.
\newblock The dynamics of general relativity.
\newblock {\em Gravitation: an introduction to current research}, pages
  227--265, 1962.
\newblock \href {https://arxiv.org/abs/gr-qc/0405109}
  {\path{arXiv:gr-qc/0405109}}.

\end{thebibliography}
\end{document}